\documentclass[11pt]{article}
\pdfoutput=1
\usepackage{jheppub}
\usepackage{tabu}
\usepackage[vcentermath]{youngtab}
\usepackage[usenames,dvipsnames,table]{xcolor}
\usepackage{graphicx,subfig}
\usepackage{amsmath,amssymb,amsthm,amsfonts,multirow,array,bm,bbm,esint}
\usepackage[mathscr]{eucal}
\usepackage[bbgreekl]{mathbbol}
\usepackage{epsf,grffile}
\usepackage{slashed}
\usepackage[numbers,sort&compress]{natbib}
\usepackage{array,tikz-cd}
\usepackage{braket}
\usepackage{multirow}
\usepackage[table]{xcolor}
\usetikzlibrary{fadings,patterns}
\usetikzlibrary{decorations.pathmorphing}

\usepackage{subfig}
\usepackage{float}
\DeclareMathAlphabet{\mathpzc}{OT1}{pzc}{m}{it}

\definecolor{labelkey}{rgb}{0.4,0.4,0.4}

\newcommand{\Tr}{{\rm {Tr}}}
\newcommand{\normord}[1]{:\mathrel{#1}:}
\newcommand{\sech}{\,\mathrm{sech}\,}
\newcommand{\csch}{\,\mathrm{csch}\,}
 
 \newcommand{\bea}{\begin{eqnarray}}
\newcommand{\eea}{\end{eqnarray}}
\newcommand{\be}{\begin{equation}}
\newcommand{\ee}{\end{equation}}
\newcommand{\ba}{\begin{align}}
\newcommand{\ea}{\end{align}}

\title{Generalized entropy of photons in AdS}

\author[a]{Sean Colin-Ellerin,}
\author[a]{Guanda Lin}

\affiliation[a]{Center for Theoretical Physics and Department of Physics,  \\
University of California, Berkeley, California 94720, USA.}

\emailAdd{scolinellerin@berkeley.edu, geoff\textunderscore guanda\textunderscore lin@berkeley.edu}

\abstract{This work analyzes the quantum corrections to holographic entanglement entropy at first subleading order in $G_{N}$ due to photon excited states in AdS. We compute the vacuum-subtracted von Neumann entropy of a $U(1)$ current excited state for a polar cap region on the cylinder in any large-$N$, strongly-coupled CFT$_{d}$ holographically dual to weakly-coupled Einstein gravity for any dimension $d>2$. We then quantize a Maxwell field in AdS$_{d+1}$ dual to the $U(1)$ current and consider a photon excited state whose vacuum-subtracted generalized entropy for the entanglement wedge is calculated. In order to factorise the Maxwell Hilbert space in AdS, we construct an extended Hilbert space and the corresponding electromagnetic edge modes. We find exact agreement between the CFT entanglement entropy and AdS generalized entropy without the inclusion of entropy of the edge modes. Finally, we show via explicit calculation that the contribution to the vacuum-subtracted von Neumann entropy from electromagnetic edge modes indeed vanishes, which is crucial for consistency with known holographic entropy formulas.}

\begin{document}

\maketitle


\section{Introduction}
\label{sec:intro}

A hallmark of any theory of quantum gravity would be to describe the entanglement of quantum gravitational degrees of freedom. One may hope that the AdS/CFT correspondence \cite{Maldacena:1997re}, as a non-perturbative theory of quantum gravity, provides an avenue to begin to answer this question. The emergence of classical spacetime from entanglement in the dual CFT has been one of the greatest successes of the AdS/CFT correspondence, as exemplified by the Hubeny-Rangamani-Ryu-Takayanagi (HRRT) formula, relating areas of surfaces in AdS to entanglement entropy in the dual CFT \cite{Ryu:2006ef,Ryu:2006bv,Hubeny:2007xt,Faulkner:2013ica,Faulkner:2017tkh}. Beyond the classical level, some progress has been made toward understanding the quantum generalization, where the entanglement entropy of quantum fields coupled to classical gravity in asymptotically anti-de Sitter spacetimes becomes important \cite{Faulkner:2013ana,Jafferis:2015del,Engelhardt:2014gca,Dong:2017xht}. 

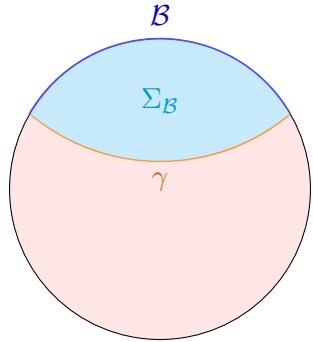
\begin{figure}[h]
\begin{center}
\begin{tikzpicture}

\filldraw[draw=black,fill=red!10!white] (0,-1) circle (2cm);
\draw[thick,color=blue!70!white] (1.73,0) arc (30:150:2);
\draw[thick,color=orange!80!white] (-1.72,0) arc (230:310:2.68);
\fill[color=cyan!20!white] (-1.72,0) .. controls (-0.94,1.32) and (0.94,1.32)  .. (1.72,0) .. controls (0.7,-0.83) and (-0.7,-0.83) .. (-1.72,0);

\node[blue!80!black] at (0,1.31) {$\mathcal{B}$};
\node[orange!80!black] at (0,-0.9) {$\gamma$};
\node[cyan!80!black] at (0,0.18) {$\Sigma_{\mathcal{B}}$};

\end{tikzpicture}
\caption{A fixed-time slice of AdS$_{d+1}$ (suppressing the spherical directions) with boundary subregion $\mathcal{B}$ (blue), quantum extremal surface $\gamma$ (orange), and quantum homology surface $\Sigma_{\mathcal{B}}$ (cyan).}
\label{fig:FLM}
\end{center}
\end{figure}

The most general holographic entanglement entropy formula \cite{Faulkner:2013ana,Engelhardt:2014gca}, known as the quantum HRRT formula, states that, in any holographic CFT dual to weakly-coupled Einstein gravity in AdS, the von Neumann entropy associated with a codimension-$1$ subregion $\mathcal{B}$ in a state $\ket{\Psi}$ in the CFT is equal to the generalized entropy in AdS associated to the dual state $\ket{\psi}$ and the bulk codimension-$1$ subregion $\Sigma_{\mathcal{B}}$ which extremizes the generalized entropy and has the minimal generalized entropy amongst such extrema
\begin{equation}\label{eqn:qHRRT}
S(\rho_{\mathcal{B}}^{\Psi}) = \underset{\Sigma_{\mathcal{B}}}{\mathrm{min\;ext}}\left\{\frac{A(\gamma)_{\psi}}{4G_{N}}+S\left(\rho_{\Sigma_{\mathcal{B}}}^{\psi}\right)\right\},
\end{equation}
where $\Sigma_{\mathcal{B}}$ is the (quantum) homology surface bounded by $\mathcal{B}$ and a codimension-$2$ surface $\gamma$ anchored on $\partial\mathcal{B}$ which is homologous to $\mathcal{B}$, as illustrated in Figure \ref{fig:FLM}. The generalized entropy is given by the area of $\gamma$, denoted by $A(\gamma)$ and labelled by the state $\psi$ (since the geometry depends on the choice of state), plus the von Neummann entropy in AdS associated with $\Sigma_{\mathcal{B}}$ in the state $\ket{\psi}$. Thus far, this formula has only been understood semi-classically, where the metric is treated as a classical field coupled to quantum matter fields. 

In this semi-classical context, some perturbative analysis in $G_{N}$ has been developed. The first quantum correction, known as the Faulkner-Lewkowycz-Maldacena (FLM) formula, has been derived from the Euclidean path integral \cite{Faulkner:2013ana} and some checks for scalar field theory have been performed \cite{Belin:2018juv,Agon:2015ftl,Agon:2020fqs,Chowdhury:2024fpd}, as well as Chern-Simons theory in AdS$_{3}$ \cite{Belin:2019mlt}. The evidence in favor of this formula at subleading orders in $G_{N}$ comes from general arguments of causality, (quantum) entanglement wedge nesting, and strong subadditivity \cite{Engelhardt:2014gca,Akers:2019lzs}, along with Euclidean path integral arguments \cite{Dong:2017xht} and some partial checks for a scalar field in AdS$_{3}$ \cite{Belin:2021htw}. By far, the most interesting results of this formula have come from applications to black hole physics whereby non-perturbative effects lead to a unitary Page curve for evaportating black holes \cite{Penington:2019npb,Almheiri:2019psf,Penington:2019kki,Almheiri:2019qdq}.

However, these results do not provide an understanding of generalized entropy for fluctuating spacetimes or even perturbative graviton fluctuations on a fixed background. 
This requires going beyond semi-classical gravity by treating the metric as a quantum variable, which poses both conceptual and technical challenges, even at the perturbative level.
One issue is how to properly define subregions because the fluctuation of the metric naturally blurs the position of the quantum extremal surface.
Even with a proposal specifying the location of the quantum extremal surface, it is still a priori mysterious what in the generalized entropy should be considered as part of the area contribution and what should be considered as part of the bulk entanglement entropy.

A further stumbling block comes from the diffeomorphism constraints of gravity. 
In any theory with constraints, such as a gauge theory, the Hilbert space does not factorise because these constraints relate modes in any subregion with those in the complement subregion.\footnote{This is even worse than the standard non-factorization of the Hilbert space in any quantum field theory coming from the large entanglement of UV modes. The standard methods of implementing a UV cutoff in the QFT in order to factorize the Hilbert space still do not lead to factorization in the presence of constraints.} 
Such a non-factorised Hilbert space structure will dramatically modify the structure of the density matrix and hence change the entanglement entropy for a subregion.

Addressing these issues is highly non-trivial. In this work, we study the generalized entropy for a $U(1)$ gauge theory in AdS, which is an important step towards our ultimate goal of understanding the contribution to holographic entanglement entropy from gravitational fluctuations for the following reasons.
Firstly, the puzzles mentioned above mostly arise from understanding the bulk graviton. 
In the dual CFT, the entanglement entropy is completely well-defined because, based on the standard extrapolate dictionary, the bulk graviton excitation is dual to excitations of the boundary stress tensor operator, which is well-defined in the CFT and has an unambiguous entanglement entropy. 
This CFT result can serve as the referee for our bulk calculation. 
Therefore, it is necessary to establish a concrete set-up in which one can explicitly compute both the boundary and bulk entropies and verify that they match, which will be demonstrated for photons (and scalars) in this work. 
Such a setup provides a tractable testing ground for the graviton calculation \cite{Colin-Ellerin24}. 

Another motivation for considering the $U(1)$ gauge theory computation in AdS is that the problem of non-factorised Hilbert spaces and gauge constraints has been fairly well-studied in Maxwell theory for a $U(1)$ gauge field \cite{Buividovich:2008gq,Donnelly:2011hn,Casini:2013rba,Radicevic:2014kqa,Donnelly:2014fua,Donnelly:2015hxa,Huang:2014pfa,Ghosh:2015iwa,Soni:2015yga,Soni:2016ogt,Blommaert:2018rsf,Ball:2024hqe}.
The standard method to obtain a factorization of the Hilbert space, given the Gauss Law constraint in Maxwell theory, is to embed the physical Hilbert space in a factorizing extended Hilbert space such that any reduced density matrix takes the form of a direct sum of superselection sectors labeled by certain center variables. 
Such a center variable is typically chosen to be the electric flux on the boundary of the subregion, which is referred to as edge modes. 
The von Neumann entropy then recieves a contribution $S_{\rm edge}$ from the Shannon entropy for the probability distribution of the different edge sectors.  
It has been shown that only after including $S_{\rm edge}$ does the entropy agree with the replica trick \cite{Donnelly:2014fua,Donnelly:2015hxa,Ghosh:2015iwa,Blommaert:2018rsf} and give the correct coefficient for the universal logarithmic divergence of entanglement entropy for a $2$-sphere in four dimensions which is determined by the $a$-type trace anomaly \cite{Donnelly:2014fua,Donnelly:2015hxa,Huang:2014pfa,Soni:2016ogt}.\footnote{
Although this is less relevant to the main topic of the current paper, we would like to mention that $S_{\rm edge}$ depends on the choice of extended Hilbert space, which amounts to a choice of center for the algebra of operators associated to $b$ \cite{Casini:2013rba}. 
The electric center is usually the most natural one with many consistency checks, but there is some dispute by the authors of \cite{Casini:2019nmu}. In this paper, we would like to argue that the edge sector gives zero contribution to the vacuum-subtracted entropy, so the choice of edge variables should not make a difference.}

Now we explain our approach to the problem of computing holographic generalized entropy for photons.
Since the FLM formula \eqref{eqn:qHRRT} serves as the guiding light for our analysis, we begin by computing the von Neumann entropy for excited primary states in a holographic $d$-dimensional CFT dual to weakly-coupled gravity in AdS$_{d+1}$. 
Entanglement entropy in large $N$, strongly coupled field theories is notoriously difficult to compute in dimensions greater than two, but we will present a simple setup where one can compute the vacuum-subtracted von Neumann entropy for a highly symmetric subregion on the cylinder in any conformal field theory satisfying certain holographic conditions. 
In particular, we will consider any conformal field theory in dimensions $d>2$ with a large central charge $C_{T}$ and a large gap $\Delta_{\mathrm{gap}}$ for the dimension of the lightest single-trace operator with spin $J > 2$ such that $1 \ll \Delta_{\mathrm{gap}} \ll C_{T}$. These are known to be necessary conditions for a CFT to be dual to weakly-coupled Einstein gravity in AdS \cite{Heemskerk:2009pn}. 

We then turn to an analysis of generalized entropy in AdS. As a prelude to the Maxwell case, we first compute the generalized entropy of a scalar field in $(d+1)$-dimensional AdS spacetime by canonically quantizing the scalar field. An excited state of the scalar field is considered, which backreacts on the spacetime to give a quantum correction to the area of the classical extremal surface. We then compute the vacuum-subtracted von Neumann entropy for the classical entanglement wedge using direct Hilbert space techniques that allow us to evaluate the desired trace. The resulting vacuum-subtracted generalized entropy is finite and agrees precisely with the vacuum-subtracted von Neumann entropy computed in the dual CFT, as expected by the FLM formula \eqref{eqn:qHRRT}.

Next, we canonically quantize a free $U(1)$ gauge field in Maxwell theory in $(d+1)$-dimensional AdS in radial gauge and consider a photon excited state. The backreaction of this excited state on the spacetime is obtained in various dimensions $d>2$ and the resulting correction to the area of the classical extremal surface is computed. Although the correction to the area is finite, it requires a very careful treatment of the IR cutoff surface, which we explain. We then compute the bulk von Neumann entropy for the photon from direct Hilbert space techniques, ignoring the edge mode contribution, and we find that the resulting vacuum-subtracted generalized entropy agrees precisely with the vacuum-subtracted von Neumann entropy for a $U(1)$ current excited state in the dual CFT. This indicates, by the FLM formula \eqref{eqn:qHRRT}, that the edge modes do not contribute to vacuum-subtracted von Neumann entropies.

Finally, we show explicitly that the vacuum-subtracted entanglement entropy of electromagnetic edge modes in AdS indeed vanishes, providing a non-trivial check of the FLM formula \eqref{eqn:qHRRT}.\footnote{When we discuss edge modes in this work, we will always be referring to those for the photon in AdS. Given that the most well-understood examples of holographic CFTs are also gauge theories, such as $\mathcal{N}=4$ SYM in four dimensions or ABJM in three dimensions,  one may worry that we should also consider edge modes in the CFT. We expect that since we are not exciting any gauge modes in the CFT, the edge modes would also not contribute to the vacuum-subtracted von Neumann entropies in those examples. In fact, our arguments in \S\ref{sec:edge} that edge modes do not contribute to such quantities in AdS can also be checked in flat space with the same conclusions.} While it had been argued previously that edge modes do not contribute to relative entropies \cite{Casini:2013rba,Moitra:2018lxn}, to our knowledge, the observation that they also do not contribute to vacuum-subtracted von Neumann entropies is new. Furthermore, many of the results obtained here for photons can be generalized to the study of perturbative gravitons in AdS dual to stress-tensor excited states in the CFT so this work serves as progress towards that ultimate goal, which will be the subject of separate work \cite{Colin-Ellerin24}.

The outline of this paper is as follows: we begin in \S\ref{sec:CFTEE_excitedstates} with a computation of the vacuum-subtracted entanglement entropy of scalar and current excited states in a putative holographic $d$-dimensional CFT. As a warm-up to $U(1)$ gauge theory, \S\ref{sec:Sgenscalar} provides a calculation of the generalized entropy for a single-particle excited state in massive scalar field theory for the classical entanglement wedge. This provides much of the technical machinery needed to study the Maxwell case of interest. Next, we give a detailed analysis of the canonical quantization of a $U(1)$ gauge field for free Maxwell theory in $(d+1)$-dimensional AdS in radial gauge in \S\ref{sec:Max}. The generalized entropy of a photon excited state is then obtained in \S\ref{sec:Sgenphoton}. Finally, \S\ref{sec:edge} contains an analysis of the vacuum-subtracted entanglement entropy for the electromagnetic edge modes in Lorenz gauge. We end in \S\ref{sec:disc} with a discussion of some open problems.

Some technical details needed for our analysis are collected in various appendices. Appendix \ref{sec:AdSRindler} contains some facts about AdS-Rindler needed for calculations of entanglement entropy in AdS. In Appendix \ref{sec:hyperbolicballeigfns}, we obtain the eigenfunctions of the Laplacian on $H^{d-1}$. Appendix \ref{sec:Bogcoeffs} gives a detailed calculation of the Bogoliubov coefficients for both scalar and Maxwell fields. Appendix \ref{sec:photonbackreact_details} provides the details of the backreaction of the photon excited state on the spacetime. Finally, Appendix \ref{sec:usefulints} contains some useful integrals needed in our computations.

\section{Entanglement entropy of excited states in CFT$_{d}$}
\label{sec:CFTEE_excitedstates}

We analyze the vacuum-subtracted von Neumann entropy $\Delta S$ for excited states in $d$-dimensional CFTs for $d>2$. We will use the perturbative expansion of the modular Hamiltonian in the subregion size, rather than the standard replica trick, as this technique turns out to be easier. 
We explain the setup of this perturbative expansion in \S\ref{sec:setup}. 
It will be found that one can compute the von Neumann entropy explicitly order-by-order in this expansion:
in \S\ref{sec:cylvacmodH} we obtain the leading order contribution from the first law of entanglement entropy $\Delta S|_{\mathcal{O}(\delta\rho)} = \Delta K_{0}$ for a scalar primary excited state \eqref{eqn:K0Odif} and $U(1)$ current excited state \eqref{eqn:K0Jdif_Max};
and in \S\ref{sec:2ndorderCFTentropy}, we compute the first sub-leading correction $\Delta S|_{\mathcal{O}((\delta\rho)^{2})}$ for the scalar \eqref{eqn:deltaS2_scalarMFT_final} and $U(1)$ current \eqref{eqn:deltaS_J_Max}. 
These boundary CFT results will be matched to the AdS computations in later sections. 

\subsection{General setup}
\label{sec:setup}

Our goal is to compute the entanglement entropy in a CFT on the cylinder $\mathbb{R} \times S^{d-1}$ with unit radius for a scalar primary excited state and the excited state of a $U(1)$ current (when such an operator exists in the theory). The region we consider is a polar cap on the spatial sphere of the cylinder given by the codimension-$1$ subregion 
\begin{equation}\label{eqn:CFTsubtregion}
\mathcal{B} = \{t=0,\,\theta \leq \theta_{0}\},
\end{equation}
where $\theta$ is the inclination angle on $S^{d-1}$ and $t$ is the time coordinate on the cylinder. The techniques to obtain excited state entanglement entropy in this setup were developed in \cite{Sarosi:2017rsq} for scalar operators which we now briefly review and readily generalize to higher-spin operators. 

\begin{figure}[h]
\begin{centering}
\begin{tikzpicture}
\draw[->] (0,-2.3) -- (0,2.5);
\shade[ball color = blue!50, opacity = 0.4] (0,0) circle (2cm);
\draw (0,0) circle (2cm);
\draw (-2,0) arc (180:360:2 and 0.6);
\draw[dashed] (2,0) arc (0:180:2 and 0.6);

\fill[red!40, opacity=0.6] (-1.525,1.3) .. controls (-0.5,0.9) and (0.5,0.9) .. (1.525,1.3) .. controls (0.75,2.25) and (-0.75,2.25) .. (-1.525,1.3);
\draw[blue!50!black!50] (0,0) -- (1.525,1.3);
\node[red!30!black!70] at (0.2,1.5) {$\mathcal{B}$};

\draw[blue!50!black!50] (0.11,0.1) arc (30:80:0.15cm);
\node[purple!50!black!50] at (0.2,0.5) {$\theta_{0}$};

\end{tikzpicture}
\caption{Polar cap entangling region (red) on the $t=0$ slice of the cylinder on which the CFT lives.}
\label{fig:entregion}
\end{centering}
\end{figure}
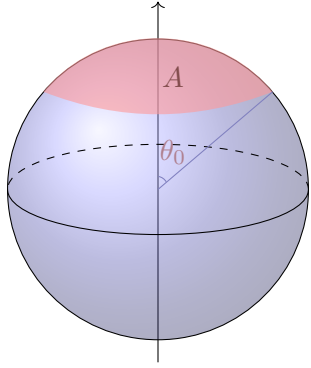

Consider the CFT state $\ket{\mathcal{O}} = \lim_{t_{E} \to -\infty} e^{\Delta_{\mathcal{O}}t_{E}}\mathcal{O}(t_{E})\ket{0}$ defined by acting with a primary operator $\mathcal{O}$, where $t_{E}$ is Euclidean time on the cylinder. This creates a primary state on the cylinder by the state-operator correspondence.\footnote{The angular position of the operator is not important because the bottom of the cylinder is mapped to a single point on the sphere in radial quantization.} This operator can be a scalar operator $O$ or an operator $\epsilon_{\mu_{1}\ldots\mu_{\ell}}\mathcal{O}^{\mu_{1}\ldots\mu_{\ell}}$ of spin $\ell$ with polarization tensor $\epsilon_{\mu_{1}\ldots\mu_{\ell}}$. The reduced density matrix for the subregion $\mathcal{B}$ is given by 
\begin{equation}\label{eqn:CFTexcitedstate_rhoA}
\rho_{\mathcal{B}}^{\mathcal{O}} = \Tr_{\mathcal{B}^{c}}|\mathcal{O}\rangle\langle\mathcal{O}|.
\end{equation}
We wish to compute the von Neumann entropy of this state
\begin{equation}\label{eqn:vNentropy}
S\left(\rho_{\mathcal{B}}^{\mathcal{O}} \right) = -\Tr\left(\rho_{\mathcal{B}}^{\mathcal{O}} \log \rho_{\mathcal{B}}^{\mathcal{O}} \right).
\end{equation}
We first map from $\mathbb{R} \times S^{d-1}$ to $\mathcal{H} = S^{1} \times H^{d-1}$ using the Casini-Huerta-Myers map \cite{Casini:2011kv} in Euclidean signature
\begin{equation}\label{eqn:hypertocyl}
\tanh t_{E} = \frac{\sin\theta_{0}\sin\tau_{E}}{\cosh u + \cos\theta_{0}\cos\tau_{E}}, \qquad \tan\theta = \frac{\sin\theta_{0}\sinh u}{\cos\tau_{E} + \cos\theta_{0}\cosh u},
\end{equation}
where the metric on $\mathcal{H}$ is given by
\begin{equation}\label{eqn:bdyhyperbolicspacemetric}
ds_{\mathcal{H}}^{2} = d\tau_{E}^{2}+du^{2}+\sinh^{2}u\,d\Omega_{d-2}^{2},
\end{equation}
with $\tau_{E} \sim \tau_{E}+2\pi$, $u \in [0,\infty)$. The image of the region $\mathcal{B}$ is $\{\tau_{E}=0,\; u \in [0,\infty)\}$ with the entangling surface $\theta=\theta_{0}$ mapped to $u=\infty$. The operator insertions are mapped to
\begin{equation}\label{eqn:opinsertionimages}
t_{E} = -\infty \to (u=0,\tau_{E}=\pi+\theta_{0}), \qquad t_{E} = \infty \to (u=0,\tau_{E}=\pi-\theta_{0}).
\end{equation}
Importantly, the vacuum state transforms to the thermal state $\rho_{0} = e^{-2\pi H_{\mathcal{H}}}$ under this map, where $H_{\mathcal{H}}$ is the Hamiltonian on $\mathcal{H}$. The vacuum modular Hamiltonian $K_{0} = 2\pi H_{\mathcal{H}}$ is thus the generator of $\tau$ translations. 

Define $\tau_{\mathcal{O}} = \pi+\theta_{0}$ and $\hat{\tau}_{\mathcal{O}} = \pi-\theta_{0}$. The (normalized) density matrix on $\mathcal{H}$ can be found by using the fact that the vacuum state on $\mathbb{R} \times S^{d-1}$ is the thermal state on $\mathcal{H}$ to obtain
\begin{equation}\label{eqn:rhoHnorm}
\rho_{\mathcal{B}}^{\mathcal{O}} = \frac{e^{-\pi H_{\mathcal{H}}}\mathcal{O}(\theta_{0})\mathcal{O}^{\dagger}(-\theta_{0})e^{-\pi H_{\mathcal{H}}}}{\langle \mathcal{O}(\tau_{\mathcal{O}})\mathcal{O}(\hat{\tau}_{\mathcal{O}})\rangle_{\mathcal{H}}},
\end{equation}
where $\langle \cdot \rangle_{\mathcal{H}} = \Tr\left(\rho^{0}_{\cal B}\cdot\right)$.
Next, we write the excited state density matrix as a perturbation of the vacuum state density matrix $\rho_{\mathcal{B}}^{0}$:
\begin{equation}\label{eqn:rhopert}
\rho_{\mathcal{B}}^{\mathcal{O}} = \rho_{\mathcal{B}}^{0} + \delta \rho
\end{equation}
where the $\mathcal{O} \times \mathcal{O}$ OPE gives an explicit form for $\delta \rho$
\begin{equation}\label{eqn:deltarho}
\delta \rho = e^{-\pi H_{\mathcal{H}}}\sum_{\substack{\mathcal{\tilde{O}} \in \mathcal{O} \times \mathcal{O} \\ \tilde{\mathcal{O}} \neq \mathbb{1}}}C_{\mathcal{O}\mathcal{O}\mathcal{\tilde{O}}}(\tau_{\mathcal{O}}-\hat{\tau}_{\mathcal{O}},\partial_{\hat{\tau}_{\mathcal{O}}})\mathcal{\tilde{O}}(\hat{\tau}_{\mathcal{O}})e^{-\pi H_{\mathcal{H}}},
\end{equation}
where the identity contribution to the OPE gives $\rho_{\mathcal{B}}^{0}$ and we are suppressing all the tensor indices on $C_{\mathcal{O}\mathcal{O}\mathcal{\tilde{O}}}$ and $\tilde{\mathcal{O}}$ which are present when $\mathcal{O}$ is a higher-spin operator. 

\begin{figure}[h]
\begin{centering}
\begin{tikzpicture}

\draw (-3,-2.5) -- (-3,2.5);
\fill[red!70!blue!30] (-3,0) circle (0.8mm);
\node at (-3.8,0) {$t_{E}=0$};
\draw[orange,thick] (-3.1,-2.6) -- (-2.9,-2.4);
\draw[orange,thick] (-3.1,-2.4) -- (-2.9,-2.6);
\draw[green,thick] (-3.1,2.6) -- (-2.9,2.4);
\draw[green,thick] (-3.1,2.4) -- (-2.9,2.6);
\node at (-3.6,-2.5) {$-\infty$};
\node at (-2.5,-2.3) {$\mathcal{O}$};
\node at (-3.6,2.5) {$+\infty$};
\node at (-2.5,2.8) {$\mathcal{O}^{\dagger}$};
\node at (-3,-3) {$\mathbb{R}$};

\draw (0,0) -- (0,2);
\shade[ball color = blue!50, opacity = 0.4] (0,0) circle (2cm);
\draw (0,0) circle (2cm);
\draw (-2,0) arc (180:360:2 and 0.6);
\draw[dashed] (2,0) arc (0:180:2 and 0.6);

\fill[red!40, opacity=0.6] (-1.525,1.3) .. controls (-0.5,0.9) and (0.5,0.9) .. (1.525,1.3) .. controls (0.75,2.25) and (-0.75,2.25) .. (-1.525,1.3);
\draw[cyan,very thick] (-1.525,1.3) .. controls (-0.5,0.9) and (0.5,0.9) .. (1.525,1.3);
\draw[blue!50!black!50] (0,0) -- (1.525,1.3);
\node[red!30!black!70] at (0.2,1.5) {$\mathcal{B}$};

\draw[blue!50!black!50] (0.11,0.1) arc (30:80:0.15cm);
\node[purple!50!black!50] at (0.2,0.5) {$\theta_{0}$};
\node at (-1.5,-3) {$\times$};
\node at (0,-3) {$\mathbb{S}^{d-1}$};

\draw [->] (2.5,0) -- (4.5,0);
\node at (3.5,0.4) {CHM};

\draw (6.8,0) ellipse (0.5cm and 2cm); 
\node at (5.5,0) {$\tau_{H}=0$};
\node at (6.8,-3) {$S^{1}$};
\fill[red!80!blue!30] (6.3,0) circle (0.8mm);
\draw[green,thick] (7.35,-1) -- (7.15,-0.8);
\draw[green,thick] (7.35,-0.8) -- (7.15,-1);
\draw[orange,thick] (7.35,0.8) -- (7.15,1);
\draw[orange,thick] (7.35,1) -- (7.15,0.8);
\node at (7.8,1.4) {$\pi+\theta_{0}$};
\node at (7.8,-1.4) {$\pi-\theta_{0}$};

\fill[red!70!blue!30] (10.5,2) .. controls (7.8,1) and (7.8,-1) ..  (10.5,-2) .. controls (10.9,-1.6) and (10.9,1.6) .. (10.5,2);
\draw[cyan,very thick] (10.5,0) ellipse (0.3cm and 2cm);
\draw[green,thick] (8.38,-0.2) -- (8.58,0);
\draw[green,thick] (8.38,0) -- (8.58,-0.2);
\draw[orange,thick] (8.38,0.1) -- (8.58,-0.1);
\draw[orange,thick] (8.38,-0.1) -- (8.58,0.1);
\node at (8.3,-3) {$\times$};
\node at (10,-3) {$H^{d-1}$};

\end{tikzpicture}
\caption{Casini-Huerta-Myers map from cylinder to thermal hyperbolic space. The subregion $B$ gets mapped to all of hyperbolic space with entangling surface $\partial B$ mapped to infinity ($u=\infty$) on the hyperboloid. The operator insertions get mapped to the origin ($u=0$) on the hyperboloid.}
\label{fig:CHM}
\end{centering}
\end{figure}

In the small cap limit $\theta_{0} \ll 1$, one finds that $\delta \rho$ is small, as can be seen from figure \ref{fig:CHM} where the two operators become close as $\theta_{0} \to 0$, so we can expand $S(\rho_{\mathcal{B}}^{\mathcal{O}})$ in $\delta \rho$ and look at the first few orders. Throughout the paper, we will use this small $\theta_{0}$ limit to simplify our computations.

We begin by expanding the excited modular Hamiltonian using the integral representation of the logarithm \cite{Sarosi:2017rsq}
\begin{equation}\label{eqn:logintrep}
K_{\mathcal{O}} = -\log \rho_{\mathcal{B}}^{\mathcal{O}} = \int_{0}^{\infty} d\beta\,\left(\frac{1}{\beta+\rho_{\mathcal{B}}^{\mathcal{O}}} - \frac{1}{\beta+1}\right) = K_{0} + \sum_{n=1}^{\infty}(-1)^{n}\delta K_{\mathcal{O}}^{(n)}
\end{equation}
where
\begin{align}\label{eqn:deltaKdef}
\begin{split}
\delta K_{\mathcal{O}}^{(n)} &= \int_{-\infty}^{\infty} ds_{1} \ldots ds_{n} \, \mathcal{K}_{n}(s_{1},\ldots,s_{n})\prod_{k=1}^{n} e^{iH_{\mathcal{H}}(s_{k}-i\pi)}\delta\rho e^{-iH_{\mathcal{H}}(s_{k}+i\pi)}
\\ \mathcal{K}_{n}(s_{1},\ldots,s_{n}) &= \frac{(2\pi)^{2}i^{n-1}}{(4\pi)^{n+1}}\frac{1}{\cosh \frac{s_{1}}{2} \cosh \frac{s_{n}}{2} \prod_{k=2}^{n}\sinh\left(\frac{s_{k}-s_{k-1}}{2}\right)}
\end{split}
\end{align}
with the contour prescription $s_{k}-s_{k-1} \to s_{k}-s_{k-1}+i\epsilon$ for any $0< \epsilon < 2\pi$, which is required for convergence of intermediate steps in the derivation of \eqref{eqn:logintrep}, \eqref{eqn:deltaKdef} (later we will need a more detailed contour prescription in \eqref{eqn:sintcontour}).

Inserting this into \eqref{eqn:vNentropy}, the vacuum-subtracted entanglement entropy can now be expressed as\footnote{In the following equation, we have a separation of $ \Delta K$ and relative entropy $S_{\rm rel}$ as the infinite sum. Although the sum starts from $n=1$, we would like to note that the leading non-zero contribution is of order $(\delta \rho)^2$ because ${\rm Tr}(\rho_0 \delta K^{(1)}_{\mathcal{O}})=0$. }
\begin{equation}\label{eqn:deltaS}
\Delta S^{\mathcal{O}} \equiv S(\rho_{\mathcal{B}}^{\mathcal{O}})-S(\rho_{\mathcal{B}}^{0}) = \Tr\left(\rho_{\mathcal{B}}^{\mathcal{O}}K_{\mathcal{O}}\right)-\Tr\left(\rho_{\mathcal{B}}^{0}K_{0}\right) = \Delta K_{0}^{\mathcal{O}} + \sum_{n=1}^{\infty}(-1)^{n}\,\Tr\left(\rho_{\mathcal{B}}^{\mathcal{O}}\delta K_{\mathcal{O}}^{(n)}\right),
\end{equation}
where the first term is given by
\begin{equation}\label{eqn:DeltaK}
\Delta K_{0}^{\mathcal{O}} = \Tr\left(\rho_{\mathcal{B}}^{\mathcal{O}}K_{0}\right)-\Tr\left(\rho_{\mathcal{B}}^{0}K_{0}\right) = \Tr\left(\delta \rho K_{0}\right)
\end{equation}
whose appearance at leading order is simply the first law of entanglement entropy \cite{Blanco:2013joa}. The results up to this point hold for any CFT, but each term appearing in \eqref{eqn:deltaS} depends somewhat sensitively on the particular theory so we must now specify the theory of interest.

We consider an arbitrary conformal field theory in dimensions $d>2$ with a large central charge $C_{T}$ and a large gap $\Delta_{\mathrm{gap}}$ for the dimension of the lightest single-trace operator with spin $J > 2$ such that $1 \ll \Delta_{\mathrm{gap}} \ll C_{T}$. The assumption that $C_{T}$ is large is related to assuming that $N$ is large since in any large-$N$ conformal field theory $C_{T} \sim N^{a}$ for some $a>0$. Large $\Delta_{\mathrm{gap}}$ is actually an unnecessary assumption for our purposes when $\mathcal{O}$ is a scalar operator, but it is crucial when $\mathcal{O}$ is a $U(1)$ current, as we will now see. 

\subsection{Expectation value of vacuum modular Hamiltonian}
\label{sec:cylvacmodH}

Let us first compute the leading contribution to $\Delta S^{\mathcal{O}}$ in \eqref{eqn:deltaS} which is given by the vacuum-subtracted expectation value of the vacuum modular Hamiltonian \eqref{eqn:DeltaK}. As previously mentioned, the vacuum modular Hamiltonian for $\mathcal{B}$ is simply the Hamiltonian on $\mathcal{H}$:
\begin{equation}\label{eqn:vacmodH}
K_{0} = 2\pi H_{\mathcal{H}} = 2\pi \int_{\Sigma_{\tau=0}}d^{d-1}x\,\sqrt{g_{\Sigma}}\,T_{\tau\tau}+c
\end{equation}
where $c$ is a constant which is fixed by the requirement $\Tr(\rho_{\mathcal{B}}^{0}) = 1$ and $\tau=i\tau_{E}$ is Lorentzian time on $\mathcal{H}$. To compute the expectation value of this operator in various excited states, we will choose to work on the cylinder. The conformal transformation of the stress-tensor under the mapping to the cylinder \eqref{eqn:hypertocyl} (in Lorentzian signature) gives
\begin{equation}
K_{0} = 2\pi \int d\Omega_{d-2}\,\int_{0}^{\theta_{0}}d\theta\,\sin^{d-2}\theta\frac{\left(\cos\theta-\cos\theta_{0}\right)}{\sin\theta_{0}}T_{tt}+c'
\end{equation}
where $c'$ is a constant that depends on $c$ and the conformal anomaly in even dimensions, but the details of which will not be needed since it does not contribute to $\Delta K_{0}$. Thus, our task is to compute the expectation value of $T_{tt}$ on the cylinder in the desired excited states and then perform the angular integrals.

For a scalar operator $O$, the three-point function on $\mathbb{R}^{d}$ is fixed by conformal Ward identities \cite{Osborn:1993cr} to be 
\begin{align}\label{eqn:OTO3ptfn}
\begin{split}
\bra{0}O(x_{1})T_{\mu\nu}(x_{2})O(x_{3})\ket{0}_{\mathbb{R}^{d}} &= C_{OOT}\left(\frac{Z_{\mu}Z_{\nu}}{Z^{2}}-\frac{1}{d}\delta_{\mu\nu}\right)\frac{1}{|x_{12}|^{d}|x_{23}|^{d}|x_{13}|^{2\Delta_{O}-d}}
\\	Z_{\mu} &\equiv \frac{x_{12,\mu}}{x_{12}^{2}}-\frac{x_{32,\mu}}{x_{32}^{2}}.
\end{split}
\end{align}
The three-point function coefficient is given by \cite{Penedones:2016voo}
\begin{equation}\label{eqn:OOT_OPEcoeff}
C_{OOT} = -\frac{\Delta_{O}}{V_{S^{d-1}}}\frac{d}{(d-1)}
\end{equation}
for the canonically normalized stress-tensor
\begin{align}\label{eqn:TTtwoptfn}
\begin{split}
\bra{0}T_{\mu\nu}(x)T_{\rho\sigma}(0)\ket{0}_{\mathbb{R}^{d}} &= \frac{C_{T}}{V_{S^{d-1}}^{2}}\frac{I_{\mu\nu,\rho\sigma}(x)}{x^{2d}}
\\	I_{\mu\nu,\rho\sigma}(x) &= \frac{1}{2}\left(I_{\mu\rho}(x)I_{\nu\sigma}(x)+I_{\mu\sigma}(x)I_{\nu\rho}(x)\right)-\frac{1}{d}\delta_{\mu\nu}\delta_{\rho\sigma}
\\	I_{\mu\nu}(x) &= \delta_{\mu\nu} - 2\frac{x_{\mu}x_{\nu}}{x^{2}},
\end{split}
\end{align}
where $V_{S^{d-1}} = 2\pi^{\frac{d}{2}}/\Gamma(\frac{d}{2})$ is the volume of $S^{d-1}$. This normalization is chosen so that the total (vacuum-subtracted) energy of the state $\ket{O}$ on the cylinder is equal to $\Delta_{O}$. We wish to map this three-point function to the cylinder under the standard map $r=e^{t_{E}}$ so we need the $T_{rr}$ three-point function. From the definition of the dual state $\bra{O} = \lim_{r \to \infty}\bra{0}O^{\dagger}(r)r^{2\Delta_{O}}$, we arrive at the desired expectation value
\begin{equation}\label{eqn:Tscalarev}
\bra{O}T_{rr}(x_{2})\ket{O} = -\frac{\Delta_{O}}{V_{S^{d-1}}}\frac{1}{|x_{2}|^{d}},
\end{equation}
which gives the vacuum-subtracted expectation value for the vacuum modular Hamiltonian
\begin{align}\label{eqn:K0Odif}
\begin{split}
\Delta K_{0}^{O} &= \frac{\bra{O}K_{0}\ket{O}}{\langle O|O\rangle}-\bra{0}K_{0}\ket{0}
\\	&=  2\pi\Delta_{O}\frac{V_{S^{d-2}}}{V_{S^{d-1}}}\,\int_{0}^{\theta_{0}}\sin^{d-2}\theta\frac{\left(\cos\theta-\cos\theta_{0}\right)}{\sin\theta_{0}}
\\	&= \frac{2\pi\Delta_{O}}{(d-1)}\frac{V_{S^{d-2}}}{V_{S^{d-1}}}\sin^{d-2}\theta_{0}\left(1-\cos\theta_{0}\,{}_{2}{F}_{1}\left(\frac{1}{2},\frac{d-1}{2},\frac{d+1}{2};\sin^{2}\theta_{0}\right)\right)
\\	&= \frac{\sqrt{\pi}\,\Gamma(\frac{d}{2})\Delta_{O}}{2\Gamma\left(\frac{d+3}{2}\right)}\theta_{0}^{d}\left(1+\mathcal{O}(\theta_{0}^{2})\right),
\end{split}
\end{align}
where the second line holds for $\theta_{0} < \pi/2$ and we expanded at small $\theta_{0}$ in the last line.\footnote{There is also an overall minus sign that one must take into account in \eqref{eqn:K0Odif} coming from the fact that \eqref{eqn:Tscalarev} gives the expectation value of $T_{t_{E}t_{E}}$, but $\Delta K_{0}^{O}$ is written in terms of $T_{tt}$ so the analytic continuation from Euclidean to Lorentzian signature gives an overall minus sign.} Observe that we have properly normalized the density matrix $\rho_{\mathcal{B}}^{O}$ in \eqref{eqn:K0Odif} which has the nice feature that, even though we evaluated this expectation value on the cylinder, it gives the same result on $\mathcal{H}$ because any conformal factor in the transformation of $O$ cancels between the numerator and denominator. 

Next, consider a $U(1)$ current excited state $\ket{\epsilon \cdot J}_{\mathbb{R}^{d}} = \epsilon^{\mu}J_{\mu}(r=0)\ket{0}_{\mathbb{R}^{d}}$ obtained by acting on the vacuum with a conserved $U(1)$ current in the CFT with arbitrary polarization vector $\epsilon^{\mu}$. The dual state is given by
\begin{equation}\label{eqn:Jdualstate}
\bra{\epsilon^{\ast} \cdot J}_{\mathbb{R}^{d}} = \lim_{|x| \to \infty}|x|^{2(d-1)}\bra{0}_{\mathbb{R}^{d}}\epsilon^{\ast\nu}{I_{\nu}}^{\mu}(x)J_{\mu}(x).
\end{equation}
We normalize $J_{\mu}$ to have the two-point function
\begin{equation}\label{eqn:JJ2ptfn}
\bra{0}J_{\mu}(x)J_{\nu}(0)\ket{0}_{\mathbb{R}^{d}} = \frac{C_{J}}{V_{S^{d-1}}^{2}}\frac{I_{\mu\nu}(x)}{x^{2(d-1)}} \implies \langle\epsilon^{\ast} \cdot J|\epsilon \cdot J\rangle_{\mathbb{R}^{d}} = |\epsilon|^{2}\frac{C_{J}}{V_{S^{d-1}}^{2}}.
\end{equation}
The three-point function of two $U(1)$ currents and the stress-tensor is fixed by conformal symmetry, the symmetries of the stress-tensor, and current conservation to be \cite{Osborn:1993cr}
\begin{equation}\label{eqn:JTJ3ptfn}
\bra{0}J_{\mu}(x_{1})T_{\rho\sigma}(x_{2})J_{\nu}(x_{3})\ket{0} = \frac{1}{x_{12}^{d}x_{23}^{d}x_{13}^{d-2}}{I_{\mu}}^{\alpha}(x_{12}){I_{\nu}}^{\beta}(x_{32})t_{\rho\sigma\alpha\beta}(Z),
\end{equation}
where
\begin{align}\label{eqn:t4def}
\begin{split}
t_{\rho\sigma\alpha\beta}(Z) &= \tilde{a}h^{1}_{\rho\sigma}(\hat{Z})\delta_{\alpha\beta}+\tilde{b}h^{1}_{\rho\sigma}(\hat{Z})h^{1}_{\alpha\beta}(\hat{Z})+\tilde{c}h^{2}_{\rho\sigma\alpha\beta}(\hat{Z})+\tilde{e}h^{3}_{\rho\sigma\alpha\beta}
\\ \hat{Z}_{\mu} &= \frac{Z_{\mu}}{|Z|}, \qquad h^{1}_{\rho\sigma}(\hat{Z}) = \hat{Z}_{\rho}\hat{Z}_{\sigma}-\frac{1}{d}\delta_{\rho\sigma}
\\	h^{2}_{\rho\sigma\alpha\beta}(\hat{Z}) &= \hat{Z}_{\rho}\hat{Z}_{\alpha}\delta_{\sigma\beta}+\hat{Z}_{\sigma}\hat{Z}_{\alpha}\delta_{\rho\beta}+\hat{Z}_{\rho}\hat{Z}_{\beta}\delta_{\sigma\alpha}+\hat{Z}_{\sigma}\hat{Z}_{\beta}\delta_{\rho\alpha} - \frac{4}{d}\hat{Z}_{\rho}\hat{Z}_{\sigma}\delta_{\alpha\beta} - \frac{4}{d}\hat{Z}_{\alpha}\hat{Z}_{\beta}\delta_{\rho\sigma} + \frac{4}{d^{2}}\delta_{\rho\sigma}\delta_{\alpha\beta}
\\	h^{3}_{\rho\sigma\alpha\beta} &= \delta_{\rho\alpha}\delta_{\sigma\beta}+\delta_{\rho\beta}\delta_{\sigma\alpha}-\frac{2}{d}\delta_{\rho\sigma}\delta_{\alpha\beta}.
\end{split}
\end{align}
There are two constraints on the coefficients:
\begin{equation}\label{eqn:JTJconservationconstraint}
\tilde{b} = d(d-2)\tilde{e}, \qquad \tilde{a} = 2(d-2)\left(\tilde{e}-\frac{\tilde{c}}{d}\right)
\end{equation}
leaving us with two constants $\tilde{c}$ and $\tilde{e}$. The Ward identity imposes one more condition
\begin{equation}\label{eqn:WardIdcondJTJ}
\tilde{c}+\tilde{e} = \frac{d}{2V_{S^{d-1}}^{3}}C_{J},
\end{equation}
leaving one overall theory-dependent constant $\tilde{e}$.\footnote{For $d=3$, there can be a parity-odd structure as well in this three-point function so there are two-theory dependent constants \cite{Afkhami-Jeddi:2018own}. Once we use our assumptions on the CFT, this term will vanish, but for general CFTs in $d=3$, our result \eqref{eqn:K0Jdif} only holds for parity-preserving theories.} We can compute the vacuum-subtracted expectation value on the cylinder of the vacuum modular Hamiltonian in terms of $\tilde{c}$, $\tilde{e}$, and the angle $\theta_{\epsilon}$ of the polarization vector:
\begin{align}\label{eqn:K0Jdif}
\begin{split}
\Delta K_{0}^{J} &\equiv \frac{\bra{\epsilon^{\ast} \cdot J}K_{0}\ket{\epsilon \cdot J}}{\langle\epsilon^{\ast} \cdot J|\epsilon \cdot J\rangle} - \bra{0}K_{0}\ket{0} 
\\	&= 2\pi V_{S^{d-2}}\frac{V_{S^{d-1}}^{2}}{C_{J}}\sin^{d-2}\theta_{0}\Bigg\{\frac{2}{d}\tilde{c}f_{+}(d,\theta_{0})-\tilde{e}\bigg[\frac{d-3-(d-2)(d+1)\cos^{2}\theta_{\epsilon}}{d-1}f_{+}(d,\theta_{0})
\\	&\qquad -\frac{(d-2)(d+1)}{d}\sin(2\theta_{\epsilon})\tan\theta_{0}\left(f_{-}(d+1,\theta_{0})-\sin^{2}\theta_{0}\right)
\\ 	&\qquad -(d-2)\sin^{2}\theta_{0}\cos(2\theta_{\epsilon})f_{+}(d+2,\theta_{0})\bigg]\Bigg\},
\end{split}
\end{align}
which holds for $\theta_{0}<\pi/2$ and we have defined the following function to simplify the above expression
\begin{equation}\label{eqn:fdef}
f_{\pm}(d,\theta_{0}) \equiv 1-\cos\theta_{0}\,{}_{2}{F}_{1}\left(\pm\frac{1}{2},\frac{d-1}{2},\frac{d+1}{2};\sin^{2}\theta_{0}\right).
\end{equation}
Note that we have not yet used our assumptions on the CFT. 

For our goal of relating this result to generalized entropy in AdS, it is important to understand how $\tilde{e}$ is related to physics of the dual AdS $U(1)$ gauge theory. All of this can be understood as follows. The two possible couplings of the gauge field to gravity that contribute to the bulk three-point vertex of two photons and one graviton are the Maxwell Lagrangian $F^{2}$ and the Lagrangian $W^{\mu\nu\rho\sigma}F_{\mu\nu}F_{\rho\sigma}$ where $F_{\mu\nu}$ is the field strength and $W^{\mu\nu\rho\sigma}$ is the Weyl tensor. Although one can add other higher-derivative terms to the bulk Lagrangian, they will not contribute to the three-point vertex as there are only two gauge-invariant vertices one can write down \cite{Hofman:2008ar}. This is the AdS analogue of the fact that the CFT three-point function only depends on two constants $\tilde{c}$, $\tilde{e}$ by current conservation. From the computation of the dual three-point Witten diagrams, one finds that the bulk couplings are related to $\tilde{c}$, $\tilde{e}$ by \cite{Hofman:2008ar,Kulaxizi:2017ixa}
\begin{equation}\label{eqn:U(1)gaugeL}
S \propto \left(\int d^{d+1}x\,\sqrt{g}F^{\mu\nu}F_{\mu\nu}+\alpha_{W}\int d^{d+1}x\,\sqrt{g}W^{\mu\nu\rho\sigma}F_{\mu\nu}F_{\rho\sigma}\right)
\end{equation}
where
\begin{equation}\label{eqn:alphaW}
\alpha_{W} \propto \frac{\tilde{c}-d(d-2)\tilde{e}}{\left(\tilde{e}+\tilde{c}\right)}.
\end{equation}
It was shown in \cite{Li:2017lmh,Afkhami-Jeddi:2018own} that for any large-$C_{T}$ CFT in $d>2$ dimensions with a large $\Delta_{\mathrm{gap}}$, crossing symmetry and causality imply that $\alpha_{W} = 0$ so such CFTs can describe only Einstein-Maxwell theory.\footnote{For $d=3$, there can be another parity-violating higher-derivative contribution to the action \eqref{eqn:U(1)gaugeL}, corresponding to the parity-odd conformal structure discussed in footnote 6, but it was shown that this will vanish using our assumptions on the CFT \cite{Afkhami-Jeddi:2018own} so we do not need to worry about it. This means \eqref{eqn:K0Jdif_Max} still applies in $d=3$.} This completely fixes the three-point function and we arrive at
\begin{align}\label{eqn:K0Jdif_Max}
\begin{split}
\left(\Delta K_{0}^{J}\right)_{1 \ll \Delta_{\mathrm{gap}} \ll C_{T}} &= \frac{\sqrt{\pi}\,\Gamma(\frac{d}{2}+1)}{(d-1)\Gamma\left(\frac{d+1}{2}\right)}\sin^{d-2}\theta_{0}\Bigg(\frac{2d^{2}-7d+7+(d-2)(d+1)\cos^{2}\theta_{\epsilon}}{d-1}f_{+}(d,\theta_{0})
\\	&\qquad -\frac{(d-2)(d+1)}{d}\sin(2\theta_{\epsilon})\tan\theta_{0}\left(f_{-}(d+1,\theta_{0})-\sin^{2}\theta_{0}\right)
\\ 	&\qquad -(d-2)\sin^{2}\theta_{0}\cos(2\theta_{\epsilon})f_{+}(d+2,\theta_{0})\Bigg).
\end{split}
\end{align}
The single-particle photon state that we will consider in AdS corresponds to choosing the polarization vector such that $\theta_{\epsilon}=0$ which gives the simpler result
\begin{align}\label{eqn:K0Jdif_Max_thetaeps=0}
\begin{split}
\left(\Delta K_{0}^{J}\right)_{1 \ll \Delta_{\mathrm{gap}} \ll C_{T}}\Big|_{\theta_{\epsilon}=0} &= \frac{\sqrt{\pi}\,\Gamma(\frac{d}{2}+1)}{(d-1)\Gamma\left(\frac{d+1}{2}\right)}\sin^{d-2}\theta_{0}
\\  &\qquad \times \Bigg((3d-5)f_{+}(d,\theta_{0})-(d-2)\sin^{2}\theta_{0}f_{+}(d+2,\theta_{0})\Bigg)
\\	&= \frac{\sqrt{\pi}(3d-5)\Gamma(\frac{d}{2}+1)}{2(d-1)\Gamma\left(\frac{d+3}{2}\right)}\theta_{0}^{d}\left(1+\mathcal{O}(\theta_{0}^{2})\right),
\end{split}
\end{align}
where in the second equality we expanded in small $\theta_{0}$ only giving the leading term for simplicity, but in relating this to the AdS calculation we actually need the subleading terms in $\theta_{0}$ which can be extracted straightforwardly from the first equality (we will need this result up to $\mathcal{O}(\theta_{0}^{2(d-1)})$).

\subsection{Second-order correction to entanglement entropy}
\label{sec:2ndorderCFTentropy}

The second-order correction and all higher-order corrections to the entanglement entropy turn out to be much simpler than the first-order correction as they are simply determined by large-$N$ factorization, i.e., they come from generalized free field correlators. This means that, unlike the first-order correction computed in the previous section, they do not know about the coupling of the matter to gravity. In practice though, one must still evaluate the integrals over modular parameters $s_{i}$ appearing in \eqref{eqn:deltaKdef} so we will content ourselves with only computing the second-order contribution.

To understand higher-order terms in \eqref{eqn:deltaS} contributing to $\Delta S^{\mathcal{O}}$, observe that
\begin{align}\label{eqn:trtocorrfn}
\begin{split}
\Tr\left(\rho_{\mathcal{B}}^{\mathcal{O}}\delta K_{\mathcal{O}}^{(n)}\right) &= \Tr\left(\frac{e^{-\pi H_{\mathcal{H}}}\mathcal{O}(\tau_{\mathcal{O}}-\pi)\mathcal{O}(\hat{\tau}_{\mathcal{O}}-\pi)e^{-\pi H_{\mathcal{H}}}}{\langle \mathcal{O}(\tau_{\mathcal{O}})\mathcal{O}(\hat{\tau}_{\mathcal{O}})\rangle_{\mathcal{H}}}\delta K_{\mathcal{O}}^{(n)}\right) 
\\	&= \Tr\left(e^{-2\pi H_{\mathcal{H}}}\frac{\mathcal{O}(\tau_{\mathcal{O}})\mathcal{O}(\hat{\tau}_{\mathcal{O}})}{\langle \mathcal{O}(\tau_{\mathcal{O}})\mathcal{O}(\hat{\tau}_{\mathcal{O}})\rangle_{\mathcal{H}}}\delta K_{\mathcal{O}}^{(n)}\right)
\\	&=  \left\langle\frac{\mathcal{O}(\tau_{\mathcal{O}})\mathcal{O}(\hat{\tau}_{\mathcal{O}})}{\langle \mathcal{O}(\tau_{\mathcal{O}})\mathcal{O}(\hat{\tau}_{\mathcal{O}})\rangle_{\mathcal{H}}}\delta K_{\mathcal{O}}^{(n)}\right\rangle_{\mathcal{H}}.
\end{split}
\end{align}
Therefore, the terms in the sum in \eqref{eqn:deltaS} can be written as the following integrated correlation functions
\begin{align}\label{eqn:deltaSnthterm}
\begin{split}
\Tr\left(\rho_{\mathcal{B}}^{\mathcal{O}}\delta K_{\mathcal{O}}^{(n)}\right) &= \int_{-\infty}^{\infty} ds_{1} \ldots ds_{n} \, \mathcal{K}_{n}(s_{1},\ldots,s_{n})
\\	&\qquad \times \bigg\langle\mathcal{T}\left(\frac{\mathcal{O}(\tau_{\mathcal{O}})\mathcal{O}(\hat{\tau}_{\mathcal{O}})}{\langle \mathcal{O}(\tau_{\mathcal{O}})\mathcal{O}(\hat{\tau}_{\mathcal{O}})\rangle_{\mathcal{H}}}\prod_{k=1}^{n}\left(\frac{\mathcal{O}(\tau_{\mathcal{O}}-\tau_{s_{k}})\mathcal{O}(\hat{\tau}_{\mathcal{O}}-\tau_{s_{k}})}{\langle \mathcal{O}(\tau_{\mathcal{O}}-\tau_{s_{k}})\mathcal{O}(\hat{\tau}_{\mathcal{O}}-\tau_{s_{k}})\rangle_{\mathcal{H}}}-1\right)\right)\bigg\rangle_{\mathcal{H}}
\end{split}
\end{align}
where $\tau_{s_{k}} = \pi-is_{k}$ and $\mathcal{T}$ indicates that this thermal correlation function must be time-ordered (up to KMS relations) in order to be well-defined. 

This time-ordering can be implemented by choosing an appropriate contour for the $s$-integrals. The desired contour is given by a constant shift in the imaginary direction for each $s$-integral: $s_{k} \to s_{k}-i\epsilon_{k}$ with
\begin{equation}\label{eqn:sintcontour}
-\pi < \epsilon_{k} < \ldots < \epsilon_{1} < \pi, \qquad -\pi+2\theta_{0} < \epsilon_{k} < \pi-2\theta_{0}, \qquad 2\theta_{0} < \epsilon_{k-1} - \epsilon_{k}.
\end{equation}
Observe that this is consistent with the earlier contour prescription. However, this contour prescription can only be satisfied for $\theta_{0} < \pi/(n+1)$ so for any fixed $\theta_{0}$ only a finite number of terms in the sum in \eqref{eqn:logintrep} are convergent. Nonetheless, it can be argued that the fully resummed answer must be convergent \cite{Sarosi:2017rsq}. We will only be interested in the four-point function contribution $(n=1,2)$ since all higher-point functions give sub-leading terms in the small $\theta_{0}$ limit so we do not have to worry about these issues. To see this, take the OPE in \eqref{eqn:deltaSnthterm} to obtain 
\begin{align}\label{eqn:KnOPE}
\begin{split}
&\Tr\left(\rho_{\mathcal{B}}^{\mathcal{O}}\delta K_{\mathcal{O}}^{(n)}\right) = \int_{-\infty}^{\infty} ds_{1} \ldots ds_{n} \, \mathcal{K}_{n}(s_{1},\ldots,s_{n})
\\	&\times \bigg\langle \sum_{\mathcal{\tilde{O}}}\frac{C_{\mathcal{O}\mathcal{O}\mathcal{\tilde{O}}}(\tau_{\mathcal{O}}-\hat{\tau}_{\mathcal{O}},\partial_{\hat{\tau}_{\mathcal{O}}})}{\langle \mathcal{O}(\tau_{\mathcal{O}})\mathcal{O}(\hat{\tau}_{\mathcal{O}})\rangle_{\mathcal{H}}}\mathcal{\tilde{O}}(\hat{\tau}_{\mathcal{O}})\bigg(\prod_{k=1}^{n}\sum_{\mathcal{\tilde{O}} \neq \mathbb{1}}\frac{C_{\mathcal{O}\mathcal{O}\mathcal{\tilde{O}}}(\tau_{\mathcal{O}}-\hat{\tau}_{\mathcal{O}},\partial_{\hat{\tau}_{\mathcal{O}}})}{{\langle \mathcal{O}(\tau_{\mathcal{O}}-\tau_{s_{k}})\mathcal{O}(\hat{\tau}_{\mathcal{O}}-\tau_{s_{k}})\rangle_{\mathcal{H}}}}\mathcal{\tilde{O}}(\hat{\tau}_{\mathcal{O}}-\tau_{s_{k}})\bigg)\bigg\rangle_{\mathcal{H}}.
\end{split}
\end{align}
Using the fact $C_{\mathcal{\mathcal{O}\mathcal{O}\tilde{O}}}(\tau_{\mathcal{O}}-\hat{\tau}_{\mathcal{O}},\partial_{\hat{\tau}_{\mathcal{O}}}) \approx C_{\mathcal{O}\mathcal{O}\mathcal{\tilde{O}}}(2\theta_{0})^{\Delta_{\mathcal{\tilde{O}}}-2\Delta_{\mathcal{O}}}$ in the small interval limit, we conclude that only the small $n$ terms contribute at leading order in $\theta_{0}$.

We now proceed to compute correlation functions appearing in \eqref{eqn:deltaSnthterm} for $n=1,2$. Large $N$ factorization implies that, at leading order in large $N$, all correlation functions are given by their generalized free field theory (GFF) values, i.e., they can be computed by Wick contractions and knowledge of the two-point function. For the scalar operator case, we can use the two-point function $\langle O(\tau_{E})O(0)\rangle = \left(2\sin\left(\frac{\tau_{E}}{2}\right)\right)^{-2\Delta_{O}}$ and $\mathcal{K}_{n}$ defined in \eqref{eqn:deltaKdef} to obtain
\begin{equation}\label{eqn:deltaS1_scalarMFT}
\Tr\left(\rho_{\mathcal{B}}^{O}\delta K_{O}^{(1)}\right)_{\mathrm{GFF}} = \frac{1}{2}\int_{-\infty}^{\infty} ds\,\frac{1}{\cosh^{4\Delta_{O}+2}\frac{s}{2}}\theta_{0}^{4\Delta_{O}}\left(1+\mathcal{O}(\theta_{0}^{2})\right)
\end{equation}
and
\begin{align}\label{eqn:deltaS2_scalarMFT}
\begin{split}
\Tr\left(\rho_{\mathcal{B}}^{O}\delta K_{O}^{(2)}\right)_{\mathrm{GFF}} &= \frac{i}{8\pi}\int_{-\infty}^{\infty} ds_{1}\,ds_{2}\,\frac{1}{\cosh\left(\frac{s_{1}}{2}\right)\cosh\left(\frac{s_{2}}{2}\right)\sinh\left(\frac{s_{2}-s_{1}}{2}\right)}
\\	&\qquad \times \frac{1}{\left(i\sinh\left(\frac{s_{1}-s_{2}}{2}\right)\right)^{4\Delta_{\mathcal{O}}}}\theta_{0}^{4\Delta_{O}}\left(1+\mathcal{O}(\theta_{0}^{2})\right)+ \mathcal{O}\left(\theta_{0}^{6\Delta_{O}}\right)
\end{split}
\end{align}
where we have taken the small $\theta_{0}$ expansion. This contribution at $\theta_{0}^{4\Delta_{O}}$ comes from the double-trace operator $[OO]_{0,0}$ in the $O \times O$ OPE and the $\mathcal{O}(\theta_{0}^{2})$ corrections come from descendants. To perform the $s_{1},s_{2}$ integrals, we make the change of variables $u=s_{1}+s_{2}$ and $v=s_{2}-s_{1}-i\pi$ and then deform the $v$ contour to the real line, which is consistent with our contour prescription, leaving us with
\begin{equation}\label{eqn:deltaS2_scalarMFT_final}
\Tr\left(\rho_{\mathcal{B}}^{O}\delta K_{O}^{(2)}\right)_{\mathrm{GFF}} = \frac{1}{4}\int_{-\infty}^{\infty} ds\,\frac{1}{\cosh^{4\Delta_{O}+2}\frac{s}{2}}\theta_{0}^{4\Delta_{O}}\left(1+\mathcal{O}(\theta_{0}^{2})\right)+ \mathcal{O}\left(\theta_{0}^{6\Delta_{\mathcal{O}}}\right).
\end{equation}
We see that the contribution from $\delta K_{O}^{(2)}$ is actually the same as that for $\delta K_{O}^{(1)}$ but with a different coefficient. Combining them in \eqref{eqn:deltaS} and evaluating this final integral, we find the vacuum-subtracted von Neumann entropy
\begin{equation}\label{eqn:DeltaS_final}
\Delta S^{O} \approx \Delta K_{0}^{O}-\frac{\sqrt{\pi}\Gamma(2\Delta_{O}+1)}{2\Gamma(2\Delta_{O}+\frac{3}{2})}\theta_{0}^{4\Delta_{O}}\left(1+\mathcal{O}(\theta_{0}^{2})\right)
\end{equation}
where we have dropped all contributions from products of double-trace operators which give contributions of order $\theta_{0}^{2k\Delta}$ with $k>2$.

All of this can be repeated for the $U(1)$ current operator using the two-point functions
\begin{align}\label{eqn:JJ2ptfn_H}
\begin{split}
\langle \epsilon^{\ast} \cdot J(\tau_{E}) \epsilon \cdot J(0) \rangle &= \frac{C_{J}}{V_{S^{d-1}}^{2}}\frac{|\epsilon|^{2}}{\left(2\sin\left(\frac{\tau_{E}}{2}\right)\right)^{2(d-1)}}
\\ \langle \epsilon \cdot J(\tau_{E}) \epsilon \cdot J(0) \rangle &= \frac{C_{J}}{V_{S^{d-1}}^{2}}\frac{\epsilon^{2}\left(1-2\cos^{2}\theta_{\epsilon}\right)}{\left(2\sin\left(\frac{\tau_{E}}{2}\right)\right)^{2(d-1)}}
\\ \langle \epsilon^{\ast} \cdot J(\tau_{E}) \epsilon^{\ast} \cdot J(0) \rangle &= \frac{C_{J}}{V_{S^{d-1}}^{2}}\frac{\left(\epsilon^{\ast}\right)^{2}\left(1-2\cos^{2}\theta_{\epsilon}\right)}{\left(2\sin\left(\frac{\tau_{E}}{2}\right)\right)^{2(d-1)}},
\end{split}
\end{align}
which gives the vacuum-subtracted von Neumann entropy
\begin{align}\label{eqn:deltaS_J_Max}
\begin{split}
\left(\Delta S^{J}\right)_{1 \ll \Delta_{\mathrm{gap}} \ll C_{T}} &\approx \left(\Delta K_{0}^{J}\right)_{1 \ll \Delta_{\mathrm{gap}} \ll C_{T}}
\\  &\qquad -\frac{\sqrt{\pi}\Gamma(2d-1)}{4\Gamma(2d-\frac{1}{2})}\left(1+\left(1-2\cos^{2}\theta_{\epsilon}\right)^{2}\right)\theta_{0}^{4(d-1)}\left(1+\mathcal{O}(\theta_{0}^{2})\right),
\end{split}
\end{align}
where the second term is the second-order contribution and again we have dropped all contributions from products of double-trace operators.

As a final comment, observe that this second-order correction to the entanglement entropy is equal to (minus) the leading-order contribution to the relative entropy due to the relation
\begin{equation}\label{eqn:relS}
S_{\mathrm{rel}}\left(\rho_{\mathcal{B}}^{\mathcal{O}}\big|\rho_{\mathcal{B}}^{0}\right) = \Delta K_{0}^{\mathcal{O}} - \Delta S^{\mathcal{O}}.
\end{equation}
We will find that this matches the relative entropy in AdS, as expected from \cite{Jafferis:2015del}. One may have worried that we need to include the subleading corrections to $\Delta K_{0}^{J}$ up to order $\theta_{0}^{4(d-1)}$ which mix with the second term in \eqref{eqn:deltaS_J_Max}, but this is not needed as we will have a more fine-grained matching between CFT and AdS quantities.

\section{Warm-up: generalized entropy of scalar field in AdS$_{d+1}$}
\label{sec:Sgenscalar}
 
As a prelude to analyzing Maxwell theory, we will first show how to compute the vacuum-subtracted generalized entropy of a scalar field for the classical entanglement wedge. This is a higher-dimensional generalization of the AdS$_{3}$ computation performed in \cite{Belin:2018juv}, which will allow us to introduce many of the tools needed for the Maxwell case in a simplified setting.

We will find that this bulk generalized entropy reproduces the vacuum-subtracted entanglement entropy in the boundary CFT, which is usually referred to as the FLM formula \cite{Faulkner:2013ana}.
More explicitly, 
for the first-order contribution to the entanglement entropy, we will show a match in \eqref{eqn:JLMSscalar} between the boundary and bulk results described by the following JLMS formula \cite{Jafferis:2015del}
\begin{equation}\label{eqn:JLMS}
\Delta K_{0}^{O} = \frac{\Delta A(\gamma_{\mathrm{ext}})_{\psi}}{4G_{N}} + \Delta K_{0}^{\psi}
\end{equation}
where $\ket{\psi}$ is the state in AdS dual to the CFT state $\ket{O}$. 
The entanglement wedge considered here is the AdS-Rindler wedge whose boundary is a Killing horizon so we can understand \eqref{eqn:JLMS} as a semiclassical statement of the gravitational Gauss Law.
Moreover, for the second-order contribution governed by the relative entropy $S_{\mathrm{rel}}$, we verify the following relation \cite{Jafferis:2015del}
\begin{equation}\label{eqn:srel}
    S_{\mathrm{rel}}^{\mathrm{CFT}}\left(\rho_{\mathcal{B}}^{O}\big|\rho_{\mathcal{B}}^{0}\right) = S_{\mathrm{rel}}^{\mathrm{AdS}}\left(\rho_{\Sigma_{\mathcal{B}}}^{\psi}\Big|\rho_{\Sigma_{\mathcal{B}}}^{0}\right)
\end{equation}
in \eqref{eqn:matchscalarsrel}. These two relations together give the FLM formula \eqref{eqn:qHRRT}.

From \eqref{eqn:JLMS} and \eqref{eqn:srel}, we can see clearly what needs to be computed. 
To begin, in \S\ref{sec:canonquant_scalar}, we quantize the scalar field (in global coordinates) in AdS and construct the state $|\psi\rangle$.
Then in \S\ref{sec:areacorrection}, with the excited state $|\psi\rangle$, we obtain the backreaction from the excitation on the AdS geometry semiclassically, and extract $\Delta A(\gamma_{\rm ext})_{\psi}$ which is the change in area of the extremal surface from the backreacted geometry. 
Finally, in \S\ref{sec:scalar_bulkEE}, we compute the bulk entanglement entropy.
We utilize the fact that the classical entanglement wedge is AdS-Rindler to construct the reduced density matrix for the global state $|\psi\rangle$ by decomposing the state into a sum over products of left and right Rindler states. This requires the Rindler quantization of the scalar field and decomposing the global modes into Rindler modes via computation of the Bogoliubov coefficients. Armed with an explicit Hilbert space construction of the reduced density matrix, the bulk entanglement entropy is then obtained via the replica trick. Combining with the area correction, we confirm at the end of \S\ref{sec:scalar_bulkEE} that both \eqref{eqn:JLMS} and \eqref{eqn:srel} are true.

\subsection{Canonical quantization in global coordinates}
\label{sec:canonquant_scalar}

Consider the AdS$_{d+1}$ spacetime whose metric in global coordinates is given by
\begin{equation}\label{eqn:AdSmetric}
ds^{2} = -(r^{2}+1)dt^{2}+\frac{dr^{2}}{(r^{2}+1)} + r^{2}d\Omega_{d-1}^{2},
\end{equation}
where $d\Omega_{d-1}^{2}$ is the metric on $S^{d-1}$ and we have chosen the AdS radius $L_{\mathrm{AdS}} = 1$. Let $\phi$ be a massive scalar field propagating on this AdS$_{d+1}$ spacetime whose dynamics are governed by the Klein-Gordon equation
\begin{equation}\label{eqn:KG}
(\nabla^{2}-m^{2})\phi=0,
\end{equation}
where the mass is related to the conformal dimension of the dual primary field $O$ in the CFT by $m^{2}=\Delta_{O}(\Delta_{O}-d)$ and we will assume $\Delta_{O}>d$. The canonical quantization of this scalar field gives
\begin{equation}\label{eqn:canonquant}
\phi(t,r,\Omega) = \sum_{n,\ell,\mathfrak{m}}\phi_{n,\ell,\mathfrak{m}}(t,r,\Omega)a_{n,\ell,\mathfrak{m}}+\phi_{n,\ell,\mathfrak{m}}(t,r,\Omega)^{\ast}a_{n,\ell,\mathfrak{m}}^{\dagger},
\end{equation}
where $\Omega = \{\phi_{i}\}$ and $\mathfrak{m}=\{m_{i}\}$ are collective coordinates and $n,\ell,\mathfrak{m}$ label the independent solutions of the Klein-Gordon equation. The solutions are
\begin{equation}\label{eqn:sepvar}
\phi_{n,\ell,\mathfrak{m}}(t,r,\Omega) = e^{-i\Omega_{n,\ell}t}f_{n,\ell,\mathfrak{m}}(r)Y_{\ell,\mathfrak{m}}^{(d-1)}(\Omega),
\end{equation}
where the quantized frequencies are given by
\begin{equation}\label{eqn:quantcondition}
\Omega_{n,\ell} = \Delta_{O}+2n+\ell,
\end{equation}
the $Y_{\ell,\mathfrak{m}}^{(d-1)}$ are $(d-1)$-dimensional spherical harmonics, and the radial wavefunctions $f_{n,\ell,\mathfrak{m}}(r)$ are known in terms of hypergeometric functions \cite{Balasubramanian:1998sn}. We will only be interested in the lowest mode of the scalar field $(n,\ell,\mathfrak{m}) = (0,0,\mathbf{0})$ for which the radial wavefunction takes the very simple form 
\begin{equation}\label{eqn:KGeqnn=0}
f_{0,0,\mathbf{0}}(r) = \frac{C_{0}}{(r^{2}+1)^{\frac{\Delta_{O}}{2}}}.
\end{equation}
The constant $C_{0}$ is fixed by requiring that $f_{0,0,\mathbf{0}}$ have unit Klein-Gordon norm:
\begin{align}\label{eqn:KGnorm}
\begin{split}
\langle f_{0,0,\mathbf{0}},f_{0,0,\mathbf{0}}\rangle &= -2\Omega_{0,0}\int dr\,d\Omega_{d-1}\,\sqrt{-g^{(0)}}\big(g^{(0)}\big)^{tt}|f_{0,0,\mathbf{0}}|^{2} = 1
\\	\implies  C_{0} &= \sqrt{\frac{\Gamma(\Delta_{O})}{V_{S^{d-1}}\Gamma(\frac{d}{2})\Gamma(\Delta_{O}+1-\frac{d}{2})}}
\end{split}
\end{align}
where $g_{\mu\nu}^{(0)}$ denotes the vacuum AdS$_{d+1}$ metric. The annihilation and creation operators satisfy the usual commutation relations 
\begin{equation}\label{eqn:globalcommrelns}
[a_{n,\ell,\mathfrak{m}},a_{n',\ell',\mathfrak{m}'}^{\dagger}] = \delta_{n,n'}\delta_{\ell,\ell'}\delta_{\mathfrak{m},\mathfrak{m}'}
\end{equation}
where the normalization of these commutation relations is determined by the canonical (equal-time) commutation relation $[\phi(x),\pi(y)] = i\delta^{d}(x-y)$. 

For our calculation of the generalized entropy, we focus on the lowest energy excited state of the scalar field
\begin{equation}\label{eqn:AdSscalarstate}
\ket{\psi} = a_{0,0,\mathbf{0}}^{\dagger}\ket{0}.
\end{equation}
This is the single-particle state in AdS dual to the scalar primary state $O$ that we considered in the CFT calculation in \S\ref{sec:CFTEE_excitedstates} because the extrapolate dictionary gives $\lim_{r \to \infty}r^{\Delta}\phi(t,r,\Omega) = \mathcal{O}(t,\Omega)$ as CFT operator.

\subsection{Correction to the area}
\label{sec:areacorrection}

The area term $A/4G_{N}$ in the generalized entropy has a leading contribution at $\mathcal{O}(G_{N}^{-1})$ coming from the classical area of the classical extremal surface in vacuum AdS, which gives the leading $\mathcal{O}(N^{2})$ entanglement entropy in the CFT by the HRRT formula. There are subleading corrections to $A/G_{N}$ which appear at $\mathcal{O}(G_{N}^{0})$ and depend on the particular perturbative state $\ket{\psi}$. We wish to compute these as they correspond to the $\mathcal{O}(N^{0})$ corrections to the entanglement entropy that we obtained in the CFT in \S\ref{sec:CFTEE_excitedstates}. Note that any correction to the position of the surface that is $\mathcal{O}(G_{N}^{1})$ will give an $\mathcal{O}(G_{N}^{2})$ correction to the area by classical extremality and thus will not contribute at the order that we are considering. Therefore, we can consider the classical extremal surface and the only contribution at the desired order comes from the correction to the area due to the change in the metric from the backreaction of the scalar field.

The backreaction can be obtained by solving the semi-classical Einstein equations, i.e., we treat the metric $g_{\mu\nu}$ as a classical field while the scalar is treated as a quantum field. That is, we wish to solve
\begin{equation}\label{eqn:Einsteineqns_scalar}
R_{\mu\nu}-\frac{1}{2}Rg_{\mu\nu}+\Lambda g_{\mu\nu} = 8\pi G_{N}\bra{\psi}T_{\mu\nu}\ket{\psi},
\end{equation}
where $\Lambda = -d(d-1)/2$ is the cosmological constant and the normal-ordered stress-tensor is given by
\begin{equation}\label{eqn:stress-tensor_scalar}
T_{\mu\nu} = \;\normord{\partial_{\mu}\phi\partial_{\nu}\phi-\frac{1}{2}g_{\mu\nu}\left((\nabla\phi)^2+m^2\phi^2\right)}.
\end{equation}
The expectation values of the stress-tensor are given by
\begin{align}\label{eqn:stress-tensor_evs_scalar}
\begin{split}
\bra{\psi}T_{tt}\ket{\psi} &= \frac{C_{0}^{2}\Delta_{O}(2\Delta_{O}-d)}{(r^2+1)^{\Delta_{O}-1}}
\\ \bra{\psi}T_{rr}\ket{\psi} &= \frac{C_{0}^{2}d\Delta_{O}}{(r^2+1)^{\Delta_{O}+1}}
\\ \bra{\psi}T_{\phi_{i}\phi_{i}}\ket{\psi} &= g_{\phi_{i}\phi_{i}}C_{0}^{2}\Delta_{O}\frac{(d-2\Delta_{O})r^2+d}{(r^2+1)^{\Delta_{O}+1}}
\end{split}
\end{align}
while all off-diagonal expectation values vanish. 

To solve the Einstein equations, we make the following metric ansatz 
\begin{equation}\label{eqn:metricansatz_scalar}
\widetilde{ds}^{2} = -\left(r^2+G_{1}(r)^2\right)dt^{2} + \frac{dr^{2}}{r^2+G_{2}(r)^2} + r^{2}d\Omega_{d-1}^{2}
\end{equation}
and expand $G_{1,2}(r)$ as a function of $G_{N}$:
\begin{equation}\label{eqn:G12}
G_{1,2}(r) = 1+G_{N}G_{1,2}^{(1)}(r)+\mathcal{O}(G_{N}^2).
\end{equation}
For the purposes of computing the correction to the area, we only need $G_{2}(r)$ so we will only present that here, although explicit expressions can be obtained for $G_{1}(r)$ as well. The $tt$ Einstein equation gives the following first-order inhomogeneous differential equation for $G_{2}(r)$:
\begin{equation}\label{eqn:ttEinsteineqn_scalar}
\partial_{r}G_{2}^{(1)}+(d-2)\frac{G_{2}^{(1)}}{r} = 8\pi\frac{C_{0}^{2}\Delta_{O}(d-2\Delta_{O})}{(d-1)}\frac{r}{(r^2+1)^{\Delta_{O}}}
\end{equation}
whose solution is
\begin{equation}\label{eqn:G21soln}
G_{2}^{(1)}(r) = \frac{\mathfrak{C}}{r^{d-2}} + 8\pi\frac{C_{0}^{2}\Delta_{O}(d-2\Delta_{O})}{d(d-1)}r^{2}\,{}_{2}{F}_{1}\left(\frac{d}{2},\Delta_{O},\frac{d}{2}+1;-r^{2}\right).
\end{equation}
Observe that the first term diverges as $r \to 0$ while the second term goes to zero so regularity of the metric demands $\mathfrak{C} = 0$, and hence
\begin{equation}\label{eqn:G21soln_final}
G_{2}^{(1)}(r) = 8\pi\frac{C_{0}^{2}\Delta_{O}(d-2\Delta_{O})}{d(d-1)}r^{2}\,{}_{2}{F}_{1}\left(\frac{d}{2},\Delta_{O},\frac{d}{2}+1;-r^{2}\right).
\end{equation}

We are now ready to compute the correction to the area due to the backreaction of the scalar excited state. To find the classical extremal surface $\gamma_{\mathrm{ext}}$, we must find the codimension-2 surface anchored on the boundary polar cap with minimal area, where we restrict to those surfaces lying in the $t=0$ slice of AdS due to time reflection symmetry. The polar cap on the boundary is symmetric along all the angular directions, except for the $\theta$ direction, so we can assume that the classical extremal surface respects the same symmetries. Thus, we parametrize our surface by $r$ so that the surface is described by $\theta(r)$. Taking $h_{\mu\nu}^{(0)}$ to be the induced metric on the surface in vacuum AdS, the area functional for such a surface is given by
\begin{equation}\label{eqn:areafunctional}
A\left[g^{(0)},\gamma\right] = \int_{\gamma}\sqrt{h^{(0)}} = 2V_{S^{d-2}}\int_{r_{\mathrm{min}}}^{\infty}dr\,(r\sin\theta)^{d-2}\sqrt{\frac{1}{r^2+1}+r^2\left(\theta'\right)^2},
\end{equation}
where $r_{\mathrm{min}}$ is the deepest point in the bulk reached by the surface. Extremizing this functional, we find the solution to the minimal surface equation to be\footnote{This solution is particularly easy to obtain in coordinates specially adapted to the symmetries of the problem \cite{Colin-Ellerin:2019vst}.}
\begin{equation}\label{eq:extremalsurf}
\theta(r) = \arccos\left(\alpha_{0}\frac{\sqrt{r^{2}+1}}{r}\right).
\end{equation}
The constant $\alpha_{0}$ can be related to $\theta_{0}$ and to $r_{\mathrm{min}}$ by
\begin{equation}\label{eqn:rmin}
\theta_{0} = \arccos(\alpha_{0}), \qquad r_{\mathrm{min}} = \frac{\alpha_{0}}{\sqrt{1-\alpha_{0}^{2}}} = \cot\theta_{0}.
\end{equation}
This data completely specifies the classical extremal surface $\gamma_{\mathrm{ext}}$. The correction to the area of the surface due to the backreaction of the scalar field is thus found to be
\begin{align}\label{eqn:areachange_scalar}
\begin{split}
\Delta A(\gamma_{\mathrm{ext}})_{\psi} &\equiv A\left[g^{(0)}+G_{N}g^{(1)},\gamma_{\mathrm{ext}}\right] - A[g^{(0)},\gamma_{\mathrm{ext}}] 
\\	&= -V_{S^{d-2}}G_{N}\int_{r_{\mathrm{min}}}^{\infty}dr\,(r\sin\theta)^{d-2}\frac{G_{2}^{(1)}(r)}{(r^{2}+1)^{\frac{3}{2}}}\sqrt{1-\frac{r_{\mathrm{min}}^{2}}{r^{2}}} + \mathcal{O}(G_{N}^{2}),
\end{split}
\end{align}
where $g_{\mu\nu}^{(1)}$ is the backreacted metric \eqref{eqn:metricansatz_scalar} at $\mathcal{O}(G_{N})$.\footnote{This is sometimes referred to as the change in the expectation value of the area operator $\hat{A}$ \cite{Jafferis:2015del,Dong:2017xht}. That is, we could think of \eqref{eqn:Einsteineqns_scalar} as solving for $\langle g_{\mu\nu}^{(1)}\rangle_{\psi}$. Since $\hat{A}[g_{\mu\nu}^{(1)}]$ is linear in $g_{\mu\nu}^{(1)}$ at leading order in $G_{N}$ as seen in \eqref{eqn:areachange_scalar}, it makes sense to think of the area we compute as an expectation value of a linear operator acting on the code subspace. This distinction is actually not important at the order we are considering for ordinary matter, but it becomes important when considering the graviton \cite{Colin-Ellerin24}.} Although each of the areas is divergent, their difference is actually finite. This cancellation of divergences turns out to be very subtle in the Maxwell case as we shall see, but for the scalar case at hand we can proceed to evaluate the integral in \eqref{eqn:areachange_scalar}. The small $\theta_{0}$ expansion corresponds to the large $r_{\mathrm{min}}$ expansion so it suffices to expand the integrand at large $r$ and evaluate the integral term-by-term in the expansion, leading to
\begin{equation}\label{eqn:areachange_scalar_final}
\frac{\Delta A(\gamma_{\mathrm{ext}})_{\psi}}{4G_{N}} = \frac{\sqrt{\pi}\,\Gamma(\frac{d}{2})\Delta_{O}}{2\Gamma\left(\frac{d+3}{2}\right)}\theta_{0}^{d}\left(1+\mathcal{O}(\theta_{0}^{2})\right)-\frac{\sqrt{\pi}\,\Gamma(\Delta_{O}+1)}{2\Gamma(\Delta_{O}+\frac{3}{2})}\theta_{0}^{2\Delta_{O}}\left(1+\mathcal{O}(\theta_{0}^{2})\right).
\end{equation}
Notice that we can distinguish the two different contributions to $\Delta A$ by their monodromy in the complex $\theta_{0}$-plane (when $\Delta_{O} \not \in \mathbb{Z}$) so it is meaningful to separate the two as distinct. It is not difficult to compute higher-order corrections in $\theta_{0}$, but the leading-order contributions will be sufficient for comparison with the CFT result. We already see that the $\theta_{0}^{d}$ contribution to $\Delta A$ agrees with $\Delta K_{0}^{O}$ computed in \eqref{eqn:K0Odif}.

\subsection{Entanglement entropy}
\label{sec:scalar_bulkEE}

Since we are only studying the leading order quantum correction to the HRRT formula, namely $\mathcal{O}(G_{N}^{0})$, we only need to consider the entanglement of a free quantum scalar field on the fixed vacuum AdS background with the entangling surface given by the classical extremal surface. Any corrections coming from coupling to gravity, such as changes in the position of the surface, changes in the background metric, or graviton loops will all be higher order effects in $G_{N}$. 

The von Neumann entropy of the excited state can be most easily studied by going to coordinates adapted to the classical entanglement wedge $\mathcal{W}[\mathcal{B}]$ defined as the domain of dependence of the homology surface $\Sigma_{\mathcal{B}}$. For the polar cap region considered here, $\mathcal{W}[\mathcal{B}]$ can be described by AdS-Rindler coordinates with the metric given by 
\begin{equation}\label{eqn:AdSRindlermetric}
ds^{2} = -(\rho^{2}-1)d\tau^{2} + \frac{d\rho^{2}}{\rho^{2}-1} + \rho^{2}\underbrace{\left(du^{2}+\sinh^{2}u\,d\Omega_{d-2}^{2}\right)}_{(dH^{d-1})^{2}},
\end{equation}
where $\tau \in \mathbb{R}$ and $\rho \in (1,\infty)$. The classical extremal surface $\gamma_{\mathrm{ext}}$ lies at $\{\tau=0$, $\rho=1\}$ and the domain of dependence of $\mathcal{B}$ in the asymptotic boundary $D[\mathcal{B}]$ lies at $\rho = \infty$. The final piece of the metric in \eqref{eqn:AdSRindlermetric} is the metric on $(d-1)$-dimensional hyperbolic space $H^{d-1}$. The relationship to global coordinates is given in App. \S\ref{sec:AdSRindler}.

We now quantize the scalar field in the AdS-Rindler wedge and relate this to the global quantization of \S\ref{sec:canonquant_scalar}.

\subsubsection{AdS-Rindler quantization}
\label{sec:AdSRindlerquant}

The massive scalar field in AdS-Rindler can be expanded in solutions to the Klein-Gordon equation
\begin{equation}\label{eqn:AdSRindlerquant}
\phi(\tau,\rho,\mathbf{x}) = \sum_{I \in \{L,R\}}\int_{0}^{\infty}\frac{d\omega}{2\pi}\sum_{\lambda}\left(e^{-i\omega\tau}g_{\omega,\lambda,I}(\rho,\mathbf{x})b_{\omega,\lambda,I} + e^{i\omega\tau}g_{\omega,\lambda,I}(\rho,\mathbf{x})^{\ast}b_{\omega,\lambda,I}^{\dagger}\right)
\end{equation}
where we refer to $\mathcal{W}[\mathcal{B}]$ as the right wedge and $\mathcal{W}[\mathcal{B}^{c}]$ as the left wedge. The wavefunctions $g_{\omega,\lambda,I}(\rho,\mathbf{x})$ are given by
\begin{equation}\label{eqn:AdSRindlerwavefn}
g_{\omega,\lambda,I}(\rho,\mathbf{x}) = \psi_{\omega,\lambda}(\rho)H_{\lambda}(\mathbf{x})
\end{equation}
where $H_{\lambda}(\mathbf{x}) \equiv H_{\lambda,\ell,\mathfrak{m}}(\mathbf{x})$ are eigenfunctions of the Laplacian on $H^{d-1}$ with eigenvalue $-\lambda$ discussed in App. \ref{sec:hyperbolicballeigfns}. The eigenvalues are continuous so the sum in \eqref{eqn:AdSRindlerquant} is actually an integral together with a sum over degeneracies ($\ell,\mathfrak{m}$). It is convenient to define $\lambda = \tilde{\lambda}^{2}+\zeta^{2}$ where $\zeta \equiv \frac{d-2}{2}$ so the spectrum of allowed eigenvalues is given by $\tilde{\lambda} \in (0,\infty)$. We also define
\begin{equation}
\omega_{\pm} = \omega \pm \tilde{\lambda}.
\end{equation}

The radial wavefunctions are
\begin{align}\label{eqn:AdSRindlerradialwavefn}
\begin{split}
\psi_{\omega,\lambda}(\rho) = \frac{\mathcal{N}_{\omega,\lambda}}{\rho^{\Delta_{O}}}\left(1-\frac{1}{\rho^{2}}\right)^{-\frac{i\omega}{2}}{}_{2}{F}_{1}\Bigg[\frac{1}{2}&\left(\Delta_{O}-i\omega_{-}-\zeta\right),\frac{1}{2}\left(\Delta_{O}-i\omega_{+}-\zeta\right),\Delta_{O}-\zeta;\frac{1}{\rho^{2}}\Bigg],
\end{split}
\end{align}
and requiring that the wavefunctions have unit (Rindler) Klein-Gordon norm gives the normalization constant
\begin{equation}\label{eqn:AdSRindlernorm}
\mathcal{N}_{\omega,\lambda} = \frac{1}{\sqrt{2\omega}}\left|\frac{\Gamma\left(\frac{1}{2}\left(\Delta_{O}-i\omega_{-}-\zeta\right)\right)\Gamma\left(\frac{1}{2}\left(\Delta_{O}-i\omega_{+}-\zeta\right)\right)}{\Gamma(\Delta_{O}-\zeta)\Gamma(i\omega)}\right|.
\end{equation}
Observe that these wavefunctions obey the standard boundary conditions for a scalar field in spacetimes with a horizon: normalizable at infinity and ingoing/outgoing at the horizon. The annihilation and creation operators satisfy the commutation relations
\begin{equation}\label{eqn:Rindlercommreln}
[b_{\omega,\lambda,I},b_{\omega',\lambda',I'}^{\dagger}] = (2\pi)^{2}\delta(\omega-\omega')\delta^{(d-1)}(\tilde{\lambda}-\tilde{\lambda}')\delta_{I,I'},
\end{equation}
where we have abused notation to write $\delta^{(d-1)}(\tilde{\lambda}-\tilde{\lambda}') \equiv \delta(\tilde{\lambda}-\tilde{\lambda}')\delta_{l,l'}\delta_{\mathfrak{m},\mathfrak{m}'}$.

Before leaving the quantization discussion, it is worthwhile to comment on the boundary conditions. 
Strictly speaking, we need to put in a UV cut-off at $\rho=1+\epsilon$ which results in a discretized spectrum for the Rindler modes. 
The reason is as follows: (1) to have a well-defined variational principle for a single wedge, one needs the boundary term in the action $(\rho^{2}-1)\partial_{\rho}\phi|_{\rm horizon}$ to vanish, however, the scalar wavefuncion oscillates near $\rho=1$ so we need a cut-off; 
(2) the introduction of a cutoff is necessary to obtain a finite von Neumann entropy which will diverge as we take $\epsilon \to 0$. 
These subtleties will be irrelevant for the analysis of the scalar, but will be important when we consider Maxwell theory.
In particular, the bulk vacuum-subtracted von Neumann entropy that we analyze in this work does not suffer from any divergences, so that it is rather trivial to take $\epsilon \rightarrow 0$ or effectively not introduce such a cutoff on the horizon and work with a continuous spectrum in practice.

\subsubsection{Density matrix construction}

To construct the reduced density matrix for the entanglement wedge corresponding to the state $|\psi\rangle$, we need to factorise the global AdS Hilbert space into the left and right Rindler wedges. 
We will consider a perturbative excitation of the global vacuum state so we first need to understand the vacuum state and the relationship between the global and Rindler operators, i.e., understand properties of the Bogoliubov coefficients. 

The global vacuum state in AdS is equal to the thermofield double state in AdS-Rindler
\begin{equation}\label{eqn:MinkvactoRindlerTFD}
\ket{0} = \bigotimes_{\omega,\lambda}\sqrt{1-e^{-2\pi\omega}}\sum_{n}e^{-\pi E_{n}}\ket{n,\omega,\lambda^{\ast}}_{L}\ket{n,\omega,\lambda}_{R},
\end{equation}
where $E_{n}=\omega n$. Henceforth, we will drop the product and simply write $\ket{0} = \sum_{n}e^{-\pi E_{n}}\ket{n^{\ast}}_{L}\ket{n}_{R}$ where $\ket{n^{\ast}}$ is the CRT conjugate of $\ket{n}$. 

We can relate the global annihilation and creation operators with the Rindler annihilation and creation operators by
\begin{equation}\label{eqn:globaltoRindlerops}
a_{n,\ell,\mathfrak{m}} = \sum_{I \in \{L,R\}}\int_{0}^{\infty}\frac{d\omega}{2\pi}\sum_{\lambda}\left(\alpha_{n,\ell,\mathfrak{m};\omega,\lambda,I}b_{\omega,\lambda,I}+\beta_{n,\ell,\mathfrak{m};\omega,\lambda,I}^{\ast}b_{\omega,\lambda,I}^{\dagger}\right)
\end{equation}
where $\alpha_{n,\ell,\mathfrak{m};\omega,\lambda,I}$, $\beta_{n,\ell,\mathfrak{m};\omega,\lambda,I}$ are the Bogoliubov coefficients. By applying \eqref{eqn:globalcommrelns} and \eqref{eqn:Rindlercommreln}, we arrive at the following condition on the Bogoliubov coefficients
\begin{align}\label{eqn:Bogcoeffcondition}
\begin{split}
\delta_{n,n'}\delta_{\ell,\ell'}\delta_{\mathfrak{m},\mathfrak{m}'} &= [a_{n,\ell,\mathfrak{m}},a_{n',\ell',\mathfrak{m}'}^{\dagger}] 
\\	&= \sum_{I \in \{L,R\}}\int_{0}^{\infty}\frac{d\omega}{2\pi}\sum_{\lambda}\left(\alpha_{n,\ell,\mathfrak{m};\omega,\lambda,I}\alpha_{n',\ell',\mathfrak{m'};\omega,\lambda,I}^{\ast}-\beta_{n,\ell,\mathfrak{m};\omega,\lambda,I}^{\ast}\beta_{n',\ell',\mathfrak{m'};\omega,\lambda,I}\right).
\end{split}
\end{align}
Let us focus on $(n,\ell,\mathfrak{m}) = (0,0,\mathbf{0})$ and we denote $\alpha_{\omega,\lambda,I} \equiv \alpha_{0,0,\mathbf{0};\omega,\lambda,I}$. The Bogoliubov coefficients must satisfy additional constraints owing to $a_{0,0,\mathbf{0}}\ket{0}=0$, which implies
\begin{equation}\label{eqn:vacann}
\alpha_{\omega,\lambda,L} = -e^{\pi\omega}\beta_{\omega,\lambda^{\ast},R}, \qquad \beta_{\omega,\lambda,L}^{\ast} = -e^{-\pi\omega}\alpha_{\omega,\lambda^{\ast},R}.
\end{equation}
This constraint is important because it allows us to write everything in terms of only the $R$ Bogoliubov coefficients. We can also rewrite left creation and annihilation operators acting on $\ket{0}$ in terms of only right operators using \eqref{eqn:MinkvactoRindlerTFD} to obtain
\begin{equation}\label{eqn:AdSRindlerops_LtoR}
b_{\omega,\lambda,L}\ket{0} = e^{-\pi\omega}b_{\omega,\lambda^{\ast},R}^{\dagger}\ket{0}, \qquad b_{\omega,\lambda,L}^{\dagger}\ket{0} = e^{\pi\omega}b_{\omega,\lambda^{\ast},R}\ket{0}.
\end{equation}
The single particle state can now be written in terms of purely right modes
\begin{equation}\label{eqn:singlepartstate_AdSRindler}
\ket{\psi} = a_{0,0}^{\dagger}\ket{0} = \int\frac{d\omega}{2\pi}\sum_{\lambda}\left((1-e^{-2\pi\omega})\alpha_{\omega,\lambda,R}^{\ast}b_{\omega,\lambda,R} + (1-e^{2\pi\omega})\beta_{\omega,\lambda,R}^{\ast}b_{\omega,\lambda,R}^{\dagger}\right)\ket{0}.
\end{equation}
Since we have succeeded in writing everything in terms of $R$ objects, we will drop the $R$ labels: $\alpha_{\omega,\lambda} \equiv \alpha_{\omega,\lambda,R}$ and $b_{\omega,\lambda} \equiv b_{\omega,\lambda,R}$. From the full density matrix $\rho^{\psi} = |\psi\rangle\langle\psi|$, we now easily obtain the reduced density matrix
\begin{align}\label{eqn:AdSRindler_scalarreducedrho}
\begin{split}
\rho_{\Sigma_{\mathcal{B}}}^{\psi} = \Tr_{\Sigma_{\mathcal{B}}^{c}}\rho^{\psi} = &\int\frac{d\omega}{2\pi}\sum_{\lambda}\left((1-e^{-2\pi\omega})\alpha_{\omega,\lambda}^{\ast}b_{\omega,\lambda}^{\dagger} + (1-e^{2\pi\omega})\beta_{\omega,\lambda}b_{\omega,\lambda}\right)e^{-K_{0}^{\mathrm{bulk}}}
\\	&\int\frac{d\omega'}{2\pi}\sum_{\lambda'}\left((1-e^{-2\pi\omega'})\alpha_{\omega',\lambda'}b_{\omega',\lambda'} + (1-e^{2\pi\omega'})\beta_{\omega',\lambda'}^{\ast}b_{\omega',\lambda'}^{\dagger}\right)
\end{split}
\end{align}
where $K_{0}^{\mathrm{bulk}} = 2\pi H_{R}$ with $H_{R}$ the Rindler Hamiltonian and we label the reduced density matrix for the right Rindler wedge by the (classical) homology surface $\Sigma_{\mathcal{B}}$. In the next section, we will need the two-point functions of the Rindler annihilation and creation operators in the AdS vacuum, which are given by
\begin{align}\label{eqn:AdSRindler2ptfns}
\begin{split}
\bra{0}b_{\omega,\lambda}b_{\omega',\lambda'}\ket{0} &= \bra{0}b_{\omega,\lambda}^{\dagger}b_{\omega',\lambda'}^{\dagger}\ket{0} = 0
\\	\bra{0}b_{\omega,\lambda}b_{\omega',\lambda'}^{\dagger}\ket{0} &= \frac{(2\pi)^{2}}{1-e^{-2\pi\omega}}\delta(\omega-\omega')\delta^{d-1}(\lambda-\lambda')
\\	\bra{0}b_{\omega,\lambda}^{\dagger}b_{\omega',\lambda'}\ket{0} &= \frac{(2\pi)^{2}}{e^{2\pi\omega}-1}\delta(\omega-\omega')\delta^{d-1}(\lambda-\lambda').
\end{split}
\end{align}

We now proceed to compute the von Neumann entropy of $\rho_{\Sigma_{\mathcal{B}}}^{\psi}$ by leveraging the small $\theta_{0}$ expansion. 

\subsubsection{First-order contribution to entanglement entropy}
\label{sec:scalarEE1}

The entanglement entropy for $\rho_{\Sigma_{\mathcal{B}}}^{\psi}$ can be computed using the replica trick. We will expand in the parameter
\begin{equation}\label{eqn:rhosmallparam}
\delta \rho = \rho_{\Sigma_{\mathcal{B}}}^{\psi} - \rho_{\Sigma_{\mathcal{B}}}^{0},
\end{equation}
where $\rho_{\Sigma_{\mathcal{B}}}^{0}$ is the reduced density matrix for $\Sigma_{\mathcal{B}}$ in the vacuum state $\ket{0}$. We find that $\delta \rho$ is small in the small polar cap limit $\theta_{0} \ll 1$. The R\'enyi entropy is defined as
\begin{equation}\label{eqn:bulkRenyi}
S_{n}^{\mathrm{AdS}}(\rho_{\Sigma_{\mathcal{B}}}^{\psi}) = \frac{1}{1-n}\log\left[\Tr\left((\rho_{\Sigma_{\mathcal{B}}}^{\psi})^{n}\right)\right],
\end{equation}
from which the entanglement entropy can be computed by
\begin{equation}\label{eqn:bulkEE}
S^{\mathrm{AdS}}(\rho_{\Sigma_{\mathcal{B}}}^{\psi}) = \lim_{n \to 1}S_{n}^{\mathrm{AdS}}(\rho_{\Sigma_{\mathcal{B}}}^{\psi}) = -\partial_{n}\Tr\left((\rho_{\Sigma_{\mathcal{B}}}^{\psi})^{n}\right)\big|_{n=1}.
\end{equation}
The expansion of the trace gives
\begin{equation}\label{eqn:traceexp_1storder}
\Tr\left((\rho_{\Sigma_{\mathcal{B}}}^{\psi})^{n}\right) = \Tr\left((\rho_{\Sigma_{\mathcal{B}}}^{0}+\delta \rho)^{n}\right) = \Tr\left((\rho_{\Sigma_{\mathcal{B}}}^{0})^{n}\right)+n\Tr\left((\rho_{\Sigma_{\mathcal{B}}}^{0})^{n-1}\delta\rho\right)+\mathcal{O}\left((\delta\rho)^{2}\right).
\end{equation}
We see, as we did in the CFT, that the first order contribution to the vacuum-subtracted entanglement entropy is
\begin{equation}\label{eqn:bulkvacsubtractEE_1storder}
\Delta S^{\mathrm{AdS}}(\rho_{\Sigma_{\mathcal{B}}}^{\psi})\big|_{\mathcal{O}(\delta \rho)} = -\partial_{n}\Tr\left((\rho_{\Sigma_{\mathcal{B}}}^{0})^{n-1}\delta\rho\right)\big|_{\mathcal{O}(\delta \rho),n=1} = -\Tr\left(\delta \rho \log \rho_{\Sigma_{\mathcal{B}}}^{0}\right) = \Delta K_{0}^{\psi},
\end{equation}
where we used the fact that $\Tr(\rho_{\Sigma_{\mathcal{B}}}^{\psi})=\Tr(\rho_{\Sigma_{\mathcal{B}}}^{0})=1$, which is simply the first law of entanglement entropy. 

There are two different ways that one can compute $\Delta K_{0}^{\psi}$. First, one can use the fact that $K_{0}^{\mathrm{bulk}}$ is the generator of boosts for the classical entanglement wedge so it can written as the following integral of the stress-tensor
\begin{equation}\label{eqn:bulkmodH}
K_{0}^{\mathrm{bulk}} = 2\pi\int_{\Sigma_{\mathcal{B}}}\sqrt{g_{\Sigma_{\mathcal{B}}}}\,\xi^{\mu}n^{\nu}T_{\mu\nu}
\end{equation}
where $\xi^{\mu} = (\frac{\partial}{\partial \tau})^{\mu}$ is the corresponding Killing vector and $n^{\nu} = \frac{1}{\sqrt{\rho^{2}-1}}(\frac{\partial}{\partial \tau})^{\nu}$ is the (unit) normal vector to the homology surface. The vacuum-subtracted expectation value of $K_{0}^{\mathrm{bulk}}$ in the state $\ket{\psi}$ can be computed from \eqref{eqn:stress-tensor_evs_scalar} in the small interval limit to obtain
\begin{align}\label{eqn:bulkmodH_explicit}
\begin{split}
\Delta K_{0}^{\psi} &= 2\pi C_{0}^{2}\Delta_{O}(2\Delta_{O}-d)V_{S^{d-2}}\theta_{0}^{2\Delta_{O}}\left(1+\mathcal{O}(\theta_{0}^{2})\right)\int_{0}^{\infty} du\,\int_{1}^{\infty}d\rho\,\frac{\rho^{d-1}\sinh^{d-2}u}{(\rho\cosh u + \sqrt{\rho^{2}-1})^{2\Delta_{O}}}
\\	&= \frac{\sqrt{\pi}\Gamma(\Delta_{O}+1)}{2\Gamma(\Delta_{O}+\frac{3}{2})}\theta_{0}^{2\Delta_{O}}\left(1+\mathcal{O}(\theta_{0}^{2})\right)
\end{split}
\end{align}
with an explicit evaluation of these integrals provided in \eqref{eqn:modHscalarpos_int}. It is interesting to note that $\Delta K_{0}^{\mathrm{\psi}}$ is independent of the dimension $d$.

Alternatively, we can compute $\Delta K_{0}^{\psi}$ using the momentum space techniques developed in \S\ref{sec:AdSRindlerquant}. This requires explicit expressions for the Bogoliubov coefficients, which we compute at leading order in small $\theta_{0}$ in App. \ref{sec:Bogcoeffs_scalar}, with the result
\begin{align}\label{eqn:Bogcoeffs_scalar}
\begin{split}
\alpha_{\omega,\lambda} &\sim \frac{C_{0}\mathcal{N}_{\lambda,0}^{H}V_{S^{d-2}}}{\mathcal{N}_{\omega,\lambda}^{\ast}}\frac{2^{\Delta_{O}-2}\Gamma(\zeta+\frac{1}{2})}{\sqrt{\pi}\Gamma(\Delta_{O})\Gamma(\Delta_{O}-\zeta)}\left|\Gamma\left(\frac{\Delta_{O}-\zeta+i\omega_{+}}{2}\right)\Gamma\left(\frac{\Delta_{O}-\zeta+i\omega_{-}}{2}\right)\right|^{2}
\\	&\qquad \times \theta_{0}^{\Delta_{O}}\left(1+\mathcal{O}(\theta_{0}^{2})\right)
\\ \beta_{\omega,\lambda} &\sim -\alpha_{\omega,\lambda}
\end{split}
\end{align}
where we use `$\sim$' to indicate that we are missing non-perturbative contributions in $\theta_{0}$ of the form $e^{-a\theta_{0}\omega}$ for some constant $a$ which have been set to $1$. The constant $\mathcal{N}_{\lambda,0}^{H}$ is the normalization of the eigenfunctions of the Laplacian on $H^{d-1}$ given in \eqref{eqn:hyperbolicev_norm}.

The Rindler Hamiltonian in momentum space is given by
\begin{equation}\label{eqn:RindlerH}
K_{0}^{\mathrm{bulk}} = 2\pi\int \frac{d\omega}{2\pi}\sum_{\lambda}\omega b_{\omega,\lambda}^{\dagger}b_{\omega,\lambda}.
\end{equation}
The first-order contribution to the entanglement entropy is thus
\begin{align}\label{eqn:bulkvacsubtractEE_1storder_explicit}
\begin{split}
\Delta K_{0}^{\psi} &= \Tr\bigg[\rho_{\Sigma_{\mathcal{B}}}^{0}\int\frac{d\omega}{2\pi}\frac{d\omega'}{2\pi}\frac{d\omega''}{2\pi}\sum_{\lambda,\lambda',\lambda''}\left((1-e^{-2\pi\omega'})\alpha_{\omega',\lambda'}b_{\omega',\lambda'} + (1-e^{2\pi\omega'})\beta_{\omega',\lambda'}^{\ast}b_{\omega',\lambda'}^{\dagger}\right)
\\	&\qquad \times \left(2\pi\omega b_{\omega',\lambda'}^{\dagger}b_{\omega',\lambda'}\right)\left((1-e^{-2\pi\omega''})\alpha_{\omega'',\lambda''}^{\ast}b_{\omega'',\lambda''}^{\dagger} + (1-e^{2\pi\omega''})\beta_{\omega'',\lambda''}b_{\omega'',\lambda''}\right)\bigg] 
\\	&\qquad -\Tr\left[\rho_{\Sigma_{\mathcal{B}}}^{0}\int\frac{d\omega}{2\pi}\sum_{\lambda}\omega b_{\omega,\lambda}^{\dagger}b_{\omega,\lambda}\right]
\\	&= \int d\omega\,\omega\sum_{\lambda}\left(|\alpha_{\omega,\lambda}|^{2}+|\beta_{\omega,\lambda}|^{2}\right)
\\	&=\frac{\sqrt{\pi}\Gamma(\Delta_{O}+1)}{2\Gamma(\Delta_{O}+\frac{3}{2})}\theta_{0}^{2\Delta_{O}}\left(1+\mathcal{O}(\theta_{0}^{2})\right),
\end{split}
\end{align}
where we performed the trace using Wick contractions and the two-point functions in \eqref{eqn:AdSRindler2ptfns}. The integrals over $\omega$ and $\lambda$ are computed in \eqref{eqn:modHscalarmom_int}. We find exact agreement with our position space calculation \eqref{eqn:bulkmodH_explicit} which provides a nice check on our Bogoliubov coefficients.

Observe that $\Delta K_{0}^{\psi}$ is equal to the second term in \eqref{eqn:areachange_scalar_final} up to a sign. Therefore, our results agree with the JLMS formula \eqref{eqn:JLMS}:
\begin{align}\label{eqn:JLMSscalar}
\begin{split}
\frac{\Delta A(\gamma_{\mathrm{ext}})_{\psi}}{4G_{N} }+\Delta K_{0}^{\psi} &= \frac{\sqrt{\pi}\,\Gamma(\frac{d}{2})\Delta_{O}}{2\Gamma\left(\frac{d+3}{2}\right)}\theta_{0}^{d}\left(1+\mathcal{O}(\theta_{0}^{2})\right)-\frac{\sqrt{\pi}\,\Gamma(\Delta_{O}+1)}{2\Gamma(\Delta_{O}+\frac{3}{2})}\theta_{0}^{2\Delta_{O}}\left(1+\mathcal{O}(\theta_{0}^{2})\right)
\\  &\qquad +\frac{\sqrt{\pi}\Gamma(\Delta_{O}+1)}{2\Gamma(\Delta_{O}+\frac{3}{2})}\theta_{0}^{2\Delta_{O}}\left(1+\mathcal{O}(\theta_{0}^{2})\right)
\\  &= \frac{\sqrt{\pi}\,\Gamma(\frac{d}{2})\Delta_{O}}{2\Gamma\left(\frac{d+3}{2}\right)}\theta_{0}^{d}\left(1+\mathcal{O}(\theta_{0}^{2})\right)
\\  &= \Delta K_{0}^{O},
\end{split}
\end{align}
as desired.

\subsubsection{Second-order contribution to entanglement entropy}
\label{sec:scalarEE2}

We can go beyond what can be accessed from the first law of entanglement by computing the $(\delta \rho)^{2}$ contribution in the expansion of the R\'enyi entropy \eqref{eqn:traceexp_1storder}. At second-order, one finds
\begin{align}\label{eqn:traceexp_2ndorder}
\begin{split}
\Tr\left[(\rho_{\Sigma_{\mathcal{B}}}^{\psi})^{n}\right]\Big|_{\mathcal{O}(\delta\rho^2)} &= \frac{n}{2}\sum_{a=0}^{n-2}\Tr\left(\delta\rho(\rho_{\Sigma_{\mathcal{B}}}^{0})^{a}\delta\rho(\rho_{\Sigma_{\mathcal{B}}}^{0})^{n-a-2}\right)
\\	&= \frac{n}{2}\Tr\left(\rho_{\Sigma_{\mathcal{B}}}^{\psi}\tilde{\rho}(n)(\rho_{\Sigma_{\mathcal{B}}}^{0})^{n-2}\right)+n(n-1)\left[\frac{1}{2}\Tr\left((\rho_{\Sigma_{\mathcal{B}}}^{0})^{n}\right)-\Tr\left(\rho_{\Sigma_{\mathcal{B}}}^{\psi}(\rho_{\Sigma_{\mathcal{B}}}^{0})^{n-1}\right)\right]
\end{split}
\end{align}
where we have defined
\begin{equation}\label{eqn:tilderhodef}
\tilde{\rho}(n) \equiv \sum_{a=0}^{n-2}(\rho_{\Sigma_{\mathcal{B}}}^{0})^{a}\rho_{\Sigma_{\mathcal{B}}}^{\psi}(\rho_{\Sigma_{\mathcal{B}}}^{0})^{-a}.
\end{equation}
This can be evaluated using the Baker-Campbell-Hausdorff formula to obtain
\begin{align}\label{eqn:tilderho}
\begin{split}
\tilde{\rho}(n) &= \int \frac{d\omega_{1}}{2\pi}\,\frac{d\omega_{2}}{2\pi}\,\sum_{\lambda_{1},\lambda_{2}}\bigg[\frac{\mathfrak{E}\left((n-1)(\omega_{2}-\omega_{1})\right)\mathfrak{E}(-\omega_{1})\mathfrak{E}(-\omega_{2})}{\mathfrak{E}(\omega_{2}-\omega_{1})}\alpha_{1}^{\ast}\alpha_{2}b_{1}^{\dagger}\rho_{\Sigma_{\mathcal{B}}}^{0}b_{2}
\\	&\qquad +\frac{\mathfrak{E}\left((n-1)(-\omega_{1}-\omega_{2})\right)\mathfrak{E}(-\omega_{1})\mathfrak{E}(\omega_{2})}{\mathfrak{E}(-\omega_{1}-\omega_{2})}\alpha_{1}^{\ast}\beta_{2}^{\ast}b_{1}^{\dagger}\rho_{\Sigma_{\mathcal{B}}}^{0}b_{2}^{\dagger}
\\	&\qquad +\frac{\mathfrak{E}\left((n-1)(\omega_{1}+\omega_{2})\right)\mathfrak{E}(\omega_{1})\mathfrak{E}(-\omega_{2})}{\mathfrak{E}(\omega_{1}+\omega_{2})}\beta_{1}\alpha_{2}b_{1}\rho_{\Sigma_{\mathcal{B}}}^{0}b_{2}
\\	&\qquad +\frac{\mathfrak{E}\left((n-1)(\omega_{1}-\omega_{2})\right)\mathfrak{E}(\omega_{1})\mathfrak{E}(\omega_{2})}{\mathfrak{E}(\omega_{1}-\omega_{2})}\beta_{1}\beta_{2}^{\ast}b_{1}\rho_{\Sigma_{\mathcal{B}}}^{0}b_{2}^{\dagger}\bigg]
\end{split}
\end{align}
where we have abbreviated $\alpha_{k} \equiv \alpha_{\omega_{k},\lambda_{k}}$ and the same for $\beta$, and to make equations cleaner we have defined
\begin{equation}\label{eqn:Edef}
\mathfrak{E}(x) = 1-e^{2\pi x}.
\end{equation}
The function $\tilde{\rho}(n)$ is now manifestly analytic in $n$ and has a simple $n$ derivative. It also has the nice property that $\tilde{\rho}(1)=0$. Therefore, evaluating the trace using Wick contractions, the second-order contribution to the vacuum-subtracted von Neumann entropy is
\begin{align}\label{eqn:bulkvacsubtractEE_2ndorder}
\begin{split}
\Delta S^{\mathrm{AdS}}(\rho_{\Sigma_{\mathcal{B}}}^{\psi})\big|_{\mathcal{O}(\delta\rho^2)} &= -\partial_{n}\Tr\left((\rho_{\Sigma_{\mathcal{B}}}^{\psi})^{n}\right)\Big|_{\mathcal{O}(\delta\rho^2),n=1}
\\	&= \frac{1}{2}-\frac{1}{2}\Tr\left(\rho_{\Sigma_{\mathcal{B}}}^{\psi}\tilde{\rho}'(1)(\rho_{\Sigma_{\mathcal{B}}}^{0})^{-1}\right)
\\	&= \pi \int \frac{d\omega_{1}}{2\pi}\,\frac{d\omega_{2}}{2\pi}\,\sum_{\lambda_{1},\lambda_{2}}\left((\omega_{1}+\omega_{2})\frac{\mathfrak{E}(\omega_{1})\mathfrak{E}(\omega_{2})}{\mathfrak{E}(\omega_{1}+\omega_{2})}+(\omega_{1}-\omega_{2})\frac{\mathfrak{E}(-\omega_{1})\mathfrak{E}(\omega_{2})}{\mathfrak{E}(\omega_{2}-\omega_{1})}\right)
\\	&\qquad \times \left(|\alpha_{1}|^{2}|\beta_{2}|^{2}+|\beta_{1}|^{2}|\alpha_{2}|^{2}+2\alpha_{1}^{\ast}\beta_{1}\alpha_{2}\beta_{2}^{\ast}\right)
\\	&= -\frac{\sqrt{\pi}\Gamma(2\Delta_{O}+1)}{2\Gamma(2\Delta_{O}+\frac{3}{2})}\theta_{0}^{4\Delta_{O}},
\end{split}
\end{align}
where the integrals over $\omega_{1,2}$ and $\tilde{\lambda}_{1,2}$ can be evaluated in a similar way to the integrals in \eqref{eqn:modHscalarmom_int}.\footnote{Observe that the second-order contribution is also independent of the dimension $d$ so this is true for the total bulk vacuum-subtracted von Neumann entropy to the order in $\theta_{0}$ considered here. It would be interesting to determine if that is true to all orders in $\theta_{0}$, and if so, why it is true.} This second-order contribution gives the relative entropy
\begin{align}
\begin{split}
-S_{\mathrm{rel}}^{\mathrm{AdS}}\left(\rho_{\Sigma_{\mathcal{B}}}^{\psi}\Big|\rho_{\Sigma_{\mathcal{B}}}^{0}\right) &= \Delta S^{\mathrm{AdS}}(\rho_{\Sigma_{\mathcal{B}}}^{\psi})-\Delta K_{0}^{\psi}
\\  &= \Delta S^{\mathrm{AdS}}(\rho_{\Sigma_{\mathcal{B}}}^{\psi}) - \Delta S^{\mathrm{AdS}}(\rho_{\Sigma_{\mathcal{B}}}^{\psi})|_{\mathcal{O}(\delta \rho)}
\\  &= \Delta S^{\mathrm{AdS}}(\rho_{\Sigma_{\mathcal{B}}}^{\psi})\big|_{\mathcal{O}(\delta\rho^2)} + \mathcal{O}((\delta \rho)^{3}).
\end{split}
\end{align}

Comparing \eqref{eqn:bulkvacsubtractEE_2ndorder} with the second-order contribution to the CFT vacuum-subtracted von Neumann entropy \eqref{eqn:DeltaS_final}, we see 
\begin{equation}\label{eqn:matchscalarsrel}
    -S_{\rm rel}^{\rm AdS}\left(\rho_{\Sigma_{\mathcal{B}}}^{\psi}\Big|\rho_{\Sigma_{\mathcal{B}}}^{0}\right) = -\frac{\sqrt{\pi}\Gamma(2\Delta_{O}+1)}{2\Gamma(2\Delta_{O}+\frac{3}{2})}\theta_{0}^{4\Delta_{O}} = - S_{\rm rel}^{\rm CFT}\left(\rho_{\mathcal{B}}^{O}\big|\rho_{\mathcal{B}}^{0}\right) ,
\end{equation}
and hence the FLM formula holds given  \eqref{eqn:matchscalarsrel} and \eqref{eqn:JLMSscalar}. To wit,
\begin{equation}
\begin{aligned}
    \Delta S^{\rm CFT}  &\equiv  \Delta S^{O} = \Delta K^{ O}_0 - S_{\rm rel}^{\rm CFT}\left(\rho_{\mathcal{B}}^{O}\big|\rho_{\mathcal{B}}^{0}\right)  \\
    & = \frac{\Delta A(\gamma_{\mathrm{ext}})_{\psi}}{4 G_N} + \Delta K^{\psi}_0 - S_{\rm rel}^{\rm AdS}\left(\rho_{\Sigma_{\mathcal{B}}}^{\psi}\Big|\rho_{\Sigma_{\mathcal{B}}}^{0}\right) =  \frac{\Delta A(\gamma_{\mathrm{ext}})_{\psi}}{4 G_N} + \Delta S^{\rm AdS}.
\end{aligned}
\end{equation}
This completes our analysis of generalized entropy for the massive scalar field.

\section{Maxwell theory in AdS$_{d+1}$}
\label{sec:Max}

Having understood the scalar case, we now turn to a detailed analysis of Maxwell theory in AdS$_{d+1}$ which is the focus of this work. 
The free Maxwell theory of a $U(1)$ gauge field $\mathbb{A}_{\mu}$ has the action
\begin{equation}\label{eqn:Maxwellaction}
S_{\mathrm{Maxwell}} = -\frac{1}{4}\int d^{d+1}x\,\sqrt{-g}F_{\mu\nu}F^{\mu\nu}
\end{equation}
where $F_{\mu\nu} = \nabla_{\mu}\mathbb{A}_{\nu}-\nabla_{\nu}\mathbb{A}_{\mu}$. The $U(1)$ gauge symmetry acts on the gauge field by $\mathbb{A}_{\mu} \to \mathbb{A}_{\mu}+\partial_{\mu}\lambda$ for any function $\lambda$.

The goal of this section is to understand the quantization of this theory, which is parallel to \S\ref{sec:canonquant_scalar} in the scalar case.
However, the existence of the gauge symmetry dramatically complicates the quantization procedure because one has to properly eliminate the gauge redundancy. 
In general, there are two ways to quantize this gauge theory: one can first gauge-fix to obtain the reduced classical phase space and then quantize the physical degrees of freedom or one can first quantize the theory and then project onto the physical Hilbert space.
In this section, we will take the first approach due to practical simplicity.
In \S\ref{sec:edge}, we will have to consider the second approach to incorporate edge modes. 

To perform the gauge-fixing-first procedure, we choose the radial gauge $A_{r}=0$. This gauge choice will be one of the constraints on phase space and we must first understand the reduced classical phase space in \S\ref{sec:Dirac}. 
We describe the solution to the Maxwell equations in \S\ref{sec:MaxwellEOM} since it is rather non-trivial to obtain. 
These two subsections are rather detailed because our target is to also give a complete discussion on Maxwell quantization in AdS.
Finally, we canonically quantize the gauge field in \S\ref{sec:Maxwellquant} and construct the lowest energy excited state, which is the simplest to consider for the entanglement entropy computation. We explain how such states are related to the $U(1)$ current states in the CFT and provide all necessary expressions associated with such states for computation of the entropy.

\subsection{Reduced phase space: Dirac's method}
\label{sec:Dirac}

The simplest way to construct the reduced phase space of Maxwell theory is to use Dirac's method for analyzing Hamiltonian systems with constraints \cite{Dirac}. Via the construction of the Dirac bracket, this method allows one to obtain the correct phase space variables and Poisson bracket on the constrained phase space without needing to explicitly solve the constraints. For nice reviews of this method, see \cite{Henneaux:1992ig,Hanson:1976cn}.

The full, unconstrained phase space is $2(d+1)$-dimensional with canonical coordinates $\mathfrak{P} = \{\mathbb{A}_{\mu},\pi^{\nu}\}$. The equal-time Poisson brackets of these canonical coordinates are
\begin{equation}\label{eqn:PB}
\{\mathbb{A}_{\mu}(x),\pi^{\nu}(x')\}_{\mathrm{P.B}} = {\delta_{\mu}}^{\nu}\delta^{d}(x-x'),
\end{equation}
where $\delta^{d}(x-x') = \prod_{i=1}^{d}\delta(x_{i}-x_{i}')$. The conjugate momenta to the gauge field $\mathbb{A}_{\mu}$ constructed from the Maxwell Lagrangian appearing in \eqref{eqn:Maxwellaction} are given by
\begin{align}\label{eqn:Aconjmom}
\begin{split}
\pi^{i} &= \frac{\delta \mathcal{L_{\mathrm{Maxwell}}}}{\delta (\partial_{t}\mathbb{A}_{i})} = \sqrt{-g}F^{it}, \qquad i=1,\ldots,d
\\ \pi^{t} &= \frac{\delta \mathcal{L_{\mathrm{Maxwell}}}}{\delta (\partial_{t}\mathbb{A}_{t})} \approx 0.
\end{split}
\end{align}
The condition $\pi^{t} \approx 0$ is a primary constraint on the theory since it follows directly from the Lagrangian. We use the standard notation $\approx 0$ to indicate that a function on phase space is weakly zero, meaning that it vanishes on the primary constraint subspace, but may have non-zero Poisson bracket. 

The canonical Hamiltonian is obtained via a Legendre transformation of the Maxwell Lagrangian \eqref{eqn:Maxwellaction} leading to
\begin{equation}\label{eqn:Hamiltoniancanon_EM}
H_{c} = \int d^{d}x\,\left(\frac{\sqrt{-g}}{4}F_{ij}F^{ij}-\frac{1}{2\sqrt{-g}}g_{tt}\pi_{i}\pi^{i}-\mathbb{A}_{t}\partial_{i}\pi^{i}\right).
\end{equation}
However, at this point, the Hamiltonian is not unique because one can construct the primary Hamitlonian $H_{p}$ from the canonical Hamiltonian by adding the primary constraints of the theory multiplied by a Lagrange multiplier field. In practice, this subtlety will turn out to not affect our subsequent discussion (the Lagrange multiplier will never explicitly appear) and once we gauge-fix, the primary constraints will be set strongly to zero so the two Hamiltonians agree. Nevertheless, for completeness, we will work with the primary Hamiltonian.

The requirement that the primary constraint $\pi^{t} \approx 0$ hold for all times gives the secondary constraint
\begin{equation}\label{eqn:secondaryconstraint}
\partial_{i}\pi^{i} = \{\pi^{t},H_{p}\}_{\mathrm{P.B}} = \partial_{t}\pi^{t} \approx 0,
\end{equation}
which we recognize as the Gauss Law.\footnote{A short calculation shows that $\sqrt{-g}\nabla_{\mu}F^{\mu t}=\partial_{i}\pi^{i}$.} The requirement that this constraint be time independent does not impose any new constraint because $\{\partial_{i}\pi^{i},H_{p}\}_{\mathrm{P.B}} \propto \partial_{i}\partial_{j}\left(\sqrt{-g}F^{ij}\right)=0$. The two constraints
\begin{equation}\label{eqn:1stclassconstr}
\psi_{1} = \pi^{t} \approx 0 \qquad \mathrm{and} \qquad \psi_{2} = \partial_{i}\pi^{i} \approx 0
\end{equation}
are first-class constraints since $\{\psi_{1},\psi_{2}\}_{\mathrm{P.B}} = 0$. These constraints generate gauge transformations on a fixed time slice $\Sigma_{t}$. In particular, for $\psi_{\epsilon} = \int_{\Sigma_{t}} d^{d}v\,\epsilon^{1}(v)\psi_{1}(v)+\epsilon^{2}(v)\psi_{2}(v)$, the equal-time Poisson bracket with $\mathbb{A}_{\mu}$ gives
\begin{equation}
\{\mathbb{A}_{t},\psi_{\epsilon}\}_{\mathrm{P.B}} = \epsilon^{1}, \qquad \{\mathbb{A}_{i},\psi_{\epsilon}\}_{\mathrm{P.B}} = -\partial_{i}\epsilon^{2}
\end{equation}
where, for a general gauge transformation $\mathbb{A}_{\mu} \to \mathbb{A}_{\mu}+\partial_{\mu}\lambda(t,v)$, we identify $\epsilon^{1}(v) = \partial_{t}\lambda(0,v)$ and $\epsilon^{2}(v) = -\lambda(0,v)$.

We will now choose a gauge and this provides the final two constraints on phase space. In the AdS/CFT correspondence, the dual operator in the CFT is the $U(1)$ current $J_{\mu}$ so the natural choice of gauge is radial gauge $\mathbb{A}_{r} = 0$,\footnote{Strictly speaking, this choice of gauge is singular at $r=0$ because vector fields are ill-defined there, but this is simply an artifact of working in global coordinates and it does not lead to any issues in our analysis.} also known as holographic gauge, which can be obtained by the gauge transformation
\begin{equation}\label{eqn:hologauge}
\mathbb{A}_{\mu} \to \mathbb{A}_{\mu}' = \mathbb{A}_{\mu} - \partial_{\mu}\int_{0}^{r} dr'\,\mathbb{A}_{r}(r').
\end{equation}
The choice that this gauge-fixing hold for all times is equivalent to requiring $\pi_{r} = \sqrt{-g}g^{tt}\partial_{r}\mathbb{A}_{t}$ by \eqref{eqn:Aconjmom}. Therefore, we have two gauge-fixing constraints
\begin{equation}\label{eqn:gaugeconstraint}
\psi_{3} = \mathbb{A}_{r} \approx 0, \qquad \psi_{4} =  \pi_{r} - \sqrt{-g}g^{tt}\partial_{r}\mathbb{A}_{t} \approx 0.
\end{equation}
The reduced phase space $\mathfrak{C}$ is defined as the submanifold of the full phase space $\mathfrak{P}$ satsifying the constraints $\{\psi_{i}=0\,|\,i=1,\ldots,4\}$. Since the total number of constraints is equal to $4$, the reduced phase space $\mathfrak{C}$ is $2(d-2)$-dimensional, as expected from the number of physical polarizations of Maxwell theory.

There is a residual gauge freedom, leaving both of the gauge-fixing constraints invariant, given by any $r$-independent gauge transformation $\mathbb{A}_{\mu} \to \mathbb{A}_{\mu}+\partial_{\mu}\Lambda(t,\Omega_{d-1})$. These are `large' gauge transformations, in the sense that they do not fall off at the asymptotic boundary, which are fixed by the choice of boundary conditions.\footnote{This means that these gauge transformations do not play a role in the Dirac bracket analysis above.} The extrapolate dictionary for AdS/CFT gives the following boundary conditions on $\mathbb{A}_{\mu}$: 
\begin{equation}\label{eqn:Abdycond}
\lim_{r \to \infty}r^{d-2}\mathbb{A}_{\mu}(t,r,\Omega_{d-1}) \propto J_{\mu}(t,\Omega_{d-1}).
\end{equation}
We can use the residual gauge freedom to fix the boundary condition $\nabla_{\mu}^{\partial}(\lim_{r \to \infty}r^{d-2}\mathbb{A}^{\mu}) = 0$, where $\nabla^{\partial}$ is the boundary covariant derivative, which corresponds to conservation of the dual CFT $U(1)$ current $\nabla^{\partial}_{\mu}J^{\mu}=0$.\footnote{Our wavefunctions will turn out to satisfy this conservation so no residual gauge freedom will be needed.}

The introduction of gauge-fixing constraints means that all constraints are now second-class because $\{\psi_{i},\psi_{j}\}_{\mathrm{P.B}} \neq 0$ for $i \in 1,2$ and $j \in 3,4$. One can explicitly check this, but a more elegant way to see it is to observe that because we have completely gauge-fixed it must be that only the identity gauge transformation preserves the gauge-fixing conditions. So, using that the first-class constraints $\psi_{1,2}$ generate gauge transformations, we find
\begin{equation}\label{eqn:gaugetransgaugefix}
0 = \delta_{\epsilon}\psi_{a}(x) = \{\psi_{a}(x),\psi_{\epsilon}\}_{\mathrm{P.B}} = \int_{\Sigma_{t}}d^{d}u\,\sum_{i=1}^{2}\epsilon^{b}(u)\{\psi_{a}(x),\psi_{b}(u)\}_{\mathrm{P.B}}, \qquad a = 3,4
\end{equation}
implies $\epsilon^{b}(x) = 0$ for $b=1,2$. We conclude that the $2 \times 2$ matrix $(\mathfrak{a}_{ab}) = \{\psi_{a},\psi_{b+2}\}_{\mathrm{P.B}}$ with $a,b\in\{1,2\}$ is invertible, and hence we must introduce the Dirac bracket in order to set these second-class constraints strongly to zero \cite{Blaschke:2020nsd}.

One way to understand the raison d'\^etre of the Dirac bracket is that if one would solve the constraints explicitly to obtain a reduced set of variables, then the Poisson bracket on the reduced phase space would agree with the Dirac bracket on the unconstrained phase space. To construct the Dirac bracket, we first need the matrix of equal-time Poisson brackets of the constraints, which is given explicitly by
\begin{align}\label{eqn:Poissonbracketmatrix}
\begin{split}
C_{ab}(x_{1},x_{2}) &= \{\psi_{a}(x_{1}),\psi_{b}(x_{2})\}_{\mathrm{P.B}} 
\\	&= \left(\begin{matrix} 0 & 0 & 0 & -\sqrt{-g}g^{tt}\big|_{x_{2}}\partial_{r_{1}}\delta^{d}(x_{12}) \\ 0 & 0 & -\partial_{r_{1}}\delta^{d}(x_{12}) & 0 \\ 0 & -\partial_{r_{1}}\delta^{d}(x_{12}) & 0 & g_{rr}\delta^{d}(x_{12}) \\ -\sqrt{-g}g^{tt}\big|_{x_{2}}\partial_{r_{1}}\delta^{d}(x_{12}) & 0 & -g_{rr}\delta^{d}(x_{12}) & 0 \end{matrix}\right),
\end{split}
\end{align}
where $x_{12} = x_{1}-x_{2}$.\footnote{Notice that this matrix is symmetric rather than antisymmetric, which is simply due to the fact the transpose exchanges $a \leftrightarrow b$, but not $x_{1} \leftrightarrow x_{2}$. If we made both swaps, then the matrix would indeed be antisymmetric.} The inverse matrix is defined by
\begin{equation}\label{eqn:Poissonbracketmatrixinvrel}
\int d^{d}x_{2}\, C_{ab}^{-1}(x_{1},x_{2})C_{bc}(x_{2},x_{3}) = \delta_{ac}\delta^{d}(x_{13}),
\end{equation}
where there is an implicit sum over $b$. Observe that $C$ takes the form $\left(\begin{smallmatrix} 0 & Q^{T} \\ Q & P \\ \end{smallmatrix}\right)$ so we find
\begin{equation}\label{eqn:Poissonbracketmatrixinv}
C^{-1} = \left(\begin{matrix} -Q^{-1}P(Q^{T})^{-1} & Q^{-1} \\ (Q^{T})^{-1} & 0 \end{matrix}\right)
\end{equation}
where
\begin{equation}\label{eqn:Ainv}
Q^{-1}(x_{1},x_{2}) = \left(\begin{matrix} 0 & f_{1}(x_{1},x_{2}) \\ f_{2}(x_{1},x_{2}) & 0 \\ \end{matrix}\right)
\end{equation}
with $f_{1}$ and $f_{2}$ obtained by
\begin{align}\label{eqn:Poissonbracketmatrixinv_entries}
\begin{split}
\\ \partial_{r_{1}}\left(\sqrt{-g}g^{tt}\big|_{x_{1}}f_{1}(x_{1},x_{2})\right) &= \delta^{d}(x_{12}) \implies f_{1}(x_{1},x_{2}) = \frac{g_{tt}}{\sqrt{-g}}\bigg|_{x_{1}}\Theta(r_{1}-r_{2})\prod_{i=1}^{d-1}\delta(\phi_{1,i}-\phi_{2,i})
\\ \partial_{r_{2}}f_{2}(x_{1},x_{2}) &= \delta^{d}(x_{12}) \implies f_{2}(x_{1},x_{2}) = -\Theta(r_{1}-r_{2})\prod_{i=1}^{d-1}\delta(\phi_{1,i}-\phi_{2,i}).
\end{split}
\end{align}
We have integrated by parts to determine $f_{1,2}$ with all boundary terms vanishing due to the presence of a delta-function.

We can now use this matrix to construct the equal-time Dirac bracket, with respect to which the constraints are strongly zero, i.e., they have vanishing bracket. For any two functions $\mathscr{P}$, $\mathscr{Q}$ on phase space, the Dirac bracket is defined by
\begin{equation}\label{eqn:Diracbracket}
\{\mathscr{P}(x),\mathscr{Q}(y)\}_{\mathrm{D.B}} = \{\mathscr{P}(x),\mathscr{Q}(y)\}_{\mathrm{P.B}} - \int d^{d}z\,d^{d}w\,\{\mathscr{P}(x),\psi_{a}(z)\}_{\mathrm{P.B}}C_{ab}^{-1}(z,w)\{\psi_{b}(w),\mathscr{Q}(y)\}_{\mathrm{P.B}},
\end{equation}
where there is an implicit sum over $a$ and $b$. The spherical components of the gauge field now obey canonical equal-time brackets
\begin{equation}\label{eqn:Diracbracketcanon}
\{\mathbb{A}_{\phi_{i}}(x),\pi^{\phi_{j}}(y)\}_{\mathrm{D.B}} = {\delta_{i}}^{j}\delta^{d}(x-y), \qquad \{\mathbb{A}_{\phi_{i}}(x),\mathbb{A}_{\phi_{j}}(y)\}_{\mathrm{D.B}} = \{\pi^{\phi_{i}}(x),\pi^{\phi_{j}}(y)\}_{\mathrm{D.B}} = 0,
\end{equation}
where the vanishing of $\{\mathbb{A}_{\phi_{i}}(x),\mathbb{A}_{\phi_{j}}(y)\}_{\mathrm{D.B}}$ follows from the fact that $C_{22}^{-1}=0$ owing to the off-diagonal structure of $Q^{-1}$ and $P$. However, $\mathbb{A}_{t}$ and $\pi^{t}$ do not obey canonical commutation relations since $\{\mathbb{A}_{t}(x),\pi^{t}(y)\}_{\mathrm{D.B}} \neq \delta^{d}(x-y)$ and $\mathbb{A}_{t}$ has non-zero Dirac bracket with other components of $\mathbb{A}_{\mu}$ so we treat them as dependent variables, in particular, we can use the two constraints $\psi_{2}=\psi_{4}=0$ to solve for $\mathbb{A}_{t}$ as a function of the $\mathbb{A}_{\phi_{i}}$:\footnote{Notice that the residual gauge freedom that we used to fix boundary conditions $\nabla^{\partial}_{\mu}\mathbb{A}^{\mu} = 0$ fixes the boundary dynamics of $\mathbb{A}_{t}$ in terms of the boundary values of the other components of $\mathbb{A}_{\mu}$.}
\begin{equation}\label{eqn:A0eqn}
0 = \partial_{i}\pi^{i} = \partial_{r}\left(\sqrt{-g}g^{tt}g^{rr}\partial_{r}\mathbb{A}_{t}\right) + \partial_{\phi_{i}}\left(\sqrt{-g}g^{tt}g^{\phi_{i}\phi_{i}}\left(\partial_{\phi_{i}}\mathbb{A}_{t}-\partial_{t}\mathbb{A}_{\phi_{i}}\right)\right).
\end{equation}

Therefore, the reduced phase space $\mathfrak{C}$ which we canonically quantize has coordinates $(\mathbb{A}^{\phi_{i}},\pi^{\phi_{j}})$ and Dirac bracket \eqref{eqn:Diracbracket}. The equations of motion of the theory are Hamilton's equations for the reduced phase space variables produced by the Dirac brackets\footnote{Now that we have constructed the Dirac bracket, it does not matter whether we use $H_{c}$ or $H_{p}$ so we will use $H_{c}$ for simplicity.}
\begin{equation}\label{eqn:DiracbracketEOM}
\partial_{t}\pi^{\phi_{i}} = \{\pi^{\phi_{i}},H_{c}\}_{\mathrm{D.B}}, \qquad \partial_{t}\mathbb{A}^{\phi_{i}} = \{\mathbb{A}^{\phi_{i}},H_{c}\}_{\mathrm{D.B}}.
\end{equation}
One finds that the equations of motion for $\pi^{\phi_{i}}$ give the Maxwell equations for the spherical directions: $\nabla_{\mu}F^{\mu\phi_{i}}=0$. The equations of motion for $\mathbb{A}^{\phi_{i}}$ simply give the relation between $\mathbb{A}^{\phi_{i}}$ and $\pi^{\phi_{i}}$. 

Finally, we must construct the symplectic form on this phase space. This can be done in the standard way \cite{Lee:1990nz} by obtaining the presymplectic density from the Maxwell Lagrangian, from which the presymplectic form follows via integration, leading to
\begin{equation}\label{eqn:symplecticform}
\Omega_{\mathfrak{P}} = \int_{\Sigma_{t}}d^{d}x\,n_{\mu}\delta \mathbb{A}_{\nu} \wedge \sqrt{-g}\,\delta F^{\mu\nu},
\end{equation}
where $n^{\mu}$ is the normal vector to $\Sigma_{t}$. The presymplectic form $\Omega_{\mathfrak{P}}$ is independent of the choice of Cauchy slice $\Sigma_{t}$ because the presymplectic density is conserved. One can show that this presymplectic form vanishes for any vector tangent to a gauge orbit so it is degenerate, and hence is not a symplectic form. However, the induced form $\Omega_{\mathfrak{C}}$ on the reduced phase space $\mathfrak{C}$ is non-degenerate because there are no gauge variations within this space. So $\Omega_{\mathfrak{C}}$ indeed defines a symplectic form. In practice, we can extract $\Omega_{\mathfrak{C}}$ from $\Omega_{\mathfrak{P}}$ by solving the constraints $\{\psi_{i}=0\,|\,i=1,\ldots,4\}$ for $\{\mathbb{A}_{t},\pi^{t},\mathbb{A}_{r},\pi^{r}\}$ in terms of $\{\mathbb{A}_{\phi_{i}},\pi^{\phi_{i}}\}$, which is what we will do in the next section when we normalize our wavefunctions.

\subsection{Equations of motion}
\label{sec:MaxwellEOM}

We need to solve the equations of motion \eqref{eqn:DiracbracketEOM} and constraints \eqref{eqn:1stclassconstr} for the gauge-fixed theory. This amounts to finding the solutions of the Maxwell equations $\nabla_{\mu}F^{\mu\nu}=0$ for all $\nu$ except $\nu=r$ since $\mathbb{A}_{r}$ is not part of the reduced phase space. The $\nu=t$ component is the Gauss Law given by $\psi_{2} = 0$. These equations are quite non-trivial to solve, but a general formalism was first developed in \cite{Ishibashi:2004wx}, which makes the problem tractable. We now briefly summarize the method, for which we refer the reader there for more details, and then apply the method to our choice of gauge.

The idea is to decompose $\mathbb{A}_{\mu}$ into a vector part and scalar part, each of which can be solved for in terms of a scalar field. Due to the elliptic regularity of any compact Riemannian manifold $\mathcal{M}$, any one-form $\eta_{i}$ on $\mathcal{M}$ has a unique decomposition as
\begin{equation}\label{eqn:vectordecomp}
\eta_{i} = \gamma_{i} + \left(\nabla_{\mathcal{M}}\right)_{i}\zeta, \qquad \nabla_{\mathcal{M}}^{i}\gamma_{i} = 0,
\end{equation}
where $\nabla_{\mathcal{M}}$ is the covariant derivative on $\mathcal{M}$. The case of interest is when $\mathcal{M} = S^{d-1}$ (as this forms part of vacuum AdS) for which the Hilbert space of all $L^{2}$ functions has a basis given by the scalar spherical harmonics $Y_{\mathbf{k}_{S}}^{(d-1)}$ satisfying
\begin{equation}\label{eqn:scalarsphharm}
\left(\nabla_{S^{d-1}}^{2}+k_{S}^{2}\right)Y_{\mathbf{k}_{S}}^{(d-1)} = 0, \qquad k_{S}^{2} = \ell(\ell+d-2), \; \ell = 0,1,2,\ldots
\end{equation}
where $\mathbf{k}_{S} = (\ell,\mathfrak{m})$, and the Hilbert space of all $L^{2}$ vector fields has a basis given by the vector spherical harmonics $\mathbf{Y}_{\phi_{i},\mathbf{k}_{V}}^{(d-1)}$ satisfying
\begin{equation}\label{eqn:vecsphharm}
\left(\nabla_{S^{d-1}}^{2}+k_{V}^{2}\right)\mathbf{Y}_{\phi_{i},\mathbf{k}_{V}}^{(d-1)} = 0, \qquad \nabla_{S^{d-1}}^{\phi_{i}}\mathbf{Y}_{\phi_{i},\mathbf{k}_{V}}^{(d-1)} = 0, \qquad k_{V}^{2} = \ell(\ell+d-2)-1, \; \ell = 1,2,\ldots
\end{equation}
Observe that the Laplacian is invariant under rotations so each eigenspace of the Laplacian forms a representation of $SO(d)$ and the divergence-free condition on vector fields is also invariant under rotations so divergence-free vector fields also form a representation of $SO(d)$. One can show that these representations are independent. Since the vacuum Maxwell equations are rotationally invariant\footnote{In $d=3$, one must use parity to argue that the vector and scalar parts form independent representations, but the Maxwell equations are also parity invariant so the argument still holds.}, the vector and scalar parts must decouple in these equations so we can solve separately for each of them.

This leads to following decomposition of the gauge field under $SO(d)$ representations:
\begin{equation}\label{eqn:gaugefielddecomp}
\mathbb{A}_{\mu} = \mathbb{A}_{\mu}^{V} + \mathbb{A}_{\mu}^{S}
\end{equation}
where, using $i,j,k,\ldots$ to label spherical components and $a,b,c,\ldots$ to label components along $(t,r)$, we have the expansions
\begin{equation}\label{eqn:gaugefieldcomps}
\mathbb{A}_{\mu}^{V}\,dx^{\mu} = \sum_{\mathbf{k}_{V}}\mathbb{A}_{\mathbf{k}_{V}}\mathbf{Y}_{\phi_{i},\mathbf{k}_{V}}^{(d-1)}\,d\phi^{i}, \qquad \mathbb{A}_{\mu}^{S}\,dx^{\mu} = \sum_{\mathbf{k}_{S}}\mathbb{A}_{a,\mathbf{k}_{S}}Y_{\mathbf{k}_{S}}^{(d-1)}\,dx^{a}+\mathbb{A}_{\mathbf{k}_{S}}\left(\nabla_{S^{d-1}}\right)_{\phi_{i}}Y_{\mathbf{k}_{S}}^{(d-1)}\,d\phi^{i}.
\end{equation}
The $U(1)$ gauge transformations $\mathbb{A}_{\mu} \to \mathbb{A}_{\mu} + \nabla_{\mu}\lambda$ only affect $\mathbb{A}_{\mu}^{S}$ so $\mathbb{A}_{\mu}^{V}$ is gauge-invariant. We will now focus only on the scalar part of the gauge field $\mathbb{A}_{\mu}^{S}$ as the lowest energy excited state generated by this part will correspond to our CFT excited state $\ket{\epsilon \cdot J}$, although one can also solve the equations of motion for the vector part in a similar way. Notice that only the modes with $\ell > 0$ appear in the expansion of the gauge field \eqref{eqn:gaugefieldcomps} so all photon modes have non-zero angular momentum.

By manipulating the spherical $\nu = \phi_{i}$ Maxwell equations, one finds that the scalar modes can be related to a single scalar field $\phi_{S}$ by the following set of equations\footnote{There is a typo in \cite{Ishibashi:2004wx} which we have corrected in the second equation of \eqref{eqn:phiSdifeq}.}
\begin{align}\label{eqn:phiSdifeq}
\begin{split}
F_{\mathbf{k}_{S},ab}^{S} &= \widehat{\nabla}_{c}\left(\frac{1}{r^{d-3}}\widehat{\nabla}^{c}\phi_{S,\mathbf{k}_{S}}\right)\epsilon_{ab}
\\ \epsilon_{ab}r^{d-3}\left(\widehat{\nabla}^{b}\mathbb{A}_{\mathbf{k}_{S}}-\mathbb{A}_{\mathbf{k}_{S}}^{b}\right) &= \widehat{\nabla}_{a}\phi_{S,\mathbf{k}_{S}}
\end{split}
\end{align}
where we have expanded $\phi_{S}$ in spherical harmonics, and $\widehat{\nabla}_{a}$ and $\epsilon_{ab}$ are the covariant derivative and Levi-Civita tensor on the $2$-dimensional part of the geometry with coordinates $(t,r)$, respectively. Furthermore, we can use the constraints $\psi_{2}=\psi_{4}=0$ to obtain
\begin{equation}\label{eqn:constrimplication}
\sqrt{g_{S^{d-1}}}\partial_{r}\left(r^{d-1}\partial_{r}\mathbb{A}_{t}\right) + \frac{r^{d-1}}{(r^{2}+1)}\partial_{\phi_{i}}\left(\sqrt{g_{S^{d-1}}}g^{\phi_{i}\phi_{i}}\left(\partial_{\phi_{i}}\mathbb{A}_{t}-\partial_{t}\mathbb{A}_{\phi_{i}}\right)\right) = 0
\end{equation}
which implies 
\begin{equation}\label{eqn:constraintA_t}
\partial_{r}\left(r^{d-1}\partial_{r}\mathbb{A}_{t,\mathbf{k}_{S}}\right)+\frac{r^{d-3}}{(r^{2}+1)}k_{S}^{2}\left(\partial_{t}\mathbb{A}_{\mathbf{k}_{S}}-\mathbb{A}_{\mathbf{k}_{S},t}\right) = 0.
\end{equation}
We plug in \eqref{eqn:phiSdifeq} to obtain a simple differential equation for $\phi_{S}$:
\begin{equation}\label{eqn:phiSgaugefixeqn}
\partial_{r}\left[r^{d-1}\widehat{\nabla}_{d}\left(\frac{1}{r^{d-3}}\widehat{\nabla}^{d}\phi_{S,\mathbf{k}_{S}}\right)-k_{S}^{2}\phi_{S,\mathbf{k}_{S}}\right] = 0 \implies r^{d-1}\widehat{\nabla}_{d}\left(\frac{1}{r^{d-3}}\widehat{\nabla}^{d}\phi_{S,\mathbf{k}_{S}}\right)-k_{S}^{2}\phi_{S,\mathbf{k}_{S}} = \mathfrak{F}(t)
\end{equation}
where $\mathfrak{F}_{\mathbf{k}_{S}}(t)$ is an arbitrary function of $t$. We now use separation of variables
\begin{equation}\label{eqn:PhiSsepofvars}
\phi_{S,n,\mathbf{k}_{S}}(t,r) = e^{-i\Omega_{n,\mathbf{k}_{S}}^{S}t}f_{n,\mathbf{k}_{S}}^{S}(r)
\end{equation}
which allows us to move all $t$-dependence in \eqref{eqn:phiSgaugefixeqn} to the righthand side and all $r$-dependence to the lefthand side so each side must be equal to a constant $\mathfrak{F}_{n,\mathbf{k}_{S}}$. Finally, we perform the rescaling $\Phi_{S,\mathbf{k}_{S}} = r^{-\frac{(d-3)}{2}}\phi_{S,\mathbf{k}_{S}}$ to arrive at
\begin{equation}\label{eqn:Maxwellscalarfinal}
\widehat{\nabla}_{a}\widehat{\nabla}^{a}\Phi_{S,\mathbf{k}_{S}}-\left(\frac{(d-1)(d-3)}{4}+k_{S}^{2}\right)\frac{1}{r^{2}}\Phi_{S,\mathbf{k}_{S}}-\frac{(d-3)(d-5)}{4}\Phi_{S,\mathbf{k}_{S}} = \mathfrak{F}_{n,\mathbf{k}_{S}}.
\end{equation}
This is a second-order inhomogeneous differential equation. We can first solve the homogeneous equation ($\mathfrak{F}_{n,\mathbf{k}_{S}}=0$) and then obtain the solution for the inhomogeneous equation in the standard way. One of the solutions to the homogeneous equation makes $\mathbb{A}_{n,\mathbf{k}_{S}}$ not regular at the origin ($r=0$) and the same is true for the part of solution coming from the inhomogeneity, leaving us with the final result
\begin{equation}
f_{n,\mathbf{k}_{S}}^{S}(r) \propto y^{\frac{(d-2+\ell)}{2}}\,{}_{2}{F}_{1}\left(\frac{d-2+\ell+\Omega_{n,\mathbf{k}_{S}}^{S}}{2},\frac{d-2+\ell-\Omega_{n,\mathbf{k}_{S}}^{S}}{2},\ell+\frac{d}{2};y\right), \quad y \equiv \frac{r^{2}}{r^{2}+1}.
\end{equation}
where we have defined the new coordinate $y$ to simplify some expressions. We can now extract the gauge field using the $a=t$ component of the second equation in \eqref{eqn:phiSdifeq}, viz.,
\begin{equation}\label{eqn:fStoA}
\partial_{r}\mathbb{A}_{n,\mathbf{k}_{S}} = \frac{(r^{2}+1)}{r^{d-3}}\partial_{t}\phi_{S,n,\mathbf{k}_{S}}.
\end{equation}
Before we integrate this up to obtain the explicit form of $\mathbb{A}_{n,\mathbf{k}_{S}}$, let us pause to observe that the normalizability of $\mathbb{A}_{\mu}$ can be understood from the large $r$ behavior of $f_{n,\mathbf{k}_{S}}^{S}$ from which we infer that the frequencies must be quantized, taking the values
\begin{equation}\label{eqn:AscalE}
\Omega_{n,\mathbf{k}_{S}}^{S} = d-2+2n+\ell, \qquad \ell = 1,2,\ldots, \; n = 0,1,2,\ldots.
\end{equation}
We now use \eqref{eqn:fStoA} to obtain the wavefunctions
\begin{equation}\label{eqn:ASfinal}
\mathbb{A}_{n,\mathbf{k}_{S}}(t,r) = \mathcal{N}_{n,\ell}^{S}e^{-i\Omega_{n,\mathbf{k}_{S}}^{S}t}\left(y^{\frac{\ell}{2}+1}{}_{3}{F}_{2}\left(2-\frac{d}{2}-n,\ell+\frac{d}{2}+n,\frac{\ell}{2}+1;\ell+\frac{d}{2},\frac{\ell}{2}+2;y\right)+\mathfrak{A}_{n,\ell}\right),
\end{equation}
where the constant of integration $\mathfrak{A}_{n,\ell}$ is fixed by ensuring that these wavefunctions satisfy the boundary conditions determined by the extrapolate dictionary \eqref{eqn:Abdycond}, which gives
\begin{equation}\label{eqn:intconst}
\mathfrak{A}_{n,\ell} = -\frac{\Gamma\left(\frac{\ell}{2}+2\right)\left(\frac{\ell+d-2}{2}\right)_{n}\Gamma\left(\frac{d}{2}+n-1\right)}{\left(\ell+\frac{d}{2}\right)_{n}\Gamma\left(\frac{d}{2}+\frac{\ell}{2}+n\right)}.
\end{equation}
Using the $a=r$ component of the second equation in \eqref{eqn:phiSdifeq}, we can also find $\mathbb{A}_{t}$:
\begin{equation}\label{eqn:At}
\mathbb{A}_{t,n,\mathbf{k}_{S}}(t,r) = -i\Omega_{n,\mathbf{k}_{S}}^{S}\mathbb{A}_{n,\mathbf{k}_{S}}(t,r)-\frac{i}{\Omega_{n,\mathbf{k}_{S}}^{S}}\frac{1}{r^{d-3}(r^{2}+1)}\,\partial_{r}\left(\frac{r^{d-3}}{(r^{2}+1)}\,\partial_{r}\mathbb{A}_{n,\mathbf{k}_{S}}(t,r)\right).
\end{equation}
The final step is to normalize the gauge field modes to have unit norm with respect to the generalized Klein-Gordon inner product, which can be obtained from the presymplectic form $\Omega_{\mathfrak{P}}$ \eqref{eqn:symplecticform} (projected onto $\mathfrak{C}$ to obtain $\Omega_{\mathfrak{C}}$), leading to
\begin{equation}\label{eqn:MaxIP}
\left<B,C\right> = -i\Omega_{\mathfrak{C}}(B^{\ast},C) = -i\int_{\Sigma_{t}}d^{d}x\,\sqrt{-g}\left[B_{\mu}^{\ast}(\nabla^{t}C^{\mu}-\nabla^{\mu}C^{t})-C_{\mu}(\nabla^{t}B^{\ast\mu}-\nabla^{\mu}B^{\ast t})\right]
\end{equation}
where the overall sign is fixed such that the inner product is positive-definite on positive frequency solutions. Notice that the vector and scalar parts of the gauge field are orthogonal with respect to this inner product. We find the normalization constant for the scalar modes\footnote{One may worry that $\mathcal{N}_{n,\mathbf{k}_{S}}^{S}$ has a pole for $d$ even coming from the $\Gamma$-function out front, but this pole is cancelled by a corresponding pole coming from $\mathfrak{S}$.}
\begin{equation}\label{eqn:Anorm}
\mathcal{N}_{n,\ell}^{S} = \frac{\Gamma\left(1-\ell-\frac{d}{2}\right)}{(\ell+2)}\left[\frac{(d-2+2n+\ell)\Gamma(d-1+2n+\ell)}{V_{S^{d-1}}\ell(\ell+d-2)\mathfrak{S}}\right]^{\frac{1}{2}},
\end{equation}
where $\mathfrak{S}$ is the following finite double-sum
\begin{align}
\begin{split}
\mathfrak{S} &= \sum_{k,m=0}^{n}\Bigg[(-1)^{k+m}\Gamma\left(1-\ell-\frac{d}{2}-k\right)\Gamma\left(1-\ell-\frac{d}{2}-m\right)\left(2-\frac{d}{2}-n\right)_{k}\left(2-\frac{d}{2}-n\right)_{m}
\\	&\qquad \times {n \choose k}{n \choose m}\Gamma\left(\frac{d}{2}+k+m+\ell\right)\Gamma\left(\frac{d}{2}+2n-k-m-1\right)\Bigg].
\end{split}
\end{align}
This completes our solution of the equations of motions so we can now canonically quantize this classical theory.

\subsection{Construction of lowest excited states}
\label{sec:Maxwellquant}

With the solution to the equations of motion, it is straightforward to quantize the gauge field by defining creation/annihilation operators, which are needed to define excited states in Fock space, and to compute their commutation relations.
Since the vector and scalar modes of the gauge field decouple, the Hilbert space becomes a tensor product $\mathbb{H} = \mathbb{H}^{S} \otimes \mathbb{H}^{V}$ for the two sets of modes. The gauge field is expanded in annihilation and creation operators for these two Fock spaces
\begin{equation}\label{eqn:gaugefieldexp}
\hat{\mathbb{A}}_{\mu}(x) = \sum_{n,\mathbf{k}_{V}}\mathbb{A}_{\mu,n,\mathbf{k}_{V}}^{V}(x)a_{n,\mathbf{k}_{V}}^{V} + \sum_{n,\mathbf{k}_{S}}\mathbb{A}_{\mu,n,\mathbf{k}_{S}}^{S}(x)a_{n,\mathbf{k}_{S}}^{S} + \mathrm{h.c.}.
\end{equation}
The unit normalization of the wavefunctions with respect to the Klein-Gordon inner product along with the commutation relation $[\mathbb{A}_{\phi_{i}}(x),\pi^{\phi_{j}}(y)] = i\delta_{i}^{j}\delta(x-y)$ coming from the Dirac bracket \eqref{eqn:Diracbracketcanon} implies that these operators satisfy the following commutation relations
\begin{equation}\label{eqn:Maxladderops_CCR}
[a_{n,\mathbf{k}_{S}}^{S},a_{n',\mathbf{k}_{S}'}^{S\dagger}] = \delta_{nn'}\delta_{\mathbf{k}_{S},\mathbf{k}_{S}'}, \qquad [a_{n,\mathbf{k}_{V}}^{V},a_{n',\mathbf{k}_{V}'}^{V\dagger}] = \delta_{nn'}\delta_{\mathbf{k}_{V},\mathbf{k}_{V}'}
\end{equation}
with all other commutators equal zero.

We need to understand the relationship between the primary states $\ket{\epsilon \cdot J}$ that we studied in the CFT and these Fock space states, which was provided in \cite{Terashima:2017gmc}. All excitations of the scalar modes for $\ell > 1$ or $\ell=1$, $n>0$ and all the vector modes correspond to descendant states in the CFT so we focus on the scalar modes for $\ell=1$, $n=0$ denoted by
\begin{equation}\label{eqn:photonstate}
\ket{\mathcal{P}_{\mathfrak{m}}} \equiv a_{0,1,\mathfrak{m}}^{S\dagger}\ket{0}.
\end{equation}
The easiest way to find the dual state in the CFT is to use the extrapolate dictionary \eqref{eqn:Abdycond}. Recall that 
\begin{equation}\label{eqn:CFTstate}
\ket{\epsilon \cdot J} = \lim_{t_{E} \to -\infty}e^{(d-1)t_{E}}\epsilon^{\mu}J_{\mu}(t_{E},\Omega)\ket{0}
\end{equation}
so to extract a given $\mathfrak{m}$ we must choose $\epsilon_{\mu}$ appropriately. To do this, we write the spherical harmonics as 
\begin{equation}\label{eqn:sphericalharmonicRd}
Y_{1,\mathfrak{m}}^{(d-1)} = s_{(1,\mathfrak{m})}^{\mu}\frac{x_{\mu}}{|x|}, \qquad x^{\mu} \in \mathbb{R}^{d}  
\end{equation}
where $s_{(1,\mathfrak{m})}^{\mu}$ is a complex vector. Using orthogonality of spherical harmonics (or by direct verification), one can show that 
\begin{equation}\label{eqn:svecrelation}
s_{(1,\mathfrak{m})}^{\mu\ast}\partial_{\mu}\left(s_{(1,\mathfrak{m}')}^{\nu}x_{\nu}\right) \propto \delta_{\mathfrak{m}\mathfrak{m}'}.
\end{equation}
Observe that 
\begin{equation}\label{eqn:largerAvec}
\partial_{\mu}\left(s_{(1,\mathfrak{m}')}^{\nu}x_{\nu}\right) = \partial_{\mu}(|x|Y_{1,\mathfrak{m}'}^{(d-1)}) = \left(Y_{1,\mathfrak{m}'}^{(d-1)},\partial_{\phi_{i}}Y_{1,\mathfrak{m}'}^{(d-1)}\right) \propto \lim_{r \to \infty}r^{d-2}\mathbb{A}_{\mu,0,1,\mathfrak{m}'}^{S}
\end{equation}
so the desired choice for our polarization vector to extract a given $\mathfrak{m}$ is
\begin{equation}\label{eqn:epsilonlm}
\epsilon_{(1,\mathfrak{m})}^{\mu} = s_{(1,\mathfrak{m})}^{\mu\ast}.
\end{equation}
With this choice, after Wick rotation of the gauge field $A_{\mu}$ to Euclidean signature $t=i\tau_{E}$, we find
\begin{equation}\label{eqn:bulktobdystate}
\ket{\epsilon_{(1,\mathfrak{m})} \cdot J} = \lim_{r \to \infty}r^{d-2}\lim_{\tau_{E} \to -\infty}e^{(d-1)\tau_{E}}\epsilon_{(1,\mathfrak{m})}^{\mu}\mathbb{A}_{\mu}(\tau_{E},r,\Omega)\ket{0} \propto a_{0,1,\mathfrak{m}}^{S\dagger}\ket{0},
\end{equation}
where the proportionality constant is not important since we always normalize our density matrix.

These are the lowest energy states in the bulk Fock space $\mathbb{H}$ with energy above the vacuum $\Delta E \equiv \bra{\mathcal{P}_{\mathfrak{m}}}H\ket{\mathcal{P}_{\mathfrak{m}}}-\bra{0}H\ket{0}  = d-1$. From \eqref{eqn:ASfinal}, we find the wavefunction for these states to be
\begin{equation}\label{eqn:lowestenergywavefn}
\begin{aligned}
\mathbb{A}^{S}_{t,0,1,\mathfrak{m}}(t,r,\Omega)&=- i \mathcal{N}_{0,1}^{S}e^{-i\Omega_{0,1}^{S}t} \left( \Omega^S_{0,1} \mathcal{A}^{S}_{0,1}(r)+\frac{3 r}{(1+r^2)^{\frac{d-1}{2}}}\right)Y_{1,\mathfrak{m}}^{(d-1)}(\Omega) \\
\mathbb{A}_{\phi_{i},0,1,\mathfrak{m}}^{S}(t,r,\Omega) & = 
\mathcal{N}_{0,1}^{S}e^{-i\Omega_{0,1}^{S}t}\mathcal{A}^{S}_{0,1}(r) \partial_{\phi_{i}}Y_{1,\mathfrak{m}}^{(d-1)}(\Omega),
\end{aligned}
\end{equation}
where
\begin{equation}
\mathcal{A}_{0,1}(r)=\left(\frac{r^2}{1+r^2}\right)^{\frac{3}{2}}\,{}_{2}{F}_{1}\left(2-\frac{d}{2},\frac{3}{2},\frac{5}{2};\frac{r^2}{1+r^2}\right)-\frac{3\sqrt{\pi}\Gamma\left(\frac{d}{2}-1\right)}{4\Gamma\left(\frac{d+1}{2}\right)},
\end{equation}
and from \eqref{eqn:Anorm} we find the normalization
\begin{equation}\label{eqn:Anorm1}
\mathcal{N}_{0,1}^{S} = \frac{1}{3}\sqrt{\frac{\Gamma(d)}{V_{S^{d-1}}\Gamma\left(\frac{d}{2}-1\right)\Gamma\left(\frac{d}{2}+1\right)}}.
\end{equation}
Moreover, the field strength which will be needed in the backreaction computation is given by
\begin{equation}
    \Big(F^S_{tr,0,1,\mathfrak{m}}, F^S_{t\phi_i,0,1,\mathfrak{m}}, F^S_{r\phi_i,0,1,\mathfrak{m}}\Big)= 3 \mathcal{N}_{0,1}^S \frac{e^{- i \Omega^S_{0,1} t} }{(r^2+1)^{\frac{d-1}{2}}} \left({-}i Y^{(d-1)}_{1,\mathfrak{m}}, i r  \partial_{\phi_i}Y^{(d-1)}_{1,\mathfrak{m}},  \frac{r^2 \partial_{\phi_i}Y_{1,\mathfrak{m}}^{(d-1)}}{r^2+1}\right). 
\end{equation}
These are the states, in particular the $\mathfrak{m}=0$ one, whose generalized entropy we shall compute in the next section.

\section{Generalized entropy of photon in AdS$_{d+1}$}
\label{sec:Sgenphoton}

We now present the calculation of the vacuum-subtracted generalized entropy for a single-photon excited state. 
In  \S\ref{sec:photonbackreact}, we solve for the backreacted geometry for this photon state, which is much more complicated compared with the scalar case so we relegate the explicit details to App.~\ref{sec:photonbackreact_details}. 
Then in \S\ref{sec:photonarea}, we compute the area correction from the backreaction and address an important subtlety which is not present in the scalar case. 

The bulk entanglement entropy calculation, to the first and second order in $\delta \rho$ obtained in \S\ref{sec:photonEE}, requires again the AdS-Rindler quantization (in Rindler radial gauge) and the computation of the Bogoliubov coefficients.
Interestingly, we find that even \textbf{without} considering the possible edge mode contribution, the bulk generalized entropy matches the boundary entanglement entropy, namely
\begin{equation}
     \Delta S^{\rm CFT}  =  \frac{\Delta A_{\cal P}}{4 G_N} + \Delta S^{\rm AdS}_{\rm bulk}.
\end{equation}
We will complete the story by showing that the edge mode entropy vanishes in \S\ref{sec:edge}.

We would like to highlight some new features compared to the scalar case in \S\ref{sec:Sgenscalar} before diving into the details. 
Firstly, since the photon is spin-1, it must carry non-zero angular momentum. We will see that this complicates the backreaction calculation, and one has to be very careful in regulating infrared divergences in the area to find the correct area change, even though the final answer is finite. 
Secondly, the fact that the transverse direction is the non-compact $H^{d-1}$ will make it non-trivial to decompose the gauge field into ``scalar" and ``vector" pieces as we did in \S\ref{sec:MaxwellEOM}. 
Thirdly, due to the existence of gauge symmetry, a gauge-invariant way of computing the Bogoliubov coefficients is required.
Finally, there is a complication regarding the gauge constraints and boundary conditions on the Rindler horizon, but this problem will be addressed in \S\ref{sec:edge} so we omit it here.

\subsection{Backreaction of photon}
\label{sec:photonbackreact}
We consider the simplest photon excited state $|\mathcal{P}\rangle \equiv \ket{\mathcal{P}_{\mathbf{0}}}=a_{0,1,\mathbf{0}}^{\dagger}|0\rangle_{\rm{AdS}_{d+1}}$, with angular momentum $\ell=1$ and zero magnetic quantum number.
The angular dependence of the wavefunction for such a state is given by\footnote{We normalize the spherical harmonics so that $\int d\Omega_{d-1}\,\sqrt{g_{S^{d-1}}}\, Y_{\ell,\mathfrak{m}}^{\ast}(\Omega)Y_{\ell',\mathfrak{m}'}(\Omega) = \delta_{\ell,\ell'}\delta_{\mathfrak{m},\mathfrak{m}'}V_{S^{d-1}}$.}
\begin{equation}\label{eqn:sphharm_m=0}
Y_{1,\mathbf{0}}(\Omega) = \sqrt{d}\cos\theta.
\end{equation}
As a result, the Maxwell stress-tensor
\begin{equation}\label{eqn:stresstensor_Max}
T_{\mu\nu}^{\mathrm{Max.}} = -\frac{2}{\sqrt{-g}}\frac{\delta S_{\mathrm{Maxwell}}}{\delta g^{\mu\nu}} = {F_{\mu}}^{\beta}F_{\nu\beta}-\frac{1}{4}g_{\mu\nu}F_{\alpha\beta}F^{\alpha\beta} 
\end{equation}
has the following expectation value in the state $|\mathcal{P}\rangle$ 
\begin{align}\label{eqn:stresstensor_Max_evm=0}
\begin{split}
\bra{\mathcal{P}}T_{tt}^{\mathrm{Max.}}\ket{\mathcal{P}} &= 9d|\mathcal{N}_{0,1}^{S}|^{2}\frac{1}{(r^{2}+1)^{d-2}}
\\ \bra{\mathcal{P}}T_{rr}^{\mathrm{Max.}}\ket{\mathcal{P}} &= -9d|\mathcal{N}_{0,1}^{S}|^{2}\frac{1}{(r^{2}+1)^{d}}\cos(2\theta)
\\ \bra{\mathcal{P}}T_{r\theta}^{\mathrm{Max.}}\ket{\mathcal{P}} &= 9d|\mathcal{N}_{0,1}^{S}|^{2}\frac{r}{(r^{2}+1)^{d}}\sin(2\theta)
\\ \bra{\mathcal{P}}T_{\theta\theta}^{\mathrm{Max.}}\ket{\mathcal{P}} &= 9d|\mathcal{N}_{0,1}^{S}|^{2}\frac{r^{2}}{(r^{2}+1)^{d}}\left(r^{2}+\cos(2\theta)\right)
\\ \bra{\mathcal{P}}T_{\phi_{i}\phi_{i}}^{\mathrm{Max.}}\ket{\mathcal{P}} &= 9d|\mathcal{N}_{0,1}^{S}|^{2}\frac{1}{(r^{2}+1)^{d}}\left(r^{2}\cos(2\theta)+1\right)g_{\phi_{i}\phi_{i}}, \quad 2 \leq i \leq d-1.
\end{split}
\end{align}
With the expectation values of the stress-tensor in hand, we can now solve the Einstein equations
\begin{equation}\label{eqn:Einsteineqns_Maxwell}
R_{\mu\nu}-\frac{1}{2}Rg_{\mu\nu}+\Lambda g_{\mu\nu} = 8\pi G_{N}\bra{\mathcal{P}}T_{\mu\nu}^{\mathrm{Max.}}\ket{\mathcal{P}}.
\end{equation}
Inspecting the Einstein equations, one notices two complications due to the state having non-trivial spin which did not occur in the scalar case. Firstly, the stress-tensor, and hence also the metric, has $\theta$ dependence so the equations are now second-order PDEs rather than the second-order ODEs found for the scalar backreaction. In fact, it is rather remarkable that a simple analytic solution even exists for these PDEs, as we shall see. Secondly, the stress-tensor now has off-diagonal components.

We make the following metric ansatz, inspired by the form of the stress-tensor expectation values \eqref{eqn:stresstensor_Max_evm=0},
\begin{equation}\label{eqn:Maxwellbackreact_ansatz}
\begin{aligned}
    \widetilde{ds}_{M}^{2} & = (ds^{(0)})^{2} + G_N ds_{M}^{2,(1)}\\
    (ds_{M}^{(1)})^{2} & = - g_{tt}^{(1)} dt^2 + 2g_{tr}^{(1)} dt\,dr + g_{tt}^{(1)} dr^{2} + 2g_{t\theta}dt\,d\theta + 2g_{r\theta}dr\,d\theta + g_{\theta\theta}^{(1)} d\Omega_{d-1},
\end{aligned}
\end{equation}
where the vacuum AdS metric $(ds^{(0)})^{2}$ is given in \eqref{eqn:AdSmetric} and we make a further ansatz for the diagonal components\footnote{The $\theta$ dependence of the ansatz can be justified as follows: due to the linearality of the Einstein equations at leading order, we can assume $ g^{(1)}$ has $\theta$ dependence $\sin 2\theta$ or $\cos 2\theta$. Moreover, the $\theta\leftrightarrow \pi-\theta$ symmetry of the background geometry and the stress tensor prohibits the $\sin 2\theta$ term. There is indeed a $\sin 2\theta$ term in $T_{r\theta}$, but it is expected to change sign when we do reflection $\theta\leftrightarrow \pi-\theta$. The metric components, on the other hand, are all diagonal components, so they should only have a $\cos 2\theta$ piece.}
\begin{equation}\label{eqn:metricansatz_maxwell}
\begin{aligned}
    g_{tt}^{(1)}(r,\theta)&= (r^2+1)\left(\mathfrak{g}_{1,0}(r) + \mathfrak{g}_{1,2}(r) \cos(2\theta) \right) \\
    g_{rr}^{(1)}(r,\theta) &=- \frac{1}{r^2+1}\left(\mathfrak{g}_{2,0}(r) + \mathfrak{g}_{2,2}(r) \cos(2\theta) \right) \\
   g_{\theta\theta}^{(1)}(r,\theta)&=r^2\left(\mathfrak{g}_{3,0}(r) + \mathfrak{g}_{3,2}(r) \cos(2\theta) \right).
\end{aligned}
\end{equation}
We parametrize $g_{tr}$, $g_{r\theta}$, and $g_{r\theta}$ similarly, which are zero at $\mathcal{O}(G_{N}^{0})$, but can be non-zero at $\mathcal{O}(G_{N}^{1})$.  

We solve the Einstein equations explicitly in App. \ref{sec:photonbackreact_details} and we find a solution such that the off-diagonal elements are equal to zero, and the diagonal elements in general take the form 
\begin{equation}\label{eqn:gij}
    \mathfrak{g}_{i,j} = (-1)^{d-1}\frac{2}{V_{S^{d-2}}(d-1)^2}\left\{\begin{aligned}
        &\frac{p_{i,j}(r^2)}{(1+r^2)^d}   & \text{if $d$ is even} \\
        &\frac{p_{i,j}(r^2)}{r^{d-1}(1+r^{2})^d} -  \frac{q_{i,j}(r^2)\tan^{-1}(r)}{r^d(1+r^{2})}   & \text{if $d$ is odd}
    \end{aligned} \right.
\end{equation}
where the $p$s and $q$s are polynomials. While a general integral representation of the solutions is given in \eqref{eqn:g32odd} and \eqref{eqn:g32even}, we cannot perform the integral as a general function of $d$, but for any given $d$ it is straightforward to evaluate. 
We list the results in various dimensions in Table \ref{tab:backreactphotonmetric}.
\begin{table}[h!]
\begin{center}
\rowcolors{2}{cyan!40!white!50}{cyan!20!white!40}
      $d$ \ {odd} \\
      \vspace{1.5mm}
     \begin{tabular}{|c|c|c|c|c}
     \hline
       \rowcolor{yellow!40!white!50} $d$  & 3 & 5  \\
       \hline
        $p_{1,0}$ & $2(1 + 10 r^2 + 13  r^4)$ & $27 + 336  r^2 + 386  r^4 + 128  r^6 - 237  r^8$ \\ 
        $p_{1,2}$ & $ 2(3 + 11 r^2 + 21 r^4 + 29 r^6)$ & $45 + 225  r^2 + 454  r^4 - 686  r^6 + 1933  r^8 - 15  r^{10}$  \\ 
         $p_{2,0}$ & $2(3 + 26 r^2 + 43 r^4)$ & $ 45 + 420  r^2 + 1154  r^4 + 3116  r^6 - 255  r^8$ \\ 
        $p_{2,2}$ & $2(3 + 11 r^2 - 11 r^4 + 61 r^6)$ & $15  + 115  r^2 + 338  r^4 - 1478  r^6 + 12063  r^8 - 45  r^{10}$  \\ 
        $p_{3,2}$ & $2(-3 - 2 r^2 + 29 r^4)(1+r^{2})$ & $(-15 - 40  r^2 - 18  r^4 + 2008  r^6 - 15  r^8)(1+r^{2})$ \\ 
        $q_{1,0}$ & $2(1 + 19 r^2)$ & $3(9+79r^2)$ \\
        $q_{1,2}$ & $ 6  (1 + r^2)^2$ & $15(3+r^2)(1 + r^2)$   \\
         $q_{2,0}$ & $6  (1 + 7  r^2)$ & $15(3 + 17  r^2)$  \\
        $q_{2,2}$ & $ 6 (1 + r^2)^2$ & $15(1+3r^2)(1 + r^2)$ \\
         $q_{3,2}$  & $6 (r^2-1)(1+r^{2})$  & $15(r^2-1)(1+r^{2})$ \\
         \hline
    \end{tabular}\\
    \vspace{3mm}
    $d$ \ {even} \\
    \vspace{1.5mm}
    \begin{tabular}{|c|c|c|}
        \hline
       \rowcolor{yellow!40!white!50} $d$  & 4  & 6 \\
       \hline
        $p_{1,0}$ & $16 (9 + 16 r^2 + 11  r^4)$  & $ \frac{64}{3}(100 + 180 r^2 + 159 r^4 + 31 r^6)$  \\ 
         $p_{1,2}$ & $16 r^2 (-4 - 21 r^2 + r^4)$ & $ 32 r^2 (-20 - 211 r^2 + 10 r^4 + r^6)$   \\ 
         $p_{2,0}$ & $ 64 r^{2}(-4 + 3 r^2)$ & $64 r^2 (-130 + 57 r^2 + 11 r^4)$  \\ 
        $p_{2,2}$ & $16 r^2 (25 - 99 r^2 + 2 r^4)$ & $32 r^2 (245 - 1934 r^2 + 25 r^4 + 4 r^6)$  \\ 
        $p_{3,2}$  & $16 r^2 (-25 + r^2)(1+r^{2})$ & $32 r^2 (-245 + 4 r^2 + r^4)(1+r^{2})$  \\
        \hline
    \end{tabular}
\end{center}
\caption{Backreacted metric due to lowest energy photon excited state in $d=3,4,5,6$.}
\label{tab:backreactphotonmetric}
\end{table}

Before moving on to the area correction, it is helpful to analyze the large $r$ behavior of these metric perturbations. 
We notice that these functions $\mathfrak{g}_{i,j}$ in general take the form $r^{-(d-2)}$ for large $r$. 
Such an observation can be understood directly from the (homogeneous) linearized Einstein equation with more details provided in App.~\ref{sec:photonbackreact_details}, in particular see \eqref{eqn:k=2relfin} and the subsequent discussion.

As a nice check of our backreacted metric, we computed the change in the ADM mass $\Delta M$ between the vacuum AdS and backreacted spacetimes as obtained from the Brown-York stress tensor of the respective spacetimes \cite{Balasubramanian:1999re,deHaro:2000vlm} and we found that $\Delta M = d-1$ which is equal to the energy of the dual CFT state $\ket{\epsilon \cdot J}|_{\theta_{\epsilon}=0}$ computed from the CFT stress-tensor. This is the expected result because the backreacted spacetime is dual to the CFT state.\footnote{One can also compute $\Delta M$ for the backreacted metric due to the scalar single-particle state obtained in \S\ref{sec:areacorrection} and reproduce the energy $\Delta_{\mathcal{O}}$ of the dual CFT scalar primary state.}

\subsection{Correction to the area}
\label{sec:photonarea}

We now turn to a detailed discussion of the correction to the area of the classical extremal surface $\gamma_{\mathrm{ext}}$ due to the backreaction of the state $\ket{\mathcal{P}}$. As we will see, there is an important subtlety coming from the regularization of the area, which happens to be absent for the scalar calculation in \S\ref{sec:areacorrection}.

The problem originates from the divergence in the area of $\gamma_{\mathrm{ext}}$. Being anchored to the asymptotic boundary of AdS, the area of this codimension-$2$ surface in both pure AdS $A[g^{(0)}]$ and in the backreacted spacetime $A[g^{(0)}+G_N g^{(1)}]$ is IR-divergent due to the infinite proper distance to the boundary.
As we will see, if we naively take the difference of these two infinite areas, an IR finite answer is obtained. However, the answer is incorrect because subtracting two infinities is ill-defined unless a consistent regularization procedure is implemented.
The proper way to compute the area correction is to regulate the divergence of both areas, and then take the difference, after which the regulator is removed. We present two different methods for achieving such a regularization and prove that they are equivalent. 

We first discuss the regularization using a cut-off surface. In the unperturbed metric $g_{\mu\nu}^{(0)}$, we choose a constant radius cut-off $r=r_c$, and the induced metric on such a surface is 
\begin{equation}\label{eq:cutoffmetric}
    d s_c^2 = -(1+r_c^2) dt^2 + r_c^2 (d\theta^2+\sin^2\theta d\Omega_{d-2}^2).
\end{equation}
We consider this cut-off surface as the surface on which the CFT lives, in particular, it is where we define the boundary conditions for our variational problem for area. In order to compare the area of the classical extremal surface in the vacuum versus backreacted metrics, we must ensure that we have the same boundary subregion $\mathcal{B}$ on which $\gamma_{\mathrm{ext}}$ is anchored. Thus, in the perturbed metric, we should require the same induced metric for the cut-off surface, at least on a fixed time slice which is where we define $\mathcal{B}$. Since the perturbed metric has angular dependence, the new cut-off surface should as well so it must be described by a wiggly surface $r=r(r_c,\theta)$ such that 
\begin{equation}\label{eq:newcutoffmetric}
    ds_c^2=-g_{tt}(r(r_c,\theta),\theta) dt^2 + g_{rr}(r(r_c,\theta),\theta) \left(\partial_{\theta}{r}(r_c,\theta)\right)^2 d\theta^2 + g_{\theta\theta}(r(r_c,\theta),\theta)(d\theta^2+\sin^2\theta d\Omega_{d-2}).
\end{equation}
We require that the new induced metric \eqref{eq:newcutoffmetric} match the old induced metric \eqref{eq:cutoffmetric} up to an error of $\mathcal{O}(G_N^2)$. 
This condition determines the wiggly cut-off surface $r=r(r_c,\theta)$. 

Let us first argue that it is necessary to consider the wiggly cut-off surface because it gives non-vanishing contribution to the area difference between the backreacted and pure AdS metrics. 
As a concrete example, we will start with the area change for the $d=4$ case, where the back-reacted metric is given by
\begin{equation}
\begin{aligned}
    \widetilde{ds}^2_{M} &=  {-} (r^2+1)\left(1 - 8G_N \frac{(11 r^4 +16 r^2 +9)+r^2(r^4 -21 r^2 -4)\cos(2\theta)}{9V_{S^2}(r^2+1)^4}\right) dt^2 \\
    &+ \frac{1}{(r^2+1)}\left(1 - 8G_N \frac{r^2(4(3r^2-4)+(2 r^4 -99 r^2 +25)\cos(2\theta))}{9V_{S^2}(r^2+1)^4}\right)^{-1} dr^2 \\
    & + r^2\left(1 - 8G_N\frac{r^2(r^2-25)\cos(2\theta)}{9V_{S^2}(r^2+1)^3}\right)d\Omega_2^2.
\end{aligned}
\end{equation}
We see that the metric has $\theta$-dependence so if we insist on using the $r=r_c$ cut-off surface, the induced metric on the cut-off surface would be $\theta$-dependent, which is not what we want. 
To obtain the same induced metric \eqref{eq:cutoffmetric} on a fixed time slice of the cut-off surface, we need to make the radius $\theta$-dependent as 
\begin{equation}\label{eq:newcutoff}
    r^2= r_c^2\left(1 + G_N \frac{8 r_c^2(r_c^2-25)\cos(2\theta)}{9V_{S^2}(r_c^2+1)^4} + \mathcal{O}(G_N^2)\right)\approx  r_c^2\left(1 + G_N \frac{8\cos(2\theta)}{9V_{S^2} r_c^2}+\mathcal{O}(G_N^2,r_{c}^{-4})\right) ,
\end{equation}
so that the induced metric, up to $\mathcal{O}(G_N^2)$, is now 
\begin{equation}
    \widetilde{ds}^2_c=(r_c^2+1)\left( 
  1-8G_N\frac{ (9 + 16 r_c^2 + 11 r_c^4) + 4 r_c^2 (-1 + r_c^2) \cos(2 \theta)}{9 V_{S^2}(1 + r_c^2)^{4}}\right) dt^2 + r_c^2 d\Omega_2^2.
\end{equation}
As stated previously, it not a problem that the $g_{tt}$ part of the metric is different from \eqref{eq:cutoffmetric} because we are only interested in the spatial parts of the metric since the classical extremal surface and boundary subregion $\mathcal{B}$ live on a fixed time slice of the bulk and boundary, respectively, and we only care that we have a consistent treatment of the area for such codimension-2 surfaces anchored on $\mathcal{B}$.

Now we compute the correction to the area using this cut-off regularization. 
As in the scalar case, the extremal surface for the unperturbed metric is given by\footnote{The expression of the extremal surface is slightly modified compared to \eqref{eq:extremalsurf} to satisfy the boundary condition $\theta(r_c)=\theta_0$.}
\begin{equation}
    \gamma_{\rm ext}: \ \theta=\cos^{-1}\left(\alpha_0 \frac{r_c}{\sqrt{r_c^2+1}} \frac{\sqrt{r^2+1}}{r}\right),
\end{equation}
and we compute the areas $A[g^{(0)},\gamma_{\rm ext}]$ and $A[g^{(0)}+G_N g^{(1)},\gamma_{\rm ext}]$\footnote{Actually, $\gamma_{\rm ext}$ is not the extremal surface for the perturbed metric $g^{(0)}+G_N g^{(1)}$. However, the extremal surface for $g^{(0)}+G_N g^{(1)}$ differs from $\gamma_{\rm ext}$ by an $\mathcal{O}(G_N)$ amount so that their area difference is of $\mathcal{O}(G_N^2)$ due to extremality.} as 
\begin{equation}
    A^{(d=4)}[g^{(0)},\gamma_{\rm ext}] =  V_{S^{2}} \left(\frac{1}{2} r_c^2 \sin^2\theta_0 +\frac{1}{4}\left(1-\log(4r_c\sin^2\theta_0)\right)\right),
\end{equation}
and 
\begin{equation}\label{eq:newarea}
    A^{(d=4)}[g^{(0)}+G_N g^{(1)},\gamma_{\rm ext}] = A^{(d=4)}[g^{(0)},\gamma_{\rm ext}] +  \frac{4}{9}G_N \sin^2\theta_0 \left( \cos( 2\theta_0) -\frac{1}{5}(12\cos( 2\theta_0) -7) \right) ,
\end{equation}
up to $\mathcal{O}(r_c^{-2}) $ corrections which vanish in the $r_c\rightarrow \infty$ limit. The key observation is that the areas are IR-divergent as $\mathcal{O}(r_c^2)$ when sending $r_c$ to $\infty$, while the new cut-off surface in \eqref{eq:newcutoff} differs from the constant $r=r_c$ surface by $O(r_c^{-2})$. 
This means having a new cut-off surface results in an $\mathcal{O}(1)$ area change, which is exactly the first $\cos(2\theta_0)$ term in parentheses in \eqref{eq:newarea}. The last term in parentheses is due to the change of metric in the bulk, which would be the only area correction contribution if one used the $r=r_c$ cut-off for both metrics. 

To summarize, we have obtained the area difference
\begin{equation}\label{eqn:d=4areadiff}
    \Delta A^{(d=4)}(\gamma_{\mathrm{ext}})_{\mathcal{P}} = 4G_N \frac{56}{45} \sin^4\theta_0,
\end{equation}
which receives contributions from both the perturbation of the metric in the bulk and the anchoring points on the boundary.
We see that, at least at the order of $\theta_0^4$, $\Delta A/4G_N$ matches the CFT modular Hamiltonian in \eqref{eqn:K0Jdif_Max_thetaeps=0}, which evaluates to $56/45$ at this leading order in $\theta_{0}$. 
\begin{figure}[h]
\begin{center}
\begin{tikzpicture}[path fading=south]

\draw[thick,color=blue!70!white] (2.5,0) arc (30:150:3);
\draw[thick,color=orange!80!white] (-2.7,0) arc (230:310.5:4.02);
\draw[path fading] (2.5,0) arc (30:2:3);
\draw[path fading] (-2.7,0) arc (150:178:3);
\node[blue!80!black] at (0,1.8) {$\mathcal{B}$};
\node[orange!80!black] at (0,-1.3) {$\gamma_{\mathrm{ext}}$};

\draw[thick,color=red!50!white,decorate,decoration={snake,segment length=6mm,amplitude=.4mm}] (2.54,-0.5) arc (30:150:3);
\draw[thick,color=red!50!white,decorate,decoration={snake,segment length=7mm,amplitude=.4mm},path fading] (2.54,-0.5) -- (2.74,-1.4);
\draw[thick,color=red!50!white,decorate,decoration={snake,segment length=7mm,amplitude=.4mm},path fading] (-2.65,-0.5) -- (-2.85,-1.25);
\node at (0,0.7) {$\color{red}{r(r_{c},\theta)}$};

\end{tikzpicture}
\end{center}
\caption{Fixed-time Cauchy slice of AdS$_{d+1}$ (with $S^{d-2}$ directions suppressed) zoomed in on boundary subregion $\mathcal{B}$ (blue) and classical extremal surface $\gamma_{\mathrm{ext}}$ (orange) with wiggly cutoff surface $r(r_{c},\theta)$ (red).}
\label{fig:wigglycutoff}
\end{figure}
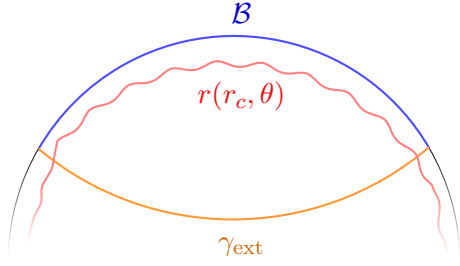

These subtleties in the calculation of the correction to the area can be argued in general dimensions. 
The codimension-2 extremal surface has an area given by 
\begin{align}\label{eqn:areafunctional_correction_Max}
\begin{split}
A[g,\gamma_{\mathrm{ext}}] = \int_{\gamma_{\mathrm{ext}}}\sqrt{h} &= V_{S^{d-2}}\int_{r_{\mathrm{min}}}^{r_c^{\prime}}dr\,\sin^{d-2}\theta\left(g_{\theta\theta}(r,\theta)\right)^{\frac{(d-2)}{2}}\left(g_{rr}(r,\theta) + \left(\theta'\right)^2 g_{\theta\theta} (r,\theta)\right)^{\frac{1}{2}}
\end{split}
\end{align}
where for $\theta$ we plugged in $\theta=\theta_{\rm ext}(r)$. One can argue that the leading IR divergence is given by 
\begin{equation}
    A[g,\gamma_{\mathrm{ext}}]\Big|_{\rm leading\ IR-div.}\approx \sin^{d-2}\theta_0\int_{r_{\mathrm{min}}}^{r_c^{\prime}} dr\, r^{d-3} \propto  \sin^{d-2}\theta_0 \left(r_c^{\prime}\right)^{d-2},
\end{equation}
where $r_c^{\prime}$ denotes the radius at which the extremal surface intersects the cut-off surface.  
Now, we would like to argue that it is important to consider the wiggled cut-off surface $r=r(r_c,\theta)$ for the perturbed metric, as illustrated in Figure \ref{fig:wigglycutoff}. 
Let us parameterize the $\theta$ component of the metric as 
\begin{equation}
g_{\theta\theta}(r,\theta) = r^2(1+G_N g^{(1)}_{\theta\theta})
\end{equation}
so the perturbed cut-off surface must take the form 
\begin{equation}
r = r(r_{c},\theta) = r_c\left(1-\frac{G_N}{2} g^{(1)}_{\theta\theta}(r_c,\theta)\right)
\end{equation}
to leave the induced metric invariant, i.e., so that $g_{\theta\theta}(r(r_{c},\theta),\theta) = r_{c}^2+\mathcal{O}(G_{N}^{2})$. 
Based on the discussion below Table \ref{tab:backreactphotonmetric}, we have $ g^{(1)}_{\theta\theta}\sim r^{-(d-2)}$, and hence the difference between $r_c^{\prime}$ and $r_c$ causes an area correction of $\mathcal{O}(1)$ in powers of $r_{c}$ (all area corrections here are of course $\mathcal{O}(G_{N})$). 
This completes our justification for the necessity of introducing the wiggled cut-off surface for the perturbed metric. 

Using this observation, we can write down the explicit expression for the area difference using the perturbed metric
\begin{align}\label{eqn:areafinalmaxwell1}
    \Delta A = G_N \frac{V_{S^{d-2}}}{2} \sin^{d-2}\theta_0 \bigg(\lim _{r_c\rightarrow \infty}r^{d-2}  g_{\theta\theta}^{(1)} \big|_{r=r_c} 
    & -\int_{r_{\min }}^{\infty} d r\left(r^2-r_{\min }^2\right)^{\frac{(d-3)}{2}} \frac{1}{r \sqrt{r^2+1}}  \\
    & \hskip -8pt \times \left[ g_{rr}^{(1)}\left(r^2-r_{\min }^2\right)- g_{\theta\theta}^{(1)} \left((d-2) r^2+r_{\min }^2\right)\right]\bigg) \nonumber 
\end{align}
where  the functions $g^{(1)}(r,\theta) $ are again evaluated on the extremal surface $\gamma_{\rm ext}$. The first term comes, as discussed above, from the perturbed cut-off surface. 

We propose that there is an alternate way of understanding this result by using dimensional regularization, where no cutoff surface is required. 
Consider a specific (integer) dimension $d+1$ and an infinitesimal parameter $\epsilon>0$ with $\tilde{d}=d-\epsilon$. 
To be more precise, we still consider a spherically symmetric state with respect to $S^{d-2}$ (as is true for the state of interest), so that the spacetime is described by the same $(t,r,\theta)$ direction with a transverse $S^{\tilde{d}-2}$ sphere.
The Einstein equations in general $\tilde{d} + 1$ dimensions is more non-trivial to solve,\footnote{By this, we mean using $d+1$ coordinates for Einstein equations to obtain differential equations for the various metric components and then taking $d \to \tilde{d}$ in the resulting equations.} but we will see that it is sufficient to use the $(d+1)$-dimensional solution obtained in the last section. 
Now let us consider the area correction using this dimensional regularization. 
The area in the unperturbed metric is obtained by simply replacing $d$ with $\tilde{d}$ in \eqref{eqn:areafunctional_correction_Max} (and set $r_{c}'=\infty$), 
and for the area in the backreacted metric we use the solutions $g_{\mu\nu}^{(1)}$ obtained in $(d+1)$ dimensions and replace all other $d$ in the area formula with $\tilde{d}$ to obtain
\begin{align}
    A^{(\tilde{d})}[g^{(0)}+G_N g^{(1)},\gamma_{\mathrm{ext}}]= A^{(\tilde{d})}[g^{(0)},\gamma_{\mathrm{ext}}]   & -G_N \frac{V_{S^{d-2}}}{2} \sin^{d-2}\theta_0 \int_{r_{\min }}^{\infty} d r  \frac{\left(r^2-r_{\min }^2\right)^{\frac{(\tilde{d}-3)}{2}}}{r \sqrt{r^2+1}}  \\
    & \hskip 30pt \times \left[ g_{rr}^{(1)}\left(r^2-r_{\min }^2\right)- g_{\theta\theta}^{(1)} \left((\tilde{d}-2) r^2+r_{\min }^2\right)\right]. \nonumber
\end{align}
The area difference can thus be written as 
\begin{align}\label{eqn:areadiff_dimreg}
    \Delta A(\gamma_{\mathrm{ext}}) &= -G_N \frac{V_{S^{d-2}}}{2} \sin^{d-2}\theta_0 \bigg(\epsilon \int_{r_{\rm min}}^{\infty} d r \frac{ \left(r^2-r_{\min }^2\right)^{\frac{(\tilde{d}-3)}{2}}}{r\sqrt{r^2+1}} r^2  g_{\theta\theta}^{(1)} \\ 
    &\qquad+\int_{r_{\rm min}}^{\infty} d r \frac{\left(r^2-r_{\min }^2\right)^{\frac{(\tilde{d}-3)}{2}}}{r \sqrt{r^2+1}} \left[ g_{rr}^{(1)}\left(r^2-r_{\min }^2\right)- g_{\theta\theta}^{(1)} \left((d-2) r^2+r_{\min }^2\right)\right] \bigg). \nonumber 
\end{align}
We will show that these two terms can be mapped to exactly the two terms appearing in \eqref{eqn:areafinalmaxwell1}.
It is easier to discuss the second term: since the integral is convergent, the correction to the $d=\tilde{d}$ answer will be of order $\epsilon$ which will drop out in the end so we reproduce the second term in \eqref{eqn:areafinalmaxwell1}. 
The first term in \eqref{eqn:areadiff_dimreg}, on the other hand, involves a formally divergent integral without regularization, and the divergence comes from the large-$r$ region. The dimensionally-regularized integral will therefore take the form 
\begin{equation}\label{eqn:areadiff_dimreg_term1}
\epsilon \int_{r_{\rm min}}^{\infty} d r \frac{ \left(r^2-r_{\min }^2\right)^{\frac{(\tilde{d}-3)}{2}}}{r\sqrt{r^2+1}} r^2  g_{\theta\theta}^{(1)} \approx \left(\lim_{r\rightarrow \infty}r^{d-2}g_{\theta\theta}^{(1)}\right)\int_{r_{\min}}^{\infty} dr\, \frac{\epsilon}{r^{1+\epsilon}} = -\left(\lim_{r\rightarrow \infty}r^{d-2}g_{\theta\theta}^{(1)}\right)+\mathcal{O}(\epsilon),
\end{equation}
matching exactly the first term in \eqref{eqn:areafinalmaxwell1}.

\begin{table}[h!]
\begin{center}
     \begin{tabular}{|c|c|}
     \hline
       \rowcolor{orange!40!white!50} $d$  & ${\Delta A_{{\mathcal{P}}}}/{4G_{N}}$  \\
       \hline
          \rowcolor{green!50!white!50} 3 &  $   -\pi  S_{\theta_{0}} + \frac{4} {3} S_{\theta_{0}}^2 + \frac{\pi}{8}S_{\theta_{0}}^3 + 2\pi T_{\theta_{0}/2} + 4 (\theta_0 T_{\theta_{0}}^{-1}-1)$\\ 
        \rowcolor{green!30!white!40} 4 &  $\frac{56}{45}S_{\theta_{0}}^{4}$\\ 
       \rowcolor{green!50!white!50} 5 &  $  - 2\pi S_{\theta_{0}} + 
 \frac{8}{3} S_{\theta_{0}}^2 - \frac{\pi}{2}  S_{\theta_{0}}^3 + 
 \frac{16}{15} S_{\theta_{0}}^4  + \frac{9\pi}{64} S_{\theta_{0}}^5 +
 \frac{64}{105} S_{\theta_{0}}^6 + 4 \pi T_{\theta_{0}/2} + 8 (\theta_0 T_{\theta_{0}}^{-1}-1)$ \\
        \rowcolor{green!30!white!40} 6 &  $\frac{208}{175}S_{\theta_{0}}^6 + \frac{32}{63}S_{\theta_{0}}^8$ \\
        \hline
    \end{tabular}
\end{center}
\caption{Correction to area of classical extremal surface due to backreaction of lowest energy photon excited state in $d=3,4,5,6$ with $S_{x} \equiv \sin x$ and $T_{x} \equiv \tan x$.}
\label{tab:areacorrection}
\end{table}

In conclusion, we have shown that it is crucial to regularize the area to correctly compute the area correction, and the two regularizations have been proven to be equivalent. The area correction results are listed for various dimensions in Table \ref{tab:areacorrection} which are exact in $\theta_{0}$. We shall see that these carefully computed area differences lead to vacuum-subtracted generalized entropies that agree with the CFT result, as already hinted at below \eqref{eqn:d=4areadiff} for $d=4$. To show this, we need to compute the bulk entanglement entropy for the photon state.

\subsection{Entanglement entropy}
\label{sec:photonEE}

We now turn to a computation of the vacuum-subtracted von Neumann entropy for the excited state $\ket{\mathcal{P}}$. As explained in the Introduction, in order to factorise the Hilbert space, we need to introduce an extended Hilbert space. This leads to two contributions to the von Neumann entropy, one from the edge modes and one from the bulk modes:
\begin{equation}
\Delta S^{\mathrm{AdS}}\left(\rho_{\Sigma_{\mathcal{B}}}^{\mathcal{P}}\right) = \Delta S^{\mathrm{bulk}}\left(\rho_{\Sigma_{\mathcal{B}}}^{\mathcal{P}}\right)+\Delta S^{\mathrm{edge}}\left(\rho_{\Sigma_{\mathcal{B}}}^{\mathcal{P}}\right),
\end{equation}
where $\rho_{\Sigma_{\mathcal{B}}}^{\mathcal{P}}$ is the reduced density matrix on $\Sigma_{\mathcal{B}}$. We will defer all discussion of the edge modes to section \ref{sec:edge}, here we simply remark that we will choose the electric center so we fix the normal component of the electric field for the Rindler modes, known as perfect magnetic conductor (PMC) boundary conditions. The von Neumann entropy for the bulk modes is actually independent of the precise value of this electric flux so it suffices to set it to zero in constructing the bulk Rindler Hilbert space.\footnote{For the purposes of computing vacuum-subtracted von Neumann entropies $\Delta S^{\mathrm{bulk}}\left(\rho_{\Sigma_{\mathcal{B}}}^{\mathcal{P}}\right)$, we will see that it is actually completely independent of choice of center, i.e., choice of boundary conditions, so long as the bulk modes are regular at the entangling surface.} 

The first step is to quantize the gauge field $\mathbb{A}_{\mu}$ in Rindler coordinates to relate the global modes to the Rindler modes, from which we obtain the Bogoliubov coefficients, and thence the reduced density matrix for the right Rindler wedge.

\subsubsection{AdS-Rindler quantization}
\label{sec:photonAdSRindler}

Consider Maxwell theory in AdS-Rindler spacetime. The reduced phase space can constructed in the same way as was done in global coordinates in \S\ref{sec:Dirac}. It is too difficult to work with the radial gauge $\mathbb{A}_{r}=0$ in Rindler coordinates so we will use the Rindler radial gauge $\mathpzc{A}_{\rho}=0$. This does not affect the final result for the von Neumann entropy as the Bogoliubov coefficients will be computed in a gauge-invariant way by using only the field strength $F_{\mu\nu}$. 

We will not provide all of the details, such as the Dirac bracket and so forth, as they are completely analogous. The pertinent details are the reduced phase space variables
\begin{equation}\label{eqn:Rindlerreducedphasespace}
\mathfrak{P} = \{\mathpzc{A}_{u},\mathpzc{A}_{\phi_{i}},\pi^{u},\pi^{\phi_{i}},|\,2 \leq i \leq d-1\},
\end{equation}
and the constraints
\begin{equation}\label{eqn:Rindlerconstraints}
\pi_{\tau} = 0, \qquad \nabla_{i}\pi^{i} = 0, \qquad \mathpzc{A}_{\rho} = 0, \qquad \partial_{\tau}\mathpzc{A}_{\rho} \approx 0,
\end{equation}
where $i$ labels the spatial directions $\{u,\phi_{j}\}$ and again we use $\approx 0$ to mean that a function on phase space is weakly zero. In addition to the constraint coming from the gauge choice $\mathpzc{A}_{\rho}=0$, we will also choose boundary conditions such that this holds on the horizon $\rho=1$. The extrapolate dictionary in these coordinates takes the form
\begin{equation}\label{eqn:Rindlerbdycondtionsinf}
\lim_{\rho \to \infty}\rho^{d-2}\mathpzc{A}_{\mu} \propto J_{\mu}.
\end{equation}

The equations of motion are Maxwell's equations $\nabla_{\mu}F^{\mu\nu}=0$ for $\nu = u,\phi_{i}$ coming from Hamilton's equations on phase space, along with $\nu=\tau$ coming from the Gauss Law constraint above. They can be solved by the same techniques as in the global case, but now with a subtlety. The decomposition of the gauge field into vector and scalar parts with respect to $SO(d)$ transformations of the $S^{d-1}$ part of the AdS$_{d+1}$ metric used compactness of $S^{d-1}$. However, the AdS-Rindler wedge only has $d-2$ compact directions given by an $S^{d-2}$ so if one tries to decompose $\mathpzc{A}_{\mu}$ with respect to transformation properties under $SO(d-1)$, one finds that Maxwell's equations become simpler, but they are still written in terms of a vector field rather than a scalar field and we were not able to solve the resulting equations. So we need to find a different method.

The decomposition on $S^{d-1}$ in \eqref{eqn:vectordecomp} can be understood as the Hodge decomposition which is a consequence of Hodge's Theorem. There exists a weaker version of the Theorem due to Kodaira \cite{10.1215/S0012-7094-90-06021-1} which states that on any Riemannian manifold $M$, the $L^{2}$ $k$-forms can be decomposed as
\begin{equation}\label{eqn:Kodairathm}
L^{2}\Omega^{k}(M) = \overline{dC_{0}^{\infty}\Omega^{k-1}(M)} \oplus \overline{\delta C_{0}^{\infty}\Omega^{k+1}(M)} \oplus L^{2}\mathcal{H}^{k}(M)
\end{equation}
where $C_{0}^{\infty}$ denotes smooth and compactly supported, the closure is taken in the space of $L^2$ forms, and the last space is given by
\begin{equation}\label{eqn:coclosedclosedforms}
L^{2}\mathcal{H}^{k}(M) = \{\omega \in L^{2}\Omega^{k}(M)\,|\,d\omega=\delta\omega=0\}.
\end{equation}
All of the vector fields considered here will be $L^{2}$-normalizable after we normalize the wavefunctions so the only subtlety comes from the closure of these spaces. We will assume that no such issues arise for us and forge ahead with the assumption that such a decomposition exists.\footnote{One could try to prove the decomposition by taking the quotient of $H^{d-1}$ by a discrete subgroup of the isometry group such that the resulting space is compact so the decomposition holds, analyze Maxwell's equations on this space, and then uplift back to $H^{d-1}$. We thank Akihiro Ishibashi for correspondence on this point.}

We decompose the gauge field into the scalar and tensor part with respect to the action of $SO(d-1,1)$:
\begin{equation}\label{eqn:gaugefielddecomp_Rindler_hyperbolic}
\mathpzc{A}_{\mu} = \mathpzc{A}_{\mu}^{V_{H}} + \mathpzc{A}_{\mu}^{S_{H}}
\end{equation}
where $\mathpzc{A}_{\mu}^{V_{H}}$ is divergenceless and $\mathpzc{A}_{\mu}^{S_{H}}$ is a divergence. Using $\alpha,\beta,\ldots$ to label hyperbolic components (not $\mu,\nu,\ldots$) and $\mathfrak{v},\mathfrak{w},\ldots$ to label components along $(\tau,\rho)$, we have
\begin{equation}\label{eqn:gaugefieldcomps_Rindler_hyperbolic}
\mathpzc{A}_{\mu}^{V_{H}}\,dx^{\mu} = \sum_{\lambda_{V}}\mathpzc{A}_{\lambda_{V}}^{V_{H}}\mathbf{H}_{\alpha,\lambda_{V}}\,dz^{\alpha}, \qquad \mathpzc{A}_{\mu}^{S_{H}}\,dx^{\mu} = \sum_{\lambda}\mathpzc{A}_{\mathfrak{v},\lambda}H_{\lambda}\,dx^{\mathfrak{v}}+\mathpzc{A}_{\lambda}D_{\alpha}H_{\lambda}\,dz^{\alpha},
\end{equation}
where $\mathbf{H}_{\alpha,\lambda_{V}}$ and $H_{\lambda}$ are the vector and scalar eigenfunctions of the Laplacian on $H^{d-1}$, respectively, and $D_{\alpha}$ is the covariant derivate on $H^{d-1}$. The spectrum of the Laplacian is continuous so the sums in \eqref{eqn:gaugefieldcomps_Rindler_hyperbolic} are actually integrals.\footnote{The spectrum of scalar eigenfunctions was computed in App. \ref{sec:hyperbolicballeigfns} with the result $\lambda \in (\zeta^{2},\infty)$ and the result for the vector eigenfunctions is known for the Hodge Laplacian \cite{Donnelly:1981aa} and hence, using the simple relationship between the Hodge Laplacian and connection Laplacian on $H^{d-1}$, one finds $\lambda_{V} = \{1\} \cup [d-\frac{3}{4},\infty)$ for $d=3$ and $\lambda_{V} = \left[(\frac{d-4}{2}\right)^{2}+d-2,\infty)$ for $d > 3$.}
The scalar eigenfunctions $H_{\lambda}$ are computed in App. \ref{sec:hyperbolicballeigfns}. We will not provide the vector eigenfunctions as we will not need them for our purposes nor do we know of anywhere in the literature where they have been constructed.\footnote{Nevertheless, they must form a complete basis of $L^{2}\Omega^{1}(H^{d-1})$ because the Laplacian is a non-negative symmetric operator on the space of smooth, compactly supported, divergence-free $1$-forms so it has a Friedrichs extension to a non-negative self-adjoint operator on the space of $L^{2}$ divergence-free $1$-forms.}

One still needs to argue that the different terms in the decomposition cannot mix so that the Maxwell equations decouple into two sets of equations as they did in global AdS. The equations are linear in $\mathpzc{A}_{\mu}$ so it suffices to show that the image of the scalar and vector parts under the Maxwell differential operator consist only of the scalar and vector parts, respectively. We checked that this indeed holds.\footnote{One could alternatively check that the Maxwell equations are invariant under $SO(d-1,1)$ isometries.}

We can now solve the Maxwell equations by adapting the formalism developed in \cite{Ishibashi:2004wx} to the decomposition on $H^{d-1}$ that we are using for AdS-Rindler, and then following similar steps to those in \S\ref{sec:MaxwellEOM}. Focusing only on the scalar part $A_{\mu}^{S_{H}}$, we obtain the equations of motion
\begin{align}\label{eqn:MaxwellRindlerEOM}
\begin{split}
F_{\mathfrak{v}\mathfrak{w}} &= \sum_{\lambda}\widetilde{\nabla}_{\mathfrak{v}}\left(\frac{1}{\rho^{d-3}}\widetilde{\nabla}^{\mathfrak{v}}\phi_{S_{H},\lambda}\right)\epsilon_{\mathfrak{v}\mathfrak{w}}H_{\lambda}
\\ \epsilon_{\mathfrak{v}\mathfrak{w}}\rho^{d-3}\left(\widetilde{\nabla}^{\mathfrak{w}}\mathpzc{A}_{\lambda}-\mathpzc{A}_{\lambda}^{\mathfrak{w}}\right) &= \widetilde{\nabla}_{\mathfrak{v}}\phi_{S_{H},\lambda}
\end{split}
\end{align}
for some scalar field $\phi_{S_{H},\lambda}$, where $\widetilde{\nabla}_{\mathfrak{v}}$ and $\epsilon_{\mathfrak{v}\mathfrak{w}}$ are the covariant derivative and Levi-Civita tensor on the $2$-dimensional part of the geometry with coordinates $(\tau,\rho)$, respectively. Inserting these into the Gauss Law constraint $\nabla_{i}\pi^{i} = 0$ given in \eqref{eqn:Rindlerconstraints}, we obtain the following differential equation for $\phi_{S_{H},\lambda}$
\begin{equation}\label{eqn:phiSHdiffeq}
\rho^{d-1}\widetilde{\nabla}_{\mathfrak{v}}\left(\frac{1}{\rho^{d-3}}\widetilde{\nabla}^{\mathfrak{v}}\phi_{S_{H},\lambda}\right)-\lambda\phi_{S_{H},\lambda} = \tilde{\mathfrak{F}}(\tau)
\end{equation}
where $\tilde{\mathfrak{F}}(\tau)$ is an arbitrary function of $\tau$. Using separation of variables,
\begin{equation}\label{eqn:phiSRindler_freq}
\phi_{S_{H},\lambda}(\tau,\rho) = \int_{\omega>0}\frac{d\omega}{2\pi}e^{-i\omega\tau}g_{\omega,\lambda}^{S}(\rho),
\end{equation}
the differential equation \eqref{eqn:phiSHdiffeq} can be written in terms of $g_{\omega,\lambda}^{S}(\rho)$ with all $\rho$ dependence on one side and all $\tau$ dependence on the other side so both are equal to a constant. First, we solve the homogeneous differential equation where this constant is set to $0$, leading to
\begin{equation}\label{eqn:gomegalambdaeqn_homo}
\rho^{2}\partial_{\rho}\left((\rho^{2}-1)\partial_{\rho}\right)g_{\omega,\lambda}^{S,\mathrm{hom.}}+(3-d)(\rho^{2}-1)\rho\partial_{\rho}g_{\omega,\lambda}^{S,\mathrm{hom.}} + \frac{\rho^{2}}{(\rho^{2}-1)}\omega^{2}g_{\omega,\lambda}^{S,\mathrm{hom.}}-\lambda g_{\omega,\lambda}^{S,\mathrm{hom.}} = 0.
\end{equation}
To solve this differential equation, we change variables $\eta = \frac{1}{\rho^{2}}$ and consider $g_{\omega,\lambda}^{S,\mathrm{hom.}}(\eta) = \eta^{a}(1-\eta)^{b}\tilde{g}_{\omega,\lambda}^{S}(\eta)$, which gives the hypergeometric differential equation for $\tilde{g}_{\omega,\lambda}^{S}(\eta)$. We find that one solution of this differential equation, as well as the part of the solution coming from the inhomogeneity, do not satisfy the extrapolate dictionary \eqref{eqn:Rindlerbdycondtionsinf} so we drop them. The other part of the homogeneous solution does satisfy the dictionary and leads to
\begin{equation}\label{eqn:gS_final}
g_{\omega,\lambda}^{S}(\rho) \propto \left(1-\frac{1}{\rho^{2}}\right)^{-i\frac{\omega}{2}}{}_{2}{F}_{1}\left(\frac{1}{2}\left(\zeta-i\omega_{+}\right),\frac{1}{2}\left(\zeta-i\omega_{-}\right),\zeta;\frac{1}{\rho^{2}}\right).
\end{equation}
From the first equation in \eqref{eqn:MaxwellRindlerEOM}, we obtain $\mathpzc{A}_{\lambda,\omega}$ as the following integral
\begin{equation}\label{eqn:Alambda_soln}
\mathpzc{A}_{\omega,\lambda}(\rho) \propto \int d\rho\,\rho^{1-d}\left(1-\frac{1}{\rho^{2}}\right)^{-i\frac{\omega}{2}-1}{}_{2}{F}_{1}\left(\frac{1}{2}\left(\zeta-i\omega_{-}\right),\frac{1}{2}\left(\zeta-i\omega_{+}\right),\zeta;\frac{1}{\rho^{2}}\right).
\end{equation}
This integral can be performed by switching again to the $\eta \in (0,1)$ coordinate so that the hypergeometric function has convergent series expansion and integrate the series term-by-term to arrive at
\begin{equation}\label{eqn:ARindler_final}
\mathpzc{A}_{\omega,\lambda}(\rho) = \mathcal{N}_{\omega,\lambda}^{S,R}\sum_{n=0}^{\infty}\frac{(\frac{1}{2}(\zeta-i\omega_{-}))_{n}(\frac{1}{2}(\zeta-i\omega_{+}))_{n}}{(\zeta)_{n}\,n!}B_{\rho^{-2}}\left(\zeta+n,-i\frac{\omega}{2}\right),
\end{equation}
where we dropped the integration constant since it is non-normalizable. Using the first equation from \eqref{eqn:MaxwellRindlerEOM} again, we can obtain $\mathpzc{A}_{\tau,\omega,\lambda}$, with the result
\begin{equation}\label{eqn:Atau_soln}
\mathpzc{A}_{\tau,\omega,\lambda} = -i\left(\omega\mathpzc{A}_{\omega,\lambda}+\frac{1}{\omega}(\rho^{2}-1)\rho^{3-d}\partial_{\rho}\left(\rho^{d-3}(\rho^{2}-1)\partial_{\rho}\mathpzc{A}_{\omega,\lambda}\right)\right).
\end{equation}

The final step is to normalize the gauge field. Notice that the vector and scalar parts of the gauge field are orthogonal with respect to this inner product. The generalized Klein-Gordon inner product is given by
\begin{equation}\label{eqn:MaxIP_Rindler}
\left<B_{R},C_{R}\right>_{R} = -i\int_{\Sigma_{\mathcal{B}}}d^{d}x\,\sqrt{-g}\left(B_{R,\mu}^{\ast}(\nabla^{\tau}C_{R}^{\mu}-\nabla^{\mu}C_{R}^{\tau})-C_{R,\mu}(\nabla^{\tau}B_{R}^{\ast\mu}-\nabla^{\mu}B_{R}^{\ast \tau})\right),
\end{equation}
where we used the symplectic form $\Omega_{R}$ on the Maxwell phase space for AdS-Rindler spacetime in order to obtain this inner product. We require that the wavefunctions are delta-function normalizable:
\begin{equation}\label{eqn:Rindlergauge_norm_final}
\left<\mathpzc{A}_{\omega,\lambda}^{S_{H}},\mathpzc{A}_{\omega',\lambda'}^{S_{H}}\right>_{R} = (2\pi)^{2}\delta(\omega-\omega')\delta^{(d-1)}(\tilde{\lambda}-\tilde{\lambda}'),
\end{equation}
where $\delta^{(d-1)}(\tilde{\lambda}-\tilde{\lambda}') \equiv \delta(\tilde{\lambda}-\tilde{\lambda}')\delta_{\mathfrak{m},\mathfrak{m}'}$. Amazingly, in spite of the complicated result for the wavefunctions \eqref{eqn:ARindler_final}, we can extract the normalization constant $\mathcal{N}_{\omega,\lambda}^{S,R}$ by studying these functions near $\rho=1$ which gives the oscillatory behavior necessary for delta-function normalizability. To do so, we must use a different form of the wavefunction than that appearing in \eqref{eqn:ARindler_final} as it is not very amenable to study near $\rho=1$. Instead, we go back to \eqref{eqn:Alambda_soln} and change variables to $\tilde{\eta} \equiv 1-\eta$ so $\rho=1$ corresponds to $\tilde{\eta}=0$. We find
\begin{align}\label{eqn:Alambda_soln_final_other}
\begin{split}
\mathpzc{A}_{\omega,\lambda}(\tilde{\eta}) &= -\mathcal{N}_{\omega,\lambda}^{S,R}\int d\tilde{\eta}\,\left[\tilde{\mathcal{X}}\frac{\left(1-\tilde{\eta}\right)^{\frac{d}{2}-2}}{\tilde{\eta}^{1+i\frac{\omega}{2}}}{}_{2}{F}_{1}\left(\frac{1}{2}\left(\zeta-i\omega_{-}\right),\frac{1}{2}\left(\zeta-i\omega_{+}\right),1-i\omega;\tilde{\eta}\right)+\mathrm{c.c.}\right]
\\	&= -\mathcal{N}_{\omega,\lambda}^{S,R}\sum_{n=0}^{\infty}\left[\tilde{\mathcal{X}}\frac{\left(\frac{1}{2}\left(\zeta-i\omega_{-}\right)\right)_{n}\left(\frac{1}{2}\left(\zeta-i\omega_{+}\right)\right)_{n}}{\left(1-i\omega\right)_{n}n!}B_{\tilde{\eta}}\left(n-i\frac{\omega}{2},\zeta\right)+\mathrm{c.c.}\right] 
\\	\tilde{\mathcal{X}} &\equiv \frac{\Gamma(\zeta)\Gamma(i\omega)}{\Gamma\left(\frac{1}{2}\left(\zeta+i\omega_{-}\right)\right)\Gamma\left(\frac{1}{2}\left(\zeta+i\omega_{+}\right)\right)}.
\end{split}
\end{align}
Now we can compute the desired normalization constant with the result
\begin{equation}\label{eqn:MaxwellRindlerwavefn_normalization}
\mathcal{N}_{\omega,\lambda}^{S,R} = \sqrt{\frac{\omega}{8\lambda}}\left|\frac{\Gamma\left(\frac{1}{2}\left(\zeta-i\omega_{-}\right)\right)\Gamma\left(\frac{1}{2}\left(\zeta-i\omega_{+}\right)\right)}{\Gamma(\zeta)\Gamma(i\omega)}\right|.
\end{equation}

The Hilbert space is a tensor product of the scalar and vector modes $\mathbb{H}_{R} = \mathbb{H}_{R}^{S} \otimes \mathbb{H}_{R}^{V}$. The gauge field is an expansion in annihilation and creation operators for these Fock spaces
\begin{equation}\label{eqn:fullRindlergaugefield}
\hat{\mathpzc{A}}_{\mu}^{\mathfrak{c}}(x) = \sum_{I \in \{L,R\}}\int_{\omega>0}\frac{d\omega}{2\pi}\sum_{\lambda_{\mathfrak{c}}}\left(e^{-i\omega\tau}\mathpzc{A}_{\mu,\lambda_{\mathfrak{c}},\omega,I}^{\mathfrak{c}}b_{\omega,\lambda_{\mathfrak{c}},I}^{\mathfrak{c}}+e^{i\omega\tau}\mathpzc{A}_{\mu,\omega,\lambda_{\mathfrak{c}},I}^{\mathfrak{c} \ast}b_{\omega,\lambda_{\mathfrak{c}},I}^{\mathfrak{c}\dagger}\right),
\end{equation}
where $\mathfrak{c} \in \{S_{H},V_{H}\}$ and $\mathpzc{A}_{\mu,\lambda,\omega,R}^{S_{H}}(\rho,\mathbf{x})$ is what we called $\mathpzc{A}_{\mu,\lambda,\omega}^{S_{H}}(\rho,\mathbf{x})$ above. The Rindler annihilation and creation operators satisfy the commutation relations
\begin{equation}\label{eqn:Rindlergauge_ladderops}
\left[b_{\omega,\lambda_{\mathfrak{c}},I}^{\mathfrak{c}},\left(b_{\omega,\lambda_{\mathfrak{c}'},I'}^{\mathfrak{c}'}\right)^{\dagger}\right] = \delta_{I,I'}\delta_{\mathfrak{c},\mathfrak{c}'}(2\pi)^{2}\delta(\omega-\omega)\delta^{(d-1)}(\lambda_{\mathfrak{c}}-\lambda_{\mathfrak{c}'})
\end{equation}
while all other commutators vanish.

Let us end with a discussion of the boundary conditions on the wavefunctions at the Rindler horizon. 
One should impose proper boundary conditions to get a well-defined variational principle. 
The boundary term from varying the Lagrangian takes the form $ F_{\rho \nu}\delta A^{\nu}$ evaluated on the extremal surface. 
There are different choices of boundary conditions \cite{Casini:2013rba}, but we use the PMC boundary conditions ($E_{\perp}=0$, $B_{\parallel}=0$ on $\gamma_{\mathrm{ext}}$) given by
\begin{equation}\label{eqn:PMC}
F_{\rho\nu,I}\big|_{\gamma_{\mathrm{ext}}} = 0, \quad \nu = \tau,\alpha \qquad \mathrm{and} \qquad \mathpzc{A}_{\rho,I}\big|_{\gamma_{\mathrm{ext}}} = 0, \qquad I \in \{L,R\}.
\end{equation}
Similar to the scalar wavefunction, the gauge field $\mathpzc{A}_{\mu,\omega,\lambda}^{S_{H}}$ behaves near the horizon as $(\rho^2-1)^{\pm i\frac{\omega}{2}}$.
Due to the same argument as in \S\ref{sec:scalar_bulkEE}, to satisfy the boundary condition, we again need to implement a radial cutoff at $\rho = 1+\epsilon$ for some $\epsilon > 0$, from which the quantization of the Rindler frequencies results. However, it is sufficient to use a continuous spectrum, which is also easier technically, for computing the bulk entropy. 

\subsubsection{Contributions to bulk entanglement entropy}
\label{sec:photonEE_bulk}

The computation of the bulk entanglement entropy follows in a very similar way to the scalar case once the Bogoliubov coefficients have been obtained. The global AdS vacuum state contains the thermofield double state for the vector and scalar modes
\begin{equation}\label{eqn:MinkvactoRindlerTFD_Max}
\ket{0} \supset \bigotimes_{\mathfrak{c} \in {S_{H},V_{H}}}\bigotimes_{\omega,\lambda}\sqrt{1-e^{-2\pi\omega}}\sum_{n}e^{-\pi E_{\mathfrak{c},n}}\ket{n,\omega,\lambda_{\mathfrak{c}}^{\ast}}_{L,\mathfrak{c}}\ket{n,\omega,\lambda_{\mathfrak{c}}}_{R,\mathfrak{c}},
\end{equation}
where $\supset$ indicates that we focus only on the $E_{\rho}|_{\gamma}=0$ superselection sector, but the global vacuum is actually a direct sum over all electric center superselection sectors.

The annihilation and creation operators for the global vector and scalar modes and the Rindler scalar and vector modes can be related by
\begin{equation}\label{eqn:globaltoRindlerops_Max}
a_{n,\ell,\mathfrak{m}}^{c} = \sum_{I}\sum_{\mathfrak{c}}\int_{0}^{\infty}\frac{d\omega}{2\pi}\sum_{\lambda}\left(\alpha_{n,\ell,\mathfrak{m};\omega,\lambda,I}^{c,\mathfrak{c}}b_{\omega,\lambda,I}^{\mathfrak{c}}+\beta_{n,\ell,\mathfrak{m};\omega,\lambda,I}^{c,\mathfrak{c}\ast}b_{\omega,\lambda,I}^{\mathfrak{c}\dagger}\right), \qquad c \in \{S,V\},
\end{equation}
where $\alpha_{n,\ell,\mathfrak{m};\omega,\lambda,I}^{c,\mathfrak{c}}$, $\beta_{n,\ell,\mathfrak{m};\omega,\lambda,I}^{c,\mathfrak{c}}$ are the Bogoliubov coefficients. Note that the annihilation and creation operators for the scalar or vector mode in global coordinates can depend on both such Rindler operators, for example, we can have $\alpha_{n,\ell,\mathfrak{m};\omega,\lambda,I}^{S,V_{H}} \neq 0$. It is a non-trivial fact which we prove in App. \ref{sec:Bogcoeffs_Max} that
\begin{equation}\label{eqn:Maxwellbogocoefrelations}
\alpha_{0,1,\mathbf{0};\omega,\lambda,I}^{S,V_{H}} = \beta_{0,1,\mathbf{0};\omega,\lambda,I}^{S,V_{H}} = 0.
\end{equation}
This is very useful for us because we did not obtain the explicit solutions for the vector modes in Rindler and this means that we do not need them for our purposes. Let us focus on the global mode of interest ($n=0$, $\ell=1$, $\mathfrak{m} = \mathbf{0}$) whose Bogoliubov coefficients we denote by $\alpha_{0,1,\mathbf{0};\omega,\lambda,I}^{S,S_{H}} \equiv \alpha_{\omega,\lambda,I}^{S,S_{H}}$ and $\beta_{0,1,\mathbf{0};\omega,\lambda,I}^{S,S_{H}} \equiv \beta_{\omega,\lambda,I}^{S,S_{H}}$ to declutter notation. As in the scalar case, we have the following relations between left and right modes:
\begin{equation}\label{eqn:vacann_photon}
\alpha_{\omega,\lambda,L}^{S,S_{H}} = -e^{\pi\omega}\beta_{\omega,\lambda^{\ast},R}^{S,S_{H}}, \qquad \beta_{\omega,\lambda,L}^{S,S_{H}\ast} = -e^{-\pi\omega}\alpha_{\omega,\lambda^{\ast},R}^{S,S_{H}},
\end{equation}
which is a consequence of $a_{0,1,\mathbf{0}}\ket{0}=0$, and
\begin{equation}\label{eqn:AdSRindlerops_LtoR_Max}
b_{\omega,\lambda,L}^{S_{H}}\ket{0} = e^{-\pi\omega}b_{\omega,\lambda^{\ast},R}^{S_{H}\dagger}\ket{0}, \qquad b_{\omega,\lambda,L}^{S_{H}\dagger}\ket{0} = e^{\pi\omega}b_{\omega,\lambda^{\ast},R}^{S_{H}}\ket{0}.
\end{equation}
We can now obtain the reduced density matrix for the state $\ket{\mathcal{P}}$ for the classical entanglement wedge $\Sigma_{\mathcal{B}}$ by rewriting the operators and Bogoliubov coefficients appearing in the density matrix purely in terms of the right wedge, leading to 
\begin{align}\label{eqn:AdSRindler_photonreducedrho}
\begin{split}
\rho_{\Sigma_{\mathcal{B}}}^{\mathcal{P}} = \Tr_{\Sigma_{\mathcal{B}}^{c}}\rho^{\mathcal{P}} = &\int\frac{d\omega}{2\pi}\sum_{\lambda}\left((1-e^{-2\pi\omega})\alpha_{\omega,\lambda}^{\ast}b_{\omega,\lambda}^{\dagger} + (1-e^{2\pi\omega})\beta_{\omega,\lambda}b_{\omega,\lambda}\right)e^{-K_{0}^{\mathrm{Max.,bulk}}}
\\	&\int\frac{d\omega'}{2\pi}\sum_{\lambda'}\left((1-e^{-2\pi\omega'})\alpha_{\omega',\lambda'}b_{\omega',\lambda'} + (1-e^{2\pi\omega'})\beta_{\omega',\lambda'}^{\ast}b_{\omega',\lambda'}^{\dagger}\right).
\end{split}
\end{align}
where we have simplified notation by dropping the $R$, $S$, and $S_{H}$ labels on the Bogoliubov coefficients and the annihilation and creation operators. The vacuum modular Hamiltonian $K_{0}^{\mathrm{Max.,bulk}} = 2\pi H_{R}^{\mathrm{Max.,bulk}}$ is the Rindler Hamiltonian for the bulk Maxwell theory.

We can expand the excited state density matrix in the small region size expansion
\begin{equation}
\rho_{\Sigma_{\mathcal{B}}}^{\mathcal{P}} = \rho_{\Sigma_{\mathcal{B}}}^{0}+\delta \rho
\end{equation}
and use the replica trick to compute the vacuum-subtracted entanglement entropy as we did for the scalar \eqref{eqn:bulkvacsubtractEE_1storder}, \eqref{eqn:bulkvacsubtractEE_2ndorder} with the result
\begin{equation}\label{eqn:DeltaSbulk}
\Delta S^{\mathrm{bulk}}\left(\rho_{\Sigma_{\mathcal{B}}}^{\mathcal{P}}\right) = \underbrace{-\Tr\left(\delta \rho \log \rho_{\Sigma_{\mathcal{B}}}^{0}\right)}_{\Delta K_{0}^{\mathcal{P}}} + \underbrace{\frac{1}{2}-\frac{1}{2}\Tr\left(\rho_{\Sigma_{\mathcal{B}}}^{\psi}\tilde{\rho}'(1)(\rho_{\Sigma_{\mathcal{B}}}^{0})^{-1}\right)}_{\Delta S^{\mathrm{bulk}}|_{\mathcal{O}((\delta \rho)^{2})}}
\end{equation}
where $\tilde{\rho}(n)$ was defined in \eqref{eqn:tilderho}.

\paragraph{First order in $\delta \rho$.} 

As we have seen throughout this work, the leading order contribution to the vacuum-subtracted von Neumann entropy in the small region size expansion comes from the vacuum-subtracted expectation value of the vacuum modular Hamiltonian. We can compute this directly in global modes using the fact that this operator is the boost generator as in the scalar case \eqref{eqn:bulkmodH}:
\begin{equation}\label{eqn:bulkmodH_Max}
K_{0}^{\mathrm{Max.,bulk}} = 2\pi\int_{\Sigma_{\mathcal{B}}}d^{d-1}x\,\sqrt{g_{\Sigma_{\mathcal{B}}}}\,\xi^{\mu}n^{\nu}T_{\mu\nu}^{\mathrm{Max.}}
\end{equation}
along with the expectation values of the Maxwell stress-tensor computed in \eqref{eqn:stresstensor_Max_evm=0} to obtain
\begin{equation}\label{eqn:bulkmodH_Max_ev}
\Delta K_{0}^{\mathcal{P}} = \frac{\sqrt{\pi}\Gamma(d)}{2\Gamma(d+\frac{1}{2})}\theta_{0}^{2(d-1)}\left(1+\mathcal{O}(\theta_{0}^{2})\right).
\end{equation}

We can also compute the vacuum-subtracted expectation value using the Rindler modes. The Rindler Hamiltonian for Maxwell theory is given by
\begin{equation}\label{eqn:RindlerHamiltonian_bulk}
H_{R}^{\mathrm{Max.,bulk}} = \int_{\omega>0}\frac{d\omega}{2\pi}\,\omega \left(\sum_{\lambda}b_{\omega,\lambda,R}^{\dagger}b_{\omega,\lambda,R}+(\mathrm{vector\,part})\right)
\end{equation}
where we have again dropped the $S$ label on $b,b^{\dagger}$, and the Bogoliubov coefficients (between $S$ global and $S_{H}$ Rindler modes) at leading order in $\theta_{0}$ were computed in App. \ref{sec:Bogcoeffs_Max} with the result
\begin{align}\label{eqn:Bogcoeff_betasimpl_Max_final_main}
\begin{split}
\beta_{\omega,\lambda} &\sim \frac{\omega\mathcal{N}_{0,1}^{S}\mathcal{N}_{\lambda,0}^{H}V_{S^{d-2}}}{\mathcal{N}_{\omega,\lambda}^{S,R\ast}}\frac{2^{d-5}3\sqrt{d}}{\sqrt{\pi}}\theta_{0}^{(d-1)}\frac{\Gamma(\frac{d-1}{2})}{\Gamma(d)\Gamma(\frac{d-2}{2})}\left|\Gamma\left(\frac{\zeta+i\omega_{+}}{2}\right)\Gamma\left(\frac{\zeta+i\omega_{-}}{2}\right)\right|^{2}
\\ \alpha_{\omega,\lambda} &\sim -\beta_{\omega,\lambda}.
\end{split}
\end{align}
We reiterate that these were computed in a gauge-invariant way so any entropic quantity obtained from them is also gauge-invariant. The desired expectation value is thus
\begin{equation}\label{eqn:bulkmodH_Max_ev_mom}
\Delta K_{0}^{\mathcal{P}} = \int d\omega\,\omega\sum_{\lambda}\left(|\alpha_{\omega,\lambda}|^{2}+|\beta_{\omega,\lambda}|^{2}\right) = \frac{\sqrt{\pi}\Gamma(d)}{2\Gamma(d+\frac{1}{2})}\theta_{0}^{2(d-1)}\left(1+\mathcal{O}(\theta_{0}^{2})\right),
\end{equation}
where the integrals can be computed in a similar way to \eqref{eqn:modHscalarmom_int}, which agrees with our alternate calculation \eqref{eqn:bulkmodH_Max_ev}. This provides a nice check of our Bogoliubov coefficients \eqref{eqn:Bogcoeff_betasimpl_Max_final_main}.

Armed with the bulk modular Hamiltonian and the area correction, we can verify the JLMS formula 
\begin{equation}\label{eqn:JLMSphoton}
    \frac{\Delta A_{\mathcal{P}}}{4 G_N} + \Delta K_0^{\mathcal{P}} = \Delta K_0^{J}.
\end{equation}
We want to verify this up to $\mathcal{O}(\theta^{2(d-1)})$ because this is the leading order term in $\Delta K_0^{\mathcal{P}}$. To do so, we can rewrite the formula by computing the difference between the CFT and AdS bulk modular Hamiltonians given in \eqref{eqn:bulkmodH_Max_ev_mom} and \eqref{eqn:K0Jdif_Max_thetaeps=0} to ``predict" the area change as 
\begin{equation}
\begin{aligned}
    \frac{\Delta A_{\mathcal{P}}}{4 G_N}\Big|_{\text{JLMS}} = & \frac{\sqrt{\pi}\,\Gamma(\frac{d}{2}+1)}{(d-1)\Gamma\left(\frac{d+1}{2}\right)}\sin^{d-2}\theta_{0}\Big((3d-5)f_{+}(d,\theta_{0})-(d-2)\sin^{2}\theta_{0}f_{+}(d+2,\theta_{0})\Big)\\
    & - \frac{\sqrt{\pi}\Gamma(d)}{2\Gamma(d+\frac{1}{2})}\theta_{0}^{2(d-1)}+\mathcal{O}(\theta_{0}^{2d-1}),
\end{aligned}
\end{equation}
where $f_{+}$ is defined in \eqref{eqn:fdef}, and check that this prediction matches the results in Table \ref{tab:areacorrection} for every $d=3,4,5,6$.

\paragraph{Second order in $\delta \rho$.}

We can also use the Bogoliubov coefficeints to compute the second-order contribution to the vacuum-subtracted von Neumann entropy. It takes the same form as the scalar result \eqref{eqn:bulkvacsubtractEE_2ndorder}, viz.,
\begin{align}\label{eqn:bulkvacsubtractEE_Max_2ndorder_explicit}
\begin{split}
\Delta S^{\mathrm{bulk}}\left(\rho_{\Sigma_{\mathcal{B}}}^{\mathcal{P}}\right)|_{\mathcal{O}((\delta \rho)^{2})} &= \pi \int \frac{d\omega_{1}}{2\pi}\,\frac{d\omega_{2}}{2\pi}\,\sum_{\lambda_{1},\lambda_{2}}\left((\omega_{1}+\omega_{2})\frac{\mathfrak{E}(\omega_{1})\mathfrak{E}(\omega_{2})}{\mathfrak{E}(\omega_{1}+\omega_{2})}+(\omega_{1}-\omega_{2})\frac{\mathfrak{E}(-\omega_{1})\mathfrak{E}(\omega_{2})}{\mathfrak{E}(\omega_{2}-\omega_{1})}\right)
\\	&\qquad \times \left(|\alpha_{1}|^{2}|\beta_{2}|^{2}+|\beta_{1}|^{2}|\alpha_{2}|^{2}+2\alpha_{1}^{\ast}\beta_{1}\alpha_{2}\beta_{2}^{\ast}\right)
\\	&= \frac{V_{S^{d-2}}^{2}}{V_{S^{d-1}}^{2}}\frac{2^{2(d-1)}}{16\pi^{3}\Gamma(d)^{2}\Gamma(\frac{d}{2})^{2}\Gamma(\frac{d}{2}-1)^{2}}\theta_{0}^{4(d-1)}
\\	&\qquad \times \int_{0}^{\infty}d\omega_{1}\,\omega_{1}|\Gamma(i\omega_{1})|^{2}\int_{0}^{\infty}d\omega_{2}\,\omega_{2}|\Gamma(i\omega_{2})|^{2}
\\	&\qquad \times \left((\omega_{1}+\omega_{2})\frac{\mathfrak{E}(\omega_{1})\mathfrak{E}(\omega_{2})}{\mathfrak{E}(\omega_{1}+\omega_{2})}+(\omega_{1}-\omega_{2})\frac{\mathfrak{E}(-\omega_{1})\mathfrak{E}(\omega_{2})}{\mathfrak{E}(\omega_{2}-\omega_{1})}\right)
\\	&\qquad \times \int_{0}^{\infty}d\tilde{\lambda}_{1}\,\left|\frac{\Gamma(\zeta+1-i\tilde{\lambda}_{1})}{\Gamma(-i\tilde{\lambda}_{1})}\right|^{2}\left|\Gamma\left(\frac{\zeta+i\omega_{1+}}{2}\right)\right|^{2}\left|\Gamma\left(\frac{\zeta+i\omega_{1-}}{2}\right)\right|^{2}
\\	&\qquad \times \int_{0}^{\infty}d\tilde{\lambda}_{2}\,\left|\frac{\Gamma(\zeta+1-i\tilde{\lambda}_{2})}{\Gamma(-i\tilde{\lambda}_{2})}\right|^{2}\left|\Gamma\left(\frac{\zeta+i\omega_{2+}}{2}\right)\right|^{2}\left|\Gamma\left(\frac{\zeta+i\omega_{2-}}{2}\right)\right|^{2}
\\	&= -\sqrt{\pi}\frac{\Gamma(2d-1)}{2\Gamma(2d-\frac{1}{2})}\theta_{0}^{4(d-1)},
\end{split}
\end{align}
where again the integrals over $\omega_{1,2}$ and $\tilde{\lambda}_{1,2}$ can be evaluated in a similar way to the integrals in \eqref{eqn:modHscalarmom_int}.
Once again, we see that this agrees with the CFT calculation in \eqref{eqn:deltaS_J_Max}\footnote{Here we set $\theta_{\epsilon}=0$ again. Notice that the entropy of a photon state relative to the vacuum matches exactly the scalar result \eqref{eqn:matchscalarsrel} by setting $\Delta_O$ to be $d-1$, which is the conformal dimensional of the U(1) current.} 
\begin{equation}\label{eqn:matchmaxwellsrel}
    -S_{\rm rel}^{\rm AdS}\left(\rho_{\Sigma_{\mathcal{B}}}^{\cal P}\Big|\rho_{\Sigma_{\mathcal{B}}}^{0}\right) = -\sqrt{\pi}\frac{\Gamma(2d-1)}{2\Gamma(2d-\frac{1}{2})}\theta_{0}^{4(d-1)} = - S_{\rm rel}^{\rm CFT}\left(\rho_{\mathcal{B}}^{J}\big|\rho_{\mathcal{B}}^{0}\right),
\end{equation}
so that the FLM fomula for the photon excited state is confirmed by combining \eqref{eqn:matchmaxwellsrel} and \eqref{eqn:JLMSphoton}, viz.,
\begin{equation}\label{eqn:FLMverify}
\begin{aligned}
        \Delta S^{\rm CFT} & \equiv (\Delta S^J)_{1 \ll \Delta_{\mathrm{gap}} \ll C_{T}} = (\Delta K^{J}_0)_{1 \ll \Delta_{\mathrm{gap}} \ll C_{T}} - S_{\rm rel}^{\rm CFT}\left(\rho_{\mathcal{B}}^{J}\big|\rho_{\mathcal{B}}^{0}\right)  \\
        & = \frac{\Delta A_{\cal P}}{4 G_N} + \Delta K^{\cal P}_0 - S_{\rm rel}^{\rm AdS}\left(\rho_{\Sigma_{\mathcal{B}}}^{\cal P}\Big|\rho_{\Sigma_{\mathcal{B}}}^{0}\right) =   \frac{\Delta A_{\cal P}}{4 G_N} + \Delta S^{\rm AdS}_{\rm bulk}.
\end{aligned}
\end{equation}
Note that both $\Delta K_0^{\mathcal{P}}$ and $S_{\rm rel}^{\rm AdS}$ are computed in the bulk of the Rindler wedge so that we define their sum as $\Delta S^{\rm AdS}_{\rm bulk}$, and we ignore the possible edge mode contribution to which we now turn.

\section{Electromagnetic edge modes}
\label{sec:edge}

It remains to understand the edge mode sector of Maxwell theory in AdS and the contribution of these modes to the vacuum-subtracted  entanglement entropy considered in \S\ref{sec:photonEE}. 
In this section, we explicitly compute the edge mode contribution to the vacuum-subtracted entanglement entropy, and find 
\begin{equation}\label{eqn:DeltaSedgevanish}
    \Delta S^{\rm AdS}_{\rm edge} = 0.
\end{equation}
This means \eqref{eqn:FLMverify} is really 
\begin{equation}
     \Delta S^{\rm CFT}  =  \frac{\Delta A_{\cal P}}{4 G_N} + \Delta S^{\rm AdS} = \frac{\Delta A_{\cal P}}{4 G_N} + \Delta S^{\rm AdS}_{\rm bulk} + \Delta S^{\rm AdS}_{\rm edge} ,
\end{equation}
so the FLM formula \eqref{eqn:qHRRT} indeed holds.

Let us start by giving a physical understanding of the terminology ``edge mode". 
Consider, for example, an AdS-Rindler wedge.
To fully specify a state in the wedge, we need not only the bulk degrees of freedom, such as bulk particle excitations, but also the boundary degrees of freedom, such as the electric flux on the entangling surface. 
Therefore, the Rindler wedge Hilbert space should take the form $\mathbb{H}_{\mathrm{bulk}} \otimes \mathbb{H}_{\mathrm{edge}}$. 

The interesting effect of these edge modes on the boundary of a Rindler wedge, which we address in detail in \S\ref{sec:edgemodes}, appears when we try to glue together the two Rindler wedge Hilbert spaces to obtain the global Hilbert space.
Naively, it is tempting to take $\left(\mathbb{H}_{\mathrm{bulk},L} \otimes \mathbb{H}_{\mathrm{edge},L}\right) \otimes \left(\mathbb{H}_{\mathrm{bulk},R} \otimes \mathbb{H}_{\mathrm{edge},R}\right)$, which is typically referred to as the extended Hilbert space $\mathbb{H}_{\mathrm{ext}}$ \cite{Buividovich:2008gq,Donnelly:2011hn,Casini:2013rba,Radicevic:2014kqa,Donnelly:2014fua,Donnelly:2015hxa,Huang:2014pfa,Ghosh:2015iwa,Soni:2015yga,Soni:2016ogt,Blommaert:2018rsf,Ball:2024hqe}, to be the global Hilbert space. 
However, the physical Hilbert space is only a subspace of $\mathbb{H}_{\mathrm{ext}}$ because the Gauss Law constraint requires that the electric flux $\Phi$ must agree on the bifurcation surface, i.e., we must take the diagonal subspace of $\mathbb{H}_{\mathrm{edge},L} \otimes \mathbb{H}_{\mathrm{edge},R}$. 
To be more precise, we define the flux operator $\hat{\Phi}_{\mathbf{x},L/R}$ on each side of the Rindler bifurcation surface $\gamma_{\rm ext}$, and impose the condition 
\begin{equation}\label{eqn:GaussLaw_flux}
\hat{\Phi}_{\mathbf{x},L}\ket{\Theta}_{\mathrm{phys}}^{\mathrm{global}} = \hat{\Phi}_{\mathbf{x},R}\ket{\Theta}_{\mathrm{phys}}^{\mathrm{global}}.
\end{equation}
This implies that $\hat{\Phi}_{\mathbf{x},L/R}$ forms a center for the L/R bulk algebra since it commutes with all operators acting on the  L/R wedge bulk Hilbert space, and hence we can label states by their bulk quantum numbers $\mathfrak{n}$ and their electric flux $\Phi$
\begin{equation}\label{eqn:physicalH}
\mathbb{H}_{\text{phys}}^{\rm global} =  \text{Span} \{\ket{\mathfrak{n}_{L},\Phi}_{L}\ket{\mathfrak{n}_{R},\Phi}_{R}   \}_{\forall \mathfrak{n}_L,\mathfrak{n}_R,\Phi } . 
\end{equation}
The physical Hilbert space on a global slice does not factorise due to gauge constraints so we need the embedding in the extended Hilbert space to define reduced density matrices.

It is evident from \eqref{eqn:physicalH} why the edge modes can contribute to entanglement entropy. 
We can think of $\Phi$ as labelling superselection sectors.
For a given superselection sector, the Hilbert space still factorises as usual, so that all the computations regarding the density matrix and the entanglement entropy still make sense. 
In particular, for the entire discussion in  \S\ref{sec:photonEE}, we were simply working in the $\Phi=0$ sector of the physical Hilbert space. 
As a result, including the edge modes can be interpreted as summing over all of the superselection sectors, so that the density matrix for the right wedge in a global state $\ket{\psi}$ is in general 
\begin{equation}
   \rho_{{\Sigma}_{\mathcal{B}}}^{\psi} =  \int \mathcal{D}\Phi \, p^{\psi}(\Phi) \rho_{{\Sigma}_{\mathcal{B}}}^{\psi}(\Phi)\,,
\end{equation}
where $\rho_{{\Sigma}_{\mathcal{B}}}^{\psi}(\Phi)$ is actually independent of $\Phi$ so it is known from the previous discussion, and $p^{\psi}(\Phi)$ is the probability distribution associated to the superselection sectors. The entropy from the edge modes is then understood as the entropy of mixing given by
\begin{equation}\label{eqn:Sedgeingeneral}
     S_{\rm edge}^{\psi} = \int \mathcal{D}\Phi \, p^{\psi}(\Phi) \log p^{\psi}(\Phi) \,.
\end{equation}
We will give further details about the entropy of edge modes and compute explicitly the vacuum-subtracted edge entropy in \S\ref{sec:edgeentropy}. 

In order to more easily understand the edge mode sector, we will switch to the Lorenz gauge in this section. 
There are both formal and technical reasons for doing so, including convenience in the construction of the algebra and in the calculation of the Bogoliubov coefficients. We will elaborate more on this as we go along, from which the reader will be able to see why Lorenz gauge is indeed the natural choice.
We compute the wavefunctions in Lorenz gauge in \S\ref{sec:Lorenz} and perform the quantization using the Gupta-Bleuler procedure.

\subsection{Gupta-Bleuler quantization and Lorenz gauge}
\label{sec:Lorenz}

We wish to quantize the Maxwell field $A_{\mu}$ in the Lorenz gauge $\nabla^{\mu}A_{\mu}=0$. The quantization procedure in Lorenz gauge is the opposite of that in radial gauge. Recall that in \S\ref{sec:Max} we first gauge-fixed Maxwell theory in radial gauge at the classical level to obtain the reduced phase space, which we then quantized. However, this is not possible in Lorenz gauge because it is a covariant gauge condition. Instead, we must first quantize the full, un-gauge-fixed theory and then impose the Lorenz gauge as a physical condition on the Hilbert space to project onto the physical Hilbert space via the Gupta-Bleuler procedure.

The first step of the procedure is to require all physical states to satisfy the condition
\begin{equation}\label{eqn:Lorenzconstraint}
\nabla^{\mu}\hat{A}_{\mu}^{(+)}\ket{\Theta}_{\mathrm{phys}} = 0,
\end{equation}
where $\hat{A}_{\mu}^{(+)}$ is the positive frequency part of $\hat{A}_{\mu}$, or equivalently all physical matrix elements of the Lorenz gauge operator vanish: $\bra{\widetilde{\Theta}}_{\mathrm{phys}}\nabla^{\mu}\hat{A}_{\mu}\ket{\Theta}_{\mathrm{phys}}=0$. The second step is to obtain the quotient by all null states, i.e, those whose Gupta-Bleuler inner product is zero, which correspond to the pure gauge modes.

To compute the solutions of the Maxwell equations that satisfy the Lorenz gauge condition, we can use the fact that the radial gauge solutions have been obtained in \S\ref{sec:Sgenphoton}, so we only need to find the gauge transformation taking us between the two gauges. 
Consider the gauge transformation from the radial gauge $(\mathbb{A}^{(\mathrm{r})}_r=0)$ to the Lorenz gauge $(\nabla^{\mu}\mathbb{A}^{(\rm L)}_{\mu}=0)$
\begin{equation}
    \mathbb{A}^{(\text{r})}_{\mu}=\partial_{\mu} \Psi + \mathbb{A}^{(\text{L})}_{\mu}.
\end{equation}
We can solve for $\Psi$ by computing the divergence on both sides of the above equation to obtain 
\begin{equation}\label{eq:Lorenzgauge}
    \nabla^{\mu} \partial_{\mu} \Psi =\nabla^{\mu} \mathbb{A}^{(\text{r})}_{\mu}=\partial_t \mathbb{A}^{(\text{r})t} + \frac{1}{\sqrt{g_{S^{d-1}}}} \partial_{\phi_{i}} \left(\sqrt{g_{S^{d-1}}} \mathbb{A}^{(\text{r})\phi_{i}}\right),
\end{equation}
where we used the fact that $\mathbb{A}_{r}^{(\text{r})}=0$ to drop the $\partial_r \mathbb{A}^{(\text{r})r}$ term in the divergence. This is an inhomogeneous Laplace equation. We can find the explicit form of $\Psi$ using the standard techniques for obtaining solutions of inhomogeneous differential equations from homogeneous ones. Focusing on the scalar part of the gauge field,\footnote{The Lorenz gauge-fixing condition is invariant under $SO(d)$ rotations of $S^{d-1}$ so we can still split the gauge field into scalar and vector parts.} for the $(n,\ell,\mathfrak{m})=(0,1,\mathbf{0})$ mode in \eqref{eqn:lowestenergywavefn}, we find
\begin{equation}
    \Psi(t,r,\Omega)= \mathcal{N}_{0,1}^{S} e^{-i \Omega_{0,1}^S t} \left(\mathcal{A}^{S}_{0,1}(r)+\frac{3 r}{(d-2)\left(1+r^2\right)^{\frac{d-1}{2}}}\right)Y_{1,\mathbf{0}}^{(d-1)}(\Omega)
\end{equation}
so the wavefunction in the Lorenz gauge takes the particularly simple form
\begin{equation}\label{eq:ALorenzGlobal}    \left(\mathbb{A}_{t,0,1,\mathbf{0}}^{(\mathrm{L})S},\mathbb{A}_{r,0,1,\mathbf{0}}^{(\mathrm{L})S},\mathbb{A}_{\phi_i,0,1,\mathbf{0}}^{(\mathrm{L})S}\right)= \frac{3\mathcal{N}_{0,1}^S}{(d-2)}e^{- i \Omega_{0,1} t}\frac{r}{(1+r^2)^{\frac{d-1}{2}}}\left( -iY_{1,\mathbf{0}}^{(d-1)},\frac{Y_{1,\mathbf{0}}^{(d-1)}}{r(1+r^2)}, \partial_{\phi_i}Y_{1,\mathbf{0}}^{(d-1)}\right).
\end{equation}
Upon inspecting the gauge transformation $\Psi$, we can see that $\Psi\sim r^{-d}$ and $\partial_{\mu}\Psi=\mathbb{A}_{\mu}^{(\rm L)}-\mathbb{A}_{\mu}^{(\rm r)}$ decays at least as fast as $r^{-d}$. This reflects the fact that the bulk gauge symmetry is mapped to a boundary global symmetry via the AdS/CFT dictionary \cite{Witten:1998qj}.\footnote{Indeed, this asymptotic behavior of $\Psi$ can be proved, in the special case of a gauge transformation between the Lorenz gauge and the radial gauge, by 
\begin{equation}
    \lim_{r\rightarrow \infty} r^{d} \nabla^{2} \Psi = \lim_{r\rightarrow \infty} r^{d} \left(\partial_t g^{tt} \mathbb{A}_t^{(\rm r)} + \frac{1}{\sqrt{g_{S^{d-1}}}} \partial_{\phi_{i}} \left(\sqrt{g_{S^{d-1}}} g^{\phi_{i}\phi_{i}} \mathbb{A}_{\phi_{i}}^{(\rm r)}\right) \right) = {\nabla}^{\partial}_{\nu} J^{\nu} = 0.
\end{equation}
Since $\Psi$ is only a function of $r^2$, this means $\nabla^{2} \Psi$ decays at least as fast as $r^{-d-2}$, and hence $\Psi \sim r^{-d}$. 
Importantly, such an asymptotic behavior means $\lim_{r\rightarrow \infty}r^{d-2} (\mathbb{A}^{(\rm L)}_{\mu}-\mathbb{A}^{(\rm r)}_{\mu}) =0$ and the boundary current is not modified by the bulk gauge choice.} Furthermore, there are no residual gauge transformations that respect this boundary condition at infinity and are smooth at $r=0$.\footnote{The divergence of $\Psi$ at $r=0$ will in the end lead to non-normalised modes so we drop it.}
This translates to not having to quotient out the null states in global coordinates.

We can now canonically quantize $\mathbb{A}_{\mu}^{(L)S}$ by writing it in terms of creation and annihilation operators in the usual way. We have not provided the unphysical modes that solve the Maxwell equations, but get projected out by the condition \eqref{eqn:Lorenzconstraint}. The Hilbert space obtained from the Gupta-Bleuler quantization procedure is isomorphic to the one we obtained from quantizing the reduced phase space in radial gauge. All states obtained by acting with creation operators in either quantization procedure are distinguished by their eigenvalues under the action of isometries of AdS so the isomorphism is trivial, i.e., $a_{n,\ell,\mathfrak{m}}^{(\mathrm{L})\dagger}\ket{0} \leftrightarrow a_{n,\ell,\mathfrak{m}}^{\dagger}\ket{0}$. Therefore, the state of interest here is $a_{0,1,\mathbf{0}}^{(\mathrm{L})\dagger}\ket{0}$ which is the image of $\ket{\mathcal{P}}$ under this isomorphism.

A similar analysis can be performed for the Rindler modes in Lorenz gauge. However, it is more complicated because there are now residual gauge degrees of freedom and the PMC boundary conditions need to be imposed. The gauge transformation function $\tilde{\Psi}$ from radial Rindler gauge to Lorenz gauge is given by
\begin{equation}\label{eqn:radialtoLorenz_Rindler}
\tilde{\Psi}(\tau,\rho,u)=\mathcal{N}_{\omega,\lambda} e^{-i \omega \tau} H_{\lambda}(u) \left(\mathpzc{A}_{\omega,\lambda}(\rho)+\frac{\rho^{2-d}}{(d-2)} \tilde{h}_{\omega,\lambda}  \right)
\end{equation}
with $\mathpzc{A}_{\omega,\lambda}$ given in \eqref{eqn:Alambda_soln}. The function $\tilde{h}_{\omega,\lambda}$ is defined by
\begin{equation}
   \tilde{h}_{\omega,\lambda} = g^S_{\omega,\lambda}(\rho) + c_{\omega,\lambda} \psi_{\omega,\lambda}(\rho)
\end{equation}
where $g^S_{\omega,\lambda}$ is given in \eqref{eqn:gS_final} and $\psi_{\omega,\lambda}(\rho)$ is the residual gauge degree of freedom satisfying 
\begin{equation}
\begin{aligned}
    & \nabla^{2}\left(e^{-i\omega\tau}H_{\lambda}(u)\rho^{2-d}\psi_{\omega,\lambda}(\rho)\right)=0 \\
    & \qquad \Rightarrow \ \psi_{\omega,\lambda}(\rho) = \rho^{-2} \left(1-\frac{1}{\rho^2}\right)^{-i \frac{\omega}{2}} {}_2F_1\left(\frac{1}{2}\left(\zeta+2-i\omega_{-}\right),\frac{1}{2}\left(\zeta+2-i\omega_{-}\right),\zeta+2;\frac{1}{\rho^2}\right) .
\end{aligned}
\end{equation}
Armed with the gauge function $\tilde{\Psi}(\tau,\rho,u)$, we find the asymptotic behavior that it again decays as $\rho^{-d}$ for large $\rho$, so based on the same argument as in the global case, the gauge transformation still does not affect the boundary current. The proportionality coefficient $c_{\omega,\lambda}$ depends on the boundary conditions which we will come to next. 

From \eqref{eqn:radialtoLorenz_Rindler}, we find the Rindler gauge field in Lorenz gauge has the wavefunctions
\begin{equation}\label{eqn:RindlerMaxwellsoln_Lorenz}
\begin{aligned}
    \mathpzc{A}_{\tau,\omega,\lambda,}^{(\mathrm{L})} & = -i \mathcal{N}_{\omega,\lambda}e^{-i\omega\tau} H_{\lambda}(u) \rho^{2-d}\left(\frac{\omega}{(d-2)}\tilde{h}_{\omega,\lambda}(\rho)+\frac{2}{\omega}\left(1-\frac{1}{\rho^2}\right)\frac{d g^{S}_{\omega,\lambda}(\rho)}{d\rho^{-2}}\right),\\
    \mathpzc{A}_{\rho,\omega,\lambda}^{(\mathrm{L})} & = \mathcal{N}_{\omega,\lambda}e^{-i\omega\tau} H_{\lambda}(u) \frac{\rho^{1-d}}{1-\rho^2}\left(g^S_{\omega,\lambda}(\rho)-\frac{2}{d-2}\left(1-\frac{1}{\rho^2}\right)\frac{d \tilde{h}_{\omega,\lambda}(\rho)}{d\rho^{-2}} \right) ,\\
    \mathpzc{A}_{u,\omega,\lambda}^{(\mathrm{L})} & = \frac{\mathcal{N}_{\omega,\lambda}}{d-2}e^{-i\omega\tau} \partial_{u}H_{\lambda}(u)\rho^{2-d}\tilde{h}_{\omega,\lambda}(\rho). 
\end{aligned}
\end{equation}
We impose the PMC boundary conditions \eqref{eqn:PMC} so we must examine their behavior in the $\rho\rightarrow 1$ limit
\begin{equation}
\begin{aligned}
    \mathpzc{A}_{\omega,\lambda}^{(\mathrm{L})} \propto (\rho^2-1)^{\frac{i \omega}{2}}\frac{i\mathcal{N}_{\omega,\lambda} \Gamma(-i\omega) \Gamma(\zeta)}{4\zeta\Gamma\Big(\frac{1}{2}(\zeta-i\omega_{-})\Big)\Gamma\Big(\frac{1}{2}(\zeta-i\omega_{+})\Big) } \left((\zeta - 2 i  \omega)+ \frac{c_{\omega,\lambda}}{(\zeta-i \omega_{-})(\zeta +i \omega_{+})} \right) + \text{c.c.},
\end{aligned}
\end{equation}
so we choose $c_{\omega,\lambda}=\frac{\lambda-\omega^2}{8\zeta}+\frac{\zeta \omega^2}{2(\lambda-\omega^2)}$ such that the coefficient in the large parenthesis is purely real, which allows us to satisfy the PMC boundary conditions.

As we found for radial gauge, this leads to discretized Rindler frequencies with spacing $\frac{2\pi}{|\log({2\epsilon})|}$. The bulk Rindler Hilbert space (in the $\Phi=0$ sector) can now be constructed in the same way as in global coordinates by imposing the Lorenz gauge condition \eqref{eqn:Lorenzconstraint} on the physical Hilbert space. Henceforth, we will drop all $(\mathrm{L})$ labels as it is understood that this section is entirely in Lorenz gauge.

Notice that we used the same normalization factors for the Lorenz gauge wavefunctions as the radial gauge ones. This is because, 
thankfully, the normalization factors receive no modification from gauge transformations, which can be proved as follows.
Recall the definition of the normalization factor from the inner product 
\begin{equation}
    \mathcal{N}^2=\langle A, A \rangle = i\int_{\Sigma} d^{d}x\,\sqrt{g_{\Sigma}} \, n_{\nu}\left(A^{*}_{\mu} F_{A}^{\mu \nu} -  A_{\mu} F_{A}^{*,\mu \nu} \right),
\end{equation}
where $\Sigma$ is a Cauchy slice and  $n^{\mu}$ is the time-like unit vector normal to $\Sigma$.\footnote{In principle, there is a gauge fixing term in the inner product definition in Lorenz gauge. However, when we consider the normalization constants, we focus on the physical modes so that the gauge fixing term is always zero.}
Consider the gauge field wavefunction $A_{\mu}$ in one gauge and $\tilde{A}_{\mu}=A_{\mu}+\nabla_{\mu}\Psi$ in another gauge. Then the inner product gives 
\begin{equation}\label{eqn:normalizationchange}
    \langle A_2, A_2 \rangle = \langle A, A\rangle - i  \int_{\partial \Sigma}d^{d-1}x\, \sqrt{g_{\partial \Sigma}}  \left(\Psi^{*} F_{A}^{\mu\nu} - \Psi F^{*,\mu\nu}_{A}\right)n_{\mu}s_{\nu} = \langle A, A\rangle,
\end{equation}
where $s^{\nu}$ is the unit normal vector to $\partial\Sigma$ in $\Sigma$. This follows from the Maxwell equations and the fall-off of the wavefunctions at the asymptotic boundary, and in the Rindler case one also needs to use the PMC boundary conditions at the horizon.

\subsection{Edge mode construction}
\label{sec:edgemodes}
From now on, we will focus on the edge mode sector of the theory adapting to AdS spacetime the formalism that was developed in \cite{Blommaert:2018rsf} for edge modes in the Rindler decomposition of Minkowski spacetime. 
Let us start with some discussion at the classical level about the variational principle and boundary conditions. 
The variation of the Maxwell action \eqref{eqn:Maxwellaction} (with a gauge-fixing term) will, in general, produce a boundary term 
\begin{equation}
	\delta S_{\rm Maxwell} = \int_{\mathcal{M}}\delta A^{\nu} (\text{EOM})_{\nu} +  \int_{\partial \mathcal{M}} \delta A_{\nu}n_{\mu} (F^{\mu\nu} +g^{\mu\nu} \nabla^{\sigma} A_{\sigma} ) \,.
\end{equation}
In our special case of decomposing global AdS into left and right Rindler wedges, we require the boundary terms from the variation on both wedges to match. This is usually referred to as the Gauss Law constraint and the corresponding gluing boundary condition 
\begin{equation}\label{eqn:Gausslawgluing}
    \left. n^\mu F_{\mu t}^R\right|_{\partial \Sigma}(\mathbf{x})=\left.n^\mu F_{\mu t}^L\right|_{\partial \Sigma }(\mathbf{x})  \,.  
\end{equation}
There is also a boundary condition from the Lorenz gauge condition
\begin{equation}\label{eqn:Lorenzgluing}
    \left. n^\mu A_{\mu}^R\right|_{\partial \Sigma }(\mathbf{x})=\left.n^\mu A_{\mu}^L\right|_{\partial \Sigma }(\mathbf{x})\,.
\end{equation}
An easy choice to satisfy these boundary conditions is to set both sides of the above equations to zero. 
This is exactly the PMC boundary condition \eqref{eqn:PMC} imposed on the non-zero frequency modes, which we refer to as bulk modes, leading to a well-defined theory for a single wedge. 
From the global perspective, however, $A_{\mu}$ and $F_{\mu\nu}$ on the entangling surface, which is a surface in the bulk of global AdS, should be allowed to vary as long as \eqref{eqn:Gausslawgluing} and \eqref{eqn:Lorenzgluing} hold. 
From the Rindler boundary condition point of view, the edge modes describe an improvement of the PMC boundary conditions.
These (classical) boundary conditions \eqref{eqn:Gausslawgluing} and \eqref{eqn:Lorenzgluing}, gluing both sides of the entangling surface $\gamma_{\mathrm{ext}}$, will be promoted to operator relations valid when acting on the physical Hilbert spaces. 

\paragraph{Gauge-invariant algebra.}

To understand this at the operator level, we need to first construct the operators and their algebra. 
The simplest way to proceed is to look at the gauge-invariant algebra of observables. This algebra is generated by the Wilson lines for any curve $C$
\begin{equation}\label{eqn:Wilsonline}
\mathcal{W}_{C} = e^{i\int_{C} A}
\end{equation}
and the flux of $\pi_{\mu}$ through any codimension-2 surface $\Omega$
\begin{equation}\label{eqn:fluxop}
\Phi_{\Omega} = \int_{\Omega}d^{d-2}x\,\sqrt{g_{\Omega}}\,n^{\mu}\pi_{\mu},
\end{equation}
where $n^{\mu}$ is the normal vector to $\Omega$. Promoting these to operators $\widehat{\Phi}_{\Omega}$, $\widehat{\mathcal{W}}_{C}$, the commutation relations are
\begin{equation}\label{eqn:gaugeinvalg}
[\widehat{\mathcal{W}}_{\mathcal{C}},\widehat{\Phi}_{\Omega}] = -\widehat{\mathcal{W}}_{C}\theta(\Omega \cap C) \implies \left[\int_{C} \hat{A},\widehat{\Phi}_{\Omega}\right] = -i\theta(\Omega \cap C),
\end{equation}
where $\theta(\Omega \cap C) = 0$ if $\Omega \cap C = \emptyset$ and $\theta(\Omega \cap C) = \pm 1$ if $\Omega \cap C \neq \emptyset$ with the plus sign when $C$ and $\Omega$ have the same orientation and the minus sign otherwise. By taking $C$ and $\Omega$ small, we recover the usual non-gauge invariant algebra between $\hat{A}_{\mu}$ and $\hat{\pi}_{\mu}$. However, this algebra is not consistent with the PMC boundary conditions which set $A_{\rho}|_{\gamma_{\mathrm{ext}}} = n_{\rho}\pi^{\rho}|_{\gamma_{\mathrm{ext}}} = 0$, that is, the algebra will be violated on the horizon. These Wilson lines must have a contribution with delta-function support on the entangling surface $\gamma_{\mathrm{ext}}$ that satisfies the correct algebra, which are the boundary (edge) modes.
For later convenience, we also define the ``local" flux algebra where we consider the region $\Omega$ on $\gamma_{\mathrm{ext}}$ with infinitesimal size centered around the point $\mathbf{x}$ and the Wilson line $C$ piercing $\gamma_{\mathrm{ext}}$ at point $\mathbf{y}$,\footnote{We will abuse terminology and refer to both $\mathcal{W}_{C}$ and $\int_{C}\hat{A}$ as Wilson line operators.} then \eqref{eqn:gaugeinvalg} becomes 
\begin{equation}\label{eqn:fluxalgebra}
    \left[\widehat{\Phi}_{\mathbf{x}},\int_{C(\mathbf{y})}\hat{A}\right] = \text{sign}_{\mathcal{C}} (2\pi)^{d-1} i\delta^{(d-1)}(\mathbf{x}-\mathbf{y})\,.
\end{equation}
With these operators defined, the boundary conditions \eqref{eqn:Gausslawgluing} and \eqref{eqn:Lorenzgluing} at the quantum level should be regarded as constraints on the physical Hilbert space, namely any physical state $\ket{\Theta}_{\mathrm{phys}}^{\mathrm{global}}$ must satisfy
\begin{equation}\label{eqn:operatorbdycond}
    \left(\widehat{\Phi}_{\mathbf{x},R}-\widehat{\Phi}_{\mathbf{x},L}\right)\ket{\Theta}_{\mathrm{phys}}^{\mathrm{global}} \approx 0, \quad \left(\int_{C(\mathbf{y}),R}\hat{A}-\int_{C(\mathbf{y}),L}\hat{A}\right)\ket{\Theta}_{\mathrm{phys}}^{\mathrm{global}} \approx 0,
\end{equation}
where $\approx$ means modulo null states. These operators act on the edge Hilbert space and the constraints \eqref{eqn:operatorbdycond} define the physical Hilbert space by relating the modes in the left and right edge Hilbert space, as we shall see. In particular, we will construct the Hilbert spaces $\mathbb{H}_{\mathrm{edge},L}$ and $\mathbb{H}_{\mathrm{edge},R}$ and the operator algebras acting on them.

Let us perform a mode decomposition of the flux and Wilson line operators
\begin{equation}\label{eqn:fluxandwilsonlinemode}
\begin{aligned}
    \widehat{\Phi}_{\mathbf{x}} = \sum_{\lambda} H_{\lambda}(\mathbf{x}) q_{\lambda} , \quad \int_{C(\mathbf{y})}\hat{A} = \sum_{\lambda} H_{\lambda}(\mathbf{y}) a_{\lambda},
\end{aligned}
\end{equation}
where $a_{\lambda}$ and $q_{\lambda}$ are Hermitian, and $H_{\lambda}$ is again the eigenfunction on $H^{d-1}$. 
To see what the gluing boundary condition \eqref{eqn:Gausslawgluing} and \eqref{eqn:Lorenzgluing} implies for the mode decomposition, we express the flux and Wilson line operator in both the left and right wedges\footnote{We use the right wedge as a reference, meaning that both the Wilson line and the normal direction of the surface points into the right wedge. 
Note that we relate the left and right states by CRT conjugation. This means we use $H_{\lambda}$ on the right wedge with $H^{*}_{\lambda}$ on the left. The extra sign for the left wedge is because the CRT conjugation will switch the direction of the Wilson line.}
\begin{equation}\label{eq:fluxandWLexpansion}
\begin{aligned}
   &\widehat{\Phi}_{\mathbf{x},R} = \sum_{\lambda} H_{\lambda}(\mathbf{x}) q_{\lambda,R} , \quad \int_{C(\mathbf{y}),R}\hat{A} = \sum_{\lambda} H_{\lambda}(\mathbf{y}) a_{\lambda,R}\\
   &\widehat{\Phi}_{\mathbf{x},L} = \sum_{\lambda} H^{*}_{\lambda}(\mathbf{x}) q_{\lambda,L} , \quad \int_{C(\mathbf{y}),L}\hat{A} = \sum_{\lambda} H^{*}_{\lambda}(\mathbf{y}) a_{\lambda,L}.
\end{aligned}
\end{equation}
Then the flux algebra \eqref{eqn:fluxalgebra} leads to the following canonical commutation relations 
\begin{equation}\label{eq:qacomm}
    [a_{\lambda,R},q_{\lambda^{\prime},R}]= 2\pi i \delta^{(d-1)}(\lambda-\lambda^{\prime}) , \quad [a_{\lambda,L},q_{\lambda^{\prime},L}]= - 2\pi i \delta^{(d-1)}(\lambda-\lambda^{\prime}),
\end{equation}
and the constraints \eqref{eqn:operatorbdycond} give 
\begin{equation}\label{eq:qaleftright}
    (q_{\lambda,R} - q_{\lambda^{\ast},L})\ket{\Theta}_{\mathrm{phys}}^{\mathrm{global}}  \approx 0, \quad (a_{\lambda,R} - a_{\lambda^{\ast},L})\ket{\Theta}_{\mathrm{phys}}^{\mathrm{global}}  \approx 0.
\end{equation}

\paragraph{Edge states.}

With these mode operators, we can explicitly construct the eigenstates. We define the state $|0\rangle _{L,R}$ to be annihilated by $q_{L,R}$. Any eigenstate $\ket{\epsilon}$ of $q$ with eigenvalue $\epsilon$ takes the form $e^{i a \epsilon}\ket{0}$ due to the commutation relation \eqref{eq:qacomm}. One can show that any physical state must be a linear combination of the following basis states 
\begin{equation}
   | \Phi \rangle =  |\Phi \rangle_L | \Phi \rangle_R =  \prod_{\lambda} e^{ - i a_{\lambda^{\ast},L} \varepsilon_{\lambda}} e^{i a_{\lambda,R} \varepsilon_{\lambda} } |0\rangle_L  |0\rangle_R,
\end{equation}
which are eigenstates of the flux operator 
\begin{equation}\label{eqn:gaussoneigenstate}
\widehat{\Phi}_{\mathbf{x}} |\Phi\rangle = \Phi(\mathbf{x}) |\Phi\rangle \equiv \bigg(\sum_{\lambda} \varepsilon_{\lambda} H_{\lambda}(\mathbf{x}) \bigg)|\Phi\rangle
\end{equation}
where the function $\Phi(\mathbf{x}) = \sum_{\lambda} \varepsilon_{\lambda} H_{\lambda}(\mathbf{x})$ is nothing but the profile of electrostatic flux on the entangling surface. These states satisfy the Gauss Law constraint
\begin{equation}\label{eqn:GaussLawmodes}
q_{\lambda,R}  |\Phi\rangle = \varepsilon_{\lambda} |\Phi\rangle = q_{\lambda^{\ast},L} |\Phi\rangle,
\end{equation}
and the Lorenz gauge constraint\footnote{The approximately equal $\approx$ means this is a null state that is orthogonal to all physical states. To verify this, we can consider the inner product $\bra{\epsilon_{\lambda_{0}}}_{L}\bra{\epsilon_{\lambda_{0}}}_{R}(a_{\lambda,R} - a_{\lambda^{\ast},L})\ket{\Phi}$ and this evaluates to $\delta({\varepsilon_{\lambda}-\varepsilon_{\lambda_0}}) \langle0|_L \langle0|_R (a_{\lambda,R}-a_{\lambda^{*},R}) |0\rangle_L |0\rangle_R$ which is zero due to the left-right symmetry.}
\begin{equation}
    (a_{\lambda,R} - a_{\lambda^{\ast},L})\ket{\Phi} \approx  0\,.
\end{equation}

In summary, the edge Hilbert spaces are given by
\begin{align}
\begin{split}
\mathbb{H}_{\mathrm{edge},L} = \text{Span}& \{\ket{\Phi}_{L}\}, \qquad \mathbb{H}_{\mathrm{edge},R} = \text{Span} \{\ket{\Phi}_{R}\}, 
\\  \mathbb{H}_{\mathrm{edge,\;phys}} &= \text{Span} \{\ket{\Phi}\} \subset \mathbb{H}_{\mathrm{edge},L} \otimes \mathbb{H}_{\mathrm{edge},R},
\end{split}
\end{align}
where the physical edge Hilbert space is embedded in the extended edge Hilbert space.

As an example, the edge part of the global vacuum state is a linear combination of these eigenstates
\begin{equation}\label{eqn:TFD_edge}
    |0\rangle_{\rm global,\;edge} \sim \int \mathcal{D}\Phi \sqrt{p(\Phi)} \,|\Phi \rangle_{L}  |\Phi \rangle_{R} 
\end{equation}
where $p(\Phi)$ can be regarded as a probability distribution of different profiles of electric flux.
As will be shown more explicitly later, it turns out that $p(\Phi)$ is a Gaussian distribution obtained from the edge sector of the Rindler boost operator.  

\paragraph{Wavefunctions and boundary conditions.} 

Thus far, we have been rather abstract, framing our entire discussion in terms of electric fluxes and Wilson lines. In fact, all of the previous analysis of edge modes holds very generally and is not special to AdS or its Rindler decomposition. To be more concrete, one can ask where these electric fluxes on the entangling surface and the conjugate Wilson lines appear in our solutions to the Maxwell equations. The answer is that the edge modes are the $\omega=0$ Rindler solutions, which violate the PMC boundary conditions \eqref{eqn:PMC}. This is intuitive because the edge modes live on the bifurcation surface so they are infinitely redshifted from the perspective of any bulk observer, and hence static.

In the Lorenz gauge, the two edge modes are (1) $A^{(\Phi)}_{\mu}$ the physical mode with non-zero radial electric flux $\Phi$ on the entangling surface, and (2) $A^{(G)}_{\mu}$ the residual pure gauge mode controlling the boundary condition of gauge field on the Rindler horizon, i.e, this mode corresponds to `large' gauge transformations satisfying Lorenz gauge that do not fall off at the entangling surface.
To compute the edge entanglement entropy, we need the Bogoliubov coefficients, denoted as  $\gamma^{(\Phi)}$  and $\gamma^{(G)}$,  respectively, between the lowest excitation in the global Hilbert space that we have been considering and the two edge modes, which means we need their explicit wavefunctions.

From \eqref{eqn:RindlerMaxwellsoln_Lorenz}, we can find the two solutions to the Maxwell equations for $\omega=0$: 
\begin{equation}\label{eq:Aedgeflux}
   \mathpzc{A}_{\mu}^{(\Phi)} = \delta_{\mu\tau}\sum_{\lambda}\frac{\mathcal{N}_{\rm edge}}{\rho^{d-2}}\, {}_2F_1\left(\frac{1}{2}(\zeta-i \tilde{\lambda}),\frac{1}{2}(\zeta + i \tilde{\lambda}),1+\zeta;\frac{1}{\rho^2}\right) H_{\lambda}(\mathbf{x}),
\end{equation}
and
\begin{equation}
\begin{aligned}
    \mathpzc{A}^{(G)}_{\mu} & = \partial_{\mu} \Psi(\rho,\mathbf{x})  \\
    \Psi &= \sum_{\lambda}\frac{\lambda}{d}\frac{\mathcal{N}_{\rm edge}}{\rho^{d}} \,{}_2F_1\left(\frac{1}{2}(2+\zeta-i \tilde{\lambda}),\frac{1}{2}(2+\zeta + i \tilde{\lambda}),2+\zeta;\frac{1}{\rho^2}\right) H_{\lambda}(\mathbf{x}),
\end{aligned}
\end{equation}
where 
\begin{equation}\label{eqn:normedge}
\mathcal{N}_{\rm edge}=\frac{ \Gamma\left(\frac{1}{2}(\zeta-i \tilde{\lambda})\right)\Gamma\left(\frac{1}{2}(\zeta+i \tilde{\lambda})  \right) }{2\Gamma(1+\zeta)\log(2\epsilon)},
\end{equation}
%
which is designed to satisfy the following brick wall boundary condition for $A^{(\Phi)}$ and $A^{(G)}$ respectively
\begin{equation}\label{eq:edgebdycond}
\begin{aligned}
    F_{\rho\tau} &\xrightarrow{\rho \to 1}  \frac{\log(\rho^2-1)}{\log(2\epsilon)}  H_{\lambda}(\mathbf{x}) \ \Leftrightarrow F_{\rho\tau}(\rho=1+\epsilon) = H_{\lambda}(\mathbf{x}) ,  \\
    \Psi & \xrightarrow{\rho \to 1} \frac{\log(\rho^2-1)}{\log(2\epsilon)}  H_{\lambda}(\mathbf{x}) \ \Leftrightarrow \Psi (\rho=1+\epsilon) = H_{\lambda}(\mathbf{x}) . \\
\end{aligned}    
\end{equation}
It is worth mentioning that these are similar to the boundary conditions imposed for the flat space edge mode in \cite{Blommaert:2018rsf}. 

Before proceeding, a comment about the discrete versus continuous spectrum is in order. 
In the above discussion and as will be clarified below, we isolate a particular mode and discuss its contribution to entropy. 
Such an operation is formally not correct unless we have a discrete spectrum.
To wit, as mentioned above, the edge mode entropy requires specifying the value of the gauge field on the boundary, which can only be implemented with a regulator.
Therefore, in the discussion below, we are obligated to go from a continuous spectrum to a discrete spectrum given as
\begin{equation}\label{eqn:freq2}
\omega_{S_{H},n} = \frac{2\pi n}{|\log({2\epsilon})|}, \qquad  n \in \mathbb{Z}_{\geq 0}. 
\end{equation}
The above discussions are purely for the right wedge. 
One can perform a similar discussion on the left wedge and the constraints \eqref{eqn:Gausslawgluing} and \eqref{eqn:Lorenzgluing} will provide gluing conditions. 
Concretely, we can write down a complete expansion of (the scalar part of) the gauge field $\hat{A}$ in Lorenz gauge 
\begin{equation}\label{eqn:Aexpansion}
    \hat{A}^{S_H}= \sum_{I\in \{L,R\}} \left\{ \sum_{\lambda} \bigg[ \Big(\sum_{\omega_n} \mathpzc{A}^{S_H}_{\omega_n,\lambda,I} b_{\omega_n,\lambda,I}  + \text{c.c.} \Big)  + \mathpzc{A}^{(\Phi)}_{\lambda,I} q_{\lambda,I} + \mathpzc{A}^{(G)}_{\lambda,I} a_{\lambda,I} \bigg] \right\},
\end{equation}
and plug this expansion into the gluing boundary conditions \eqref{eqn:Gausslawgluing} and \eqref{eqn:Lorenzgluing}. 
The constraints on the bulk modes have already been taken care of by the PMC boundary conditions, so we only need to focus on the edge modes. 
We find that the $q_{\lambda,I}$ and $a_{\lambda,I}$ mode operators are exactly the same as the ones appearing in the mode decompositions of the flux/Wilson-line operators in \eqref{eqn:fluxandwilsonlinemode}:
they satisfy $q_{\lambda,R} = q_{\lambda^{*},L}$, $a_{\lambda,R} = - a_{\lambda^{*},L}$ which is a direct consequence of \eqref{eqn:Gausslawgluing} and \eqref{eqn:Lorenzgluing},
along with the commutation relations $ [a_{\lambda,R},q_{\lambda,R}]= [q_{\lambda,L},a_{\lambda,L}]= i$. 
To verify this, we need the canonical commutation relation $[\hat{A}_{\mu}(x),\hat{\pi}^{\nu}(y)]=i \delta^{\nu}_{\mu} (2\pi)^{(d-1)}\delta^{(d-1)}(x-y)$ and the inner product  $\langle  \mathpzc{A}^{(\Phi)}_{\lambda,I},  \mathpzc{A}^{(G)}_{\lambda^{\prime},I^{\prime}} \rangle = (-1)^{I}\delta_{I I^{\prime}} 2\pi \delta(\lambda- \lambda^{\prime})$ where $(-1)^{I}$ is $1$ for $I=R$ and $-1$ for $I=L$,\footnote{Such inner products between a non-gauge mode and a gauge mode are relatively easy to evaluate. We will give an example evaluating a similar inner product later.} following from  the boundary condition \eqref{eq:edgebdycond}.

Let us end with some brief comments about how this analysis would differ if one worked in radial gauge.
We have seen that after the mode decomposition, the $(q_{\lambda},a_{\lambda})$ algebra is particularly useful for constructing the edge mode states. 
For the radial gauge, one would naively get a trivial (Rindler) radial Wilson line operator, so that the $a_{\lambda}$ operator may not be easily defined. However, it is not actually trivial because we only use `small' gauge transformations which vanish at the entangling surface to go to radial gauge. This means we can not set $A_{\rho}=0$ on the entangling surface with such gauge transformations. Instead, $A_{\rho}$ can be delta-function localized on the entangling surface, leading to a non-trivial radial Wilson line \cite{Harlow:2015lma}. However, from the reduced phase space perspective, it is not obvious how to construct such an extended phase space in radial gauge or how to find the wavefunctions for these edge modes. These are the formal and technical reasons, respectively, for switching to Lorenz gauge and it would interesting to understand how to explicitly construct these edge modes in radial gauge.


\subsection{Entanglement entropy of edge modes}\label{sec:edgeentropy}

We finally have all the ingredients we need to define excited states for the edge modes and compute their entanglement entropy. The full TFD state is given by a tensor product of the bulk part, which has been already captured in \eqref{eqn:MinkvactoRindlerTFD_Max}, and the edge part which is discussed in \eqref{eqn:TFD_edge}, viz.,
\begin{equation}\label{eqn:MinkvactoRindlerTFD_Max_edge}
\ket{0} = \int \mathcal{D}\Phi\, \bigotimes_{\mathfrak{c} \in {S_{H},V_{H}}} \sqrt{p_{\mathfrak{c}}(\Phi)} \bigotimes_{\omega,\lambda_{\mathfrak{c}}}\sqrt{1-e^{-2\pi\omega}}\sum_{n}e^{-\pi E_{\mathfrak{c},n}}\ket{\Phi;n,\omega,\lambda_{\mathfrak{c}}^{\ast}}_{L,\mathfrak{c}}\ket{\Phi;n,\omega,\lambda_{\mathfrak{c}}}_{R,\mathfrak{c}}.
\end{equation}
Henceforth, we will drop the scalar/vector index $\mathfrak{c}$ because the global state $\ket{\mathcal{P}}$ we consider can written purely in terms of scalar Rindler mode excitations so we will only focus on $\mathfrak{c}=S_{H}$. In order to write the excited state $\ket{\mathcal{P}}$ in terms of left and right Rindler modes in the extended Hilbert space, we need the Bogoliubov coefficients for the edge modes.

\paragraph{Bogoliubov coefficients.}
There are two methods one can use to compute the Bogoliubov coefficients: (1) from the two-point function with one operator sent to the boundary and (2) the inner product between the global and Rindler wavefunctions. Since the computation of Bogoliubov coefficients for the edge modes is quite subtle, we will present both methods and show that they agree.

The first method was used to obtain the bulk Bogoliubov coefficients in App. \ref{sec:Bogcoeffs_Max}. 
For the second method, one can expand the Maxwell field as an operator $\hat{A}$ in both global and Rindler coordinates as follows 
\begin{equation}
\begin{aligned}
     \hat{{A}}^{\mu}_{\rm global} & = \sum_{n,\ell} \mathbb{A}^{\mu}_{n,\ell,m} a_{n,\ell,\mathfrak{m}} + \text{c.c.} , \\
     \hat{{A}}^{\mu}_{\rm Rindler} & = \sum_{I\in\{L,R\}}\sum_{\omega,\lambda} \left[\left(\mathpzc{A}^{\mu}_{\omega,\lambda,I} b_{\omega,\lambda,\mathfrak{m}} + \text{c.c.}\right) + \mathpzc{A}^{(\Phi)}_{\lambda,I} q_{\lambda,I} + \mathpzc{A}^{(G)}_{\lambda,I} a_{\lambda,I}\right].
\end{aligned}
\end{equation}
To obtain a particular global excitation, for example the one of interest $a^{\dagger}_{0,1,\mathbf{0}}$, as a linear combination of Rindler creation and annihilation operators $b,b^{\dagger}$, we perform a projection via the inner product 
\begin{align}\label{eqn:a010frominnerwithedge}
    a^{\dagger}_{0,1,\mathbf{0}} &= \langle \mathbb{A}_{0,1,\mathbf{0}},\hat{{A}}_{\rm global}\rangle = 
    \langle \mathbb{A}_{0,1,\mathbf{0}},\hat{{A}}_{\rm Rindler}\rangle 
    \\  &= \sum_{I\in \{L,R\}} \sum_{\omega,\lambda} \bigg(\underbrace{\langle \mathbb{A}_{0,1,\mathbf{0}}, \mathpzc{A}_{\omega,\lambda,I} \rangle}_{\alpha_{0,1,\mathbf{0};\omega,\lambda,I}} b_{\omega,\lambda,I} + \text{c.c}\bigg) + \underbrace{\langle \mathbb{A}_{0,1,\mathbf{0}}, \mathpzc{A}^{(\Phi)}_{\lambda,I} \rangle}_{\gamma_{0,1,\mathbf{0};\lambda,I}^{(\Phi)}}q_{\lambda,I} + \underbrace{\langle \mathbb{A}_{0,1,\mathbf{0}}, \mathpzc{A}^{(G)}_{\lambda,I} \rangle}_{\gamma_{0,1,\mathbf{0};\lambda,I}^{(G)}}a_{\lambda,I}, \nonumber
\end{align}
where we are again dropping the $S$ label since we only consider the scalar modes. This means that the Bogoliubov coefficients can be extracted from the inner product, which is defined as an integral on the $t=\tau=0$ Cauchy slice $\Sigma$.\footnote{This inner product turns out to be too difficult to compute for the bulk modes, which is why we used the two-point function method in App. \ref{sec:Bogcoeffs_Max}. However, the inner product turns out to be doable for the edge modes.} To simplify notation, we will denote $\gamma_{\lambda}^{(\Phi)} \equiv \gamma_{0,1,\mathbf{0};\lambda,R}^{(\Phi)}$ and $\gamma_{\lambda}^{(G)} \equiv \gamma_{0,1,\mathbf{0};\lambda,R}^{(G)}$, while the relationship between $L$ and $R$ edge Bogoliubov coefficients will be derived later.
Also we remark that the second step of \eqref{eqn:a010frominnerwithedge} relies heavily on the covariance of the gauge field, naturally pointing us to the Lorenz gauge.

From the two-point function method, $\gamma_{\lambda}^{(G)}$ can be computed using the same techniques as in App.~\ref{sec:Bogcoeffs_Max}, 
but with two changes coming from the normalization factor for the Rindler mode being changed to $\mathcal{N}_{\rm edge}$ and the fact that the discretization of frequencies \eqref{eqn:freq2} affects the Fourier decomposition in the $\tau$ direction.
One finds 
\begin{equation}\label{eq:edgebogoliubov}
\begin{aligned}
    \gamma_{\lambda}^{(G)} & \sim i  \frac{\mathcal{N}_{0,1}^{S} \mathcal{N}_{\tilde{\lambda},0}^{H} V_{S^{d-2}}}{ |\log(2\epsilon)| \mathcal{N}_{\rm edge} }  
    \frac{2^{d-4}3\sqrt{d}}{\sqrt{\pi}}\frac{\Gamma(\frac{d-1}{2})}{\Gamma(d)\Gamma(\frac{d-2}{2})}\left|\Gamma\left(\frac{\zeta+i\tilde{\lambda}}{2}\right)\Gamma\left(\frac{\zeta-i\tilde{\lambda}}{2}\right)\right|^{2} \theta_{0}^{d-1}\\
    & =  i \frac{3\sqrt{d} \mathcal{N}_{0,1}^{S}\mathcal{N}_{\lambda,0}^{H}V_{S^{d-2}}}{4(d-1)\Gamma(\zeta+1)} \lambda \left|\Gamma\left(\frac{\zeta+i\tilde{\lambda}}{2}\right)\Gamma\left(\frac{\zeta - i\tilde{\lambda}}{2}\right)\right| \theta_{0}^{d-1}
\end{aligned}
\end{equation}
with the details given in App.~\ref{sec:Bogcoeffs_edge}.

Next, we present the calculation of the edge Bogoliubov coefficient via the inner product. 
The edge Bogoliubov coefficients $\gamma_{\lambda}^{(G)}$ for the `large' gauge modes are actually the important ones for the edge entropy, and these turn out to be easy to compute via the inner product\footnote{The unit vectors $n_{\tau}$ and $s_{\rho}$ are not exactly well defined on the entangling surface, but we can define them on the brick wall and take the $\epsilon\rightarrow 0$ limit.}
\begin{equation}
    \gamma_{\lambda}^{(G)} = \langle \mathbb{A}_{0,1,\mathbf{0}}, \mathpzc{A}^{(G)}_{\lambda} \rangle = i  \int_{\gamma_{\mathrm{ext}}}d^{d-1}x\, \sqrt{g_{\gamma_{\mathrm{ext}}}}  \Psi_{\lambda} F_{0,1,\mathbf{0}}^{\mu\nu\ast} n_{\mu} s_{\nu}, 
\end{equation}
where $s_{\rho}$ is the unit normal vector to $\gamma_{\mathrm{ext}}$ in $\Sigma_{\mathcal{B}}$, we used that the Rindler mode is a pure gauge mode with no field strength, and integrated by parts using that the global field strength satisfies the Maxwell equations (the contribution from the asymptotic boundary vanishes). 
Using the boundary condition \eqref{eq:edgebdycond}, one can evaluate the integral over $H^{d-1}$ and obtain 
\begin{equation}
     \gamma_{\lambda}^{(G)} \sim   i \frac{3\sqrt{d} \mathcal{N}_{0,1}^{S}\mathcal{N}_{\lambda,0}^{H}V_{S^{d-2}}}{4(d-1)\Gamma(\zeta+1)} \lambda \left|\Gamma\left(\frac{\zeta+i\tilde{\lambda}}{2}\right)\Gamma\left(\frac{\zeta - i\tilde{\lambda}}{2}\right)\right| \theta_{0}^{d-1}.
\end{equation}

Curiously enough, we find that the other set of edge Bogoliubov coefficients, although harder to compute, is suppressed by $|\log(2\epsilon)|$. 
The details are explained in App.~\ref{sec:Bogcoeffs_edge}, but the idea is that we can write down the inner product definition as 
\begin{equation}
\begin{aligned}
     \gamma_{\lambda}^{(\Phi)}&=\langle \mathbb{A}_{0,1,\mathbf{0}}, \mathpzc{A}^{(\Phi)}_{\lambda} \rangle = -i \int_{\Sigma_{\mathcal{B}}}d^{d}x\, \sqrt{g_{\Sigma_{\mathcal{B}}}}\,  n_{t} \mathbb{A}_{\mu,0,1,\mathbf{0}} F^{t\mu\ast}_{\lambda},
\end{aligned}
\end{equation}
and evaluate in the AdS-Rindler coordinates. The transverse direction integral is roughly the same as the integral for $\gamma^{(G)}$ above, but the remaining $\rho$ integral is convergent even in the $\epsilon\rightarrow 0$ limit.
However, recall that in the definition of $F^{\mu\nu}_{\lambda}$ the normalization factor $\mathcal{N}_{\rm edge}$ is suppressed by $|\log(2\epsilon)|$,  giving us 
\begin{equation}
    \gamma^{(\Phi)}_{\lambda} \sim \frac{\#}{ |\log(2\epsilon)|} ,
\end{equation}
which vanishes in the $\epsilon\rightarrow 0$ limit.

The above calculations are for the right wedge, and one can obtain the left wedge coefficients similarly. 
In fact, since both Bogoliubov coefficients computations are localized on the horizon, it is not difficult to argue, by directly computing the inner product, that
\begin{equation}\label{eq:edgebogoleftright}
    \quad \gamma^{(\Phi)}_{\lambda,R}= - \gamma^{(\Phi)}_{\lambda^*,L} ,  \quad \gamma^{(G)}_{\lambda,R}= - \gamma^{(G)}_{\lambda^*,L} 
\end{equation}
given the analogous boundary condition \eqref{eq:edgebdycond} for the left wedge.\footnote{There are some subtleties here because the $\tau,\rho$ coordinates are defined differently for the left wedge compared to the left wedge. In particular, near the horizon $\partial_{\rho_L} \sim -\partial_r$ while $\partial_{\rho_R}\sim \partial_r$, and $\partial_{\tau_{L}}\sim -\partial_{t}$ while $\partial_{\tau_{R}}\sim \partial_{t}$. 
Therefore, the left boundary condition is actually $F_{\tau\rho,L}=H_{\lambda}^{\ast}(u)$ for the flux mode and $\Psi_{\tau\rho,L}=-H_{\lambda}^{\ast}(u)$ for the gauge mode.
}
Recall that for the bulk Bogoliubov coefficients, the condition $a_{0,1,\mathbf{0}}|0\rangle = 0$ was used to relate those for the left wedge to those for the right \eqref{eqn:vacann_photon}. Here we consider this relation again, but now it is used to check the validity of the edge Bogoliubov coefficient relations \eqref{eq:edgebogoleftright}.\footnote{As another check, one can consider the commutation relation $[a_{0,1,\mathbf{0}},a_{0,1,\mathbf{0}}^{\dagger}] = 1$ and make sure that the existence of the edge piece does not spoil this, which is indeed this case. }
Given the global vacuum written as the Rindler TFD state \eqref{eqn:MinkvactoRindlerTFD_Max_edge} and the Bogoliubov expansion of $a_{0,1,\mathbf{0}}$, we find 
\begin{equation}\label{eqn:a010withedge}
\begin{aligned}
    a_{0,1,\mathbf{0}} |0\rangle & = \Big(\sum_{I\in\{L,R\}}\sum_{\lambda} \gamma^{(\Phi)}_{\lambda} q_{\lambda,I} +  \gamma^{(G)}_{\lambda} a_{\lambda,I} \Big)
    \int \mathcal{D}\Phi \, \sqrt{p(\Phi)} |\Phi\rangle_L |\Phi\rangle_R \, + \text{bulk piece}  \\
   & = \int \mathcal{D}\Phi  \sqrt{p(\Phi)} \, \sum_{\lambda} \left\{\gamma^{(\Phi)}_{\lambda,R} \left( q_{\lambda,R}  -  q_{\lambda^{\ast},L} \right) +   \gamma^{(G)}_{\lambda,R} \left(a_{\lambda,R} - a_{\lambda^{\ast},L} \right) 
   \right\} |\Phi\rangle_L |\Phi\rangle_R \\
    & \approx 0 , 
\end{aligned}
\end{equation}
where the bulk piece vanishes due to the relation between bulk Bogoliubov coefficients given in \eqref{eqn:vacann_photon}, and the edge piece gives a null state due to the constraints \eqref{eq:qaleftright}. 


\paragraph{Entanglement entropy.}
We are now ready to compute the edge mode contribution to the entanglement entropy.
Let us begin with the vacuum state. 
Starting from the TFD state \eqref{eqn:MinkvactoRindlerTFD_Max_edge} and 
tracing out the left wedge, we obtain the (normalized) reduced density matrix
\begin{equation}\label{eqn:fullvac_reducedrho}
   \rho^0_{{\Sigma}_{\mathcal{B}}} =  \int \mathcal{D}\Phi \, p(\Phi) \bigotimes_{\omega,\lambda}(1-e^{-2\pi\omega})\sum_{n}e^{-2\pi E_{n}}\ket{\Phi;n,\omega,\lambda}\bra{\Phi;n,\omega,\lambda}.
\end{equation}
As explained at the beginning of \S\ref{sec:edge}, the existence of the distribution $p(\Phi)$ will lead to the (vacuum) edge entropy, see \eqref{eqn:Sedgeingeneral}. 
We can determine  $p(\Phi)$ as follows. By expressing the density matrix in terms of the corresponding modular Hamiltonian 
\begin{equation}
    \rho^0_{{\Sigma}_{\mathcal{B}}} =  \frac{e^{ - K_0}}{\mathcal{Z}} = \frac{e^{ - K_0^{\rm bulk}}}{\mathcal{Z}_{\mathrm{bulk}}} \frac{e^{ - K_0^{\rm edge}}}{{\mathcal{Z}_{\mathrm{edge}}}}
\end{equation}
where $\mathcal{Z}$ is the partition function, we see that $e^{-K_0^{\rm bulk}}/\mathcal{Z}_{\mathrm{bulk}}$ will lead to the bulk vacuum density matrix discussed for the scalar in \S\ref{sec:scalar_bulkEE} and for the photon in \S\ref{sec:photonEE_bulk}. However, for the photon, there is a new piece $e^{ - K_0^{\rm edge}}$ coming from the electric fluxes, which leads to the probability distribution 
\begin{equation}
    p(\Phi) = \frac{e^{-K_0^{\rm edge}(\Phi)}}{\mathcal{Z}_{\mathrm{edge}}}. 
\end{equation}

To compute $ K_0^{\rm edge}$ for AdS-Rindler, we start from the usual definition of $K_0$ in terms of the boost generator for Rindler, but we must be very careful about the boundary terms. In particular, we focus on the $\omega=0$ sector, since it is only this mode for which we have non-trivial boundary conditions (fluxes). 
Analogous to that given in \cite{Blommaert:2018rsf} for Minkowski space, the evaluation of the time-independent contribution to \eqref{eqn:bulkmodH_Max} will be reduced to a boundary term using the Maxwell equations
\begin{equation}
\begin{aligned}
    K_{0}^{\mathrm{edge}} & = 2\pi\int_{\Sigma_{\mathcal{B}}}d^{d-1}x\,\sqrt{g_{\Sigma_{\mathcal{B}}}}\,\xi^{\mu}n^{\nu}T_{\mu\nu}^{\mathrm{Max.}}\Big|_{\omega=0} \\
    & = \pi  \int_{{\Sigma}_{\mathcal{B}}} d^{d-1}x\,\sqrt{g} F_{\tau\mu} F^{\tau\mu}\Big|_{\omega=0} = \pi \int_{\gamma_{\mathrm{ext}}} d^{d-2}x\,\sqrt{g} A_{\tau} F^{\tau\rho}\Big|_{\omega=0}.
\end{aligned}
\end{equation}
Notice that the `large' gauge mode $\mathpzc{A}_{\mu}^{(G)}$ gives no contribution to $K_{0}^{\mathrm{edge}}$ because it has vanishing field strength. Plugging in the explicit expression \eqref{eq:Aedgeflux} of $\mathpzc{A}_{\tau}$ and $F^{\tau\rho}$ for the flux mode $\mathpzc{A}^{(\Phi)}$, we find\footnote{Since the flux profile $\Phi(\mathbf{x})$ should be real, we have $\varepsilon_{\lambda}= \varepsilon_{\lambda^{\ast}}^{*}$. }
\begin{equation}
    K_{0}^{\mathrm{edge}} = 2 \pi\sum_{\lambda} 2 \frac{q_{\lambda,R} q_{\lambda^{\ast},R}}{\lambda |\log(2 \epsilon)| } , \ \ \text{and} \ \ \bra{\Phi}K_{0}^{\mathrm{edge}}\ket{\Phi} = 2 \pi\sum_{\lambda}   \frac{ 2 |\varepsilon_{\lambda}|^2}{\lambda |\log(2 \epsilon)| }
\end{equation}
for $|\Phi\rangle$ satisfying \eqref{eqn:gaussoneigenstate} where again $\Phi=\sum_{\lambda} \varepsilon_{\lambda} H_{\lambda}$. 
As a result, the reduced density matrix for the global vacuum state \eqref{eqn:fullvac_reducedrho} can be fully specified as 
\begin{equation}
\begin{aligned}
     \rho^0_{\Sigma_{\mathcal{B}}} & =\frac{1}{\mathcal{Z}} \int \bigg(\prod_{\lambda} d\varepsilon_\lambda  d\varepsilon_{\lambda}^{*} e^{-\frac{4\pi |\varepsilon_{\lambda}|^2}{\lambda \log(2 \epsilon) } }\bigg)|\Phi\rangle\langle\Phi| \,  e^{- K_{0}^{\rm bulk}}\\
     & = \frac{1}{\mathcal{Z}_{\mathrm{edge}}} \int \bigg(\prod_{\lambda} d\varepsilon_\lambda  d\varepsilon_{\lambda}^{*} e^{-\frac{4\pi |\varepsilon_{\lambda}|^2}{\lambda \log(2 \epsilon) } } \bigg) \bigotimes_{\omega,\lambda}(1-e^{-2\pi\omega})\sum_{n}e^{-2\pi E_{n}}\ket{\Phi;n,\omega,\lambda}\bra{\Phi;n,\omega,\lambda}.
\end{aligned}
\end{equation}
This leads to a non-zero vacuum entanglement entropy for the edge modes, which is in fact divergent for $\epsilon \to 0$.

Given the result for the vacuum density matrix, we can also consider the excited state density matrix. 
Note that the creation operator is 
\begin{equation}
    a^{\dagger}_{0,1,\mathbf{0}} = a^{\dagger, \text{bulk}}_{0,1,\mathbf{0}} - \Big(\sum_{I\in\{L,R\}}\sum_{\lambda} \gamma^{(\Phi)}_{\lambda} q_{I,\lambda} +  \gamma^{(G)}_{\lambda} a_{I,\lambda} \Big) 
\end{equation}
where the bulk creation operator has been used in \S\ref{sec:photonEE_bulk}.
When acting on the vacuum state, the edge piece and the bulk piece will act differently, and we highlight the edge piece action because it will provide a new contribution correcting the probability distribution $p(\Phi)$
\begin{equation}
\begin{aligned}
	& a^{\dagger}_{0,1,\mathbf{0}}|0\rangle - a^{\dagger,\text{bulk}}_{0,1,\mathbf{0}}|0\rangle \\
 & 
 =  -\frac{1}{\sqrt{\cal Z}}
 \sum_{\lambda^{\prime}} \gamma_{\lambda^{\prime}}^{(G)} (a_{R,\lambda^{\prime}}- a_{L,\lambda^{\prime,*}})   \int \prod_{\lambda} d\varepsilon_\lambda  d\varepsilon_{\lambda}^{*} 
 e^{ - \frac{ 2 \pi |\varepsilon_{\lambda}|^2}{\lambda \log(2 \epsilon) } }   e^{i \varepsilon_{\lambda} a_{\lambda,R}} e^{- i \varepsilon_{\lambda} a_{\lambda^*,L}} \sum_{\mathfrak{n}_{L.R}}|0;\mathfrak{n}_L\rangle_{L} |0;\mathfrak{n}_R\rangle_{R} \\
& =   \frac{i}{\sqrt{\cal Z}} \sum_{\lambda^{\prime}} \gamma_{\lambda^{\prime}}^{(G)} \int \prod_{\lambda} d\varepsilon_\lambda  d\varepsilon_{\lambda}^{*} e^{ - \frac{ 2 \pi |\varepsilon_{\lambda}|^2}{\lambda \log(2 \epsilon) } } \partial_{\epsilon_{\lambda^{\prime}}} \Big( e^{i \varepsilon_{\lambda} a_{\lambda,R}} e^{- i \varepsilon_{\lambda} a_{\lambda^*,L}} \sum_{\mathfrak{n}_{L.R}}|0;\mathfrak{n}_L\rangle_{L} |0;\mathfrak{n}_R\rangle_{R} \Big) \\
 & =  -\frac{i}{\sqrt{\cal Z}} \sum_{\lambda^{\prime}}   \gamma_{\lambda^{\prime}}^{(G)}  \int \prod_{\lambda}  d\epsilon_\lambda  d\varepsilon_{\lambda}^{*} \bigg( \partial_{\lambda^{\prime}|\lambda}  e^{ - \frac{2 \pi |\varepsilon_{\lambda}|^2}{\lambda \log(2 \epsilon) } } \bigg)   e^{i \varepsilon_{\lambda} a_{\lambda,R}} e^{- i \varepsilon_{\lambda} a_{\lambda^*,L}} \sum_{\mathfrak{n}_{L.R}}|0;\mathfrak{n}_L\rangle_{L} |0;\mathfrak{n}_R\rangle_{R} \,. 
\end{aligned}
\end{equation}
where we use $\mathfrak{n}_{L,R}$ as a shorthand notation for the bulk quantum number $n,\omega,\lambda$ and an integrate-by-parts trick has been used so that  $\partial_{\lambda^{\prime}|\lambda}$ is defined so that 
\begin{equation}
\begin{aligned}
	 \partial_{\lambda^{\prime}|\lambda}  e^{-\frac{2\pi |\varepsilon_{\lambda}|^2}{\lambda \log(2 \epsilon) } }  = \begin{cases}  e^{-\frac{2\pi |\varepsilon_{\lambda}|^2}{\lambda \log(2 \epsilon) } } &  \text{ if } \lambda \neq  \lambda^{\prime} \\ 
\partial_{\varepsilon_{\lambda}} e^{-\frac{2\pi |\varepsilon_{\lambda}|^2}{\lambda \log(2 \epsilon) } }  & \text{ if } \lambda = \lambda^{\prime}
\end{cases}
\end{aligned}
\end{equation}
Note that the second case leads to a $|\log(2\epsilon)|$ supressed probability which will be important later. 

Then the density matrix for the excited state can be acquired by tracing out the left side
\begin{equation}
	\rho^{\mathcal P}_{\Sigma_{\cal B}} = \rho^{0}_{\Sigma_{\cal B}} + \delta \rho_{\rm bulk} + \delta \rho_{\rm edge} , 
\end{equation}
where $\delta \rho_{\rm edge}$ is specified to be 
\begin{equation}
	 \delta \rho_{\rm edge} =  \frac{1}{{\cal Z}} \sum_{\lambda^{\prime}}  \Big| \gamma_{\lambda^{\prime}}^{(G)} \Big|^2  \int \prod_{\lambda}  d\varepsilon_\lambda  d\varepsilon_{\lambda}^{*} \bigg(  \partial_{\lambda^{\prime}|\lambda} e^{ - \frac{2 \pi |\varepsilon_{\lambda}|^2}{\lambda \log(2 \epsilon) } } \bigg)^2   e^{i \varepsilon_{\lambda} a_{\lambda,R}} e^{-K_0^{\rm bullk}}e^{- i \varepsilon_{\lambda}^{*} a_{\lambda^{*},R}} \,. 
\end{equation}
Therefore, the edge contribution to $\Delta S$ at first-order in $\delta\rho$ is simply 
\begin{equation}
\Delta S_{\rm edge}(\rho^{\mathcal P}_{\Sigma_{\cal B}})|_{\mathcal{O}(\delta \rho)} = - \text{Tr}(\delta \rho_{\rm edge} \log \rho^0_{\Sigma_{\cal B}}),
\end{equation}
which is expressed explicitly by
\begin{align}
\begin{split}
\Delta S_{\rm edge}(\rho^{\mathcal P}_{\Sigma_{\cal B}})|_{\mathcal{O}(\delta \rho)} &= \int \frac{d\lambda}{2\pi} \Big| \gamma_{\lambda}^{(G)}\Big|^2 \frac{4 {\pi}}{|\lambda \log(2\epsilon)|} 
\\  &\propto \frac{1}{ |\log(2\epsilon)|} \int  \frac{d\lambda}{2\pi}  \lambda \left|\Gamma\left(\frac{\zeta+i\tilde{\lambda}}{2}\right)\Gamma\left(\frac{\zeta-i\tilde{\lambda}}{2}\right)\right|^2 \theta_0^{2(d-1)}.
\end{split}
\end{align}
The $\lambda$ integral is convergent because the $\Gamma$-functions decay exponentially at large $\lambda$ so we conclude that the edge contribution to the vacuum-subtracted entropy vanishes in the $\epsilon\rightarrow 0$ limit. Based on the same reasoning, it is clear that the second order contribution to $\Delta S$ is also zero as it simply involves a product of more Bogoliubov coefficients, so that
\begin{equation}
\Delta S_{\rm edge}(\rho^{\mathcal P}_{\Sigma_{\cal B}}) = 0.
\end{equation}
This feature is actually not special to the fact that we are working in AdS. For instance, we have checked that the edge modes have vanishing vacuum-subtracted entanglement entropy for Gaussian wavepacket excited states in Minkowski spacetime.

\section{Discussion}
\label{sec:disc}

Motivated by the desire to understand the generalized entropy of gravitational fluctuations in asymptotically AdS spacetimes, we studied the simpler problem of generalized entropy of photons described by Maxwell theory in pure AdS. We computed the vacuum-subtracted von Neumann entropy for a polar cap region in a $U(1)$ current excited state in any CFT dual to weakly-coupled Einstein gravity in AdS in dimensions $d>2$. Explicit expressions were obtained using an expansion in the small subregion size and known results from the CFT bootstrap for the three-point function of $U(1)$ currents with the CFT stress-tensor, along with large-$N$ factorization for four-point functions of $U(1)$ currents. Via the FLM formula, this CFT result served as the referee for our analysis of the generalized entropy of photon excited states in AdS, which requires that the two must agree.

On the AdS side, we quantized free Maxwell theory and considered the lowest energy photon excited state which is dual to the CFT primary state. We obtained the backreaction of this photon excited state on the spacetime that is needed for the change in area of the classical extremal surface due to the photon excitation. We showed that it is necessary to regularize the area of the classical extremal surface in a consistent way for both pure AdS and for the backreacted metric in order to obtain the correct area difference between the two for $\gamma_{\mathrm{ext}}$. 
Two different methods were provided, namely the introduction of a cutoff surface and dimensional regularization, which were shown to be equivalent. The vacuum-subtracted bulk entanglement entropy of the excited state was then explicitly computed using Hilbert space techniques by leveraging the fact that the classical entanglement wedge is AdS-Rindler. We found that the vacuum-subtracted generalized entropy obtained from the area difference plus the contribution from the vacuum-subtracted bulk entanglement entropy of the excited state agreed with the CFT result sans the entanglement entropy of edge modes. We then verified by explicit construction of the edge mode sector and calculation of vacuum-subtracted edge entropy that it is equal to zero, as it must for consistency with the FLM formula \eqref{eqn:qHRRT}. Along the way, we also verified that in the scalar case the FLM formula indeed holds.

\paragraph{CFT replica trick.} Our computation of the vacuum-subtracted von Neumann entropy for excited states in a CFT on the cylinder $\mathbb{R} \times S^{d-1}$ in \S\ref{sec:CFTEE_excitedstates} used the expansion of the modular Hamiltonian for the excited state in perturbations around the vacuum where the size of the perturbation was controlled by the size of our boundary subregion $\mathcal{B}$ which we took to be small. 
Alternatively, one could try to compute this vacuum-subtracted entanglement using the replica trick. This requires that we compute the $n$th moment of the density matrix $\Tr((\rho_{\mathcal{B}}^{\mathcal{O}})^{n})$ given by the path integral over the $n$-branched cover of the cylinder, branched over $\mathcal{B}$ with the operators $\mathcal{O}$ inserted at the top and bottom of every copy of the cylinder. The general formalism for computing this was developed in \cite{Sarosi:2016atx} and goes as follows. The Casini-Huerta-Myers map \eqref{eqn:hypertocyl} can be extended to this branched cover such that it maps to hyperbolic space at inverse temperature $2\pi n$, i.e., to the spacetime $\mathcal{H}_{n} = S_{n}^{1}\times H^{d-1}$. The desired trace then becomes a $2n$-point function on $\mathcal{H}_{n}$:
\begin{equation}\label{eqn:trrhon}
\Tr((\rho_{A}^{\mathcal{O}})^{n}) \propto \frac{\langle \prod_{k=0}^{n-1}\mathcal{O}(\tau_{k})\mathcal{O}(\hat{\tau}_{k})\rangle_{\mathcal{H}_{n}}}{\prod_{k=0}^{n-1}\langle\mathcal{O}(\tau_{k})\mathcal{O}(\hat{\tau}_{k})\rangle_{\mathcal{H}}},
\end{equation}
where $\tau_{k} = \pi+\theta_{0}+2\pi k$ and $\hat{\tau}_{k} = \pi-\theta_{0}+2\pi k$. The limit of small subregion $\mathcal{B}$ is an OPE limit for this correlation function which brings operators together pairwise so one might hope to compute this OPE expansion at large $N$ for the $U(1)$ current on $\mathcal{H}_{n}$, at least near $n=1$ which is relevant for the von Neumann entropy. This worked for the scalar case \cite{Sarosi:2016atx} and it would be nice to reproduce our results in \S\ref{sec:CFTEE_excitedstates} for higher-spin fields using this replica trick method.

\paragraph{No edge modes in entropy differences.}

To make all of our calculations tractable, we considered a $U(1)$ current primary state and a polar cap subregion in the CFT whose classical entanglement wedge is AdS-Rindler. However, we expect the conclusions in this work will hold for general boundary subregions and boundary current excited states, e.g., any descendant of $J_{\mu}$. In particular, our result in \S\ref{sec:edge} that the entanglement entropy of electromagnetic edge modes does not contribute to vacuum-subtracted von Neumann entropy seems to be a general fact about entanglement in Maxwell theory that follows simply from the divergence structure of edge mode entropy. We have also checked that it holds true for excited states of Maxwell theory for the left and right Rindler wedges of Minkowski spacetime. It would be nice to give an abstract quantum field theory argument that does not rely upon the specific choice of state, the subregion, or even the background spacetime. It seems likely from the arguments given in \S\ref{sec:edge} that it actually holds for the entanglement entropy difference between any two states with finite energy difference.

\paragraph{Edge modes as modular zero modes.}

The result in \S\ref{sec:edge} that the entropy of edge modes does not contribute to vacuum-subtracted von Neumann entropy for photon excited states means that the corresponding quantity in the CFT is not sensitive them. It would be very interesting to find a CFT quantity that can detect the electromagnetic edge modes of the classical extremal surface $\gamma_{\mathrm{ext}}$. One would expect that they contribute to the von Neumann entropy itself in AdS as they do in Minkowski space, however, it is not known how to compute such a quantity in the dual strongly-coupled, large-$N$ CFT in dimensions $d>2$. A possible quantity that is sensitive to edge modes in AdS is the modular zero mode considered by Faulkner and Lewkowycz \cite{Faulkner:2017vdd}. It was shown by these authors that for a boundary subregion $\mathcal{B}$ and state $\Psi$, a free scalar field in the dual asymptotically-AdS geometry that is located on the classical extremal surface can be reconstructed in the CFT by integrating over $\mathcal{B}$ against a kernel the following modular zero mode of the dual CFT operator $\mathcal{O}$:
\begin{equation}\label{eqn:modzeromode}
\mathcal{O}_{0}(x) = \int_{-\infty}^{\infty}ds\,e^{iK_{\Psi}s}\mathcal{O}(x)e^{-iK_{\Psi}s}.
\end{equation}
It was conjectured that a similar relationship should continue to hold for a $U(1)$ current in the following form
\begin{equation}\label{eqn:edgemodezeromode}
\int_{\gamma}d^{d-1}x\,\sqrt{h}\, E_{\perp} = \int_{\mathcal{B}}d^{d-1}x\, f_{J}(x)n^{\mu}(J_{\mu})_{0}(x)
\end{equation}
where $n^{\mu}$ is the normal vector to $\mathcal{B}$, $E_{\perp}$ is the electric field normal to $\gamma_{\mathrm{ext}}$ on a Cauchy slice, and $f_{J}(x)$ is a kernel built out of the bulk-to-boundary propagator. The lefthand side can be understood as the electric flux operator $\Phi_{\gamma}$ considered in \S\ref{sec:edgemodes} which acts non-trivially on the edge modes. This result was actually derived for AdS-Rindler in \cite{deBoer:2016pqk} using the fact that the causal wedge and entanglement wedge agree in that case. It would be nice to derive \eqref{eqn:edgemodezeromode} for general states and boundary subregions using the techniques in \cite{Faulkner:2017vdd}. Even better, one wishes for a local version of this formula with no integral on the lefthand side in which case the kernel $f_{J}$ would now depend on the position of $E_{\perp}$, as was found for the scalar field. This would allow one some control over the algebra of operators for the edge modes directly from the dual CFT.

\paragraph{Generalized entropy of gravitational fluctations.}
We motivated this work by our desire to understand the generalized entropy for gravitons so let us end with some final comments on this problem. 
Our calculation of CFT entanglement entropy in \S\ref{sec:CFTEE_excitedstates} readily generalizes to stress-tensor excited states dual to graviton excited states, which serves as the referee for our AdS anaylsis.
With the help of the CFT `data', we have some ideas about how to resolve the various AdS puzzles.
As mentioned in the introduction, one ambiguity comes from the definition of the quantum extremal surface: the surface can fluctuate once we allow for graviton fluctuations. 
We expect that some (at least partial) gauge fixing is required to specify the surface, but it is not yet clear how exactly this works. One may need to make choices and check whether the choice agrees with the CFT expectation. 
Another subtlety comes from how to distribute the graviton contribution in the generalized entropy between the area term and the bulk matter entropy term. Already in the JLMS paper \cite{Jafferis:2015del}, the authors proposed two, in principle equivalent, ways to distribute the graviton contribution between the two terms in \eqref{eqn:qHRRT} (in the original paper it was the modular Hamiltonian but the logic should be the same for the entropy). Our calculation for gravitons will be able to make a concrete test.
Furthermore, there is the outstanding conceptual problem of how to define graviton edge modes and, in particular, how to choose the center variables, although some progress has been made \cite{Donnelly:2016auv,Geiller:2017xad,Geiller:2017whh,Freidel:2019ees,Freidel:2020xyx,Donnelly:2020xgu,Ciambelli:2021vnn,Ciambelli:2021nmv,Ciambelli:2022cfr,Donnelly:2022kfs,Blommaert:2024cal}. 
One may hope that our result here for the vanishing of entropy for electromagnetic edge modes in vacuum-subtracted entanglement entropy continues to hold for gravitons so that this issue can be avoided in our setup.

With these puzzles clarified, one may then hope to actually derive the quantum HRRT formula \eqref{eqn:qHRRT} via the Euclidean path integral by generalizing existing techniques \cite{Lewkowycz:2013nqa,Faulkner:2013ana,Dong:2017xht}, properly taking into account gravitational fluctuations. 
We hope to report on these in the near future \cite{Colin-Ellerin24}.

\acknowledgments

The authors would like to thank Alexandre Belin, Akihiro Ishibashi, Geoffrey Penington, Pratik Rath, and Ronak Soni for useful discussions. This work was supported in part by the Berkeley Center for Theoretical Physics; by the Department of Energy, Office of Science, Office of High Energy Physics under QuantISED Award DE-SC0019380 and under contract DE-AC02-05CH11231. SC-E was supported by the National Science Foundation under Award Number 2112880. SC-E would like to thank University of Milano-Bicocca and SISSA for hospitality during the final stages of this work, and would like to thank KITP for hospitality during the workshop ``What is String Theory? Weaving Perspectives Together'' where this research was supported in part by grant NSF PHY-2309135 to the Kavli Institute for Theoretical Physics (KITP).

\appendix

\section{AdS-Rindler}
\label{sec:AdSRindler}

Here we collect some useful facts about AdS-Rindler that are used throughout this work. The transformation between global coordinates and the Rindler coordinates describing the classical entanglement wedge $\mathcal{W}[\mathcal{B}]$ is given by
\begin{align}\label{eqn:Rindlertoglobal}
\begin{split}
t &= \arctan\left(\frac{\sqrt{\rho^{2}-1}\sinh\tau}{\rho\cosh u\cosh\eta + \sqrt{\rho^{2}-1}\cosh\tau\sinh\eta}\right)
\\ r &= \sqrt{\rho^{2}\sinh^{2}u + \left(\sqrt{\rho^{2}-1}\cosh\tau\cosh\eta+\rho\cosh u \sinh\eta\right)^{2}}
\\ \theta &= \arccos\left(\frac{ \sqrt{\rho^{2}-1}\cosh\tau\cosh\eta+\rho\cosh u \sinh\eta,}{\sqrt{\rho^{2}\sinh^{2}u + \left(\sqrt{\rho^{2}-1}\cosh\tau\cosh\eta+\rho\cosh u \sinh\eta\right)^{2}}}\right)
\end{split}
\end{align}
with all angular coordinates on the common $S^{d-2}$ being the same. The parameter $\eta$ is related to the angular size of the polar cap $\theta_{0}$ by
\begin{equation}\label{eqn:etatotheta}
\cosh\eta = \frac{1}{\sin\theta_{0}}.
\end{equation}
%

\section{Eigenfunctions of Laplacian on $H^{d-1}$}
\label{sec:hyperbolicballeigfns}

Consider the hyperbolic space $H^{d-1}$ whose metric is given by
\begin{equation}\label{eqn:hyperbolicballmetric}
ds^{2} = du^{2} + \sinh^{2}u \,d\Omega_{d-2}^{2}
\end{equation}
where $u \in (0,\infty)$ and $d\Omega_{d-2}^{2}$ is the metric on $S^{d-2}$. The eigenfunctions $H_{\lambda,\ell,\mathfrak{m}}(u,\Omega)$ of the Laplacian satisfy
\begin{equation}\label{eqn:hyperbolicLapeqn}
\nabla_{H^{d-1}}^{2}H_{\lambda,\ell,\mathfrak{m}}(u,\Omega) = -\lambda H_{\lambda,\ell,\mathfrak{m}}(u,\Omega)
\end{equation}
and can be written as
\begin{equation}\label{eqn:hyperbolicballeigfn}
H_{\lambda,\ell,\mathfrak{m}}(u,\Omega) = h_{\lambda,\ell}(u)Y_{\ell,\mathfrak{m}}^{(d-2)}(\Omega),
\end{equation}
where $Y_{\ell,\mathfrak{m}}^{(d-2)}(\Omega)$ is a $(d-2)$-dimensional spherical harmonic. These eigenfunctions have been studied in \cite{10.2969/jmsj/02710082}. The eigenvalue equation takes a nicer form using the variable $z=\tanh^{2}\left(\frac{u}{2}\right)$ which gives
\begin{align}\label{eqn:hyperbolicballeigneqn_simpl}
\begin{split}
z(1-z)^{2}\partial_{z}^{2}h_{\lambda,\ell}(z)+\frac{1}{2}(1-z)&\left((d-5)z+d-1\right)\partial_{z}h_{\lambda,\ell}(z)
\\	&+\left(\tilde{\lambda}^{2}+\zeta^{2}-\frac{1}{4}\ell(\ell+d-3)\frac{(1-z)^{2}}{z}\right)h_{\lambda,\ell}(z) = 0.
\end{split}
\end{align}
In solving this eigenvalue equation, it turns out to be simpler to label solutions by $\tilde{\lambda}$ which is related to $\lambda$ by
\begin{equation}\label{eqn:lambdatotildelambda}
\lambda = \tilde{\lambda}^{2}+\zeta^{2}, \qquad \zeta \equiv \frac{d-2}{2}.
\end{equation}
The requirement that solutions be regular at the boundary $u=\infty$ implies $\tilde{\lambda} \in (0,\infty)$. The only solution to the eigenvalue equation which is regular at the origin $u=0$ is
 
\begin{equation}\label{eqn:hyperbolicballeigneqn_soln}
h_{\lambda,\ell}(u) = \mathcal{N}_{\lambda,\ell}^{H}\tanh^{\ell}\left(\frac{u}{2}\right)\sech^{2\zeta-2i\tilde{\lambda}}\left(\frac{u}{2}\right){}_{2}{F}_{1}\left(\ell+\zeta-i\tilde{\lambda},\frac{1}{2}-i\tilde{\lambda},\ell+\zeta+\frac{1}{2};\tanh^{2}\left(\frac{u}{2}\right)\right)
\end{equation}
where $\mathcal{N}_{\lambda,\ell}^{H}$ is a normalization constant. The final step is to normalize these solutions to have unit norm:
\begin{align}\label{eqn:hyperbolicip}
\begin{split}
2\pi\delta^{(d-1)}(\tilde{\lambda}-\tilde{\lambda}') &= \langle H_{\lambda,\ell,\mathfrak{m}},H_{\lambda',\ell',\mathfrak{m}'} \rangle 
\\	&= \int_{H^{d-1}}d^{d-1}x\,\sqrt{g_{H^{d-1}}}H_{\lambda,\ell,\mathfrak{m}}^{\ast}(u,\Omega)H_{\lambda',\ell',\mathfrak{m}'}(u,\Omega)
\\	&= |\mathcal{N}_{\lambda,\ell}^{H}|^{2}V_{S^{d-2}}\delta_{\ell,\ell'}\delta_{\mathfrak{m},\mathfrak{m}'}\int_{0}^{\infty}du\,|\sinh u|^{d-2}\tanh^{2\ell}\left(\frac{u}{2}\right)\sech^{2(d-2)+2i(\tilde{\lambda}-\tilde{\lambda}')}\left(\frac{u}{2}\right)
\\	&\qquad \times {}_{2}{F}_{1}\left(\ell+\zeta+i\tilde{\lambda},\frac{1}{2}+i\tilde{\lambda},\ell+\zeta+\frac{1}{2};\tanh^{2}\left(\frac{u}{2}\right)\right)
\\	&\qquad \times {}_{2}{F}_{1}\left(\ell+\zeta-i\tilde{\lambda}',\frac{1}{2}-i\tilde{\lambda}',\ell+\zeta+\frac{1}{2};\tanh^{2}\left(\frac{u}{2}\right)\right).
\end{split}
\end{align}
This integral looks very difficult to compute, but we can use the fact that the singular part of the integral comes from $u \to \infty$ so we can expand around that point and extract the coefficient of $\delta(\tilde{\lambda}-\tilde{\lambda}')$. This gives
\begin{equation}\label{eqn:hyperbolicev_norm}
\mathcal{N}_{\lambda,\ell}^{H} = \frac{2^{-\zeta}}{\sqrt{V_{S^{d-2}}}}\frac{\Gamma(\ell+\zeta-i\tilde{\lambda})\Gamma\left(\frac{1}{2}-i\tilde{\lambda}\right)}{\Gamma(\ell+\zeta+\frac{1}{2})\Gamma(-2i\tilde{\lambda})}.
\end{equation}
%

\section{Calculation of Bogoliubov coefficients}
\label{sec:Bogcoeffs}

In this appendix, we compute the Bogoliubov coefficients $\alpha_{\omega,k},\beta_{\omega,k}$ relating the annihilation and creation operators in Rindler and global coordinates \eqref{eqn:globaltoRindlerops}. We use a higher-dimensional generalization of the method used in Appendix A of \cite{Belin:2018juv}. 

\subsection{Scalar field}
\label{sec:Bogcoeffs_scalar}

Consider the following vacuum two-point function
\begin{equation}\label{eqn:Bog2ptfn}
\mathfrak{F}(t,\Omega) = \lim_{r \to \infty}r^{\Delta}\bra{0}\phi(t,r,\Omega)a_{0,0,\mathbf{0}}^{\dagger}\ket{0}.
\end{equation}
The idea is to compute it using global and Rindler coordinates and then equate the two to extract the Bogoliubov coefficients. The result in global coordinates is
\begin{equation}\label{eqn:Bog2ptfn_global}
\mathfrak{F}(t,\Omega) = C_{0}e^{-i\Delta t},
\end{equation}
and the result in Rindler coordinates is
\begin{align}\label{eqn:Bog2ptfn_Rindler}
\begin{split}
\mathfrak{F}(t,\Omega) &= \lim_{\rho \to \infty}r(\tau,\rho,\mathbf{x})^{\Delta}\bra{0}\int\frac{d\omega}{2\pi}\sum_{\lambda}\left(e^{-i\omega\tau}g_{\omega,\lambda}(\rho,\mathbf{x})b_{\omega,k}+e^{i\omega\tau}g_{\omega,\lambda}^{\ast}(\rho,\mathbf{x})b_{\omega,\lambda}^{\dagger}\right)
\\	&\qquad \qquad \int\frac{d\omega'}{2\pi}\sum_{\lambda'}\left((1-e^{-2\pi\omega'})\alpha_{\omega',\lambda'}^{\ast}b_{\omega',\lambda'}^{\dagger} + (1-e^{2\pi\omega'})\beta_{\omega',\lambda'}b_{\omega',\lambda'}\right)\ket{0}
\\	&= \lim_{\rho \to \infty}r(\tau,\rho,\mathbf{x})^{\Delta}\int\frac{d\omega}{2\pi}\sum_{\lambda}\left(e^{-i\omega\tau}g_{\omega,\lambda}(\rho,\mathbf{x})\alpha_{\omega,\lambda}^{\ast}-e^{i\omega\tau}g_{\omega,\lambda}^{\ast}(\rho,\mathbf{x})\beta_{\omega,\lambda}\right)
\end{split}
\end{align}
where we have used \eqref{eqn:AdSRindler2ptfns}. Now, let us examine the boundary limit. From \eqref{eqn:Rindlertoglobal}, one finds
\begin{equation}\label{eqn:globaltoRindler_bdylimit}
\lim_{\rho \to \infty}r(\tau,\rho,\mathbf{x})^{\Delta} = \lim_{\rho \to \infty}\rho^{\Delta}\left(\sinh^2 u + \left(\cosh u\sinh\eta+\cosh\tau\cosh\eta\right)^2\right)^{\frac{\Delta}{2}}.
\end{equation}
It is also manifest from \eqref{eqn:AdSRindlerwavefn} and \eqref{eqn:AdSRindlerradialwavefn} that
\begin{equation}\label{eqn:r_bdylimit}
\lim_{\rho \to \infty}\rho^{\Delta}g_{\omega,\lambda}(\rho,\mathbf{x}) = \mathcal{N}_{\omega,\lambda}H_{\lambda}(\mathbf{x}).
\end{equation}
Plugging these into \eqref{eqn:Bog2ptfn_Rindler} gives
\begin{align}\label{eqn:Bog2ptfn_Rindler2}
\begin{split}
\mathfrak{F}(t,\Omega) &= \left(\sinh^2 u + (\cosh u\sinh\eta+\cosh\tau\cosh\eta\big)^2\right)^{\frac{\Delta}{2}}
\\	&\qquad \int\frac{d\omega}{2\pi}\sum_{\lambda}\left(\mathcal{N}_{\omega,\lambda}e^{-i\omega\tau}H_{\lambda}(\mathbf{x})\alpha_{\omega,\lambda}^{\ast}-\mathcal{N}_{\omega,\lambda}^{\ast}e^{i\omega\tau}H_{\lambda}^{\ast}(\mathbf{x})\beta_{\omega,\lambda}\right).
\end{split}
\end{align}
We also need to take the boundary limit of $t$
\begin{equation}\label{eqn:tvarphi_bdylimit}
\lim_{\rho \to \infty}t(\tau,\rho,\mathbf{x}) = \arctan\left(\frac{\sinh\tau}{\cosh\tau\sinh\eta+\cosh u\cosh\eta}\right) \equiv t(\tau,\mathbf{x}).
\end{equation}
Equating the desired two-point function in global and Rindler coordinates gives
\begin{align}\label{eqn:Rindler=global2ptfn}
\begin{split}
C_{0}e^{-i\Delta t(\tau,\mathbf{x})} &= \left(\sinh^2 u + (\cosh u\sinh\eta+\cosh\tau\cosh\eta)^2\right)^{\frac{\Delta}{2}}
\\	&\qquad \int\frac{d\omega}{2\pi}\sum_{\lambda}\left(\mathcal{N}_{\omega,\lambda}e^{-i\omega\tau}H_{\lambda}(\mathbf{x})\alpha_{\omega,\lambda}^{\ast}-\mathcal{N}_{\omega,\lambda}^{\ast}e^{i\omega\tau}H_{\lambda}^{\ast}(\mathbf{x})\beta_{\omega,\lambda}\right).
\end{split}
\end{align}
Define
\begin{equation}\label{eqn:Bdef}
\mathcal{B}(\tau,\mathbf{x}) = \frac{C_{0}e^{-i\Delta t(\tau,\mathbf{x})}}{\left(\sinh^2 u + (\cosh u\sinh\eta+\cosh\tau\cosh\eta)^2\right)^{\frac{\Delta}{2}}}
\end{equation}
so that \eqref{eqn:Rindler=global2ptfn} can be rewritten as
\begin{equation}\label{eqn:Beq}
\mathcal{B}(\tau,\mathbf{x}) =  \int\frac{d\omega}{2\pi}\sum_{\lambda}\left(\mathcal{N}_{\omega,\lambda}e^{-i\omega\tau}H_{\lambda}(\mathbf{x})\alpha_{\omega,\lambda}^{\ast}-\mathcal{N}_{\omega,\lambda}^{\ast}e^{i\omega\tau}H_{\lambda}^{\ast}(\mathbf{x})\beta_{\omega,\lambda}\right).
\end{equation}
Using orthogonality of the eigenfunctions of the Laplacian on $H^{d-1}$, we can invert \eqref{eqn:Beq} to obtain an expression for the Bogoliubov coefficeints
\begin{align}\label{eqn:Bogcoeffs}
\begin{split}
\beta_{\omega,\lambda} &= -\frac{1}{\mathcal{N}_{\omega,\lambda}^{\ast}}\int_{-\infty}^{\infty} d\tau\,e^{-i\omega\tau}\int_{H^{d-1}}d^{d-1}x\,\sqrt{-g_{H^{d-1}}}H_{\lambda}(\mathbf{x})\mathcal{B}(\tau,\mathbf{x})
\\ \alpha_{\omega,\lambda} &= \frac{1}{\mathcal{N}_{\omega,\lambda}^{\ast}}\int_{-\infty}^{\infty} d\tau\,e^{-i\omega\tau}\int_{H^{d-1}}d^{d-1}x\,\sqrt{-g_{H^{d-1}}}H_{\lambda}(\mathbf{x})\mathcal{B}^{\ast}(\tau,\mathbf{x}).
\end{split}
\end{align}
The integral over the spherical harmonic gives
\begin{equation}\label{eqn:sphericalharmonicint}
\int d\Omega_{d-2}\,\sqrt{g_{S^{d-2}}}\,Y_{\ell,\mathfrak{m}}(\Omega) = \delta_{\ell,0}\delta_{\mathfrak{m},0}V_{S^{d-2}},
\end{equation}
leading to
\begin{align}\label{eqn:Bogcoeff_betasimpl}
\begin{split}
\beta_{\omega,\lambda} &= -\frac{C_{0}\mathcal{N}_{\lambda,0}^{H}V_{S^{d-2}}}{\mathcal{N}_{\omega,\lambda}^{\ast}}\int_{0}^{\infty}du\,\frac{\sinh^{d-2}u}{\cosh^{2\zeta-2i\tilde{\lambda}}\left(\frac{u}{2}\right)}{}_{2}{F}_{1}\left(\zeta-i\tilde{\lambda},\frac{1}{2}-i\tilde{\lambda},\zeta+\frac{1}{2};\tanh^{2}\left(\frac{u}{2}\right)\right) 
\\	&\qquad \times \int_{-\infty}^{\infty} d\tau\,e^{-i\omega\tau}\frac{e^{-i\Delta t(\tau,\mathbf{x})}}{\left(\sinh^2 u + (\cosh u\sinh\eta+\cosh\tau\cosh\eta)^2\right)^{\frac{\Delta}{2}}}
\\	&\sim -\frac{C_{0}\mathcal{N}_{\lambda,0}^{H}V_{S^{d-2}}}{\mathcal{N}_{\omega,\lambda}^{\ast}}\theta_{0}^{\Delta}\int_{0}^{\infty}du\,\frac{\sinh^{d-2}u}{\cosh^{2\zeta-2i\tilde{\lambda}}\left(\frac{u}{2}\right)}{}_{2}{F}_{1}\left(\zeta-i\tilde{\lambda},\frac{1}{2}-i\tilde{\lambda},\zeta+\frac{1}{2};\tanh^{2}\left(\frac{u}{2}\right)\right) 
\\	&\qquad \int_{-\infty}^{\infty} d\tau\,\frac{e^{-i\omega\tau}}{(\cosh u+\cosh\tau)^{\Delta}}\theta_{0}^{\Delta}\left(1+\mathcal{O}(\theta_{0}^{2})\right)
\\	&= -\frac{C_{0}\mathcal{N}_{\lambda,0}^{H}V_{S^{d-2}}}{\mathcal{N}_{\omega,\lambda}^{\ast}}\frac{2^{\Delta-2}\Gamma(\zeta+\frac{1}{2})}{\sqrt{\pi}\Gamma(\Delta)\Gamma(\Delta-\zeta)}\left|\Gamma\left(\frac{\Delta-\zeta+i(\omega+\tilde{\lambda})}{2}\right)\Gamma\left(\frac{\Delta-\zeta+i(\omega-\tilde{\lambda})}{2}\right)\right|^{2}
\\	&\qquad \times \theta_{0}^{\Delta}\left(1+\mathcal{O}(\theta_{0}^{2})\right),
\end{split}
\end{align}
where we used \eqref{eqn:Bogcoeff_int1} and \eqref{eqn:Bogcoeff_intfinal} to evaluate the integrals and we have dropped $\delta_{\ell,0}\delta_{\mathfrak{m},0}$ so it is implicit that $\ell=\mathfrak{m}=0$. The use of `$\sim$' was explained below \eqref{eqn:Bogcoeffs_scalar}. Similarly, one finds
\begin{equation}\label{eqn:Bogcoeff_alpha}
\alpha_{\omega,\lambda} \sim -\beta_{\omega,\lambda}.
\end{equation}
We can compare these results with their $d=2$ values computed in \cite{Belin:2018juv}. Since $-\lambda$ is the eigenvalue of the Laplacian which should be identified with $-k^{2}$, one finds $|k| = \sqrt{\lambda|_{d=2}} = \tilde{\lambda}$. Thus, we obtain
\begin{equation}\label{eqn:Bogcoeff_betad=2}
\beta_{\omega,\lambda}|_{d=2} = -\frac{1}{\mathcal{N}_{\omega,\lambda}^{\ast}}\frac{2^{\Delta}}{2\sqrt{\pi}\Gamma(\Delta)^{2}}\left|\Gamma\left(\frac{\Delta+i(\omega+\tilde{\lambda})}{2}\right)\right|^{2}\left|\Gamma\left(\frac{\Delta+i(\omega-\tilde{\lambda})}{2}\right)\right|^{2}\theta_{0}^{\Delta}\left(1+\mathcal{O}(\theta_{0}^{2})\right),
\end{equation}
which agrees with \cite{Belin:2018juv}, up to a phase which can be removed by rotating the fields and up to a factor of $1/\sqrt{2}$ coming from the fact that we are working with $|k|$ and not $k$.

\subsection{Maxwell field: bulk modes}
\label{sec:Bogcoeffs_Max}

We can obtain the Bogoliubov coefficients for the Maxwell field by repeating the procedure we performed for the scalar field. One problem is that the global and Rindler modes are in different gauges so to compare them we have to perform a gauge transformation on one of them so that they are in the same gauge, which makes the comparison more technically challenging. Furthermore, we want the Bogoliubov coefficients to be manifestly gauge-invariant so that the entropy obtained from them is too. Therefore, we will use the field strength which is gauge-invariant, and hence the Bogoliubov coefficients will be too.

We consider the following two vacuum two-point functions:
\begin{equation}\label{eqn:Bog2ptfn_Max_rtheta}
\mathfrak{F}^{\mathrm{Max.},r\theta}(t,\Omega) = \lim_{r \to \infty}r^{d-1}\bra{0}F^{r\theta}(t,r,\Omega)a_{0,1,\mathbf{0}}^{S\dagger}\ket{0}
\end{equation}
where we used the fact that $A_{\theta} \sim r^{-(d-2)}$ at large $r$ so $F^{r\theta} \sim r^{-(d-1)}$, and
\begin{equation}\label{eqn:Bog2ptfn_Max_rt}
\mathfrak{F}^{\mathrm{Max.},rt}(t,\Omega) = \lim_{r \to \infty}r^{d-1}\bra{0}F^{rt}(t,r,\Omega)a_{0,1,\mathbf{0}}^{S\dagger}\ket{0}
\end{equation}
where we used the fact that $A_{t} \sim r^{-(d-2)}$ at large $r$ so $F^{rt} \sim r^{-(d-1)}$. The results in global coordinates are
\begin{align}\label{eqn:Bog2ptfn_global_Max}
\begin{split}
\mathfrak{F}^{\mathrm{Max.},r\theta}(t,\Omega) &= -\mathcal{N}_{0,1}^{S}e^{-i\Omega_{0,1,\mathbf{0}}^{S}t}3\sqrt{d}\sin\theta
\\ \mathfrak{F}^{\mathrm{Max.},rt}(t,\Omega) &= \mathcal{N}_{0,1}^{S}e^{-i\Omega_{0,1,\mathbf{0}}^{S}t}3i\sqrt{d}\cos\theta.
\end{split}
\end{align}

To compute $\mathfrak{F}^{\mathrm{Max.}}$ in terms of Rindler modes, we need to know if $a_{0,1,\mathbf{0}}^{S\dagger}$ depends on both $b_{\omega,\lambda}^{S_{H}}$ and $b_{\omega,\lambda_{V}}^{V_{H}}$. This can answered by looking at the global Klein-Gordon inner product
\begin{align}\label{eqn:KGinnerproduct_globaltoRindlerV}
\begin{split}
\langle A_{0,1,\mathbf{0}}^{S},\tilde{A}_{\omega,\lambda_{V}}^{V_{H}}\rangle &= -i\int_{t=0}\sqrt{-g}\left[A_{\mu,0,1,\mathbf{0}}^{S\ast}(\nabla^{t}A_{\omega,\lambda_{V}}^{V_{H}\mu}-\nabla^{\mu}A_{\omega,\lambda_{V}}^{V_{H}t})-A_{\mu,\omega,\lambda_{V}}^{V_{H}}(\nabla^{t}A_{0,1,\mathbf{0}}^{S\mu\ast}-\nabla^{\mu}A_{0,1,\mathbf{0}}^{St\ast})\right]
\\	&= -i\int_{t=0}\sqrt{-g}\left[A_{\theta,0,1,\mathbf{0}}^{S\ast}\nabla^{t}A_{\omega,\lambda_{V}}^{V_{H}\theta}-A_{\theta,\omega,\lambda_{V}}^{V_{H}}\nabla^{t}A_{0,1,\mathbf{0}}^{S\theta\ast}\right]
\end{split}
\end{align}
where, on the first line, the second term vanishes since $A_{\omega,\lambda_{V}}^{V_{H}t}|_{t=0} \propto A_{\omega,\lambda_{V}}^{V_{H}\tau}|_{t=0} = 0$ and the fourth term vanishes after integration by parts (for the integral over $H_{d-1}$) since $A_{\mu}^{V_{H}}$ is divergenceless. Observe that 
\begin{equation}\label{eqn:Rindlervector_theta}
A_{\theta,\omega,\lambda_{V}}^{V_{H}} = \frac{\partial u}{\partial \theta}A_{u,\omega,\lambda_{V}}^{V_{H}}.
\end{equation}
Recall that we can expand the vector part of the Rindler gauge field in terms of eigenfunctions of the vector Laplacian on $H^{d-1}$:
\begin{equation}\label{eqn:Rindlervector_exp}
A_{\mu}^{V_{H}}dx^{\mu} = \sum_{\lambda_{V}}A_{\lambda_{V}}^{V_{H}}\mathbf{H}_{\alpha,\lambda_{V}}dx^{\alpha}
\end{equation}
where $\ell,\mathfrak{m}$ label the $S^{d-2}$ part. We can decompose the $\mathbf{H}_{\alpha,\lambda_{V}}$ in terms of how it transforms under the isometries of $S^{d-2}$:
\begin{equation}\label{eqn:vectoreigfn_decomp}
\mathbf{H}_{\alpha,\lambda_{V}} = \mathbf{H}_{\alpha,\lambda_{V}}^{S}+\mathbf{H}_{\alpha,\lambda_{V}}^{V}
\end{equation}
where
\begin{align}\label{eqn:vectoreigenfn_parts}
\begin{split}
\mathbf{H}_{\alpha,\lambda_{V}}^{S}dx^{\alpha} &= \sum_{\mathbf{k}_{S}}V_{u,\lambda_{V},\mathbf{k}_{S}}^{S}(u)Y_{\mathbf{k}_{S}}^{(d-2)}dx^{u}+V_{\lambda_{V},\mathbf{k}_{S}}^{S}(u)D_{\phi_{i}}\mathbf{Y}_{\mathbf{k}_{S}}^{(d-2)}dx^{\phi_{i}}
\\	\mathbf{H}_{\alpha,\lambda_{V}}^{V}dx^{\alpha} &= \sum_{\mathbf{k}_{V}}V_{\lambda_{V},\mathbf{k}_{V}}^{V}(u)\mathbf{Y}_{\alpha,\mathbf{k}_{V}}^{(d-2)}dx^{\alpha}.
\end{split}
\end{align}
Now observe that $A_{0,1,\mathbf{0}}^{S}$ is constant on $S^{d-2}$ so the integral in \eqref{eqn:KGinnerproduct_globaltoRindlerV} is only non-zero for $V_{u,\lambda_{V},\mathbf{0}}^{S}(u)$ by the orthogonality of spherical harmonics on $S^{d-2}$. Finally, we consider the divergenceless property of $A_{\mu}^{V_{H}}$, which can be applied separately to $V_{\alpha,\lambda_{V}}^{S}$ and to $V_{\alpha,\lambda_{V}}^{V}$ since the divergenceless condition is invariant under rotation of $S^{d-2}$. We are only interested in the $V_{\alpha,\lambda_{V}}^{S}$ since this is the only one with a non-zero $u$ component. Therefore, the divergenceless condition gives
\begin{align}\label{eqn:divlesscondition}
\begin{split}
0 &= \nabla_{\alpha}V_{\lambda_{V},\mathbf{0}}^{S\alpha} = \frac{1}{\sqrt{g_{H^{d-1}}}}\partial_{\alpha}\left(\sqrt{g_{H^{d-1}}}V_{\lambda_{V},\mathbf{0}}^{S\alpha}\right) = \frac{1}{\sinh^{d-2}u}\partial_{u}\left(\sinh^{d-2}u\,V_{\lambda_{V},\mathbf{0}}^{Su}\right) 
\\	&\implies V_{\lambda_{V},\mathbf{0}}^{Su} = \mathscr{A}\csch^{d-2}u,
\end{split}
\end{align}
which is not regular at the origin so we conclude that $\mathscr{A}=0$, and hence the inner product \eqref{eqn:KGinnerproduct_globaltoRindlerV} vanishes. We conclude that only the scalar part of the Rindler gauge field can appear in the expansion of $A_{0,1,\mathbf{0}}^{S}$ in Rindler modes.

We are now ready to compute $\mathfrak{F}^{\mathrm{Max.}}(t,\Omega)$ in terms of Rindler modes. The boundary limit gives
\begin{align}\label{eqn:FthetarRindler_bdylimit}
\begin{split}
\lim_{\rho \to \infty}\rho^{d-1}F^{S_{H}r\theta} &= \frac{\cosh\eta}{\sqrt{\sinh^{2}u+(\cosh\tau\cosh\eta+\cosh u\sinh\eta)^{2}}} \lim_{\rho\to\infty}\bigg[\sinh u\sinh \tau\rho^{d-1}F_{\rho\tau}^{S_{H}}
\\	&+\cosh u\sinh \tau\rho^{d-4}F_{u\tau}^{S_{H}}+\left(\cosh u\cosh \tau+\tanh\eta\right)\rho^{d-1}F_{\rho u}^{S_{H}}\bigg].
\end{split}
\end{align}
The $F_{u\tau}^{S_{H}}$ does not give a finite contribution in the limit and the other two terms give
\begin{equation}\label{eqn:Frhotau_bdylimit}
\lim_{\rho\to\infty}\rho^{d-1}F_{\rho\tau}^{S_{H}} = 2\int_{\omega>0}\frac{d\omega}{2\pi}\sum_{\lambda}\frac{\lambda}{\omega}\left(e^{-i\omega\tau}i\mathcal{N}_{\omega,\lambda}^{S,R}H_{\lambda}b_{\omega,\lambda}^{S_{H}} - e^{i\omega\tau}i\mathcal{N}_{\omega,\lambda}^{S,R\ast} H_{\lambda}^{\ast}\left(b_{\omega,\lambda}^{S_{H}}\right)^{\dagger}\right)
\end{equation}
and
\begin{equation}\label{eqn:Furho_bdylimit}
\lim_{\rho\to\infty}\rho^{d-1}F_{\rho u}^{S_{H}} = -2\int_{\omega>0}\frac{d\omega}{2\pi}\sum_{\lambda}\left(e^{-i\omega\tau}\mathcal{N}_{\omega,\lambda}^{S,R}D_{u}H_{\lambda}b_{\omega,\lambda}^{S_{H}} +e^{i\omega\tau}\mathcal{N}_{\omega,\lambda}^{S,R\ast}D_{u}H_{\lambda}^{\ast}\left(b_{\omega,\lambda}^{S_{H}}\right)^{\dagger}\right).
\end{equation}
Let us define
\begin{equation}\label{eqn:Frthetawavefn}
\mathscr{R}_{\omega,\lambda}^{(r\theta)}(\tau,u) \equiv \mathcal{N}_{\omega,\lambda}^{S,R}\left(i\frac{\lambda}{\omega}\cosh\eta\sinh u\sinh \tau H_{\lambda}-\left(\cosh\eta\cosh u\cosh \tau+\sinh\eta\right)D_{u}H_{\lambda}\right).
\end{equation}
Then \eqref{eqn:FthetarRindler_bdylimit} becomes
\begin{align}\label{eqn:FthetarRindler_bdylimit_final}
\begin{split}
\lim_{\rho \to \infty}\rho^{d-1}F^{S_{H}r\theta} &= \frac{2}{\sqrt{\sinh^{2}u+(\cosh\tau\cosh\eta+\cosh u\sinh\eta)^{2}}}
\\	&\qquad \times \int_{\omega>0}\frac{d\omega}{2\pi}\sum_{\lambda}\left(e^{-i\omega\tau}\mathscr{R}_{\omega,\lambda}^{(r\theta)}(\tau,u)b_{\omega,\lambda}^{S_{H}}+e^{i\omega\tau}\mathscr{R}_{\omega,\lambda}^{(r\theta)\ast}(\tau,u)\left(b_{\omega,\lambda}^{S_{H}}\right)^{\dagger}\right).
\end{split}
\end{align}
One of the desired two-point functions can now be computed
\begin{align}\label{eqn:Bog2ptfn_Rindler_Max_rtheta}
\begin{split}
\mathfrak{F}^{\mathrm{Max.},r\theta}(t,\Omega) &= 2\left(\sinh^2 u + (\cosh u\sinh\eta+\cosh\tau\cosh\eta\big)^2\right)^{\frac{d}{2}-1}
\\	&\qquad \times \bra{0}\int\frac{d\omega}{2\pi}\sum_{\lambda}\left(e^{-i\omega\tau}\mathscr{R}_{\omega,\lambda}^{(r\theta)}(\tau,u)b_{\omega,\lambda}^{S_{H}}+e^{i\omega\tau}\mathscr{R}_{\omega,\lambda}^{(r\theta)\ast}(\tau,u)\left(b_{\omega,\lambda}^{S_{H}}\right)^{\dagger}\right)
\\	&\qquad \qquad \int\frac{d\omega'}{2\pi}\sum_{\lambda'}\left((1-e^{-2\pi\omega'})\alpha_{\omega',\lambda'}^{S,S_{H}\ast}b_{\omega',\lambda'}^{\dagger} + (1-e^{2\pi\omega'})\beta_{\omega',\lambda'}^{S,S_{H}}b_{\omega',\lambda'}\right)\ket{0}
\\	&= 2\left(\sinh^2 u + (\cosh u\sinh\eta+\cosh\tau\cosh\eta\big)^2\right)^{\frac{d}{2}-1}
\\	&\qquad \times \int\frac{d\omega}{2\pi}\sum_{\lambda}\left(e^{-i\omega\tau}\mathscr{R}_{\omega,\lambda}^{(r\theta)}(\tau,u)\alpha_{\omega,\lambda}^{S,S_{H}\ast}-e^{i\omega\tau}\mathscr{R}_{\omega,\lambda}^{(r\theta)}(\tau,u)\beta_{\omega,\lambda}^{S,S_{H}}\right)
\end{split}
\end{align}
where we used \eqref{eqn:globaltoRindler_bdylimit}. We also need the boundary limit of $\theta$:
\begin{equation}\label{eqn:theta_bdylimit}
\lim_{\rho \to \infty}\theta(\tau,\rho,\mathbf{x}) = \arccos\left(\frac{\cosh\tau\cosh\eta+\cosh u \sinh\eta,}{\sqrt{\sinh^{2}u + \left(\cosh\tau\cosh\eta+\cosh u \sinh\eta\right)^{2}}}\right) \equiv \theta(\tau,\mathbf{x}).
\end{equation}

Equating the two-point function in global and Rindler coordinates, we have
\begin{align}\label{eqn:Rindler=global2ptfn_Max}
\begin{split}
-\mathcal{N}_{0,1}^{S}e^{-i\Omega_{0,1,\mathbf{0}}^{S}t(\tau,\mathbf{x})}3\sqrt{d}&\sin\left(\theta(\tau,\mathbf{x})\right) = 2\left(\sinh^2 u + (\cosh u\sinh\eta+\cosh\tau\cosh\eta\big)^2\right)^{\frac{d}{2}-1}
\\	&\qquad \times \int\frac{d\omega}{2\pi}\sum_{\lambda}\left(e^{-i\omega\tau}\mathscr{R}_{\omega,\lambda}^{(r\theta)}(\tau,u)\alpha_{\omega,\lambda}^{S,S_{H}\ast}-e^{i\omega\tau}\mathscr{R}_{\omega,\lambda}^{(r\theta)\ast}(\tau,u)\beta_{\omega,\lambda}^{S,S_{H}}\right).
\end{split}
\end{align}
We now define
\begin{equation}\label{eqn:Brthetadef_Max}
\mathcal{B}^{\mathrm{Max.},(r\theta)}(\tau,\mathbf{x}) = -\frac{\mathcal{N}_{0,1}^{S}e^{-i\Omega_{0,1,\mathbf{0}}^{S}t(\tau,\mathbf{x})}3\sqrt{d}\sin\left(\theta(\tau,\mathbf{x})\right)}{2\left(\sinh^2 u + (\cosh u\sinh\eta+\cosh\tau\cosh\eta\big)^2\right)^{\frac{d}{2}-1}}
\end{equation}
so that \eqref{eqn:Rindler=global2ptfn_Max} can be rewritten as
\begin{equation}\label{eqn:Beqrtheta_Max}
\mathcal{B}^{\mathrm{Max.},(r\theta)}(\tau,\mathbf{x}) = \int\frac{d\omega}{2\pi}\sum_{\lambda}\left(e^{-i\omega\tau}\mathscr{R}_{\omega,\lambda}^{(r\theta)}(\tau,u)\alpha_{\omega,\lambda}^{S,S_{H}\ast}-e^{i\omega\tau}\mathscr{R}_{\omega,\lambda}^{(r\theta)\ast}(\tau,u)\beta_{\omega,\lambda}^{S,S_{H}}\right).
\end{equation}

We can follow the same procedure for $F^{S_{H}rt}$ and once the dust settles, the desired two-point function is found to be
\begin{align}\label{eqn:Bog2ptfn_Rindler_Max_rt}
\begin{split}
\mathfrak{F}^{\mathrm{Max.},rt}(t,\Omega) &= 2\frac{\left(\sinh^2 u + (\cosh u\sinh\eta+\cosh\tau\cosh\eta\big)^2\right)^{\frac{d}{2}}}{\sinh^{2}\tau+(\cosh\tau\sinh\eta+\cosh u\cosh\eta)^{2}}
\\	&\qquad \times \int\frac{d\omega}{2\pi}\sum_{\lambda}\left(e^{-i\omega\tau}\mathscr{R}_{\omega,\lambda}^{(rt)}(\tau,u)\alpha_{\omega,\lambda}^{S,S_{H}\ast}-e^{i\omega\tau}\mathscr{R}_{\omega,\lambda}^{(rt)}(\tau,u)\beta_{\omega,\lambda}^{S,S_{H}}\right),
\end{split}
\end{align}
where
\begin{equation}\label{eqn:Frtwavefn}
\mathscr{R}_{\omega,\lambda}^{(rt)}(\tau,u) \equiv \mathcal{N}_{\omega,\lambda}^{S,R}\left(-i\frac{\lambda}{\omega}\left(\cosh\eta\cosh u\cosh \tau+\sinh\eta\right) H_{\lambda}+\cosh\eta\sinh u\sinh \tau D_{u}H_{\lambda}\right).
\end{equation}
We now define
\begin{align}\label{eqn:Brtdef_Max}
\begin{split}
\mathcal{B}^{\mathrm{Max.},(rt)}(\tau,\mathbf{x}) &= \frac{\mathcal{N}_{0,1}^{S}e^{-i\Omega_{0,1,\mathbf{0}}^{S}t(\tau,\mathbf{x})}3i\sqrt{d}\cos\left(\theta(\tau,\mathbf{x})\right)}{2\left(\sinh^2 u + (\cosh u\sinh\eta+\cosh\tau\cosh\eta\big)^2\right)^{\frac{d}{2}}}
\\	&\qquad \times \left(\sinh^{2}\tau+(\cosh\tau\sinh\eta+\cosh u\cosh\eta)^{2}\right),
\end{split}
\end{align}
and equating with the Rindler answer gives
\begin{equation}\label{eqn:Beqrt_Max}
\mathcal{B}^{\mathrm{Max.},(rt)}(\tau,\mathbf{x}) = \int\frac{d\omega}{2\pi}\sum_{\lambda}\left(e^{-i\omega\tau}\mathscr{R}_{\omega,\lambda}^{(rt)}(\tau,u)\alpha_{\omega,\lambda}^{S,S_{H}\ast}-e^{i\omega\tau}\mathscr{R}_{\omega,\lambda}^{(rt)\ast}(\tau,u)\beta_{\omega,\lambda}^{S,S_{H}}\right).
\end{equation}
The idea is to now take a linear combination of $\mathcal{B}^{\mathrm{Max.},(r\theta)}(\tau,\mathbf{x})$ and $\mathcal{B}^{\mathrm{Max.},(rt)}(\tau,\mathbf{x})$ such that we only have $H_{\lambda}$, which we can then invert using orthogonality of these eigenfunctions. We find
\begin{align}\label{eqn:Bdef_max}
\begin{split}
\mathcal{B}^{\mathrm{Max.}}(\tau,\mathbf{x}) &\equiv \frac{i}{\left(\cosh\eta\cosh u\cosh \tau+\sinh\eta\right)^{2}-\cosh^{2}\eta\sinh^{2} u\sinh^{2} \tau}
\\	&\qquad \times \left(\left(\cosh\eta\cosh u\cosh \tau+\sinh\eta\right)\mathcal{B}^{\mathrm{Max.},(rt)}+\cosh\eta\sinh u\sinh \tau\,\mathcal{B}^{\mathrm{Max.},(r\theta)}\right)
\\	&= \int\frac{d\omega}{2\pi}\sum_{\lambda}\left(\frac{\lambda}{\omega}\mathcal{N}_{\omega,\lambda}^{S,R}e^{-i\omega\tau}H_{\lambda}(\mathbf{x}) \alpha_{\omega,\lambda}^{S,S_{H}\ast}-\frac{\lambda}{\omega}\mathcal{N}_{\omega,\lambda}^{S,R\ast}e^{i\omega\tau}H_{\lambda}^{\ast}(\mathbf{x})\beta_{\omega,\lambda}^{S,S_{H}}\right).
\end{split}
\end{align}
Therefore, we obtain the Bogoliubov coefficients
\begin{align}\label{eqn:Bogcoeffs_Max}
\begin{split}
\beta_{\omega,\lambda}^{S,S_{H}} &= -\frac{1}{\mathcal{N}_{\omega,\lambda}^{S,R\ast}}\frac{\omega}{\lambda}\int_{-\infty}^{\infty} d\tau\,e^{-i\omega\tau}\int_{H^{d-1}}d^{d-1}x\,\sqrt{g_{H^{d-1}}}H_{\lambda}(\mathbf{x})\mathcal{B}^{\mathrm{Max.}}(\tau,\mathbf{x})
\\ \alpha_{\omega,\lambda}^{S,S_{H}} &= \frac{1}{\mathcal{N}_{\omega,\lambda}^{S,R\ast}}\frac{\omega}{\lambda}\int_{-\infty}^{\infty} d\tau\,e^{-i\omega\tau}\int_{H^{d-1}}d^{d-1}x\,\sqrt{g_{H^{d-1}}}H_{\lambda}(\mathbf{x})\mathcal{B}^{\mathrm{Max.}\ast}(\tau,\mathbf{x}).
\end{split}
\end{align}
Using \eqref{eqn:Brthetadef_Max}, \eqref{eqn:Brtdef_Max} and performing the integral over $S^{d-2}$ \eqref{eqn:sphericalharmonicint}, we obtain
\begin{align}\label{eqn:Bogcoeff_betasimpl_Max}
\begin{split}
&\beta_{\omega,\lambda}^{S,S_{H}} = -\frac{\omega\mathcal{N}_{0,1}^{S}\mathcal{N}_{\lambda,0}^{H}V_{S^{d-2}}}{2\lambda\mathcal{N}_{\omega,\lambda}^{S,R\ast}}\int_{0}^{\infty}du\,\frac{\sinh^{d-2}u}{\cosh^{2\zeta-2i\tilde{\lambda}}\left(\frac{u}{2}\right)}{}_{2}{F}_{1}\left(\zeta-i\tilde{\lambda},\frac{1}{2}-i\tilde{\lambda},\zeta+\frac{1}{2};\tanh^{2}\left(\frac{u}{2}\right)\right) 
\\	&\qquad \times \int_{-\infty}^{\infty} d\tau\,e^{-i\omega\tau}\frac{3i\sqrt{d}\,e^{-i\Omega_{0,1,\mathbf{0}}^{S}t(\tau,\mathbf{x})}}{\left(\sinh^2 u + (\cosh u\sinh\eta+\cosh\tau\cosh\eta\big)^2\right)^{\frac{d-1}{2}}}
\\	&\qquad \times \frac{\left(\cosh\eta\cosh u\cosh \tau+\sinh\eta\right)}{\left(\left(\cosh\eta\cosh u\cosh \tau+\sinh\eta\right)^{2}-\cosh^{2}\eta\sinh^{2} u\sinh^{2} \tau\right)}
\\	&\qquad \times \Bigg[i\frac{\left(\cosh\tau\cosh\eta+\cosh u \sinh\eta\right)\left(\sinh^{2}\tau+(\cosh\tau\sinh\eta+\cosh u\cosh\eta)^{2}\right)}{\left(\sinh^{2}u + \left(\cosh\tau\cosh\eta+\cosh u \sinh\eta\right)^{2}\right)}
\\	&\qquad \qquad -\frac{\cosh\eta\sinh^{2}u\sinh\tau}{\left(\cosh\eta\cosh u\cosh \tau+\sinh\eta\right)}\Bigg]
\\	&\sim \frac{3\sqrt{d}\omega\mathcal{N}_{0,1}^{S}\mathcal{N}_{\lambda,0}^{H}V_{S^{d-2}}}{2\lambda\mathcal{N}_{\omega,\lambda}^{S,R\ast}}\theta_{0}^{d-1}\int_{0}^{\infty}du\,\frac{\sinh^{d-2}u}{\cosh^{2\zeta-2i\tilde{\lambda}}\left(\frac{u}{2}\right)}{}_{2}{F}_{1}\left(\zeta-i\tilde{\lambda},\frac{1}{2}-i\tilde{\lambda},\zeta+\frac{1}{2};\tanh^{2}\left(\frac{u}{2}\right)\right)
\\	&\qquad \times \int_{-\infty}^{\infty} d\tau\,e^{-i\omega\tau}\frac{\left(\cosh u\cosh\tau+1\right)}{\left(\cosh u+\cosh\tau\right)^{d}}.
\end{split}
\end{align}
Using the trigonometric identity $2\cosh\tau\cosh(i\omega\tau) = \cosh(\tau(i\omega+1))+\cosh(\tau(i\omega-1))$ and then similar integrals and manipulations as in the scalar case \eqref{eqn:Bogcoeff_int1} and \eqref{eqn:Bogcoeff_intfinal}, albeit much more complicated, we find
\begin{equation}\label{eqn:Bogcoeff_betasimpl_Max_final}
\begin{aligned}
\beta_{\omega,\lambda}^{S,S_{H}} \sim & \frac{3\sqrt{d} \omega\mathcal{N}_{0,1}^{S}\mathcal{N}_{\lambda,0}^{H}V_{S^{d-2}}}{\lambda \mathcal{N}_{\omega,\lambda}^{S,R\ast}}\frac{2^{d-5}}{\sqrt{\pi}}\theta_{0}^{d-1}\frac{\Gamma(\zeta + \frac{1}{2})}{\Gamma(d)\Gamma(\zeta)}\lambda \left|\Gamma\left(\frac{\zeta+i\omega_{+}}{2}\right)\Gamma\left(\frac{\zeta+i\omega_{-}}{2}\right)\right|^{2}\\
  =  &  \frac{3\sqrt{d \omega \lambda} \,\mathcal{N}_{0,1}^{S}\mathcal{N}_{\lambda,0}^{H}V_{S^{d-2}}}{\sqrt{2} }
 \frac{ |\Gamma(i \omega)| }{(d-1)\Gamma(\frac{d}{2})}
  \left|\Gamma\left(\frac{\zeta+i\omega_{-}}{2}\right)\Gamma\left(\frac{\zeta+i\omega_{+}}{2}\right)\right|\theta_{0}^{d-1}.
\end{aligned}
\end{equation}
Similarly, we find
\begin{equation}\label{eqn:Bogcoeff_alphasimpl_Max_final}
\alpha_{\omega,\lambda}^{S,S_{H}} \sim -\beta_{\omega,\lambda}^{S,S_{H}}.
\end{equation}
\subsection{Maxwell field: edge modes}\label{sec:Bogcoeffs_edge}

\paragraph{Two-point function calculation of $\gamma^{(G)}$.}
First, observe that the two-point function calculation used above can only detect $\gamma^{(G)}$ for the following reason. 
Recall that we consider the vacuum two-point functions:
\begin{equation}
\mathfrak{F}^{\mathrm{Max.},\mu\nu}(t,\Omega) = \lim_{r \to \infty}r^{d-1}\bra{0}F^{\mu\nu}(t,r,\Omega)a_{0,1,\mathbf{0}}^{S\dagger}\ket{0}
\end{equation}
and expand both $F^{\mu\nu}$ and $a_{0,1,\mathbf{0}}^{S\dagger}$ in terms of Rindler operators. 
However, only the flux mode contributes to the edge part of the field strength so $F^{\mu\nu}$ only contains the $q_{\lambda}$ operators, hence it can only extract the $a_{\lambda}$ mode in $a_{0,1,\mathbf{0}}^{S\dagger}$.\footnote{There is indeed also $q_\lambda$ mode in $a_{0,1,\mathbf{0}}^{\dagger}$, but recall that the $q_{\lambda}$ term cancels due to the operator relation $q_{\lambda,L}=q_{\lambda,R}$ and the Bogoliubov coefficient \eqref{eq:edgebogoleftright}. }

With this clarified, the two-point function calculation follows the same steps as those above in \S\ref{sec:Bogcoeffs_Max} with one modification: to isolate the edge modes, we introduced discretized frequencies \eqref{eqn:freq2} so the Fourier decomposition should now be viewed as a decomposition on a compact circle with the size $|\log(2\epsilon)|$, or equivalently, the Rindler time $\tau$ lies in the range $\tau \in \left(-|\log(2\epsilon)|/2,|\log(2\epsilon)|/2\right)$. 
There are also minor differences coming from the fact that the edge mode has exactly zero frequency and a normalization factor $\mathcal{N}_{\rm edge}$ in \eqref{eqn:normedge}. 
One final comment is about the phase of the Bogoliubov coefficient: because we have $[a,q]=i$ to isolate a particular edge mode (instead of $[b,b^{\dagger}]=1$), the edge Bogoliubov coefficient will have an extra $i$.

To address these differences, we change the calculation as follows.
First, the zero frequency sector in \eqref{eqn:Frhotau_bdylimit} and \eqref{eqn:Furho_bdylimit} are given by 
$\lim_{\rho\rightarrow \infty} \rho^{d-1}F_{\rho\tau}|_{\omega=0} = \mathcal{N}_{\rm edge} H_{\lambda} + \text{c.c}$ and $F_{\rho u}|_{\omega=0} = 0$. 
Then the $\mathcal{B}^{\rm Max.}(\tau,\mathbf{x})$ in \eqref{eqn:Bdef_max} has a zero frequency sector $\mathcal{B}^{\rm Max.}(\tau,\mathbf{x})|_{\omega=0} = \mathcal{N}_{\rm edge} H_{\lambda} \gamma^{(\Phi)}_{\lambda}$. 
The inversion process can now be carried out, bearing in mind that the time interval is not infinite anymore as mentioned previously, leading to
\begin{equation}
\begin{aligned}
    \gamma^{(G)}_{\lambda} & = \frac{i}{|\log(2\epsilon)| \mathcal{N}_{\rm edge}}  \int_{-|\log(2\epsilon)|/2}^{|\log(2\epsilon)|/2} d\tau \int_{H^{d-1}} d^{d-1} x \sqrt{g_{H^{d-1}}} \mathcal{B}^{\rm Max.}(\tau,\mathbf{x}) H_{\lambda}^{*}(\mathbf{x})\\
    & \approx \frac{2 i \Gamma(\zeta)}{ \Gamma\left(\frac{1}{2}(\zeta-i \tilde{\lambda})\right)\Gamma\left(\frac{1}{2}(\zeta+i \tilde{\lambda}) \right) }  \int_{-\infty }^{+\infty} d\tau \int_{H^{d-1}} d^{d-1} x \sqrt{g_{H^{d-1}}} \mathcal{B}^{\rm Max.}(\tau,\mathbf{x}) H_{\lambda}^{*}(\mathbf{x}) \,.
\end{aligned}
\end{equation}
Observe that the combination $|\log(2\epsilon)| \mathcal{N}_{\rm edge}$ is finite as $\epsilon \to 0$, and one can argue that it is safe to send the $\tau$ integral bounds to $\pm\infty$ because of the convergence of the integral. 
The computation of the $\tau$ and $u$ integrals is actually identical to the long procedure that we went through in \S\ref{sec:Bogcoeffs_Max}, with the result 
\begin{equation}
\begin{aligned}
    \gamma^{(G)}_{\lambda} &  \sim  i  \frac{3\sqrt{d} \mathcal{N}_{0,1}^{S}\mathcal{N}_{\lambda,0}^{H}V_{S^{d-2}}}{|\log(2\epsilon)| \mathcal{N}_{\rm edge}}\frac{2^{d-4}}{\sqrt{\pi}} \theta_{0}^{d-1}\frac{\Gamma(\zeta + \frac{1}{2})}{\Gamma(d)\Gamma(\zeta)}\lambda \left|\Gamma\left(\frac{\zeta+i\tilde{\lambda}}{2}\right)\Gamma\left(\frac{\zeta - i\tilde{\lambda}}{2}\right)\right|^{2} \\
    & =  i \frac{3\sqrt{d} \mathcal{N}_{0,1}^{S}\mathcal{N}_{\lambda,0}^{H}V_{S^{d-2}}}{4(d-1)\Gamma(\zeta+1)} \lambda \left|\Gamma\left(\frac{\zeta+i\tilde{\lambda}}{2}\right)\Gamma\left(\frac{\zeta - i\tilde{\lambda}}{2}\right)\right| \theta_{0}^{d-1} \,.
\end{aligned}
\end{equation}

\paragraph{Inner-product calculation of $\gamma^{(G)}$ and $\gamma^{(\Phi)}$.}
We can alternatively compute the Bogoliubov coefficients for the edge modes using the inner-product (this is too difficult for the bulk modes, but tractable for the edge modes), in fact, this is the only method we have to compute $\gamma^{(\Phi)}$. It is simpler to describe the $\gamma^{(G)}$ computation first because it naturally localizes on the horizon, taking the form 
\begin{equation}
    \gamma_{\lambda}^{(G)} = \langle \mathbb{A}_{0,1,\mathbf{0}}, \mathpzc{A}^{(G)}_{\lambda} \rangle = i  \int_{\gamma_{\mathrm{ext}}}d^{d-1}x\, \sqrt{g_{\gamma_{\mathrm{ext}}}}  \Psi_{\lambda} F_{0,1,\mathbf{0}}^{\mu\nu\ast} n_{\mu} s_{\nu}   
\end{equation}
Specifying $F_{0,1,\mathbf{0}}^{t \mu \ast}n_{t} s_{\mu}$ on the horizon gives 
\begin{equation}
    F_{0,1,\mathbf{0}}^{t \mu\ast}n_{t} s_{\mu} \sqrt{g_{\gamma_{\rm ext}}}  = \frac{3 \sqrt{d} \mathcal{N}_{0,1}^{S}}{r \sqrt{r^2+1}}((1+r^2)\cos^2\theta_0 + \sin^2\theta_0) (r^2 - (r^2+1)\cos^2\theta_0)^{\frac{d-1}{2}}  \sqrt{g_{S^{d-2}}},
\end{equation}
which can be transformed to Rindler coordinates and expanded for small $\theta_0$, giving the result 
\begin{equation}
\begin{aligned}
    \gamma^{(G)}_{\lambda} & \sim  i \theta_0^{d-1} 3 \sqrt{d} \mathcal{N}_{0,1}^{S} \mathcal{N}_{\lambda,\mathbf{0}}^{H} \int d^{d-1} \mathbf{x} \, \sqrt{g_{S^{d-2}}} \sinh^{d-2} u  \cosh^{-d} u  H_{\lambda}(u) \\
    & = i \frac{3\sqrt{d} \mathcal{N}_{0,1}^{S}\mathcal{N}_{\lambda,0}^{H}V_{S^{d-2}}}{ 4 (d-1)\Gamma(\zeta + 1 )} \lambda \left|\Gamma\left(\frac{\zeta+i\tilde{\lambda}}{2}\right)\Gamma\left(\frac{\zeta - i\tilde{\lambda}}{2}\right)\right| \theta_{0}^{d-1}, 
\end{aligned}
\end{equation}
where we have used 
\begin{equation}
    \int_{0}^{\infty} du\, \sinh^{d-2}u \cosh^{-n} u  \, H_{\lambda}(u) =  \mathcal{N}_{\lambda,\mathbf{0}}^{H} \frac{2^{n-2}  \Gamma(\frac{d-1}{2})}{\sqrt{\pi} \Gamma(n)}
   \Gamma\left(\frac{-\zeta - i \lambda +   n}{2}\right)
   \Gamma\left(\frac{-\zeta + i \lambda + n}{2}\right).
\end{equation}

Next, we compute the Bogoliubov coefficient $\gamma^{(\Phi)}$ for the flux mode
\begin{equation}
\begin{aligned}
     \gamma^{(\Phi)}&=\langle \mathbb{A}_{0,1,\mathbf{0}} , \mathpzc{A}^{(\Phi)}_{\lambda} \rangle = -i \int_{\Sigma_{\mathcal{B}}} \sqrt{g_{{\Sigma}_{\mathcal{B}}}} \mathbb{A}_{\mu,0,1,\mathbf{0}}   F^{t \mu}_{\lambda}  n_{t}. \\ 
\end{aligned}
\end{equation}
Such an inner product is naturally computed in the Rindler wedge and one only needs the coordinate transformation of $A_{\mu}$ to Rindler to perform the computation.  
Then $\sqrt{g_{\Sigma}}n_\tau$ is simply $\rho^{d-1}\sinh^{d-2} u \sqrt{g_{S^{d-2}}}$, while the global gauge field and the Rindler field strength components are given near $\rho=1$ by
\begin{equation}
\begin{aligned}
    A_{u} &\xrightarrow{\rho \to 1}  -\frac{3\sqrt{d} \mathcal{N}_{0,1}^S }{(d-2)}\theta_0^{d}  \sinh u \cosh^{-d}u  \sqrt{\rho^2 -1 },  & A_{\rho} \xrightarrow{\rho \to 1}  \frac{3\sqrt{d} \mathcal{N}_{0,1}^S }{(d-2)}\theta_0^{d} \cosh^{1-d}u  \frac{1}{\sqrt{\rho^2 -1 }}, \\ 
    F^{\tau\rho} & \xrightarrow{\rho \to 1} \frac{\log(\rho^2-1)}{\log(2\epsilon)} H_{\lambda}(\mathbf{x}),  & F^{\tau u } \xrightarrow{\rho \to 1}  \frac{4}{\lambda} \frac{1}{|\log(2\epsilon)| (\rho^2-1)}\partial_{u}H_{\lambda}(\mathbf{x}).
\end{aligned}
\end{equation}
Then note that the $\rho$ integral is actually finite, but $\mathcal{N}_{\rm edge}$ is $|\log(2\epsilon)|$ suppressed, so the Bogoliubov coefficient is vanishing in the $\epsilon\rightarrow 0$ limit.

\section{Details of photon backreaction}
\label{sec:photonbackreact_details}

We provide the details of solving the Einstein equations for backreaction of the lowest energy photon excited state
\begin{equation}\label{eqn:Einsteineqns_Maxwell2}
R_{\mu\nu}-\frac{1}{2}Rg_{\mu\nu}+\Lambda g_{\mu\nu} = 8\pi G_{N}\bra{\mathcal{P}} T_{\mu\nu}^{\mathrm{Max.}}\ket{\mathcal{P}},
\end{equation}
where the state $\ket{\mathcal{P}}$ is the $\mathfrak{m}=\mathbf{0}$ state given in Section~\ref{sec:photonbackreact} and the stress tensor is given in \eqref{eqn:stresstensor_Max_evm=0}. The calculation is very non-trivial so we will give all the details.

Given that the expectation values of the stress-tensor along all of the $S^{d-2}$ directions $\phi_{i}$ ($i=2,\ldots,d-1$) vanish, and the non-zero expectation values only have $r,\theta$ dependence, we make the following metric ansatz
\begin{align}\label{eqn:Maxwellbackreact_ansatz2}
\begin{split}
\widetilde{ds}_{M}^{2}\big|_{\mathfrak{m}=\mathbf{0}} = -F_{1}(r,\theta)dt^{2} + 2g_{tr}(r,\theta)&dt\,dr + \frac{dr^{2}}{F_{2}(r,\theta)} 
\\  &+ 2g_{t\theta}(r,\theta)dt\,d\theta + 2g_{r\theta}(r,\theta)dr\,d\theta + F_{3}(r,\theta)d\Omega_{d-1}.
\end{split}
\end{align}
This leads to the Ricci tensor at $\mathcal{O}(G_{N})$
\begin{align}\label{eqn:Riccitensor_Maxwell1}
\begin{split}
R_{tt} &= \frac{1}{4}F_{1}'F_{2}'+\frac{1}{2}F_{2}F_{1}''-\frac{1}{4}\frac{F_{2}(F_{1}')^2}{F_{1}}+\frac{(d-1)}{4}\frac{F_{2}F_{1}'F_{3}'}{F_{3}}+\frac{1}{2}\left(\frac{\partial_{\theta}^{2}F_{1}}{r^{2}}-\frac{2(r^{2}+1)}{r}\partial_{\theta}g_{\theta r}\right)
\\	&\qquad +\frac{(d-2)}{2}\cot\theta\left(\frac{\partial_{\theta}F_{1}}{r^{2}}-\frac{2(r^{2}+1)}{r}g_{\theta r}\right)
\\ R_{tr} &= -dg_{rt}+\partial_{\theta}\left[\frac{1}{2r^{2}}\left(\partial_{r}g_{t\theta}-\partial_{\theta}g_{tr}\right)-\frac{g_{t\theta}}{r(r^{2}+1)}\right]
\\ R_{t\theta} &= -2g_{t\theta}+\frac{(r^{2}+1)}{2}\partial_{r}\left(\partial_{\theta}g_{tr}-\partial_{r}g_{t\theta}\right) + \frac{(d-3)}{2}\frac{(r^{2}+1)}{r}\left(\partial_{\theta}g_{tr}-\partial_{r}g_{t\theta}\right)+r\partial_{\theta}g_{tr}
\\ R_{rr} &= \frac{1}{4}\frac{(F_{1}')^2}{(F_{1})^{2}}-\frac{1}{2}\frac{F_{1}''}{F_{1}}-\frac{(d-1)}{4}\frac{F_{2}'F_{3}'}{F_{2}F_{3}}-\frac{1}{4}\frac{F_{2}'F_{1}'}{F_{2}F_{1}}+\frac{(d-1)}{4}\frac{(F_{3}')^2}{(F_{3})^{2}}-\frac{(d-1)}{2}\frac{F_{3}''}{F_{3}}
\\	& \quad  + \partial_{\theta}\left[\frac{1}{2r^{2}}\left(2\partial_{r}g_{r\theta}+\frac{\partial_{\theta}F_{2}}{(r^{2}+1)^{2}}\right)+\frac{g_{\theta r}}{r(r^{2}+1)}\right]	 
\\ & \quad+\frac{(d-2)}{2}\cot\theta\left[\frac{1}{r^{2}}\left(2\partial_{r}g_{r\theta}+\frac{\partial_{\theta}F_{2}}{(r^{2}+1)^{2}}\right)+\frac{2g_{\theta r}}{r(r^{2}+1)}\right]
\\ R_{r\theta} & = -\left(d + \frac{(d-2)}{r^{2}}\right)g_{r\theta}+\frac{(2r^{2}+1)}{2r(r^{2}+1)^{2}}\partial_{\theta}F_{1}-\frac{\partial_{\theta}F_{1}'}{2(r^{2}+1)}+\left(\frac{1}{2r(r^{2}+1)^{2}}-\frac{(d-1)}{2r(r^{2}+1)}\right)\partial_{\theta}F_{2}
\\	&\qquad -\frac{(d-2)}{2}\left(\frac{\partial_{\theta}F_{3}'}{r^{2}}-2\frac{\partial_{\theta}F_{3}}{r^{3}}\right)
\\ R_{\theta\theta} & = -\frac{1}{4}\frac{F_{1}'F_{2}F_{3}'}{F_{1}}-\frac{1}{4}F_{2}'F_{3}'-\frac{(d-3)}{4}\frac{F_{2}(F_{3}')^{2}}{F_{3}}-\frac{1}{2}F_{2}F_{3}''+\frac{1}{2}\frac{\partial_{\theta}^{2}(F_{2}-F_{1})}{(r^{2}+1)}-\frac{(d-2)}{2}\frac{\partial_{\theta}^{2}F_{3}}{r^{2}}
\\	&\qquad -\frac{(d-2)}{2}\cot\theta\frac{\partial_{\theta}F_{3}}{r^{2}}+(d-2)+(r^{2}+1)\partial_{\theta}g_{r\theta}'+\left(2r+\frac{(d-2)(r^{2}+1)}{r}\right)\partial_{\theta}g_{r\theta}
\\	&\qquad +(d-2)\frac{(r^{2}+1)}{r}\cot\theta\, g_{r\theta}
\\ R_{\phi_{i}\phi_{i}} &= \Bigg[(r^{2}+1)\cot\theta\,g_{\theta r}'+\frac{(r^{2}+1)}{r}\partial_{\theta}g_{\theta r}+\left((2d-3)r+\frac{(2d-5)}{r}\right)\cot\theta\,g_{\theta r}
\\	&\qquad -\frac{1}{2r^{2}}\partial_{\theta}^{2}F_{3}+\left(\frac{1}{2}\frac{\partial_{\theta}(F_{2}-F_{1})}{(r^{2}+1)}-(2d-5)\frac{\partial_{\theta}F_{3}}{2r^{2}}\right)\cot\theta
\\	&\qquad -\left(\frac{1}{4}F_{2}'F_{3}'+\frac{1}{4}\frac{F_{1}'F_{2}F_{3}'}{F_{1}}+\frac{1}{2}F_{2}F_{3}''+\frac{(d-3)}{4}\frac{F_{2}(F_{3}')^{2}}{F_{3}}\right)+(d-2)\Bigg]\frac{g_{\phi_{i}\phi_{i}}^{(0)}}{r^{2}},
\end{split}
\end{align}
where prime labels radial derivatives and $2 \leq i \leq d-1$. The Ricci scalar at $\mathcal{O}(G_{N})$ is 
\begin{align}\label{eqn:Ricciscalar_Maxwell}
\begin{split}
R &= \frac{1}{2}\frac{(F_{1}')^{2}F_{2}}{F_{1}^2}-\frac{1}{2}\frac{F_{2}'F_{1}'}{F_{1}}-\frac{(d-1)}{2}\frac{F_{2}F_{1}'F_{3}'}{F_{1}F_{3}}-\frac{F_{2}F_{1}''}{F_{1}}-\frac{(d-1)}{2}\frac{F_{2}'F_{3}'}{F_{3}}-\frac{(d-1)(d-4)}{4}\frac{F_{2}(F_{3}')^{2}}{(F_{3})^{2}}
\\	&\qquad -(d-1)\frac{F_{2}F_{3}''}{F_{3}}+\frac{(d-1)(d-2)}{F_{3}} + \frac{1}{r^{2}(r^{2}+1)}\left(\partial_{\theta}^{2}(F_{2}-F_{1})+(d-2)\cot\theta\partial_{\theta}(F_{2}-F_{1})\right)
\\	&\qquad - \frac{(d-2)}{r^{4}}\left(\partial_{\theta}^{2}F_{3}+(d-2)\cot\theta\partial_{\theta}F_{3}\right)+2\frac{(r^{2}+1)}{r^{2}}\partial_{\theta}g_{\theta r}'+\frac{2}{r}\left(d+\frac{(d-2)}{r^{2}}\right)\partial_{\theta}g_{\theta r}
\\	&\qquad + 2(d-2)\frac{(r^{2}+1)}{r^{2}}\cot\theta\,g_{\theta r}'+\frac{2(d-2)}{r}\left(d+\frac{(d-2)}{r^{2}}\right)\cot\theta\,g_{\theta r}.
\end{split}
\end{align}

With all the pieces in hand, we now turn to solving the Einstein equations \eqref{eqn:Einsteineqns_Maxwell2}. One finds that $R_{t\phi_{i}} = R_{r\phi_{i}} = R_{\theta\phi_{i}} = 0$ for $2 \leq i \leq d-1$ and $R_{\phi_{i}\phi_{j}}=0$ for $i \neq j$ so those sets of equations are satisfied. Note that $g_{tr}$ and $g_{t\theta}$ only appear in the $tr$ and $t\theta$ components of the Einstein equations (up to $\mathcal{O}(G_N)$) so these two equations can be solved by  $g_{tr}=g_{t\theta}=0$ and this solution does not affect the other equations. For the remaining equations, we write 
\begin{align}\label{eqn:F12M}
\begin{split}
F_{1,2}(r,\theta) &= (r^2+1)\left(1+G_{N}G_{1,2}^{M(1)}(r,\theta)+\mathcal{O}(G_{N}^{2})\right)
\\  F_{3}(r,\theta) &= r^2\left(1+G_{N}G_{3}^{M(1)}(r,\theta)+\mathcal{O}(G_{N}^{2})\right).
\end{split}
\end{align}
This particular choice for the $F_{i}$ turns out to make the Einstein equations tractable for solving for the $G_{i}^{M}$.
Let us also assume that $g_{r\theta}=0$, which is not justified a priori, but will lead to a non-trivial solution for the other components. 

As mentioned in \S\ref{sec:photonbackreact}, we posit the following form of the solution
\begin{equation}\label{eqn:Giexp}
G_{i}^{M(1)}(r,\theta) = \mathfrak{g}_{i,0}(r)+\mathfrak{g}_{i,2}(r)\cos(2\theta).
\end{equation}
Denote the Einstein tensor appearing on the lefthand side of \eqref{eqn:Einsteineqns_Maxwell2} by $E_{\mu\nu}$. Consider the following combination of the Einstein equations (with no sum on $i$):
\begin{equation}\label{eqn:Etheta+Ephi}
\left({E_{\theta}}^{\theta}-{E_{\phi_{i}}}^{\phi_{i}}\right)\Big|_{\mathcal{O}(G_{N})} = 8\pi \left(\bra{\mathcal{P}}({T_{\theta}}^{\theta})\ket{\mathcal{P}} - \bra{\mathcal{P}}{T_{\phi_{i}}}^{\phi_{i}}\ket{\mathcal{P}}\right),
\end{equation}
which gives
\begin{equation}\label{eqn:Etheta+Ephi2}
\sin\theta\partial_{\theta}\left(\csc\theta\partial_{\theta}U(r,\theta)\right) = 288\pi d\,|\mathcal{N}_{0,1}^{S}|^{2}\frac{r^{2}(r^{2}-1)}{(r^{2}+1)^{d}}\sin^{2}\theta,
\end{equation}
where
\begin{equation}\label{eqn:Udef}
U(r,\theta) \equiv G_{2}^{M(1)}(r,\theta)-G_{1}^{M(1)}(r,\theta)-(d-3)G_{3}^{M(1)}(r,\theta).
\end{equation}
Plugging in the metric ansatz \eqref{eqn:Giexp}, we find that the solution satisfies
\begin{equation}\label{eqn:frakg_rel1}
\mathfrak{g}_{2,2}(r)-\mathfrak{g}_{1,2}(r)-(d-3)\mathfrak{g}_{3,2}(r) = 72\pi d\,|\mathcal{N}_{0,1}^{S}|^{2}\frac{r^{2}(r^{2}-1)}{(r^{2}+1)^{d}}
\end{equation}
with no constraints on $\mathfrak{g}_{i,0}$ or $\mathfrak{g}_{i,1}$. Next, the $r\theta$ equation is
\begin{align}\label{eqn:Einstein-Maxwell_rtheta}
\begin{split}
R_{r\theta}+dg_{r\theta} &= 8\pi G_{N}\bra{\mathcal{P}}T_{r\theta}^{\mathrm{Max.}}\ket{\mathcal{P}}
\\ \implies G_{1}^{M(1)}-r(r^{2}+1){G_{1}^{M(1)}}'&+\left(r^{2}+2-d(r^{2}+1)\right)G_{2}^{M(1)}-(d-2)r(r^{2}+1){G_{3}^{M(1)}}'
\\	&= -72\pi d\,|\mathcal{N}_{0,1}^{S}|^{2}\frac{r^{2}}{(r^{2}+1)^{d-1}}\cos(2\theta)+\mathfrak{f}(r),
\end{split}
\end{align}
where we have integrated both sides with respect to $\theta$ in the second line so $\mathfrak{f}(r)$ is an arbitrary function. 
For the diagonal Einstein equations, we find that the $tt$ equation is
\begin{align}\label{eqn:Einstein-Maxwell_tt}
\begin{split}
&-(d-2+dr^{2})G_{2}^{M(1)}-r(r^{2}+1){G_{2}^{M(1)}}'-(d-2)G_{3}^{M(1)}-r\left(d+(d+1)r^{2}\right){G_{3}^{M(1)}}'
\\	&-r^{2}(r^{2}+1){G_{3}^{M(1)}}''+\frac{1}{(d-1)}\left((d-2)\cot\theta\partial_{\theta}+\partial_{\theta}^{2}\right)\left(G_{2}^{M(1)}-(d-2)G_{3}^{M(1)}\right) 
\\  & \qquad\qquad\qquad\qquad\qquad\qquad\qquad\qquad\qquad\qquad =\frac{144\pi d}{(d-1)}|\mathcal{N}_{0,1}^{S}|^{2}\frac{r^{2}}{(r^{2}+1)^{d-1}},
\end{split}
\end{align}
and the $rr$ equation is
\begin{align}\label{eqn:Einstein-Maxwell_rr}
\begin{split}
&r(r^{2}+1){G_{1}^{M(1)}}'+(d-2+dr^{2})G_{2}^{M(1)}+(d-2)\left(rG_{3}^{M(1)}\right)'+(d-1)r^{3}{G_{3}^{M(1)}}'
\\	&+\frac{1}{(d-1)}\left((d-2)\cot\theta\partial_{\theta}+\partial_{\theta}^{2}\right)\left(G_{1}^{M(1)}+(d-2)G_{3}^{M(1)}\right) 
\\  & \qquad\qquad\qquad\qquad\qquad\qquad\qquad\qquad\qquad\qquad = -\frac{144\pi d}{(d-1)}|\mathcal{N}_{0,1}^{S}|^{2}\frac{r^{2}}{(r^{2}+1)^{d-1}}\cos(2\theta),
\end{split}
\end{align}
and the $\theta\theta$ equation is
\begin{align}\label{eqn:Einstein-Maxwell_thetatheta}
\begin{split}
&\left((d-2)(d-3)+d(d-1)r^{2}\right)G_{2}^{M(1)}+(d-2)(d-3)G_{3}^{M(1)}+r\left(d-2+(d+1)r^{2}\right){G_{1}^{M(1)}}'
\\	&+r\left(d-2+(d-1)r^{2}\right){G_{2}^{M(1)}}'+r(d-2)\left(d-1+(d+1)r^{2}\right){G_{3}^{M(1)}}'
\\	&+r^{2}(r^{2}+1)\left(G_{1}^{M(1)}+(d-2)G_{3}^{M(1)}\right)''+(d-2)\cot\theta\partial_{\theta}\left(G_{1}^{M(1)}-G_{2}^{M(1)}+(d-3)G_{3}^{M(1)}\right) 
\\  & \qquad\qquad\qquad\qquad\qquad\qquad\qquad\qquad\qquad\qquad = 144\pi d\,|\mathcal{N}_{0,1}^{S}|^{2}\frac{r^{2}}{(r^{2}+1)^{d}}\left(r^{2}+\cos(2\theta)\right),
\end{split}
\end{align}
and the $\phi_{i}\phi_{i}$ equation (for any $2 \leq i \leq d-1$) is
\begin{align}\label{eqn:Einstein-Maxwell_phiphi}
\begin{split}
&\left((d-2)(d-3)+d(d-1)r^{2}\right)G_{2}^{M(1)}+(d-2)(d-3)G_{3}^{M(1)}+r\left(d-2+(d+1)r^{2}\right){G_{1}^{M(1)}}'
\\	&+r\left(d-2+(d-1)r^{2}\right){G_{2}^{M(1)}}'+r(d-2)\left(d-1+(d+1)r^{2}\right){G_{3}^{M(1)}}'
\\	&+r^{2}(r^{2}+1)\left(G_{1}^{M(1)}+(d-2)G_{3}^{M(1)}\right)''+\left(\partial_{\theta}^{2}+(d-3)\cot\theta\partial_{\theta}\right)\left(G_{1}^{M(1)}-G_{2}^{M(1)}+(d-3)G_{3}^{M(1)}\right) 
\\  &\qquad\qquad\qquad\qquad\qquad\qquad\qquad\qquad\qquad\qquad = 144\pi d\,|\mathcal{N}_{0,1}^{S}|^{2}\frac{r^{2}}{(r^{2}+1)^{d}}\left(r^{2}\cos(2\theta)+1\right).
\end{split}
\end{align}

Let us start by looking at the $\cos(2\theta)$ terms in these equations, i.e., we want solve for the $\mathfrak{g}_{i,2}$. We obtain from \eqref{eqn:Einstein-Maxwell_rtheta}, \eqref{eqn:Einstein-Maxwell_tt}, \eqref{eqn:Einstein-Maxwell_rr}, \eqref{eqn:Einstein-Maxwell_thetatheta}, \eqref{eqn:Einstein-Maxwell_phiphi} the following set of equations
\begin{align}\label{eqn:k=2terms}
\begin{split}
\mathfrak{g}_{1,2}-r(r^{2}+1)\mathfrak{g}_{1,2}'+\left(r^{2}+2-d(r^{2}+1)\right)\mathfrak{g}_{2,2}-(d-2)&r(r^{2}+1)\mathfrak{g}_{3,2}' 
\\  \qquad &= -72\pi d\,|\mathcal{N}_{0,1}^{S}|^{2}\frac{r^{2}}{(r^{2}+1)^{d-1}}
\\ -\left(d^{2}-d+2+(d-1)dr^{2}\right)\mathfrak{g}_{2,2}-(d-1)r(r^{2}+1)\mathfrak{g}_{2,2}'+&(d-2)(d+1)\mathfrak{g}_{3,2}
\\	-r(d-1)\left(d+(d+1)r^{2}\right)\mathfrak{g}_{3,2}'-(d-1)r^{2}(r^{2}+1)\mathfrak{g}_{3,2}'' &= 0
\\ -2d\mathfrak{g}_{1,2}+(d-1)r(r^{2}+1)\mathfrak{g}_{1,2}'+(d-1)(d-2+dr^{2})\mathfrak{g}_{2,2}&-(d-2)(d+1)\mathfrak{g}_{3,2}
\\	+r(d-1)\left((d-2)+(d-1)r^{2}\right)\mathfrak{g}_{3,2}' &= -144\pi d\,|\mathcal{N}_{0,1}^{S}|^{2}\frac{r^{2}}{(r^{2}+1)^{d-1}}
\\ -2(d-2)\mathfrak{g}_{1,2}+r\left(d-2+(d+1)r^{2}\right)\mathfrak{g}_{1,2}'+r^{2}(r^{2}+1)\mathfrak{g}_{1,2}''&
\\ +(d-1)\left(d-2+dr^{2}\right)\mathfrak{g}_{2,2}+r\left(d-2+(d-1)r^{2}\right)\mathfrak{g}_{2,2}'&-(d-2)(d-3)\mathfrak{g}_{3,2}
\\ +r(d-2)\left(d-1+(d+1)r^{2}\right)\mathfrak{g}_{3,2}'+(d-2)r^{2}(r^{2}+1)\mathfrak{g}_{3,2}'' &= 144\pi d\,|\mathcal{N}_{0,1}^{S}|^{2}\frac{r^{2}}{(r^{2}+1)^{d}}
\\ -2(d-1)\mathfrak{g}_{1,2}+r\left(d-2+(d+1)r^{2}\right)\mathfrak{g}_{1,2}'+r^{2}(r^{2}+1)\mathfrak{g}_{1,2}''&
\\ +\left(d^{2}-3d+4+(d-1)dr^{2}\right)\mathfrak{g}_{2,2}+r\left(d-2+(d-1)r^{2}\right)\mathfrak{g}_{2,2}'&-(d-3)d\mathfrak{g}_{3,2}
\\ +r(d-2)\left(d-1+(d+1)r^{2}\right)\mathfrak{g}_{3,2}'+(d-2)r^{2}(r^{2}+1)\mathfrak{g}_{3,2}'' &= 144\pi d\,|\mathcal{N}_{0,1}^{S}|^{2}\frac{r^{4}}{(r^{2}+1)^{d}}
\end{split}
\end{align}
We can add the second and third equations to obtain
\begin{equation}\label{eqn:k=2rel1}
2d\left(\mathfrak{g}_{1,2}+\mathfrak{g}_{2,2}\right)+(d-1)r(r^{2}+1)\left(\mathfrak{g}_{2,2}'-\mathfrak{g}_{1,2}'+\frac{1}{r}\left(r^{2}\mathfrak{g}_{3,2}'\right)'\right) = 144\pi d\,|\mathcal{N}_{0,1}^{S}|^{2}\frac{r^{2}}{(r^{2}+1)^{d-1}},
\end{equation}
and we can add the third equation to $(d-1)$ times the first equation to obtain
\begin{equation}\label{eqn:k=2rel2}
(d+1)\mathfrak{g}_{1,2}-(d-1)r^{2}\mathfrak{g}_{2,2}+(d-2)(d+1)\mathfrak{g}_{3,2}-r^{3}(d-1)\mathfrak{g}_{3,2}' = 72\pi d(d+1)\,|\mathcal{N}_{0,1}^{S}|^{2}\frac{r^{2}}{(r^{2}+1)^{d-1}}.
\end{equation}
We can now solve \eqref{eqn:frakg_rel1} for $\mathfrak{g}_{1,2}$ and plug the result into \eqref{eqn:k=2rel1} to obtain
\begin{align}\label{eqn:g22rel}
\begin{split}
\mathfrak{g}_{2,2} &= \frac{(d-3)}{2}\mathfrak{g}_{3,2}-\frac{(d-1)}{4d}r(r^{2}+1)\left((d-1)\mathfrak{g}_{3,2}'+r\mathfrak{g}_{3,2}''\right)
\\	&\qquad \qquad \qquad +36\pi d\,|\mathcal{N}_{0,1}^{S}|^{2}(d-2)\frac{r^{4}\left(-(d+1)+(d-1)r^{2}\right)}{(r^{2}+1)^{d}}.
\end{split}
\end{align}
Next, we use \eqref{eqn:g22rel} in \eqref{eqn:k=2rel2} (again using \eqref{eqn:frakg_rel1} for $\mathfrak{g}_{1,2}$) to obtain a differential equation for $\mathfrak{g}_{3,2}$:
\begin{align}\label{eqn:k=2relfin}
\begin{split}
&\frac{(d-1)}{2}\left(-(d+1)+(d-3)r^{2}\right)\mathfrak{g}_{3,2}+\frac{(d-1)}{4d}r\left((d-1)(d+1)+2(3d-1)r^{2}-(d-1)^{2}r^{4}\right)\mathfrak{g}_{3,2}'
\\	&-\frac{(d-1)}{4d}r^{2}(r^{2}+1)\left(-(d+1)+(d-1)r^{2}\right)\mathfrak{g}_{3,2}'' 
\\	&+8\pi9d|\mathcal{N}_{0,1}^{S}|^{2}\frac{r^{4}}{(r^{2}+1)^{d}}\left(2(d+1)+\frac{(d-2)}{2d}\left(-(d+1)+(d-1)r^{2}\right)^{2}\right) = 0.
\end{split}
\end{align}
The two solutions to the homogeneous version of this equation, i.e., without the source term on the last line, can be found with the resulting general solution given by their linear combination, viz.,
\begin{equation}\label{eqn:g32hom1} 
\mathfrak{g}_{3,2}^{\mathrm{hom.}} = C_{1}\frac{(r^{2}-1)}{r^{d}}+C_{2}r^{2}\left[1+(1-r^{2})\frac{d}{(d+2)}{}_{2}{F}_{1}\left(1,\frac{d}{2}+1,\frac{d}{2}+2;-r^{2}\right)\right],
\end{equation}
where $C_{1}$ and $C_{2}$ are constants to be fixed later by smoothness and boundary conditions. This is a fairly simple answer, but obtaining the solution to the full, inhomogeneous differential equation involves integration of the two linearly independent solutions in \eqref{eqn:g32hom1} (times the reciprocal of the Wronskian) times the source term in \eqref{eqn:k=2relfin} which is a rational polynomial in $r$. Such an integral turns out to be intractable for arbitrary $d$, but for any fixed $d$ we will be able to straightforwardly compute the solution to the full, inhomogeneous equation. 

First, we rewrite \eqref{eqn:g32hom1} in terms of simpler functions plus a finite sum:
\begin{equation}\label{eqn:g32hom2}
\mathfrak{g}_{3,2}^{\mathrm{hom.}}(r) = C_{1}\frac{(r^{2}-1)}{r^{d}}+C_{2}r^{2}\begin{cases} 1-(-1)^{\frac{d+1}{2}}\frac{d(r^{2}-1)}{r^{d+3}}\left(r\tan^{-1}r+\sum_{k=1}^{\frac{d+1}{2}}\frac{(-1)^{k}}{(2k-1)}r^{2k}\right) & d\mathrm{\;odd} \\  1-(-1)^{\frac{d}{2}}\frac{d}{2}\frac{(r^{2}-1)}{r^{d+2}}\left(\log(1+r^{2})+\sum_{k=1}^{\frac{d}{2}}\frac{(-1)^{k}}{k}r^{2k}\right) & d\mathrm{\;even}. \end{cases}
\end{equation}
Decomposing the full solution as
\begin{equation}\label{eqn:g32}
\mathfrak{g}_{3,2}(r) = \mathfrak{g}_{3,2}^{\mathrm{hom.}}(r)+\mathfrak{g}_{3,2}^{\mathrm{inhom.}}(r),
\end{equation}
we have the inhomogeneous term in even and odd dimensions
\begin{align}\label{eqn:g32odd}
\begin{split}
&\mathfrak{g}_{3,2}^{\mathrm{inhom.,odd}}(r) = \frac{(r^{2}-1)}{r^{d}}\Bigg\{\left[\frac{r^{d+2}}{(r^{2}-1)}-(-1)^{\frac{d+1}{2}}\frac{d}{r}\left(r\tan^{-1}r+\sum_{k=1}^{\frac{d+1}{2}}\frac{(-1)^{k}}{(2k-1)}r^{2k}\right)\right]
\\  &\qquad\qquad\qquad\qquad\qquad\qquad\qquad\qquad\qquad \times \int dr\,\frac{(r^{2}-1)(r^{2}+1)}{2r\left(-(d+1)+(d-1)r^{2}\right)}\mathfrak{s}(r)
\\	&-\int dr\,\frac{(r^{2}+1)r^{d+1}}{2\left(d(r^{2}-1)-(r^{2}+1)\right)}\left[1-(-1)^{\frac{d+1}{2}}d\frac{(r^{2}-1)}{r^{d+3}}\left(r\tan^{-1}r+\sum_{k=1}^{\frac{d+1}{2}}\frac{(-1)^{k}}{(2k-1)}r^{2k}\right)\right]\mathfrak{s}(r)\Bigg\}
\end{split}
\end{align}
and
\begin{align}\label{eqn:g32even}
\begin{split}
&\mathfrak{g}_{3,2}^{\mathrm{inhom.,even}}(r) = \frac{(r^{2}-1)}{r^{d}}\Bigg\{\left[\frac{r^{d+2}}{(r^{2}-1)}-(-1)^{\frac{d}{2}}\frac{d}{2}\frac{(r^{2}-1)}{r^{d}}\left(\log(1+r^{2})+\sum_{k=1}^{\frac{d}{2}}\frac{(-1)^{k}}{k}r^{2k}\right)\right]
\\  &\qquad\qquad\qquad\qquad\qquad\qquad\qquad\qquad\qquad \times\int dr\,\frac{(r^{2}-1)(r^{2}+1)}{2r\left(-(d+1)+(d-1)r^{2}\right)}\mathfrak{s}(r)
\\	&-\int dr\,\frac{(r^{2}+1)}{2\left(-(d+1)+(d-1)r^{2}\right)}\left[r^{d+1}-(-1)^{\frac{d}{2}}\frac{d}{2}\frac{1}{r}\left(\log(1+r^{2})+\sum_{k=1}^{\frac{d}{2}}\frac{(-1)^{k}}{k}r^{2k}\right)\right]\mathfrak{s}(r)\Bigg\},
\end{split}
\end{align}
where 
\begin{equation}\label{eqn:Sdef}
\mathfrak{s}(r) = 72\pi d\,|\mathcal{N}_{0,1}^{S}|^{2}\frac{r^{2}}{(r^{2}+1)^{d}}\left(2(d+1)r^{2}+\frac{(d-2)}{2d}r^{2}\left(-(d+1)+(d-1)r^{2}\right)^{2}\right)
\end{equation}
is the source term in \eqref{eqn:k=2relfin}. As mentioned previously, these integrals which give $\mathfrak{g}_{3,2}$ are difficult to compute for arbitrary $d$, but the answer is simple to compute for a given $d$.

The constants $C_{1}$ and $C_{2}$ appearing in \eqref{eqn:g32hom2} can be determined by looking at the small $r$ and large $r$ behavior of $\mathfrak{g}_{3,2}^{\mathrm{hom.}}$:
\begin{equation}\label{eqn:g32hom}
\mathfrak{g}_{3,2}^{\mathrm{hom.}}(r \to 0) = -\frac{C_{1}}{r^{d}}+\mathcal{O}(r^{2-d}), \qquad \mathfrak{g}_{3,2}^{\mathrm{hom.}}(r \to \infty) = 2C_{2}\frac{(d-1)}{(d-2)} + \mathcal{O}(r^{-2}).
\end{equation}
Thus, the first solution diverges at the origin ($r=0$) and the second solution does not satisfy asymptotically AdS boundary conditions. One finds that neither of these unwanted behaviors can be cancelled by $\mathfrak{g}_{3,2}^{\mathrm{inhom.}}$ so we must set $C_{1}=C_{2}=0$. Therefore, the entire solution to the full, inhomogeneous differential equation is given solely by $\mathfrak{g}_{3,2}^{\mathrm{inhom.}}$. This completes our computation of $\mathfrak{g}_{3,2}$.

Armed with the solution for $\mathfrak{g}_{3,2}$, we can then obtain $\mathfrak{g}_{2,2}$ from \eqref{eqn:g22rel} and obtain $\mathfrak{g}_{1,2}$ from \eqref{eqn:k=2rel2}. We list the results for $d=3,4,5,6$ in \S\ref{sec:photonbackreact} (see \eqref{eqn:gij} and Table \ref{tab:backreactphotonmetric}).
One can check that these solutions satisfy the fourth and fifth equations in \eqref{eqn:k=2terms}, which verifies that we have indeed found a solution to the Einstein equations.

Finally, we look at the $\mathfrak{g}_{i,0}$ terms. We can eliminate $\mathfrak{g}_{3,0}$ either using a residual diffeomorphism gauge freedom\footnote{For instance, one can use the diffeomorphism $x^{\mu} \to x^{\mu} + \xi^{\mu}$ where $\xi^{r} = -G_{N}r\mathfrak{g}_{3,0}(r)$, $\xi^{\mu \neq r} = 0$ and redefine $\mathfrak{g}_{1,0}$, $\mathfrak{g}_{2,0}$ by their shifted functions after the diffeomorphism.} or simply making it part of our ansatz. We thus find from \eqref{eqn:Einstein-Maxwell_tt}, \eqref{eqn:Einstein-Maxwell_rr}, \eqref{eqn:Einstein-Maxwell_thetatheta},
\begin{align}\label{eqn:k=0terms}
\begin{split}
-\frac{\left(d-2+dr^{2}\right)}{2}\mathfrak{g}_{2,0}+\frac{(d-2)}{(d-1)}\left((d-2)\mathfrak{g}_{3,2}-\mathfrak{g}_{2,2}\right)-\frac{r(r^{2}+1)}{2}\mathfrak{g}_{2,0}' &= \frac{72\pi d}{(d-1)}|\mathcal{N}_{0,1}^{S}|^{2}\frac{r^{2}}{(r^{2}+1)^{d-1}}
\\ -2\frac{(d-2)}{(d-1)}\mathfrak{g}_{1,2}+r(r^{2}+1)\mathfrak{g}_{1,0}'+(d-2+dr^{2})\mathfrak{g}_{2,0}-2\frac{(d-2)^{2}}{(d-1)}\mathfrak{g}_{3,2} &= 0
\\ 2\left(\mathfrak{g}_{2,2}-\mathfrak{g}_{1,2}\right)+r\left(1+\frac{(d+1)}{(d-2)}r^{2}\right)\mathfrak{g}_{1,0}'+\frac{1}{(d-2)}r^{2}(r^{2}+1)\mathfrak{g}_{1,0}''&
\\ +\left((d-3)+\frac{(d-1)d}{(d-2)}r^{2}\right)\mathfrak{g}_{2,0}+r\left(1+\frac{(d-1)}{(d-2)}r^{2}\right)\mathfrak{g}_{2,0}'&
\\ -2(d-3)\mathfrak{g}_{3,2} &= \frac{144\pi d}{(d-2)}|\mathcal{N}_{0,1}^{S}|^{2}\frac{r^{4}}{(r^{2}+1)^{d}}.
\end{split}
\end{align}
The first equation in \eqref{eqn:k=0terms} can be solved for $\mathfrak{g}_{2,0}$, and then the second equation in \eqref{eqn:k=0terms} can be used to obtain
%
$\mathfrak{g}_{1,0}$. Since these solutions depend on our solution for $\mathfrak{g}_{3,2}$, they also require that we fix a dimension $d$ and the solutions for $d=3,4,5,6$ are also given in \S\ref{sec:photonbackreact} (see \eqref{eqn:gij} and Table \ref{tab:backreactphotonmetric}).
One can check that these solutions satisfy the third equation in \eqref{eqn:k=0terms}. This completes our solution for the backreaction of the photon on the metric.

\section{Useful integrals}
\label{sec:usefulints}

Here we provide the details of some integrals needed for our computations. For \eqref{eqn:bulkmodH_explicit}, we used the following integral
\begin{align}\label{eqn:modHscalarpos_int}
\begin{split}
&\int_{0}^{\infty} du\,\int_{1}^{\infty}d\rho\,\rho^{d-2\Delta-1}\sinh^{d-2} u\left(\cosh u + \frac{\sqrt{\rho^{2}-1}}{\rho}\right)^{-2\Delta} 
\\	&= \frac{\Gamma(2\Delta+2-d)\Gamma(\frac{d-1}{2})}{2^{2\Delta+2-d}\Gamma(2\Delta+\frac{3-d}{2})}\int_{1}^{\infty}d\rho\,\rho^{d-2\Delta-1}{}_{2}{F}_{1}\left[2\Delta,2\Delta+2-d,\frac{3-d}{2}+2\Delta;\frac{1}{2}\left(1-\frac{\sqrt{\rho^{2}-1}}{\rho}\right)\right]
\\	&= 2^{\frac{d-3}{2}}\Gamma(2\Delta+2-d)\Gamma\left(\frac{d-1}{2}\right)\int_{0}^{1}dy\,y(1-y^{2})^{-\frac{(d+5)}{4}}P_{\frac{d-3}{2}}^{\frac{d-1}{2}-2\Delta}\left(y\right), \qquad y \equiv \frac{\sqrt{\rho^{2}-1}}{\rho}
\\	&= \frac{\Gamma(\frac{d-1}{2})\Gamma(\Delta-\frac{d}{2})}{8\Gamma(\Delta+\frac{3}{2})},
\end{split}
\end{align}
where we used eq. (3.518.5) from \cite{GradRyz} in the second line and on the last line we used eq. (18.1.6) of \cite{Bateman1954}.

For \eqref{eqn:bulkvacsubtractEE_1storder_explicit}, we used the integral
\begin{align}\label{eqn:modHscalarmom_int}
\begin{split}
\\	\int_{0}^{\infty}&d\omega\,|\Gamma(1+i\omega)|^{2}\int_{0}^{\infty}d\tilde{\lambda}\,\left|\frac{\Gamma(\zeta-i\tilde{\lambda})}{\Gamma(-i\tilde{\lambda})}\Gamma\left(\frac{\Delta-\zeta+i(\omega+\tilde{\lambda})}{2}\right)\Gamma\left(\frac{\Delta-\zeta+i(\omega-\tilde{\lambda})}{2}\right)\right|^{2}
\\	&= \frac{2^{2\zeta-2}}{\pi}\int_{0}^{\infty}d\omega\,|\Gamma(1+i\omega)|^{2}\int_{0}^{\infty}d\tilde{\lambda}\,\left|\frac{\Gamma(\frac{\zeta-i\tilde{\lambda}}{2})\Gamma(\frac{\zeta+1-i\tilde{\lambda}}{2})}{\Gamma(-i\tilde{\lambda})}\right|^{2}
\\	&\qquad \times \left|\Gamma\left(\frac{\Delta-\zeta+i(\omega+\tilde{\lambda})}{2}\right)\Gamma\left(\frac{\Delta-\zeta+i(\omega-\tilde{\lambda})}{2}\right)\right|^{2}
\\	&= \frac{4\pi\Gamma(\frac{d-1}{2})\Gamma\left(\Delta+1-\frac{d}{2}\right)}{2^{2\Delta-2\zeta}\Gamma(\Delta+\frac{1}{2})}\int_{0}^{\infty}d\omega\,|\Gamma(1+i\omega)|^{2}|\Gamma(\Delta+i\omega)|^{2}
\\	&= \frac{\pi^{2}\Gamma(\frac{d-1}{2})\Gamma\left(\Delta+1-\frac{d}{2}\right)\Gamma(\Delta)\Gamma(\Delta+1)}{2^{2\Delta-2\zeta}\Gamma(\Delta+\frac{3}{2})},
\end{split}
\end{align}
where we used the de Branges-Wilson Beta integral (eq. (5.3.15) in https://dlmf.nist.gov/5.13) to evaluate the $\tilde{\lambda}$ integral and we used eq. (6.413.1) in \cite{GradRyz} to evaluate the $\omega$ integral.

The integral required for the calculation of the scalar Bogoliubov coefficients \eqref{eqn:Bogcoeff_betasimpl} is very involved so we will explain it in some detail here. We find
\begin{align}\label{eqn:Bogcoeff_int1}
\begin{split}
&\int_{0}^{\infty}du\,\frac{\sinh^{d-2}u}{\cosh^{2\zeta-2i\tilde{\lambda}}\left(\frac{u}{2}\right)}{}_{2}{F}_{1}\left(\zeta-i\tilde{\lambda},\frac{1}{2}-i\tilde{\lambda},\zeta+\frac{1}{2};\tanh^{2}\left(\frac{u}{2}\right)\right)\int_{-\infty}^{\infty} d\tau\,\frac{e^{-i\omega\tau}}{(\cosh u+\cosh\tau)^{\Delta}}
\\	&= \frac{\sqrt{2\pi}\Gamma(\Delta+i\omega)\Gamma(\Delta-i\omega)}{\Gamma(\Delta)}\int_{0}^{\infty}du\,\frac{\sinh^{d-\Delta-\frac{3}{2}}u}{\cosh^{2\zeta-2i\tilde{\lambda}}\left(\frac{u}{2}\right)}P_{i\omega-\frac{1}{2}}^{\frac{1}{2}-\Delta}(\cosh u)
\\	&\qquad \times {}_{2}{F}_{1}\left(\zeta-i\tilde{\lambda},\frac{1}{2}-i\tilde{\lambda},\zeta+\frac{1}{2};\tanh^{2}\left(\frac{u}{2}\right)\right)
\\	&= \frac{\sqrt{2\pi}\Gamma(\Delta+i\omega)\Gamma(\Delta-i\omega)}{\Gamma(\Delta)}\int_{0}^{\infty}du\,\sinh^{d-\Delta-\frac{3}{2}}u\cosh u \, P_{i\omega-\frac{1}{2}}^{\frac{1}{2}-\Delta}(\cosh u)
\\	&\qquad \times {}_{2}{F}_{1}\left(\frac{\zeta-i\tilde{\lambda}+1}{2},\frac{\zeta+i\tilde{\lambda}+1}{2},\zeta+\frac{1}{2};-\sinh^{2}u\right) 
\\	&= 2^{\zeta}\frac{\sqrt{\pi}\Gamma(\zeta+\frac{1}{2})\Gamma(\Delta+i\omega)\Gamma(\Delta-i\omega)}{\Gamma(\Delta)}\int_{0}^{\infty}du\,\sinh^{\frac{d}{2}-\Delta}u \, P_{i\omega-\frac{1}{2}}^{\frac{1}{2}-\Delta}(\cosh u)P_{i\tilde{\lambda}-\frac{1}{2}}^{\frac{1}{2}-\zeta}(\cosh u)
\end{split}
\end{align}
where to perform the $\tau$ integral in the second line we used eq. (3.517.1) from \cite{GradRyz}. This $u$ integral can be evaluated by using the following identity given by eq. (7.8.5) in \cite{Bateman1953}:
\begin{equation}\label{eqn:Legendreidentity}
P^{\mu}_{\nu}(z) = \sqrt{\frac{2}{\pi}}\frac{1}{\Gamma(-\mu-\nu)\Gamma(\nu-\mu+1)}(z^{2}-1)^{-\frac{\mu}{2}}\int_{0}^{\infty}dt\,e^{-tz}t^{-\mu-\frac{1}{2}}K_{\nu+\frac{1}{2}}(t),
\end{equation}
which gives
\begin{align}\label{eqn:Bogcoeff_intfinal}
\begin{split}
&\int_{0}^{\infty}du\,\frac{\sinh^{d-2}u}{\cosh^{2\zeta-2i\tilde{\lambda}}\left(\frac{u}{2}\right)}{}_{2}{F}_{1}\left(\zeta-i\tilde{\lambda},\frac{1}{2}-i\tilde{\lambda},\zeta+\frac{1}{2};\tanh^{2}\left(\frac{u}{2}\right)\right)\int_{-\infty}^{\infty} d\tau\,\frac{e^{-i\omega\tau}}{(\cosh u+\cosh\tau)^{\Delta}}
\\	&= 2^{\zeta+\frac{1}{2}}\frac{\Gamma(\zeta+\frac{1}{2})}{\Gamma(\Delta)}\int_{0}^{\infty} dt\,t^{\Delta-1}K_{i\omega}(t)\int_{0}^{\infty}du\,e^{-t\cosh u}\sinh^{\frac{d-1}{2}}u\, P_{i\tilde{\lambda}-\frac{1}{2}}^{\frac{1}{2}-\zeta}(\cosh u)
\\	&= 2^{\zeta+1}\frac{\Gamma(\zeta+\frac{1}{2})}{\sqrt{\pi}\Gamma(\Delta)}\int_{0}^{\infty} dt\,t^{\Delta-\zeta-1}K_{i\omega}(t)K_{i\tilde{\lambda}}(t)
\\	&= 2^{\Delta-2}\frac{\Gamma(\zeta+\frac{1}{2})}{\sqrt{\pi}\Gamma(\Delta)\Gamma(\Delta-\zeta)}\left|\Gamma\left(\frac{\Delta-\zeta+i(\omega+\tilde{\lambda})}{2}\right)\Gamma\left(\frac{\Delta-\zeta+i(\omega-\tilde{\lambda})}{2}\right)\right|^{2},
\end{split}
\end{align}
where in the second and third equalities we used 7.141.5 and 6.576.4 from \cite{GradRyz}, respectively.


\bibliography{GeneralizedEntropyPhoton-refs} 
\bibliographystyle{JHEP}


\end{document}